\documentclass[showpacs,aps,rmp,twocolumn,preprintnumbers,amsmath,amssymb,longbibliography]{revtex4-1}

\usepackage{graphicx,epsfig,epsf,color,hhline}
\usepackage{dcolumn}
\usepackage{bm}
\usepackage{bbm}
\usepackage{dsfont}
\usepackage{tensor,pbox}
\usepackage{hyperref}
\usepackage{amsthm,bbm}
\usepackage[english]{babel}

\newcommand{\reddd}[1]{{#1}}
\newcommand{\red}[1]{{#1}}
\newcommand{\colb}[1]{{#1}}

\newcommand{\ket}[1]{\left|#1\right>}
\newcommand{\bra}[1]{\left<#1\right|}

\newcommand{\bea}{\begin{eqnarray}}
\newcommand{\ea}{\end{eqnarray}}
\newcommand{\eea}{\end{eqnarray}}

\newcommand{\sumint}[1]

\begin{document}
\newcommand{\ri}{ i}
\newcommand{\re}{ e}
\newcommand{\bx}{{\bm x}}
\newcommand{\bd}{{\bm d}}
\newcommand{\be}{{\bm e}}
\newcommand{\br}{{\bm r}}
\newcommand{\bk}{{\bm k}}
\newcommand{\bA}{{\bm A}}
\newcommand{\bE}{{\bm E}}
\newcommand{\bB}{{\bm B}}
\newcommand{\bI}{{\bm I}}
\newcommand{\bH}{{\bm H}}
\newcommand{\bR}{{\bm R}}
\newcommand{\bZero}{{\bm 0}}
\newcommand{\bM}{{\bm M}}
\newcommand{\bX}{{\bm X}}
\newcommand{\bn}{{\bm n}}
\newcommand{\bs}{{\bm s}}
\newcommand{\bv}{{\bm v}}
\newcommand{\tbs}{\tilde{\bm s}}
\newcommand{\rSi}{{\rm Si}}
\newcommand{\beps}{{\bm \epsilon}}
\newcommand{\bsig}{{\bm \sigma}}
\newcommand{\bGa}{{\bm \Gamma}}
\newcommand{\rg}{{\rm g}}
\newcommand{\tr}{{\rm tr}}
\newcommand{\xmax}{x_{\rm max}}
\newcommand{\xb}{\overline{x}}
\newcommand{\pb}{\overline{p}}
\newcommand{\ra}{{\rm a}}
\newcommand{\rx}{{\rm x}}
\newcommand{\rs}{{\rm s}}
\newcommand{\rP}{{\rm P}}
\newcommand{\up}{\uparrow}
\newcommand{\down}{\downarrow}
\newcommand{\hc}{H_{\rm cond}}
\newcommand{\kb}{k_{\rm B}}
\newcommand{\cI}{{\cal I}}
\newcommand{\tit}{\tilde{t}}
\newcommand{\cE}{{\cal E}}
\newcommand{\cC}{{\cal C}}
\newcommand{\Ubs}{U_{\rm BS}}
\newcommand{\sech}{{\rm sech}}
\newcommand{\qq}{{\bf ???}}
\newcommand*{\etal}{\textit{et al.}}

\renewcommand{\red}{}
\newcommand{\redd}{}
\def\vec#1{\bm{#1}}
\def\ket#1{|#1\rangle}
\def\bra#1{\langle#1|}
\def\braket#1#2{\langle#1|#2\rangle}
\def\ketbra#1#2{|#1\rangle|\langle#1|}

\title{Quantum-enhanced measurements without entanglement}

\author{Daniel Braun$^{1}$, Gerardo Adesso$^{2}$, Fabio Benatti$^{3,4}$, Roberto Floreanini$^{4}$, Ugo Marzolino$^{5,6}$, Morgan~W.~Mitchell$^{7,8}$, Stefano Pirandola$^{9}$}
\affiliation{$^1\mbox{Institute of Theoretical
Physics, University
  T\"ubingen, Auf der Morgenstelle 14, 72076 T\"ubingen, Germany}$}
\affiliation{$^2\mbox{Centre for the Mathematics and Theoretical Physics of Quantum Non-Equilibrium Systems (CQNE),}$\\ $\mbox{School of Mathematical Sciences, University of Nottingham, University Park,
Nottingham NG7 2RD, United Kingdom}$}
\affiliation{$^3\mbox{Department of Physics, University of Trieste, 34151 Trieste, Italy}$}
\affiliation{$^4\mbox{Istituto Nazionale di Fisica Nucleare, Sezione di
Trieste, 34151 Trieste, Italy}$}
\affiliation{$^5\mbox{Department of
Physics, Faculty of Mathematics and Physics, University of
Ljubljana, 1000 Ljubljana, Slovenia}$}
\affiliation{$^6\mbox{Division of Theoretical Physics, Ru\dj er Bo\v{s}kovi\'c Institute, 10000 Zagreb, Hrvatska}$}
\affiliation{$^7\mbox{ICFO -- Institut de Ciencies Fotoniques, The
Barcelona Institute of Science and Technology,}$\\ $\mbox{08860 Castelldefels
(Barcelona), Spain}$}
\affiliation{$^8\mbox{ICREA -- Instituci\'{o}
Catalana de Recerca i Estudis Avan\c{c}ats, 08015 Barcelona,
Spain}$}
\affiliation{$^9\mbox{Computer Science and York Centre for
Quantum Technologies, University of York, York YO10 5GH, United
Kingdom}$}


\begin{abstract}
Quantum-enhanced measurements exploit quantum mechanical effects for
increasing the sensitivity of measurements of certain physical
parameters and have great potential for both
fundamental science and concrete
applications. Most of the research has so far focused on using highly
entangled states, which are, however, difficult to produce and
to stabilize for a large number of constituents.  In the following we review
alternative mechanisms, notably the use of more general quantum
correlations such as quantum discord, identical particles,
or non-trivial Hamiltonians; the estimation of thermodynamical
parameters or parameters characterizing non-equilibrium states; and the use of
quantum phase transitions. We describe both theoretically achievable
enhancements and enhanced sensitivities, not primarily based on
entanglement, that have already been demonstrated
experimentally, and indicate some possible future research directions.
\end{abstract}
\maketitle

\tableofcontents
\section{Introduction}
\label{sec:IntroI.1}\label{sec.Intro}
\subsection{Aim and scope}
Quantum-enhanced measurements  aim at improving
measurements of physical parameters by using quantum effects.  The
improvement sought is
an enhanced sensitivity for a given amount of resources such
as mean or maximum energy used, number of probes, number of
measurements, and integration time. Ideas in this direction go back at
least to the late 1960s when the effect of quantum noise on
the estimation of classical parameters started to be studied
in a systematic way using appropriate mathematical tools
\cite{helstrom_quantum_1969,Holevo1982}. In the early
1980s first detailed proposals appeared on how to
enhance the sensitivity of gravitational wave detectors by using
squeezed light
\cite{caves_measurement_1980,caves_quantum-mechanical_1981}. Nowadays,
squeezed light is routinely produced in many labs, and used for
instance to enhance sensitivity in
gravitational wave observatories
\cite{AasiNP2013others,chua_quantum_2015}.

Quantum-enhanced measurements have the potential of enabling many
important applications, both scientific and technological.  Besides
gravitational wave detection, there are
proposals or demonstrations for the improvement of
time- or frequency-standards,  navigation, remote sensing, measurement of
very small magnetic fields
 (with applications to
medical brain- and heart-imaging), measurement of the
parameters of space-time,
thermometry, and
many more. The literature  on the topic of quantum metrology is vast
and for a general introduction we refer to the available reviews
\cite{wiseman_quantum_2009,Paris2009,GiovannettiPRL2006,SmerziR,toth_quantum_2014,pezze_non-classical_2016,degen_quantum_2016}.

From the theoretical side, the standard tool for evaluating a possible
quantum enhancement has become the so-called quantum Cram\'er-Rao
bound
\cite{helstrom_quantum_1969,Holevo1982,Braunstein94,braunstein_generalized_1996}. It
provides a lower bound on the variance ${\rm Var}(\theta_{\rm est})$
of any unbiased estimator
function $\theta_{\rm est}$ that maps observed data obtained from
arbitrary quantum
measurements  to an estimate
of the parameter $\theta$. The bound is optimized over all possible 
measurements 
and data analysis schemes, in a sense made precise below.
  In
the limit of an infinite number of measurements the bound can be
saturated. It thus represents a valuable benchmark that can in
principle be achieved once all technical noise problems have been
solved, such that only the unavoidable noise  inherent
in the
quantum state itself remains. 

A standard classical method of noise reduction is to average
measurement results from $N$ independent, identically prepared
systems.
In a quantum mechanical formulation with
pure states, the
situation corresponds to having the $N$ quantum systems in an initial
product state,
$|\psi\rangle=\otimes_{i=1}^N\ket{\phi}_i$. Suppose that the parameter
is encoded in the state through a unitary evolution with a Hamiltonian
$H(\theta)=\theta\sum_{i=1}^N
h_i$, i.e.~$|\psi(\theta)\rangle=\exp(-iH(\theta))|\psi\rangle$. Based
on the quantum Cram\'er-Rao bound one can show that
with $M$ final measurements, the smallest achievable
variance $
\text{Var}(\theta_\text{est})$ of the estimation of $\theta$ is
\begin{equation} \label{dxminUs}
\text{Var}(\theta_\text{est})_\text{min}=\frac{1}{NM(\Lambda-\lambda)^2}\,,
\end{equation}
where $\Lambda$ and
$\lambda$ are the largest and smallest eigenvalue of $h_i$, respectively,
taken for simplicity here as identical for all subsystems
\cite{GiovannettiPRL2006}.
In fact, this $1/\sqrt{N}$ scaling can be easily understood as a
consequence of the central limit theorem in the simplest case that one
measures the systems independently. But since  (\ref{dxminUs})
is optimized over all measurements of the full system, it also implies
that entangling measurements of all systems after the parameter has
been encoded in the state cannot improve the $1/\sqrt{N}$ scaling.\\
Unfortunately, there is no unique definition of the Standard Quantum Limit in the
literature.  Whereas in the described $1/\sqrt{N}$ scaling $N$ refers
to the number of distinguishable sub-systems, the term Standard Quantum Limit is used for
example in quantum optics typically for a scaling as $1/\sqrt{\bar n}$
with the average number of photons $\bar n$, which in the same mode
are to be considered as indistinguishable (see Sec.\ref{sec.IdPar}).
In this context, the $1/\sqrt{\bar n}$ scaling is
also called ``shot-noise limit'', referring to the quantum noise that
arises from the fact that the electromagnetic energy is quantized in
units of photons. Furthermore, the prefactor in these scaling
behaviors is not fixed. We therefore may define quite generally Standard Quantum Limit
as the best scaling that can be achieved when employing only ``classical'' resources.

While this is not yet a mathematical definition
either, it becomes precise once the classical resources are specified
in the problem at hand.
\red{This may be achieved adopting a resource-theory framework, in
  which classical states of some specific sort are identified and
  formalised as ``free'' states (i.e.~given at no cost), and any other
  state is seen as
  possessing a resource content which may allow us to outperform free
  states in practical applications, leading specifically to quantum-enhanced measurements beyond
  the Standard Quantum Limit scaling. For instance, separable states are the free
  (classical) states in the resource-theory of entanglement
  \cite{Horodecki09}, while states diagonal in a reference basis are
  the free (classical) states in the resource theory of quantum
  coherence \cite{Streltsov2016C}. In quantum optics, Glauber's
  coherent states and their mixtures are regarded as the free
  (classical) states \cite{MandelRMP}, and any other state can yield a
  nonclassical scaling. In the latter example, considering the mean
  photon number $\bar n$ as an additional resource, one can fix the
  prefactor of the Standard Quantum Limit scaling, so that quantum enhancements are
  possible not only by improving the scaling
law, but also by changing the prefactor.}

\red{However, basing our review exclusively on a resource-theory picture
would be too restrictive, as cases of enhanced sensitivity are
readily available for which no resource theory has been worked
out yet (see \cite{PhysRevLett.115.070503} and references therein for
a recent overview of existing resource theories).  Examples are the
use of quantum phase transitions, {\redd for which} one
can compare the sensitivity at the phase transition with the sensitivities
away from the phase transition, or instances of Hamiltonian
engineering, {\redd for which} one can evaluate the effect of added terms in the
Hamiltonian. Rather than developing resource theories for all these
examples, which would be beyond the scope of this review, we point out
the enhancements achievable compared to the sensitivity without the
use of the mechanism under consideration.  } 

Based again on the quantum Cram\'er-Rao bound one can show 
that
initially entangled states can improve the scaling to $1/N$
\reddd{(see e.g.~\cite{GiovannettiPRL2006})}, known as the
``Heisenberg-limit''. 
Similarly to the Standard Quantum Limit, there is no unique
definition of the Heisenberg-limit in the literature (see the remarks in sec.\ref{sec.genH}).
Nevertheless, 
achieving ``the Heisenberg-limit'' has been the goal of large experimental and
theoretical efforts over the last two decades. However, only few
experiments achieved the $1/N$ scaling of the Heisenberg-limit and only for very
small numbers of sub-systems, where  the scaling advantage is
still far from allowing one to beat the best possible classical
measurements. This has several reasons: first of all, it is
already very difficult to achieve even the Standard Quantum Limit, as all
non-intrinsic noise sources have to be eliminated. Secondly,
resources such as photons are cheap, such that classically one can
operate with very large photon numbers, whereas entangled states
with large photon numbers are difficult to produce. Thirdly, and
most fundamentally, quantum-enhanced measurements schemes are plagued
by decoherence.
Indeed, it has been shown that a small amount of Markovian
decoherence brings the $1/N$ scaling for certain highly entangled
states back to the $1/\sqrt{N}$ scaling of the Standard Quantum Limit
\cite{huelga_improvement_1997,Kolodynski10,escher_general_2011}.
{The reduction to the Standard Quantum Limit also affects the
estimation of noise in programmable and teleportation-covariant
channels \cite{Laurenza17}}. Recent research has focussed on
finding optimal states in the presence of decoherence, and at
least for non-Markovian noise, a certain improvement can still be
obtained from entangled states
\cite{PhysRevLett.109.233601,PhysRevA.84.012103}. Also, niche
applications are possible, {\redd for which} the light-intensity must be very
small, as in some biological applications.  Nevertheless, it
appears worthwhile to think about alternative possible quantum
enhancement principles other than the use of highly entangled
states, and this is the focus of the
present survey.\\

Many results have been obtained over the last years for such
alternative schemes that are worth a comprehensive and exhaustive
review that compares their usefulness with respect to the main-stream
research focused on highly entangled states.
We structure the review by different ways of
breaking the conditions that are known to lead to Standard Quantum Limit scaling of the
sensitivity.  Firstly, by going away from pure states, more general
forms of quantum correlations such as quantum discord become
possible. \red{These become naturally important once we look at
  estimation of loss parameters, quantum illumination problems, and
  other applications that typically involve the loss of probes}.
Secondly, in the derivation of the Standard Quantum Limit the quantum systems are
distinguished by an index $i$, which supposes that they are
distinguishable. Cold atoms, on the other hand, have to be considered in general
as indistinguishable particles, and the same is true for photons,
which have been used for quantum-enhanced
measurements from the very beginning. Hence, statements about the
necessity of entanglement have to be re-examined for indistinguishable
particles. \red{It turns out that the permutational symmetry of the
  quantum states required due to
indistinguishability of the particles leads immediately to the level
of quantum-enhanced sensitivity that for distinguishable particles
would require to entangle them.}
Thirdly, the structure of the Hamiltonian is rather
restrictive: a.) many hamitonians do not have a bound spectrum
characterized by largest and smallest eigenvalues $\Lambda$ and
$\lambda$ as assumed in eq.(\ref{dxminUs}).  Indeed, one of the most common
systems used in quantum metrology, the harmonic oscillator that
represents e.g.~a single mode of an electro-magnetic field, has an
unbound spectrum. And b.), the Hamiltonian $H(\theta)=\theta\sum_{i=1}^N
h_i$ does not
allow for any interactions.  \red{Taking into account these freedoms opens
the path to many new forms of enhanced sensitivity}.
Fourthly, unitary evolutions with a
Hamiltonian that depends on the parameter are not the only way of
coding a parameter in a state. In statistical mechanics, for example,
there are parameters that describe the statistical ensemble, such as
temperature or chemical potential for systems in thermal equilibrium,
but which are not of Hamiltonian
origin.  The same is true for non-equilibrium states. For many of these
situations, the corresponding QCRs have been obtained only recently,
and it often turned out that improvements beyond the Standard Quantum Limit should be
possible.  \red{Furthermore, it is known even in classical statistical
physics that phase transitions can lead to diverging susceptibilities
and hence greatly enhanced sensitivities. The same is true for quantum
phase transitions, and we therefore review as well the use of quantum
phase transitions for quantum-enhanced measurements.}\\

{\redd While a growing number of researchers are investigating possibilities
of breaking the Standard Quantum Limit without using entanglement \reddd{(see
e.g.~\cite{TilmaPRA2010})},
these still appear to be a minority.}
The situation is comparable to
other aspects and fields of quantum information treatment, where
previously it was thought that entanglement is necessary. For example,
for a long time entanglement has been considered as necessary for
non-locality, until it was realized that certain aspects of
non-locality can arise
without entanglement \cite{Bennett99}. 
Recent reviews
of quantum-enhanced measurements schemes using entanglement
\cite{Paris2009,GiovannettiNPhot2011,SmerziR,toth_quantum_2014,pezze_non-classical_2016,degen_quantum_2016}
are available and we do not survey this vast literature here,
but focus rather exclusively on quantum-enhanced measurements schemes that are not essentially
based on the use of entanglement,
hoping that our review will stimulate research in these
directions.
Before reviewing these schemes, we give a short
introduction to parameter estimation theory and the precise
definition of the quantum Cram\'er-Rao bound. A more elaborate pedagogical introduction
to classical and quantum parameter estimation theory can be found in
\cite{JulienThesis2017}.

{  \subsection{Parameter estimation theory}
Consider the following task in classical statistical analysis: Given a
probability distribution $p_\theta(x)$ of a random variable $x$ that
continuously varies as function of a single real parameter $\theta$, estimate
$\theta$ as precisely as possible from $M$ samples drawn, i.e.~a set
of random values $\{x_i\},\,\,i=1,2,\ldots,M$.  We denote this
$M$-sample as $\bx$ for short, and denote the probability to find the
drawn samples
in the intervals $x_i\ldots x_i+dx_i$ as $p_\theta(\bx)d^M\bx$, with
$d^M\bx=dx_1\ldots dx_M$. For independently drawn, identically
distributed samples, $p_\theta(\bx)=p_\theta(x_1)\cdot\ldots \cdot
p_\theta(x_M)$, but the formalism allows for arbitrary
joint-probability distributions $p_\theta(\bx)$ i.e.~also correlations
between different samplings of the distribution. For simplicity we
take the support of $x$ to be the real numbers.

The task is accomplished by using an estimator function $\theta_{\rm
  est}(x_1,\ldots,x_M)$ that takes as input the drawn
random values and nothing else, and outputs an estimate of the
parameter $\theta$. Many
different estimator functions are possible, some more useful than
others. Through its random arguments the estimator will itself
fluctuate from one sample to another. One would like to have an
estimator that on average gives the true value of $\theta$,
$E(\theta_{\rm est})=\theta$, where $E(\ldots)=\int d^M\bx\,
p_\theta(\bx)(\ldots)$ is the mean value of a quantity over the
distribution. This should hold at least in an infinitesimal
interval about the true value of
$\theta$; such an estimator is called
``unbiased''. Secondly, one would like the estimator to fluctuate as
little as possible. The latter request makes only sense together with
the first one, as otherwise we could just choose a constant estimator,
which of course would not reproduce the correct value of $\theta$ in
most cases. Now consider the following chain of equalities, valid for
an unbiased estimator:
\begin{eqnarray}
  \label{eq:QCRBderiv}
  1&=&\frac{\partial}{\partial \theta}E(\theta_{\rm est})=\int d^M\bx\,
  \frac{\partial}{\partial \theta}p_\theta(\bx)\theta_{\rm
  est}(\bx)\nonumber\\
&=&\int d^M\bx\, p_\theta(\bx)\left(\frac{\partial}{\partial \theta}\ln
  p_\theta(\bx)\right)\theta_{\rm est}(\bx)\nonumber\\
&=&\int d^M\bx\, p_\theta(\bx)\left(\frac{\partial}{\partial \theta}\ln
  p_\theta(\bx)\right)(\theta_{\rm est}(\bx)-\theta)\nonumber\\
&=&\big\langle \frac{\partial \ln p_\theta}{\partial\theta},\theta_{\rm est}-\theta\big\rangle\,.
\end{eqnarray}
In the step before the last one we used that
$\theta(\partial/\partial_\theta)E(1)=0$ due to the normalization of
the probability distribution valid for all values of $\theta$. The
scalar product in the last step is defined for any two real functions
$a(\bx), b(\bx)$ as $\langle a,b\rangle=\int d^M\bx\,
p_\theta(\bx)a(\bx)b(\bx)$. Using the Cauchy-Schwarz inequality for
this scalar product, we immediately arrive at the (classical)
Cram\'er-Rao (lower) bound for the variance of the estimator
\begin{equation}
  \label{eq:CRB}
  \text{Var}(\theta_{\rm est})\ge \frac{1}{J_\theta^{(M)}}\,,
\end{equation}
where the (classical) Fisher information $J_\theta$ is defined as
\begin{eqnarray}
  \label{eq:FI}
  J_\theta^{(M)}&=&\int d^M\bx p_\theta(\bx)\left(\frac{\partial \ln
      p_\theta(\bx)}{\partial \theta}\right)^2\nonumber\\
&=&\int d^M\bx
  \frac{1}{p_\theta(\bx)}\left(\frac{\partial
      p_\theta(\bx)}{\partial \theta}\right)^2\,.
\end{eqnarray}
The bound can be saturated iff the two vectors in the scalar product
are parallel, i.e.~for $\partial \ln p_\theta(\bx)/\partial
\theta=A(\theta)(\theta_{\rm est}(\bx)-\theta)$, where $A(\theta)$ is a
possibly $\theta-$dependent proportionality factor. If one
differentiates this condition once more and then integrates it over
with $p_\theta(\bx)$, one finds that $A(\theta)=J_\theta^{(M)}$. Hence, an
unbiased estimator exists iff there is a function $f(\bx)$ independent
of $\theta $ such that $\partial \ln p_\theta/\partial
\theta=J_\theta^{(M)}(f(\bx)-\theta)$.  In that case one can choose
$\theta_{\rm est}(\bx)=f(\bx)$. One can show that for many (but not
all) members of the family of exponential probability distributions,
i.e.~distributions that can be written in the form
$p_\theta(x)=a(x)\exp(b(\theta)c(x)+d(\theta))$ with some functions
$a(x),b(\theta),c(x),d(\theta)$,
this condition is satisfied, meaning that in such cases the
Cram\'er-Rao bound can be saturated even for finite $M$. For
$M\to\infty$  and identically, independently distributed samples, the
so-called maximum-likelihood estimator saturates the bound. One easily
shows from eq.\eqref{eq:FI} that the Fisher information is additive,
such that for independently drawn, identically distributed samples
$J_\theta^{(M)}=M J_\theta$ with $J_\theta\equiv J_\theta^{(1)}$. }

{\redd \subsection{Quantum parameter estimation theory}
\label{sec:Fisherinfo}
In quantum mechanics the state of a system is given by a density
matrix $\rho_\theta $, i.e.~a positive hermitian operator with trace
equal one that can depend on the parameter $\theta$, which we assume
to be a classical parameter. Random data are
created when measuring some observable of the system whose statistics
will depend on $\theta$ through the quantum state $\rho_\theta$. Again
we would like to estimate $\theta$ as precisely as possible based on
the measurement data (see FIG.\ref{fig1}).
\begin{figure}[h!]
  \begin{center}
    \includegraphics[width=0.45\textwidth]{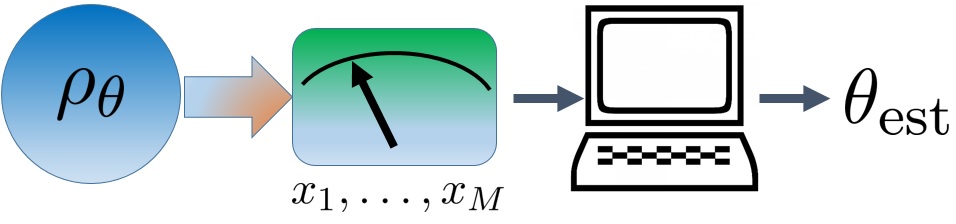}
    \caption{(Color online) General setup of quantum parameter
      estimation. A quantum 
      measurement of a
      system in quantum state $\rho_\theta$ that depends on a
      classical parameter $\theta$ is performed (general POVM
      measurement) and produces data $x_1,\ldots,x_M$. The data are
      analysed with an estimator function that outputs an estimate
      $\theta_\text{est}$ of $\theta$. The goal is to obtain an
      unbiased estimate with as small as possible statistical fluctuations. }
    \label{fig1}
  \end{center}
\end{figure}

The most general measurements are
so-called POVM measurements (POVM=positive-operator-valued measure).
These are measurements that generalize and include projective von Neumann
measurements and are relevant in particular when the system is
measured through
an ancilla system to which it is coupled \cite{Peres93}.  They
consist of a set of positive operators $M_x$, where $x$ labels
possible measurement outcomes that we take again as $x\in\mathbb R$
for simplicity. They obey a completeness relation, $\int_\mathbb{R}M_x
dx=\mathbb{I}$, where $\mathbb{I}$ is the identity operator on the
Hilbert-space of the system. The probability-density to find outcome $x$ is
given by $p_\theta(x)=\tr(\rho_\theta M_x)$, and it is through this
equation that the contact with the classical parameter
estimation theory can be made: Plugging in $p_\theta(x)$ in eq.~\eqref{eq:FI}
with $M=1$, we
are led to the Fisher information
\begin{eqnarray}
  \label{eq:Jqm}
  J_\theta&=&\int dx\frac{1}{\tr(\rho_\theta
  M_x)}\left(\tr\left(\frac{\partial\rho_\theta}{\partial
  \theta}M_x\right)\right)^2\nonumber\\
&=&\int dx\frac{1}{\tr(\rho_\theta
  M_x)}\left(\tr\left(\frac{1}{2}(\rho_\theta
    L_{\rho_\theta}+L_{\rho_\theta}\rho_\theta) M_x\right)\right)^2\,,\label{eq:IJ2}
\end{eqnarray}
where in the last step we have introduced the so-called symmetric logarithmic
derivative $L_{\rho_\theta}$, defined indirectly through
\begin{equation}
  \label{eq:L}
  \frac{\partial\rho_\theta}{\partial_\theta}=\frac{1}{2}(\rho_\theta
  L_{\rho_\theta}+L_{\rho_\theta}\rho_\theta)\,,
\end{equation}
in analogy to the classical logarithmic derivative $\partial \ln
\rho_\theta/\partial_\theta$.
Compared to the classical case, one has in the quantum mechanical
setting the
additional freedom to choose a suitable measurement in order to obtain
a distribution $p_\theta(x)$ that contains
as much information as possible on the parameter $\theta $. Based on
eq.~\eqref{eq:IJ2}, one can find a similar chain of inequalities
as in the classical case based on the Cauchy-Schwarz inequality that
leads to the bound
\begin{equation}
  \label{eq:JI3}
  J_\theta\le I_\theta\equiv\tr(\rho_\theta L_{\rho_\theta}^2)\,,
\end{equation}
where $I_\theta $ is known as the ``quantum Fisher information''.

Similarly as for the classical Fisher information, the quantum Fisher information of
uncorrelated states is additive,
\cite{fujiwara_additivity_2002}:
\begin{equation}\label{eq:additivity_qfi}
I_\theta(\rho(\theta)\otimes
\sigma(\theta))=I_\theta(\rho(\theta))+I_\theta( \sigma(\theta))\,,
\end{equation}
such that for $M$ independent identical POVM measurements of the same
system, prepared always in the same state, the total quantum Fisher information satisfies
$I_\theta^{(M)}=M I_\theta$ with $I_\theta\equiv
I_\theta^{(1)}$.
Inequalities \eqref{eq:JI3} and
\eqref{eq:CRB} then lead to the so-called quantum Cram\'er-Rao bound,
 \begin{equation}\label{qfi_inequality0}
  \text{Var}(\theta_\text{est}) \geq \frac{1}{M I_\theta}\,.
 \end{equation}
Additivity of the quantum Fisher information also immediately implies the $1/\sqrt{N}$ scaling
in eq.\eqref{dxminUs}, as
the quantum Fisher information of the $N$ uncorrelated subsystems is just $N$
times the quantum Fisher information of a single subsystem.
Inequality \eqref{eq:JI3} can be saturated with a POVM that
consists of projectors onto eigenstates of $L_\theta$
\cite{helstrom_quantum_1969,Holevo1982,Braunstein94}.
As the quantum Cram\'er-Rao bound is already optimized, no
measurement of the whole system, even if entangling the individual systems, can
improve the sensitivity when the parameter was already imprinted on a
product state. \\

The quantum Cram\'er-Rao bound has become the most widely used quantity for establishing the
ultimate
sensitivity of measurement schemes. It derives its power from the facts
that firstly it is already optimized over all possible data analysis schemes (unbiased
estimator functions) and all possible (POVM-) measurements, and that
secondly it can be saturated at least in the limit of infinitely many
measurements and using the optimal POVM consisting of projectors onto the
eigenstates of $L_{\rho_\theta}$.
} 

In \cite{Braunstein94} it was
shown that $I_\theta$ is a geometric measure on how much
$\rho(\theta)$ and $\rho(\theta+d\theta)$ differ, where $d\theta$ is
an infinitesimal increment of $\theta$.
The geometric measure is given by the Bures-distance,
\begin{equation} \label{db2}
ds_{\rm Bures}^2(\rho,\sigma)\equiv 2\left(1-\sqrt{F(\rho,\sigma)}\right)\,,
\end{equation}
where the fidelity $F(\rho,\sigma)$ is defined as
\begin{equation}
  \label{eq:fidel}
  F(\rho,\sigma)=||\rho^{1/2}\sigma^{1/2}||^2_1\,,
\end{equation}
and $||A||_1\equiv {\rm
  tr}\sqrt{AA^\dagger}$ denotes the trace norm \cite{Miszczak09}. With
this \cite{Braunstein94} showed that
\begin{equation}\label{IBures}
I_\theta=4ds_{\rm
  Bures}^2(\rho(\theta),\rho(\theta+d\theta))/d\theta^2\,,
\end{equation}
unless the rank of $\rho(\theta)$ changes with $\theta$ and thus
produces removable singularities
\cite{Banchi15,safranek_discontinuities_2016}, a situation which
we do not consider in this review. The quantum Cram\'er-Rao bound thus offers the
physically intuitive picture that the parameter $\theta$ can be
measured the more precisely the more strongly the state
$\rho(\theta)$ depends on it. For pure states
$\rho(\theta)=\ket{\psi(\theta)}\bra{\psi(\theta)}$, the quantum Fisher information
reduces to the overlap of the derivative of the state
$\ket{\partial_\theta\psi(\theta)}$ with itself and the original
state \cite{braunstein_generalized_1996,Paris2009},
\begin{equation}
  \label{eq:Ipure}
I_\theta(\ketbra{\psi_\theta})=4(\braket{\partial_\theta\psi_\theta}{\partial_\theta\psi_\theta}+\braket{\partial_\theta\psi_\theta}{\psi_\theta}^2)   \,.
\end{equation}
If the parameter is imprinted on a pure state via a unitary transformation
with hermitian generator $G$ as $\ket{\psi_\theta}=\exp(i
\theta G)\ket{\psi}$, eq.\eqref{eq:Ipure} gives $I_\theta=4
\text{Var}(G)\equiv 4(
\braket{\psi_\theta}{G^2|\psi_\theta}-\braket{\psi_\theta}{G|\psi_\theta}^2)$.
With a maximally entangled state of the $N$
subsystems and a suitable measurement, one can reach a scaling of the
quantum Fisher information proportional to $N^2$ \cite{GiovannettiPRL2006}, the mentioned
Heisenberg-limit. This can be seen most easily for a pure state of the
form $\ket{\psi}=(\ket{\Lambda}^{\otimes N}+\ket{\lambda}^{\otimes
  N})/\sqrt{2}$, where $\ket{\Lambda}$ and $\ket{\lambda}$ are
two eigenstates of $G$ to two different
eigenvalues $\Lambda,\lambda$.

For mixed states, the Bures-distance is in general
difficult to calculate, but $I_\theta(\rho(\theta))$ is a convex
function of $\rho(\theta)$, i.e.~for two
density matrices $\rho(\theta)$ and $\sigma(\theta)$ and  $0\leq
\lambda \leq 1$ we have  \cite{fujiwara_quantum_2001}
\begin{equation}\label{eq:convexity_QFI}
I_\theta(\lambda \rho(\theta) +(1-\lambda) \sigma(\theta)) \leq \lambda  I_\theta(\rho(\theta))+(1-\lambda) I_\theta( \sigma(\theta))\,.
\end{equation}
This can be used to obtain an upper bound for the quantum Fisher information. Convexity also
implies that the precision of measurements cannot be
increased by classically mixing states with mixing probabilities
independent of the parameter \cite{Braun10}.   \\

In principle, the optimal measurement that saturates the quantum Cram\'er-Rao bound can be
constructed by diagonalizing $L(\theta)$. The projectors onto its
eigenstates form a POVM that yields the optimal measurement. However,
such a construction requires that the precise value of the parameter
$\theta$ is already known.  If that was the case, one could skip the
measurement altogether and choose the estimator as $\theta_{\rm
  est}=\theta$, with vanishing
uncertainty, i.e.~apparently violating the quantum Cram\'er-Rao bound in most cases
\cite{chapeau-blondeau_optimized_2015} 
(note, however, that for a
state that depends on $\theta$, the condition
$\theta_\text{est}=\theta$ for an unbiased estimator
cannot be fulfilled in a whole $\epsilon$-interval about $\theta$,
such that there is no formal contradiction). 
If
$\theta$ is not
known, the more common
approach is therefore to use the quantum Cram\'er-Rao bound as a benchmark as function
of $\theta$, and then
check whether physically motivated measurements can achieve it.
More general schemes have been proposed to mitigate the
problem of prior knowledge of the parameter. This includes the van
Trees inequality
\cite{van_trees_detection_2001,gill_applications_1995}, Bayesian approaches
\cite{RivasNJP2012,macieszczak_optimal_2014}, adaptive measurements
\cite{serafini_feedback_2012,okamoto_experimental_2012,PhysRevLett.104.093601,PhysRevA.73.063824,PhysRevA.87.019901,PhysRevA.65.043803,berry_optimal_2000,armen_adaptive_2002,wiseman_adaptive_1995,fujiwara_strong_2006,HigginsNJP2009},
and
approaches specialized to
particular parameter estimation problems such as phase estimation
\cite{hall_universality_2012}.
Another point to be kept in mind is that the quantum Cram\'er-Rao bound can be reached
asymptotically for a large number of
measurements, but not necessarily for a finite number of
measurements. The latter case is clearly relevant for experiments and
subject of active current research (see
e.g.~\cite{1367-2630-18-9-093009}
).\\
These limitations not withstanding, we base this review almost
exclusively on the quantum Cram\'er-Rao bound (with the exception of Sec.\ref{QCD} on
quantum channel discrimination and parts of Sec.\ref{sec.NLO}, where a
signal-to-noise ratio is used), given that the overwhelming majority of
results have been obtained for it and allow an in-depth
comparison of different strategies. \red{A certain number of results
  have been obtained as well for the quantum Fisher information optimized over all input
  states \cite{Fujiwara01,fujiwara_quantum_2003}, a quantity sometimes
  called channel quantum Fisher information. We do not review this
  literature here, as in this type of work sensitivity is typically not
  separately optimized over entangled or
  non-entangled initial states.}

The Fisher information can be generalized to multi-parameter
estimation \cite{helstrom_quantum_1969,Paris2009},
$\bm{\theta}=(\theta_1,\theta_2,\dots)$. The Bures distance between
two infinitesimal close states then reads
\begin{equation} \label{Bures.infinitesimal}
ds^2_{\rm Bures}(\rho_{\bm{\theta}},\rho_{\bm{\theta}+d\bm{\theta}})=2\left(1-\tr\sqrt{\sqrt{\rho_{\bm{\theta}}} \, \rho_{\bm{\theta}+d\bm{\theta}}\sqrt{\rho_{\bm{\theta}}}}\right).
\end{equation}
An expansion of
$ds^2_{\rm Bures}(\rho_{\bm{\theta}},\rho_{\bm{\theta}+d\bm{\theta}})$
leads to the quantum Fisher information matrix \cite{Sommers2003,Paris2009},
\begin{equation}\label{eq:ds2multi}
ds^2_{\rm
  Bures}(\rho_{\bm{\theta}},\rho_{\bm{\theta}+d\bm{\theta}})=\frac{I_{\theta_k,\theta_{k'}}}{4}d\theta_k
d\theta_{k'}\,,
\end{equation}
where
$I_{\theta_k,\theta_{k'}}=\tr\rho_{\bm{\theta}}(L_{\theta_k}L_{\theta_{k'}}+L_{\theta_k'}L_{\theta_{k}})/2$,
and $L_{\theta_k}$ is the symmetric logarithmic derivative with
respect to parameter $\theta_k$. The quantum Cram\'er-Rao bound
generalizes to a lower bound on the co-variance matrix
$Cov[\bm{\theta}]$ of the
parameters $\theta_i$  \cite{helstrom_quantum_1969,Helstrom1976,Paris2009},
\begin{equation}
  \label{eq:QCRBm}
  Cov[\bm{\theta}]\ge\frac{1}{M}(I(\bm{\theta}))^{-1}\,,
\end{equation}
where $Cov[\bm{\theta}]_{ij}=\langle \theta_i\theta_j\rangle-\langle
\theta_i\rangle\langle\theta_j\rangle$, and $A\ge B$ means that $A-B$
is a positive-semidefinite matrix.  Contrary to the
single parameter
quantum Cram\'er-Rao bound, the bound \eqref{eq:QCRBm} can in general not be saturated, even
in the limit of infinitely many measurements.
The Bures metric has also been called fidelity
susceptibility in the framework of quantum phase transitions \cite{Gu2010}.


\section{Quantum correlations beyond entanglement}

\subsection{Parallel versus sequential strategies in unitary quantum
metrology}\label{sec:parvseq}

One of the most typical applications of quantum metrology is the
task of unitary parameter estimation, exemplified in particular by
phase estimation \cite{GiovannettiPRL2006, GiovannettiNPhot2011}.
Let $U_\theta = \exp{(-i \theta H)}$ be a unitary transformation,
with $\theta$ the unknown parameter to be estimated, and $H$ a
selfadjoint Hamiltonian operator which represents the generator of
the transformation. The typical estimation procedure then consists
of the following steps: a) preparing an input probe in a state
$\rho$; b) propagating the state with the unitary transformation
$U_\theta$; c) measuring the output state $\rho_\theta = U_\theta
\rho U_\theta^\dagger$; d) performing classical data analysis to
infer an estimator $\theta_{\mathrm{est}}$ for the parameter
$\theta$.

Let us now assume one has the availability of $N$ utilizations of the
transformation $U_\theta$. Then, the use of $N$ uncorrelated probes in a
global initial state $\rho^{\otimes N}$, each of which is undergoing the
transformation $U_\theta$ in parallel, yields an estimator whose minimum variance
scales
as $1/N$ (Standard Quantum Limit). On the other hand, by using an initial
entangled state $\rho$ of the $N$ probes, and propagating each  with the
unitary $U_\theta$ in parallel, one can in principle achieve the Heisenberg-limit, meaning
that an optimal estimator $\theta_{\mathrm{est}}$ can be constructed whose
asymptotic variance, in the limit $N \gg 1$, scales as $1/N^2$. However, it
is not difficult to realize that the very same precision can be reached
without the use of entanglement, by simply preparing a single input probe in
a superposition state with respect to the eigenbasis of the generator $H$,
and letting the probe undergo $N$ sequential iterations of the
transformation $U_\theta$.

For instance, thinking of each probe as a qubit for simplicity, and fixing
the generator $H$ to be the Pauli matrix $\sigma_z$, one can either consider
a parallel scheme with $N$ input probes in the
Greenberger-Horne-Zeilinger (GHZ, or cat-like) maximally
entangled state $\ket{\Psi} = (\ket{00\ldots0}+\ket{11\ldots1})/\sqrt{2}$,
or a sequential scheme with a single probe in the superposition
$\ket{\psi} = (\ket{0}+\ket{1})/\sqrt2$. {\redd In the first case, the
  state after imprinting the parameter reads $U_\theta^{\otimes N}\ket{\Psi}=(e^{-iN
  \theta}\ket{00\ldots 0}+e^{iN\theta}\ket{11\ldots 1})/\sqrt{2}$, while in the second
  case $U_\theta^{N}\ket{\psi}=(e^{-iN\theta
  }\ket{0}+e^{iN\theta}\ket{1})/\sqrt{2}$. Hence, in both schemes one
  achieves an $N$-fold increase of the phase between two orthogonal
  states,} and  both schemes reach therefore the Heisenberg-limit scaling in
the estimation of the phase shift $\theta$, meaning that the quantum Cram\'er-Rao bound
can be asymptotically saturated in both cases by means of an
optimal measurement, associated to a quantum Fisher information scaling quadratically
with $N$. The equivalence between entanglement in parallel schemes
and \textit{coherence} (namely, superposition in the eigenbasis of
the generator) \cite{Baumgratz2014,Streltsov2016C,Marvian2016} in
sequential schemes further extends to certain schemes of quantum
metrology in the presence of noise, namely when the unitary
encoding the parameter to be estimated and the noisy channel
commute with each other (e.g. in the case of phase estimation
affected by dephasing) \cite{Boixo2012,Maccone2014}, although in
more general instances entanglement is shown to provide an
advantage
\cite{huelga_improvement_1997,escher_general_2011,Maccone2014}. In
general, sequential schemes such that individual probes are
initially correlated with an ancilla (on which the parameter is
not imprinted) and assisted by feedback control (see
Sec.~\ref{sec.qfeed}) can match or outperform any parallel scheme
for estimation of single or multiple parameters encoded in unitary
transformations even in the presence of noise
\cite{Maccone2014,Sekatski2016,Maccone2016,Yuan2015,Yuan2016,Nichols2016,Youse2017}.
While probe and ancilla typically need to be entangled for such
sequential schemes to achieve maximum quantum Fisher information, this observation
removes the need for large-scale multiparticle entangled probes in
the first place.

Similarly, in continuous variable optical interferometry
\cite{caves_quantum-mechanical_1981}, equivalent performances can
be reached (for unitary phase estimation) by using either a
two-mode entangled probe, such as a N00N state, or a single-mode
non-classical state, such as a squeezed state. These are
elementary examples of quantum-enhanced measurements achievable without entanglement, yet
exploiting genuinely quantum effects such as nonclassicality and
superposition. Such features can be understood by observing that
both optical nonclassicality in infinite-dimensional systems and
coherence (superposition) in finite-dimensional systems can be
converted to entanglement within a well-defined resource-theoretic
framework
\cite{Asboth2005,Vogel2014,Singh2015,killoran2016converting}, and
can be thought-of as equivalent resources to entanglement for
certain practical purposes, as is evidently the case for unitary
metrology. \footnote{An additional scenario in which the quantum limit can be
reached without entanglement is when a multipartite state is used to
measure multiple parameters, where each parameter is encoded locally
onto only one subsystem --- it has recently been shown that
entanglement between the subsystems is not advantageous, and can even
be detrimental, in this setting \cite{knott2016,proctor2018}}. 

\subsection{General results on the usefulness of entanglement}
More generally, for unitary metrology with multipartite probes in a
parallel setting, a quite general formalism has been developed to
identify the metrologically useful correlations in the probes in order
to achieve quantum-enhanced measurements \cite{Smerzi0} (see \cite{SmerziR,toth_quantum_2014}
for recent reviews). Specifically, let us consider an input state
$\rho$ of $N$ qubits and a linear interferometer with Hamiltonian
generator given by $H=J_l=\frac12 \sum_{i=1}^N \sigma_l^{(i)}$, i.e.~a
component of the collective (pseudo-)
angular momentum of the $N$ probes in the direction $l=x,y,z$,
with $\sigma_l^{(i)}$ denoting the $l$th Pauli matrix for qubit
$i$. If $\rho$ is $k$-producible, i.e., it is a convex mixture of
pure states which are tensor products of at most $k$-qubit states,
then the quantum Fisher information is bounded above as follows \cite{Toth12,Hyllus12},
\begin{equation}\label{ew:QFIkprod}
I_\theta(\rho, J_l) \leq n k^2 + (N-n k)^2\,,
\end{equation}
where $n$ is the integer part of $N/k$. This means that genuine multipartite entangled probes ($k=N$) are required to reach the maximum sensitivity, given by the Heisenberg-limit $I_\theta \propto N^2$, even though partially entangled states can still result in quantum-enhanced measurements beyond the Standard Quantum Limit.

A similar conclusion has been reached in \cite{Augusiak15} considering the geometric measure of entanglement, which quantifies how far $\rho$ is from the set of fully separable ($1$-producible) states according to the fidelity. Namely, for unitary metrology with $N$ parallel probes initialized in the mixed state $\rho$, in the limit $N \rightarrow \infty$ a nonvanishing value of the geometric measure of entanglement of $\rho$ is necessary for the exact achievement of the Heisenberg-limit. However, a sensitivity arbitrarily close to the Heisenberg-limit, $I_\theta \propto N^{2-\epsilon}$ for any $\epsilon > 0$, can still be attained even if the geometric measure of entanglement of $\rho$ vanishes asymptotically for $N \rightarrow \infty$. In deriving these results, the authors proved an important continuity relation for the quantum Fisher information in unitary dynamics \cite{Augusiak15}.


\subsection{Role of quantum discord in parameter estimation with mixed probes}\label{sec:discordmodi}

Here we will focus our attention on possible advantages stemming from the
use of quantum correlations more general than entanglement in the (generally
mixed) state of the input probes for a metrological task. Such correlations
are usually referred to under the collective name of \textit{quantum discord}
\reddd{
\cite{Ollivier2001,Henderson2001}, see also \cite{Modi2012,ABC} for recent reviews}. {\redd{The name quantum discord originates from a mismatch between two possible quantum generalizations of the classical mutual information, a measure of correlations between two (or more) variables described by a joint probability distribution \cite{Ollivier2001}. A direct generalization leads to the so-called quantum mutual information $I(\rho)=S(\text{Tr}_A \rho) +
S(\text{Tr}_B \rho)-S(\rho)$, that quantifies total correlations in the state $\rho$ of a bipartite system $AB$, with $S(\rho)=-\text{Tr}(\rho \log \rho)$ being the von Neumann entropy. An alternative generalization leads instead to $J^{(A)}(\rho)=\sup_{\{\Pi^A\}} I(\Pi^A[\rho])$,
 a measure of one-sided classical correlations  that quantifies how much the marginal entropy of, say,  subsystem $B$ is decreased (i.e., how much additional information is acquired) by performing a minimally disturbing measurement on subsystem $A$ described by a POVM
$\{\Pi^A\}$, with $\Pi^A[\rho]$ being the conditional state of the system $AB$ after such measurement  \cite{Henderson2001}. The difference between the former and the latter quantity is precisely the quantum discord,
\begin{equation}  \label{eq:DA}
D^{(A)}(\rho) = I(\rho) - J(\rho)\,,
\end{equation}
that quantifies therefore just the quantum portion of the total correlations in the state $\rho$ from the perspective of subsystem $A$.
It is clear from the definition above that the state $\rho$ of a bipartite
system $AB$ has nonzero discord (from the point of view of  $A$)
if and only if it is altered by all possible local measurements
performed on subsystem $A$: disturbance by measurement is a genuine quantum feature which is captured by the concept of discord, see
\cite{Modi2012,ABC} for more details. Every entangled state is also discordant, but
the converse is not true; in fact, almost all separable states still exhibit
nonzero discord \cite{FerraroAcin2010}. The only bipartite states with zero discord, from the point
of view of subsystem $A$, are so-called classical-quantum states, which take
the form
\begin{equation}  \label{eq:chiA}
\chi^{(A)} = \sum_i p_i \ket{i}\bra{i}^A \otimes \tau_i^B,
\end{equation}
where the states $\{\ket{i}^A\}$ form an orthonormal basis for
subsystem $A$, and $\{\tau_i^B\}$ denote a set of arbitrary states
for subsystem $B$, while $\{p_i\}$ stands for a probability
distribution. These states are left invariant by measuring $A$ in the basis $\{\ket{i}^A\}$, which entails that $D^{(A)}(\chi^{(A)})=0$.
}}

In a multipartite setting, one can define fully classical states
as the states with zero discord with respect to all possible
subsystems, or alternatively as the states which are left
invariant by a set of local measurements performed on all
subsystems. Such states take the form $\chi = \sum_{i_1, \ldots,
i_N} p_{i_1,\ldots,i_N} \ket{i_1}\bra{i_1}^{A_1} \otimes \cdots
\otimes \ket{i_N}\bra{i_N}^{A_N}$ for an $N$-particle system
$A_1\ldots A_N$; i.e., they are diagonal in a local product basis.
One can think of these states as the only ones which are
completely classically correlated, that is, completely described
by a classical multivariate probability distribution
$\{p_{i_1,\ldots,i_N}\}$, embedded into a density matrix
formalism. An alternative way to quantify discord in a (generally
multipartite) state $\rho$ is then by taking the distance between
$\rho$ and the set of classically correlated states, according to
a suitable (quasi)distance function. For instance, the relative
entropy of discord \cite{Modi10} is defined as
\begin{equation}\label{eq:DR}
D_R(\rho) = \inf_{\chi} S(\rho \| \chi)\,,
\end{equation}
where the minimization is over all classically correlated states
$\chi$, and $S(\rho\|\chi) = \text{Tr}(\rho \log \rho - \rho \log
\chi)$ denotes the quantum relative entropy. For a dedicated
review on different measures of discord-type quantum correlations
\reddd{we refer the reader to} \cite{ABC}.

{\redd{Let us now discuss the role of quantum discord in
metrological contexts. \cite{Modi2011} investigated the estimation
of a unitary phase $\theta$ applied to each of $N$ qubit probes,
initially prepared in mixed states with either \reddd{(a)} no correlations;
\reddd{(b)} only classical correlations; or \reddd{(c)} quantum correlations
(discord and/or entanglement). All the considered families of
$N$-qubit probe states were chosen with the same spectrum, i.e.~in
particular the same degree of mixedness (which is a meaningful
assumption if one is performing a metrology experiment in an
environment with a fixed common temperature), and were selected
due to their relevance in recent nuclear magnetic resonance (NMR) experiments
\cite{Jones2009}.}} In particular, given an initial thermal state
$\rho_0(p)=\left(\frac{1+p}{2}\ket{0}\bra{0} +
\frac{1-p}{2}\ket{1}\bra{1}\right)$ for each single qubit (with
purity parameter $0\leq p \leq 1$), the product states
$\rho_N^{\text{\reddd{(a)}}}(p) = \left[H \rho_0(p) H \right]^{\otimes N}$
were considered for case \reddd{(a)}, and the GHZ-diagonal states
$\rho_N^{\text{\reddd{(c)}}}(p) = C H_1 C \rho_0(p)^{\otimes N} C H_1 C$
were considered for case \reddd{(c)}, with $H$ denoting the single-qubit
Hadamard gate (acting on each qubit in the first case, and only on
the first qubit in the second case), and $C=
\otimes_{j=2}^N$~\textsc{C-Not}${}_{1j}$ a series of Control-Not
operations acting on pairs of qubits $1$ and $j$. {\redd{These two
classes of states  give rise to quantum Fisher information
$I_\theta^{\text{\reddd{(a)}}} = p^2
N$ and $I_\theta^{\text{\reddd{(c)}}} \gtrapprox p^2 N^2$, respectively. By
comparing the two cases, the authors of \cite{Modi2011} concluded
that a quantum enhancement, scaling as
 $I_\theta^{\text{\reddd{(c)}}}/I_\theta^{\text{\reddd{(a)}}} \approx N $, is possible using pairs of mixed probe states with arbitrary (even infinitesimally small) degree of
purity. This advantage persists even when the states in
strategy \reddd{(c)} are fully separable, which occurs for $p \lesssim a + b / N$ (with $a$ and $b$ determined numerically for each value of $N$), in which case both strategies are unable to beat the Standard Quantum Limit, yet the quadratic enhancement of \reddd{(c)} over \reddd{(a)} is maintained, being independent of $p$. The authors then argue that multipartite quantum discord --- which increases with $N$
according to the relative entropy measure of eq.~(\ref{eq:DR}) and vanishes only at $p=0$ --- may
be responsible for this enhancement.}}
Let us remark that, even
though the quantum Fisher information is convex (which means that for every separable but
discordant mixed state there exists a pure product state with a higher or equal
quantum Fisher information), the analysis in \cite{Modi2011} was performed at fixed
spectrum (and thus degree of purity) of the input probes,  a constraint which allowed the
authors to still identify an advantage in using correlations weaker than
entanglement, as opposed to no correlations. However, it is  presently
unclear whether these conclusions are special to the selected
classes of states, or can be further extended to more general
settings, including noisy metrology.

In a more recent work, \cite{Cable2015} analyzed a model of
unitary quantum metrology inspired by the computational algorithm
known as deterministic quantum computation with one quantum bit
(DQC1) or ``power of one-qubit'' \cite{knill_power_1998}. Using
only one pure qubit supplemented by a register of $l$ maximally
mixed qubits, all individually subject to a local unitary phase
shift $U_\theta$, their model was shown to achieve the Standard Quantum Limit for the
estimation of $\theta$, which can be conventionally obtained using
the same number of qubits in pure uncorrelated states. They found
that the Standard Quantum Limit can be exceeded by using one additional qubit, which
only contributes a small degree of extra purity, which, however,
for any finite amount of extra purity leads to an entangled state
at the stage of parameter encoding. In this model, incidentally,
the output state after the unitary encoding was found to be always
separable but discordant, with its discord vanishing only in the
limit of vanishing variance of the estimator for the parameter
$\theta$. It is not quite clear if and how the discord in the
final state can be interpreted in terms of a resource for
metrology, but the achievement of the Standard Quantum Limit with all but one probes
in a fully mixed state was identified as a quantum enhancement
without the use of entanglement. This suggests that further
investigation on the role of quantum discord (as well as state
purity) in metrological algorithms with vanishing entanglement may
be in order. A protocol for multiparameter estimation using DQC1
was  studied in \cite{BoixoSomma}, although the resource role of
correlations was not discussed there. In \cite{PhysRevA.93.023805}
a detailed investigation of a DQC1-based protocol was made based
on coherently controlled Rydberg interactions between a single
atom and an atomic ensemble containing $N$ atoms. The protocol
allows one to estimate a phase shift assumed identical for all
atoms in the atomic ensemble with a sensitivity that interpolates
smoothly between Standard Quantum Limit and Heisenberg-limit when the purity of the atomic ensemble
increases from a fully mixed state to pure states. It leads to a
cumulative phase shift proportional to $N$, and the scheme can in
fact also be seen as an implementation of ``coherent averaging'',
with the control qubit playing the role of the ``quantum bus''
(see Sec.~\ref{sec.cohav}).

\subsection{Black-box metrology and the interferometric power}

As explicitly discussed in Sec.~\ref{sec:parvseq}, for unitary parameter
estimation, if one has full prior information on the generator $H$ of the
unitary transformation $U_\theta$ imprinting the parameter $\theta$, then no
correlations are required whatsoever, and probe states with coherence in the
eigenbasis of $H$ suffice to achieve quantum-enhanced measurements in a sequential scheme. Recently,
\cite{Girolami2013,Girolami2014,Adesso2014} investigated quantum metrology
in a so-called black-box paradigm, according to which the generator $H$ is assumed not fully known
a priori. {\redd{In such a case, suppose one selects a fixed (but arbitrary) input single-particle probe $%
\rho$, then it is impossible to guarantee a precision in the estimation of $%
\theta$ for {\it all} possible nontrivial choices of $H$. This is because, in the
worst case scenario, the black-box unitary transformation may be generated
by a $H$ which commutes with the input state $\rho$, resulting in no
information imprinted on the probe, hence in a vanishing quantum Fisher information. It is clear
then that, to be able to estimate parameters independently of the choice of the generator,
one needs an ancillary system correlated with the probe. But what type of correlations are needed? It is in this
 context that discord-type correlations, rather than entanglement or
classical correlations, are found to play a key resource role.

Consider a standard two-arm interferometric configuration, and let us retrace
the steps of parameter estimation in the black-box scenario \cite%
{Girolami2014}: a) an input state $\rho$ of two particles, the probe $A$ and the
ancilla $B$, is prepared; b)  particle $B$ is transmitted with no interaction, while particle $A$ enters a
black-box where it undergoes a unitary transformation $U_\theta = \exp(- i
\theta H)$ generated by a Hamiltonian $H$, whose spectrum is known but whose eigenbasis is unknown at this stage;  c) the
agent controlling the black-box announces the full specifics of the generator $H$, so that parties
$A$ and $B$ can jointly perform the best possible measurement on the  two-particle output state $\rho_\theta = (U_\theta \otimes
\mathbb{I}) \rho (U_\theta \otimes \mathbb{I})^\dagger$; d) the whole process is iterated $N$ times, and an optimal
unbiased estimator $\theta_{\mathrm{est}}$ is eventually constructed for the parameter $%
\theta$. In the limit $N \gg 1$, for any specific black-box setting $H$, the corresponding quantum Fisher information $I_\theta(\rho, H)$
determines the maximal precision enabled by the input state $\rho$ in
estimating the parameter $\theta$ generated by $H$, as prescribed by  the quantum Cram\'er-Rao bound.

One can then introduce a figure of merit quantifying the worst case
precision guaranteed by the state $\rho $ for the estimation of $\theta$ in this black-box protocol. This is done naturally by
minimizing the quantum Fisher information over all generators $H$ within the given spectral class (the
 spectrum is assumed nondegenerate, with a canonical choice being that of
equispaced eigenvalues) \cite{Girolami2013,Girolami2014}. This defines (up to a normalization constant) the \textit{%
interferometeric power}  of the bipartite state $\rho $ with respect to
the probing system $A$,
\begin{equation}
P^{(A)}(\rho )=\frac{1}{4}\min_{H}I_{\theta }(\rho ,H)\,.  \label{eq:IP}
\end{equation}%
Remarkably, as proven in \cite{Girolami2014}, the interferometric power turns out to be a measure of discord-type
correlations in the input state $\rho $. In particular, it vanishes if and
only if $\rho $ takes the form of a classical-quantum state, eq.~(\ref%
{eq:chiA}).}} This entails that states with zero discord cannot guarantee a
precision in parameter estimation in the worst case scenario, while any
other bipartite state (entangled or separable) with nonzero discord
is suitable for estimating
parameters encoded by a unitary transformation (acting on one subsystem) no
matter the generator, with minimum guaranteed precision quantified by the interferometric power
of the state. This conclusion holds both for parameter estimation in
finite-dimensional systems \cite{Girolami2014}, and for continuous-variable
optical interferometry \cite{Adesso2014}. Recently, it has been shown more formally that entanglement accounts only for a portion of the quantum correlations relevant for bipartite quantum interferometry. In particular, the interferometric power, which is by definition a lower bound to the quantum Fisher information (for any fixed generator $H$), is itself bounded from below in bipartite systems of any dimension by a measure of entanglement aptly named the interferometric entanglement, which simply reduces to the squared concurrence for two-qubit states \cite{thereismore}.
The interferometric power can be evaluated in closed
form, solving analytically the minimization in eq.~(\ref{eq:IP}), for all
finite-dimensional states such that subsystem $A$ is a qubit \cite{Girolami2014}, and for all two-mode Gaussian states when the minimization
is restricted to Gaussian unitaries \cite{Adesso2014}. An experimental
demonstration of black-box quantum-enhanced measurements relying on discordant states as opposed to
classically correlated states has been reported using a two-qubit NMR
ensemble realized in chloroform \cite{Girolami2014}.

{\redd{We finally notice that, while (quantum) correlations with an ancilla are required to achieve a nonzero worst case precision when minimizing the quantum Fisher information over the choice of the generator $H$ within a fixed spectral class, as in the scenario considered here, single-probe (non-maximally mixed) states may however suffice to be useful resources in the arguably more practical case in which the average precision, rather than the minimal,  is considered instead as a figure of merit. This scenario is further discussed in Sec.~\ref{sub:AVQ}.}}

\subsection{Quantum estimation of bosonic loss}\label{sec.BL}

Any quantum optical communication, from fibre-based to free-space
implementations, is inevitably affected by energy dissipation. The
fundamental model to describe this scenario is the lossy channel.
This attenuates an incoming bosonic mode by transmitting a
fraction $\eta \leq 1$ of the input photons, while sending the
other fraction $1-\eta $ into the environment. The maximum number
of bits per channel use at which we can transmit quantum
information, distribute entanglement or generate secret keys
through such a lossy channel are all equal to $-\log
(1-\eta)$~\cite{Pirandola15}, a fundamental rate-loss tradeoff
that only quantum repeaters may surpass~\cite{Pirandola16}. For
these and other implications to quantum communication, it is of
paramount importance to estimate the transmissivity of a lossy
channel in the best possible way.

Quantum estimation of bosonic loss was first studied in
\cite{Monras07} by using single-mode pure Gaussian states
(see also \cite{Pinel13}). In this setting, the performance
of the coherent state probes at fixed input energy provides the
shot-noise limit or classical benchmark, which has to be beaten by
truly quantum probes. Let us denote by $\bar{n}$ the mean number
of photons, then the shot-noise limit is equal to $I_{\eta }\simeq
\eta ^{-1}\bar{n}$~\cite{Monras07,Pinel13}. The use of squeezing
can beat this limit, following the original intuition for phase
estimation of \cite{caves_quantum-mechanical_1981}. In fact,
\cite{Monras07} showed that, in the regime of small loss
$\eta \simeq 1$ and small energy $\bar{n}\simeq 0$, a squeezed
vacuum state can beat the Standard Quantum Limit.
 The use of squeezing for estimating the interaction parameter in
bilinear bosonic Hamiltonians (including beam-splitter
interactions) was also discussed in \cite{Gaiba09}, showing
that unentangled single-mode squeezed probes offer equivalent
performance to entangled two-mode squeezed probes for practical
purposes.


The optimal scaling $I_{\eta }\simeq [\eta (1-\eta)]^{-1} \bar{n}$
can be achieved by using Fock states at the
input~\cite{AdessoAnno09}. 
 Note that, because Fock states can only be used when the input
energy corresponds to integer photon numbers, in all the other
cases one needs to engineer superpositions, e.g., between the
vacuum and the one-photon Fock state if we want to explore the
low-energy regime $\bar{n}\lesssim 1$. Non-Gaussian qutrit and
quartet states can be designed to beat the best Gaussian
probes~\cite{AdessoAnno09}. It is still an open question to
determine the optimal probes for estimating loss at any energy
regime. It is certainly known that the bound $I_{\eta } \leq [\eta
(1-\eta)]^{-1} \bar{n}$ holds for any $\bar{n}$, as it can be proven by
dilating the lossy channel into a beam-splitter unitary and then
performing parameter estimation~\cite{Monras07}. Note that this
bound is computed by considering $N$ uncorrelated probes in
parallel. It is therefore an open question to find the best
performance that is achievable by the most general (adaptive)
strategies.

Interestingly, the problem of estimating the loss parameters of a
pair of lossy bosonic channels has been proven formally equivalent
to the problem of estimating the separation of two incoherent optical
point-like sources~\cite{Cosmo2016}. In this
context~\cite{Mankei2016} showed that a pair of weak thermal
sources can be resolved independently from their separation if one
adopts quantum measurements based on photon counting, instead of
standard intensity measurements. Thus, quantum detection
strategies enables one to beat the so-called ``Rayleigh's curse''
which affects classical imaging~\cite{Mankei2016}. This curse is
reinstated in the classical limit of very bright thermal
sources~\cite{Cosmo2016,Nair2016}. On the other hand,
\cite{Cosmo2016} showed that quantum-correlated sources can be
super-resolved at the sub-Rayleigh scale. In fact, it is possible
to engineer quantum-correlated point-like sources that are not
entangled (but discordant) which displays super-resolution, so
that the closer the sources are the better their distance can be
estimated.


The estimation of loss becomes complicated in the presence of
decoherence, such as thermal noise in the environment and non-unit
efficiency of the detectors. From this point of view,
\cite{Spedalieri16} considered a very general model of
Gaussian decoherence which also includes the potential presence of
non-Markovian memory effects. In such a scenario,
\cite{Spedalieri16} showed the utility of asymmetrically
correlated thermal states (i.e., with largely different photon
numbers in the two modes), fully based on discord and void of
entanglement. These states can be used to estimate bosonic loss
with a sensitivity that approaches the shot noise limit and may
also surpass it in the presence of correlated noise and memory
effects in the environment. This kind of thermal quantum metrology
has potential applications for practical optical instruments
(e.g., photometers) or at different wavelengths (e.g., far
infrared, microwave or X-ray) {\redd for which} the generation of quantum
features, such as coherence, number states, squeezing or
entanglement, may be challenging.

\subsection{Gaussian quantum metrology}

Clearly we may also consider the estimation of other parameters
beyond loss. In general, Gaussian quantum metrology aims at
estimating any parameter or multiple parameters encoded in a
bosonic Gaussian channel. As shown in~\cite{PirandolaLupo16}, the
most general adaptive estimation of noise parameters (such as
thermal or additive noise) cannot beat the Standard Quantum Limit. This is because
Gaussian channels are teleportation-covariant, i.e., they suitably
commute with the random operations induced by quantum
teleportation, a property which is shared by large class of
quantum channels at any dimension~\cite{Pirandola15}. The joint
estimation of specific combinations of parameters, such as loss
and thermal noise, or the two real components of a displacement,
has been widely studied in the
literature \cite{Monras11,Bellomo2009,Bellomo2010-1,Bellomo2010-2,PhysRevA.87.012107,Gao2014,PhysRevA.95.012305,Gagatsos2016},
but the ultimate performance achievable by adaptive (i.e.,
feedback-assisted) schemes is still unknown.

If we employ Gaussian states at the input of a Gaussian channel,
then we have Gaussian states at the output and we may exploit
closed formulas for the quantum Fisher information. These formulas can be derived by
direct evaluation of the symmetric logarithmic derivative~\cite{Monras13,Jiang2014,Safranek2015,Liuzzo2017} or by considering the
infinitesimal expression of the quantum
fidelity~\cite{Banchi15,Pinel13,Pinel12}. The latter approach may
exploit general and handy formulas. In fact, for two arbitrary
multi-mode Gaussian states, $\rho _{1}$ and $\rho _{2} $, with mean
values $u_{1}$\ and $u_{2}$, and covariance matrices $V_{1}$ and
$V_{2}$, we may write the Uhlmann-Jozsa fidelity~\cite{Banchi15}
\begin{eqnarray}
\!\!\!\!\!\!F(\rho _{1},\rho _{2}) &=&\frac{F_{\mathrm{tot}}e^{-\frac{1}{2}%
(u_{2}-u_{1})^{T}(V_{1}+V_{2})^{-1}(u_{2}-u_{1})}}{\sqrt{\det (V_{1}+V_{2})}}%
,  \label{genFIDmain} \\
\!\!\!\!\!\!F_{\mathrm{tot}}^{2} &=&\det \left[ 2\left( \sqrt{\openone+\frac{(V_{\mathrm{%
aux}}\Omega )^{-2}}{4}}+\openone\right) V_{\mathrm{aux}}\right],
\end{eqnarray}%
where we set $V_{\mathrm{aux}}:=\Omega
^{T}(V_{1}+V_{2})^{-1}\left( \Omega /4+V_{2}\Omega V_{1}\right) $
with $\Omega $ being the symplectic form~\cite{Banchi15}. Specific
expressions for the fidelity were previously given for single-mode
Gaussian states~\cite{Scutaru98}, two-mode Gaussian
states~\cite{Marians12}, multi-mode Gaussian states assuming that
one of the states is pure~\cite{Spedalieri13}, and multi-mode
squeezed thermal Gaussian states with vanishing first
moments~\cite{Paraoanu2000}.

From eq.~(\ref{genFIDmain}) we may derive the Bures metric
$ds^{2}_{\rm Bures}$. In fact, consider two infinitesimally-close Gaussian
states $\rho $, with statistical moments $u$ and $V$, and $\rho
+d\rho $, with statistical moments $u+du$ and $V+dV$. Expanding at
the second order in $du$ and $dV$,
one finds~\cite{Banchi15}%
\begin{equation}
ds^{2}_{\rm Bures}=2[1-\sqrt{F(\rho,\rho+d\rho)}]=\frac{du^{T}V^{-1}du%
}{4}+\frac{\delta }{8}~,  \label{BuresM}
\end{equation}%
where $\delta
:=4\mathrm{Tr}[dV(4\mathcal{L}_{V}+\mathcal{L}_{\Omega
})^{-1}dV]$, $\mathcal{L}_{A}X:=AXA$, and the inverse of the superoperator $4%
\mathcal{L}_{V}+\mathcal{L}_{\Omega }$ refers to the
pseudo-inverse. A similar expression was also computed
by~\cite{Monras13} using the symmetric logarithmic derivative, with
further refinements
in~\cite{Safranek2015}. From the Bures metric in
eq.~\eqref{BuresM} we may derive the quantum Fisher information (see eq.~\eqref{IBures})
for the estimation of any parameter encoded in a multi-mode (pure
or mixed) Gaussian state directly in terms of the statistical
moments. 
{ Eq.~(\ref{BuresM}) is written in a compact basis-independent and parametrization-independent form, valid for any
multi-mode Gaussian state. For an explicit parametrization via
multiple parameters ${\bm \theta} = (\theta_1, \theta_2, \dots)$, one can expand the differential
and write $dV= \sum_k \partial_{\theta_k} V \,
d\theta_k$, and similarly for $du$. In this way, $  ds^2_{\rm Bures} = \sum_{k,k'} \frac{1}{4} I_{\theta_k,\theta_{k'}} d\theta_k d\theta_{k'}$ as in Eq.~(\ref{eq:ds2multi}),
with
\begin{align}
    I_{\theta_k,\theta_{k'}} &=
{(\partial_{\theta_k}u^T)V^{-1}(\partial_{\theta_{k'}}u)%
} \nonumber
\\& \quad +
2\mathrm{Tr}[(\partial_{\theta_k}V)(4\mathcal{L}_{V}+\mathcal{L}_{\Omega
})^{-1}(\partial_{\theta_{k'}}V)]~.
\label{BuresMB}
\end{align}
Eqs.~(\ref{BuresM}) and~(\ref{BuresMB}) have been derived
following eq.~(\ref{IBures}), namely explicitly computing the
fidelity function for two most general multi-mode Gaussian states, and then
taking the limit of two infinitesimally close states. A
similar approach was used for fermionic Gaussian states
in~\cite{Banchi2014}.}

{\redd{
An alternative derivation of the bosonic quantum Fisher information for multi-mode Gaussian states, based on the
use of the symmetric logarithmic derivative, has been recently
obtained in \cite{Liuzzo2017}.
        Furthermore, \cite{Liuzzo2017}  derived a necessary and sufficient compatibility condition such that the quantum Cram\'er-Rao bound eq.~(\ref{eq:QCRBm}) is asymptotically achievable in multiparameter Gaussian quantum metrology, meaning that a single optimal measurement exists which is able to extract the maximal information on all the parameters simultaneously. For any pair of parameters $\theta_k,\theta_{k'} \in {\bm \theta}$,  in terms of  the  symmetric logarithmic derivatives $L_{\rho_{{\theta_k}}}$ and $L_{\rho_{{\theta_{k'}}}}$, the corresponding quantum Fisher information matrix element is defined as $I_{{\theta_k},{\theta_{k'}}} \equiv \text{Re}\left[ \text{Tr}\left(\rho_{{\bm \theta}} L_{\rho_{{\theta_k}}} L_{\rho_{{\theta_{k'}}}}\right)\right]$, while the measurement compatibility condition amounts to $Y_{{\theta_k},{\theta_{k'}}} \equiv \text{Im}\left[ \text{Tr}\left(\rho_{{\bm \theta}} L_{\rho_{{\theta_k}}} L_{\rho_{{\theta_{k'}}}}\right)\right] = 0$ \cite{Ragy2016}. In terms of the first and second  statistical moments $u$ and $V$ of a $m$-mode Gaussian state $\rho_{\boldsymbol \theta}$, we have then \cite{Liuzzo2017}:
\begin{eqnarray}
\!\!\!\! I_{{\theta_k},{\theta_{k'}}}&=&(\partial_{\theta_k}{u}^T) V^{-1}(\partial_{\theta_{k'}}{u})+2\text{Tr}(\partial_{\theta_{k'}} V K_{{\theta_k}}), \label{eq:gqfim}\\
\!\!\!\! Y_{{\theta_k},{\theta_{k'}}}&=&\frac{1}{2}(\partial_{\theta_k}{u}^T)V^{-1}\Omega V^{-1}(\partial_{\theta_{k'}}{u})+ 16\text{Tr}\left(\Omega K_{{\theta_{k'}}} V  K_{{\theta_k}} \right), \nonumber \\ && \label{eq:gcompatibility}
\end{eqnarray}
with $K_{\theta}= \sum_{i,j=1}^m \sum_{l=0}^{3} \frac{(a_{\theta})^{ij}_l}{\nu_i\nu_j-(-1)^l}{S^T}^{-1}{M}_l^{ij}S^{-1}$,
where  ${(a_{\theta})^{ij}_l}=\text{Tr}\left(S^{-1}\partial_\theta V  {S^T}^{-1}{M}_l^{ij}\right)$,  $\{\nu_i\}$ are the symplectic eigenvalues of the covariance matrix $V$, $S^{-1}$ is the symplectic transformation that brings $V$ into its diagonal form, $S^{-1}V{S^T}^{-1}=\bigoplus_{i=1}^m \nu_i\openone$, and the set of matrices ${M}_l^{ij}$ have all zero entries except for the $2 \times 2$ block in position $ij$ which is given by
$  \big\{ (M)_l^{ij}\big\}_{l\in\{0,\dots,3\}} = \frac{1}{\sqrt{2}}\big\{i \sigma_y,~\sigma_z,~\openone,~\sigma_x\big\}$. 
}} {Note that eq.~(\ref{eq:gqfim}) can also be obtained from
(\ref{BuresM}) by explicitly writing all the operators in the
basis {\redd in which} $V$ is diagonal, observing that $(4\mathcal{L}_{V}+\mathcal{L}_{\Omega
})^{-1}(\partial_{\theta}V)=K_{{\theta}}$. On the other hand,
Eq.~(\ref{eq:gcompatibility}) cannot be obtained from the limit of
the fidelity formula. }

In the context of this review, \cite{Pinel12} studied in particular the quantum Cram\'er-Rao bound for
estimating a parameter $\theta$ which is  encoded in a pure
multi-mode Gaussian state. It was realized that, in the limit of
large photon number, no entanglement nor correlations between
different modes are necessary for obtaining the optimal
sensitivity. Rather, a detection mode can be used based on the
derivative of the mean photon field with respect to the parameter
$\theta$, into which all the resources in terms of photons and
squeezing should be put. The mean photon field is defined as
$\bar{a}_\theta(\br,t)=\langle\psi_\theta|a(\br,t)|\psi_\theta\rangle$,
with all parameter dependence in the pure Gaussian quantum state
$\ket{\psi}_\theta$, $a(\br,t)=\sum_i a_iv_i(\br,t)$, 
where $v_i(\br,t)$ are orthonormal mode functions found from
solving Maxwell's equation with appropriate boundary conditions,
$a_i$ is the annihilation operators of mode $i$, and the sum is
over all modes. The mean field can be normalized,
$u_\theta=\bar{a}_\theta(\br,t)/||\bar{a}_\theta||$, where the
norm $||f(\br,t)||= (\int |f(\br,t)|^2 d^2\br dt)^{1/2}$ contains
spatial integration over a surface perpendicular to the light beam
propagation and temporal integration over the detection time. The
detection mode is then defined as
$\tilde{v}_1(\br,t)=\frac{\bar{a}_\theta'(\br,t)}{||\bar{a}_\theta'||}$,
where $'$ means derivative with respect to $\theta$. The detection mode
can be complemented by other, orthonormal modes to obtain a full
basis, but these other modes need not be excited for achieving maximum quantum Fisher information.
The quantum Fisher information reads then
\begin{equation}
  \label{eq:QFIGauss}
  I_\theta=N_\theta\left(4||u'_\theta||^2+\left(\frac{N'_\theta}{N_\theta}\right)^2
  \right){V}^{-1}_{\theta,[1,1]}\,,
\end{equation}
where $N_\theta$ is the mean photon number, and
$V^{-1}_{\theta,[1,1]}$ the matrix element of the
inverse covariance matrix of the Gaussian state corresponding to
the detection mode $\tilde{v}_1(\br,t)$.  All other modes are chosen
orthonormal to it.
The Standard Quantum Limit corresponds to a quantum Fisher information of a coherent state, in which case
$V^{-1}_{\theta,[1,1]}=1$. Hence, an improvement over
the Standard Quantum Limit is possible with pure Gaussian states by squeezing the
detection mode.  The scaling with $N_\theta$ can be modified if
$V^{-1}_{\theta,[1,1]}$ depends on $N_\theta$. For a
fixed total energy a scaling $I_\theta\propto N_\theta^{3/2}$ can
be achieved. This was proposed in \cite{refId0} for measuring a
beam displacement. The quantum Cram\'er-Rao bound in eq.\eqref{eq:QFIGauss} can be reached
by homodyne detection with the local oscillator in this detection
mode.

By using compact expressions of the quantum Fisher information for multi-mode Gaussian
states, \cite{Safranek2016} developed a practical method to find
optimal Gaussian probe states for the estimation of parameters
encoded by Gaussian unitary channels. Applications of the method
to the estimation of relevant parameters in single-mode and
two-mode unitary channels, such as phase, single-mode squeezing,
two-mode squeezing, and transmissivity of a beam splitter,
confirmed that separable probes can achieve exactly the same
precision as entangled probes, leading the authors of
\cite{Safranek2016} to remark how entanglement does not play any
significant role in achieving the Heisenberg-limit for unitary Gaussian quantum
metrology.

The same conclusion has been reached by considering the estimation of any small parameter $\theta$ encoded in Bogoliubov transformations, i.e., Gaussian unitary channels corresponding to arbitrary linear transformations of a set of $n$ canonical mode operators \cite{Friis2015}. In the limit of infinitesimal transformations ($\theta \ll 1$), and considering an arbitrary (Gaussian or not) pure $n$-mode probe state with input mean photon number $N_\theta$, \cite{Friis2015} showed by means of a perturbative analysis that the maximal achievable quantum Fisher information scales as $I_\theta\propto N_\theta^{2}$, that is, at the Heisenberg-limit. Remarkably, such a quantum-enhanced scaling requires nonclassical (e.g., squeezed) but not necessarily
entangled states.

Further results on the use of bosonic probes and the role of mode entanglement in Gaussian and non-Gaussian quantum metrology are presented in Sec.~\ref{sec.bosons}.

\subsection{Quantum channel discrimination} \label{QCD}

A fundamental protocol which is closely related to quantum
metrology is quantum channel
discrimination~\cite{Childs00,Acin01,Sacchi05,Lloyd08,Tan08,Pirandola11,Invernizzi11},
which may be seen as a sort of digitalized version of quantum
metrology. Its basic formulation is binary and involves the task
of distinguishing between two quantum channels, $\mathcal{E}_{0}$
or $\mathcal{E}_{1}$, associated with two a priori probabilities
$\pi _{0}:=\pi $ and $\pi _{1}=1-\pi $. During the encoding phase,
one of such channels is picked by Alice and stored in a box which
is then passed to Bob. In the decoding phase, Bob uses a suitable
state at the input of the box and performs a quantum measurement
of its output. Bob may also use ancillary systems which are
quantum correlated with the input probes and are directly sent to
the measurement. For the specific tasks of discriminating bosonic
channels, the input is assumed to be constrained in energy, so
that we fix the mean number of photons $\bar{n}$ per input probe,
or more strongly, the mean total number of photons which are
globally irradiated through the box~\cite{Weedbrook12}.

Quantum channel discrimination is an open problem in general. However, when
we fix the input state, it is translated into an easier problem to solve,
i.e., the quantum discrimination of the output states. In the binary case,
this conditional problem has been fully solved by the so-called Helstrom
bound which provides the minimum mean error probability $\bar{p}$ in the
discrimination of any two states $\rho _{0}$\ and $\rho _{1}$. Assuming
equiprobable states ($\pi =1/2$), this bound is simply given by their trace
distance $D$, i.e., we have~\cite{Helstrom1976}%
\begin{equation}
\bar{p}=\frac{1}{2}\left[ 1-D(\rho _{0},\rho _{1})\right] .
\end{equation}

In the case of multi-copy discrimination, in which we probe the box
$N$ times and we aim to distinguish the two outputs $\rho
_{0}^{\otimes N}$ and $\rho _{1}^{\otimes N}$, the mean error
probability $\bar{p}(N)$ may be not so easy to compute and,
therefore, we resort to suitable bounds. Using the quantum
fidelity $F(\rho_{0},\rho_{1})$ from \eqref{eq:fidel}, and setting
\begin{equation}
Q(\rho_{0},\rho_{1}):=\inf_{0\leq
s\leq1}\text{Tr}(\rho_{0}^{s}\rho_{1}^{1-s}),
\end{equation}
we may then write~\cite{Fuchs99,Audenaert2007,Banchi15}
\begin{equation}
\frac{1-\sqrt{1-F(\rho_{0},\rho_{1})^{N}}}{2}\leq\bar{p}(N)\leq\frac{Q^{N}(\rho_{0},\rho_{1})}{2},\label{QCBdef}
\end{equation}
where $Q^{N}(\rho_0,\rho_1)/2$ is the quantum Chernoff bound
(QCB)~\cite{Audenaert2007}. In particular, the QCB is
asymptotically tight for large $N$. Furthermore, it can be easily
computed for arbitrary multi-mode Gaussian
states~\cite{PirandolaLloyd08}.

Since the conditional output states can be optimally
distinguished, the non-trivial part in quantum channel
discrimination is the optimization of the mean error probability
$\bar{p}$ over the input states. For this reason, it is an
extremely rich problem and depending on the types of quantum
channels, quantum correlations may play an important role or not.
We now discuss some specific cases in more detail.


Quantum channel discrimination has various practical applications.
One which is very well known is quantum
illumination~\cite{Lloyd08,Tan08} which forms the basis for a
``quantum radar''~\cite{Barzanjeh15}. Despite the fact that
entanglement is used at the input between the signal (sent to
probe a potential target) and the idler (kept at the radar state
for joint detection), entanglement is completely absent at the
output between reflected and idler photons. Nevertheless the
scheme assures a superior performance with respect to the use of
coherent states; in particular, an increase by a factor 4 of the
exponent $-(\ln P(e))/M$ of the asymptotic error probability
$P(e)$ (where $M$ is the number of transmissions)~\cite{Tan08}.
For this reason, the quantum illumination advantage has been
studied in relation with the consumption of other discord-type
quantum correlations beyond
entanglement~\cite{Weedbrook2016,Bradshaw2016}. More precisely,
the enhanced performance of quantum illumination (with respect to
signal probing not assisted by an idler) corresponds to the amount
of discord which is expended to resolve the target (i.e., to
encode the information about its presence or absence). Quantum
illumination was demonstrated experimentally
in~\cite{Zhang13,Lopaeva13,Zhang15}.

Another application of quantum channel discrimination is quantum
reading~\cite{Pirandola11}. Here the basic aim is to discriminate
between two different channels which are used to encode an
information bit in a cell of a classical memory. In an optical
setting, this means to discriminate between two different
reflectivities, generally assuming the presence of decoherence
effects, such as background stray photons. The maximum amount of
bits per cell that can be read is called ``quantum reading
capacity''~\cite{Pirandola11b}. This model has also been studied in
the presence of thermal and correlated decoherence, as that
arising from optical diffraction~\cite{Lupo13}. In all cases, the
classical benchmark associated with coherent states can be largely
beaten by non-classical states, as long as the mean number of
photons hitting the memory cells is suitably low.

Depending on the regime, we may choose a different type of
non-classical states. In the presence of thermal decoherence
induced by background photon scattering, two-mode squeezed vacuum
states between signal modes (reading the cells) and idler modes
(kept for detection) are nearly-optimal. However, in the absence
of decoherence, the sequential readout of an ideal memory (where
one of the reflectivities is exactly $100\%$) is optimized by
number states at the input~\cite{Nair11}. \cite{Roga2015}
showed that, in specific regimes, the quantum advantage can be
related with a particular type of quantum correlations, the
discord of response, which is defined as the trace, or Hellinger,
or Bures minimum distance from the set of unitarily perturbed
states~\cite{Roga2014}. \cite{Roga2015} also identified
particular regimes in which strongly discordant states are able to
outperform pure entangled transmitters.

Let us consider the specific case of unitary channel discrimination. Suppose
that the task is to decide whether a unitary $U_{\theta }$ was applied or
not to a probing subsystem $A$ of a joint system $(A,B)$. In other words,
the aim is to discriminate between the two possible output states $\rho
_{\theta }=(U_{\theta }\otimes \mathbb{I})\rho (U_{\theta }\otimes \mathbb{I}%
)^{\dagger }$ (when the unitary $U_{\theta }$ has acted on $A$) or $\rho $
(equal to the input, when the identity has acted on $A$ instead). In the
limit of an asymptotically large number $N\gg 1$ of copies of $\rho $, the
minimal probability of error in distinguishing between $\rho _{\theta }$ and
$\rho $, using an optimal discrimination strategy scales approximately as
the QCB $Q(\rho ,\rho _{\theta })^{N}/2$.

It is clear that the quantity $1-Q(\rho ,\rho _{\theta })$ plays a similar
role in the present discrimination context as the quantum Fisher information in the parameter
estimation scheme. One can therefore introduce an analogous figure of merit
quantifying the worst case ability to discriminate, guaranteed by the state $%
\rho $. The \textit{discriminating strength}  of the bipartite state $%
\rho $ with respect to the probing system $A$ is then defined as~\cite%
{Farace2014}
\begin{equation}
D^{(A)}(\rho )=\min_{H}[1-Q(\rho ,\rho _{\theta })]\,,
\label{eq:DS}
\end{equation}%
where the minimization is performed once more over all generators
$H$ within a given non-degenerate spectrum.

As proven in~\cite{Farace2014}, the discriminating strength is another measure of discord-type
correlations in the input state $\rho $, which vanishes if and only if $\rho
$ is a classical-quantum state as in eq.~(\ref{eq:chiA}). The discriminating strength is also
computable in closed form for all finite-dimensional states such that
subsystem $A$ is a qubit. In the latter case, the discriminating strength turns out to be
proportional to the local quantum uncertainty~\cite{Girolami2013}, a further
measure of discord-type correlations defined as in eq.~(\ref{eq:IP}), but
with the quantum Fisher information replaced by the Wigner-Yanase skew information~\cite%
{Girolami2013}. The discriminating strength has also been extended to continuous-variable
systems, and evaluated for special families of two-mode Gaussian states
restricting the minimization in eq.~(\ref{eq:DS}) to Gaussianity-preserving
generators (i.e., quadratic Hamiltonians)~\cite{Rigovacca2015}.

Finally, we notice that the
presence and use of quantum correlations beyond entanglement has also been investigated in
other tasks related to metrology and illumination, such as ghost imaging with (unentangled) thermal source beams
\cite{Ragy2012}. Adopting a coarse-grained two-mode description of the
beams, quantum discord was found to be relevant for the implementation
of ghost imaging in the regime of low illumination, while more
generally total correlations in the thermal source beams were shown to
determine the quality of the imaging, as quantified by the
signal-to-noise ratio.

\subsection{\red{Average precision in black-box settings}\label{sub:AVQ}}

\red{The results reviewed so far in this Section highlight a clear resource role for quantum discord,
specifically measured by operational quantifiers such as the interferometric power and the discriminating strength,
in black-box metrology settings, elucidating in particular how quantum
correlations beyond entanglement manifest themselves as coherence in all local bases
for the probing subsystem. Discordant states, i.e., all states but those of
eq.~(\ref{eq:chiA}), are not only disturbed by all possible local
measurements on $A$, but are also modified by --- hence sensitive to --- all
nontrivial unitary evolutions on subsystem $A$. This is exactly the
ingredient needed for the estimation and discrimination tasks described
above.}

In practice, however, one might want to assess the general purpose
performance of probe states, rather than their worst case scenario only. One
can then introduce alternative figures of merit quantifying how suitable a
state is, on average, for estimation or discrimination of unitary
transformations, when the average is performed over all generators of a
fixed spectral class. This can be done by replacing the minimum with an
average according to the Haar measure, in Eqs.~(\ref{eq:IP}) and (\ref{eq:DS}),
respectively. Such a study has been carried out in \cite{Farace2016} by
defining the local average Wigner-Yanase skew information, which corresponds
to the average version of the discriminating strength in case the probing subsystem $A$ is a
qubit \cite{Farace2014}.

Unlike the minimum, the average skew information is found not to
be a measure of discord anymore. In particular, it vanishes only
on states of the form $\frac{\mathbb{I}^{A}}{d_{A}}\otimes \tau
^{B}$, that is, tensor product states between a maximally mixed
state on $A$, and an arbitrary state on $B$ \cite{Farace2016}.
\red{This entails that, to ensure a reliable discrimination of local
unitaries on average, the input states need to have either one of
these two (typically competing) ingredients: nonzero local purity
of the probing subsystem, or nonzero correlations (of any nature)
with the ancilla. The interplay between the average performance
and the minimum one, which instead relies on discord, as well as a
study of the role of entanglement, are detailed in
\cite{Farace2016}. A similar study has been recently performed in continuous variable systems, in which the average quantum Fisher information for estimating
the amount of squeezing applied to an input single-mode probe, without previous knowledge on the phase of the applied squeezing, was investigated with and without the use of a correlated ancilla \cite{Rigovacca2017}.}

\section{Identical particles} \label{sec.IdPar}

%
%
%
%
%
%
%
%
%
%
%
%

Measuring devices and sensors operating with many-body systems are
among the most promising instances {\redd for which} quantum-enhanced
measurements can be actually experimented; indeed, their large
numbers of elementary constituents play the role of resources according to which the accuracy of parameter estimation can be scaled.
Typical instances in which the quantum-enhanced measurement paradigm has been studied are in fact interferometers
based on ultracold atoms confined in optical lattices
\cite{Leggett1,Stringari0,Pethick0,Knight0,Haroche0,Leggett2,Kohl0,Inguscio0,Giorgini0,Cronin0,Yukalov0}
where a precise control on the state preparation and on the dynamics can nowadays be obtained.
These systems are made of spatially confined bosons or fermions,
{\it i.e.}~of constituents behaving as identical particles,
a fact that has not been properly taken into account
in most of the literature.

In systems of distinguishable particles, the notion of separability and entanglement is well-established~\cite{Horodecki09}: it is strictly associated with the natural tensor
product structure of the multi-particle Hilbert space and expresses the fact that one is able to identify each one of the constituent subsystems with their corresponding single-particle Hilbert spaces.
On the contrary, in order to describe identical particles one must extract from the tensor product structure of the whole Hilbert space either the symmetric (bosonic) or the anti-symmetric (fermionic) sector \cite{Feynman0,Sakurai0}.
This fact demands a more general approach to the notions of non-locality
and entanglement based not on the particle aspect proper \colb{for} first quantization, rather on the mode description typical of second quantization~\cite{Zanardi0,Narnhofer0,Viola1,Viola2,Benatti1,Benatti2,Argentieri0,Benatti3,Benatti4,Marzolino0,Benatti5,Benatti6,Benatti7,Benatti8,Werner1,Werner2,Werner3}.

The notion of entanglement in many-body systems has already
been addressed and discussed in the literature: for instance,
see~\cite{Schliemann0,Paskauskas0,Li0,Ghirardi0,Eckert0,Micheli0,Milburn0,Wiseman0,Verstraete0,Dowling0,Cirac0,Kraus0,Grabowski0,Calabrese0,Song0,Buchleitner0,Balachandran0,Shi0,Lewenstein0,Bloch0,Amico0,Modi2012}
and references therein. Nevertheless, only limited results actually apply to the case of
identical particles and their applications to quantum-enhanced measurements. From the existing
literature on possible metrological uses of identical particles, there
emerges as a controversial issue the distinction between particle and
mode entanglement.  Before illustrating the general approach developed
in~\cite{Benatti1,Benatti3,Benatti6} within which this matter can be
settled, we shortly overview the main aspects of the problem.
Readers who feel that the discussion whether the states in
question are to be considered as entangled or not is rather academic
may be reassured by the very pragmatical result that independently of
this discussion, systems of
indistinguishable bosons offer a metrological advantage over
distinguishable particles, in the sense that for certain measurements
one would have to massively entangle the latter for obtaining the same
sensitivity as one obtains ``for free'' from the symmetrized states of
the former.  This we show explicitly in Sec.\ref{sec.bosons}.  \\

Entanglement based on the particle description proper for first
quantization has been discussed for pure states in
\cite{Schliemann0,Paskauskas0,Li0}. In the fermionic case, Slater
determinants are identified as the only non-entangled fermionic pure
states, for, due to the Pauli exclusion principle, they are the least
correlated many-particle states. In the case of bosons, two
inequivalent notions of particle entanglement have been put forward:
in \cite{Paskauskas0}, a pure bosonic state is declared non-entangled
if and only if all particles are prepared in the same single particle
state, thus leading to a bosonic product state.
In \cite{Li0}, bosonic non-entangled pure states are identified as those corresponding to
permanents (the bosonic analogue of Slater determinants); in other words, a pure bosonic state is non-entangled not only when, as in the first approach mentioned above, it is a product state, but also when all bosons are prepared in pairwise orthogonal single particle states.
The particle-based entanglement was then studied in \cite{Eckert0} both for bosons and fermions, and then generalized within a more mathematical and abstract setting in
\cite{Grabowski0,Grabowski1}.

From the point of view of particle description, a different
perspective was provided in \cite{Ghirardi0,Ghirardi1}, based  on the
fact that non-entangled pure states should possess a complete set of
local properties, identifiable by local measurements. For a more
recent non standard approach, still based on the particle description,
see~\cite{Compagno}.

In all these approaches, only bipartite systems consisting of two
identical particles are discussed; however, in the fermionic case,
simple necessary and sufficient criteria of many
particle-entanglement, based on the single-particle reduced density
matrix, have been elaborated in~\cite{Plastino0}.

A change of perspective occurred in
\cite{Zanardi0,Vedral0,Viola1,Viola2} where the focus moved from
particles to orthonormal modes; within this approach, states that are
not mode-entangled are characterized by correlations that can be
explained in terms of joint classical occupation probabilities of the
modes. It then follows that entanglement and non-locality depend on
the mode description which has been chosen. In all cases, however,
identical-particle states
represented by fermionic Slater determinants or bosonic permanents can
be neither particle nor mode-entangled, in direct conflict with claims
that these states are particle-entangled and hence metrologically
useful \cite{Demkowicz0}.
In an attempt to resolve the conflict, in \cite{Plenio0} it is shown
that such a pseudo, or ``fluffy-bunny'' entanglement as called in
\cite{wiseman_ferreting_2003,Beenakker06}, which is due to bosonic
state symmetrization, can
be turned into the entanglement of distinguishable, and thus
metrologically accessible, modes; however, the operations needed for
such a transformation are non-local in the mode picture
and thus ultimately \colb{responsible for the achieved sub shot-noise
  accuracies.}

The variety of approaches regarding entanglement in identical particle
systems and their use in metrological applications, in particular
whether entanglement is necessary or not to achieve sub-shot-noise
accuracies, can be looked at from the unifying point of view provided
by the algebraic approach to quantum many-body systems proper to
second quantization~\cite{Bratteli0,Strocchi0}.
Within this scheme, we first address the relations between mode and
particle entanglement. Then, we focus on
specific quantum metrological issues and show that neither entangled
states, nor preliminary state preparation as spin-squeezing are
necessary in order to achieve sub shot-noise accuracies using systems
of identical particles.

\subsection{Particle and mode entanglement}

In quantum mechanics, indistinguishable particles cannot be
identified by specific labels, whence their states must be either completely symmetrized (bosons) or anti-symmetrized (fermions).
This makes second quantization a most suited approach to deal with
them, while the particle representation proper for first quantization
results to be
too restrictive for a consistent treatment of entangled  identical particles.

As a simple example of the consequences of indistinguishability, consider two qubits, each of them described by the Hilbert space $\mathcal{H}=\mathbb{C}^2$, where we select the orthonormal basis $\{\vert 0\rangle,\vert 1\rangle\}$ of eigenvectors of $\sigma_z$: $\sigma_z\vert0\rangle=\vert 0\rangle$, $\sigma_z\vert1\rangle=-\vert 1\rangle$.
If the two qubits describe distinguishable particles, the tensor product structure of the common Hilbert space $\mathbb{C}^4=\mathbb{C}^2\otimes\mathbb{C}^2$ exhibits the fact that one knows which is qubit $1$ and which one qubit $2$. Instead, if the two qubits are indistinguishable, the corresponding Hilbert space loses its tensor product structure and becomes, in the bosonic case, the three-dimensional subspace $\mathbb{C}^3$ spanned by the
orthogonal \colb{two-particle symmetric vectors
$\vert 00\rangle$, $\vert11\rangle$ and $\vert\psi^{(2)}_+\rangle:=(\vert 01\rangle\,+\,\vert 10\rangle)/\sqrt{2}$}, whereas,
in the Fermionic case, it reduces to the anti-symmetric \colb{two-particle vector $\vert\psi^{(2)}_-\rangle:=(\vert 01\rangle\,-\,\vert 10\rangle)/\sqrt{2}$}.

An important 
consequence of this fact is that
generic mixed states of  indistinguishable particles must be
represented by density matrices that arise from convex linear
combinations of projectors onto symmetrized or anti-symmetrized pure
states. Indeed, while operators must be symmetrized in order to comply
with particle indistinguishability, by symmetrizing density matrices
of distinguishable particles, say sending
$\rho=\rho_1\otimes\rho_2$ into
$\big(\rho_1\otimes\rho_2+\rho_2\otimes\rho_1\big)/2$
one cannot in general obtain an appropriate bosonic or fermionic
state, unlike
sometimes stated in the literature (see e.g.~\cite{Demkowicz0}).
For instance, the two qubit symmetric mixed state $\rho\otimes \rho$ cannot be a
fermionic density matrix since the only fermionic state is
\colb{$\vert\psi^{(2)}_-\rangle\langle\psi^{(2)}_-\vert$}.
On the other hand, in the bosonic case, $\rho\otimes\rho$ can be a
bona-fide bosonic state only if $\rho$ is pure. Indeed, as
\colb{$\rho\otimes\rho\vert\psi^{(2)}_-\rangle={\rm det}(\rho)\vert\psi^{(2)}_-\rangle$, if
the determinant of the density matrix}, ${\rm det}(\rho)\neq 0$, then $\rho\otimes\rho$ has support also in
the anti-symmetric component of the  Hilbert space. A careful discussion
of the relationship between permutationally invariant density matrices
and symmetric/anti-symmetric states can be found in
\cite{PhysRevA.94.033838}.

Instead of on the standard particle picture, second quantization is based on the so-called mode picture.
In general, a mode is any of the normalized vectors $\vert\psi\rangle$ of the same single particle Hilbert space $\mathcal{H}$ that is used to describe each one of a system of identical particles.
In practice, one fixes an orthonormal basis $\{\vert\psi_i\rangle\}$ in $\mathcal{H}$ and populates the $i$-th mode by acting on the so-called vacuum vector with (powers of) the creation operator $a^\dag_i$, $a^\dag_i\vert 0\rangle=\vert\psi_i\rangle$, while the adjoint operators $a_i$ annihilate the vacuum, $a_i\vert 0\rangle=0$.
For bosons, $[a_i\,,\,a^\dag_j]=a_ia^\dag_j-a^\dag_ja_i=\delta_{ij}$ and one can find arbitrarily many particles in a given mode, while for fermions $\{a_i\,,\,a^\dag_j\}=a_i\,a^\dag_j\,+\,a^\dag_j\,a_i=\delta_{ij}$ and each mode can be occupied by one fermion, at most.

Typical modes are given by the eigenvectors of a given single particle Hamiltonian, orthogonal polarization directions, the left and right position in a double-well potential, or the atomic positions in an optical lattice.
In the case of free photons, typical modes are plane waves labeled by
wave vector and polarization, arising from the quantization of
classical electrodynamics in terms of independent harmonic
oscillators. Within this picture, saying that there are $n$ photons in
a certain energy mode can also be interpreted as a
quantum oscillator being promoted to its $n$-th excited state.

In dealing with distinguishable particles, quantum entanglement \cite{Horodecki09} is basically approached by referring to the tensor product structure of the total Hilbert space which embodies the fact that particles can be identified. From the previous discussion, it follows that, in the case of identical particles, one ought to consider the entanglement of modes rather than the entanglement of particles.
The relations between the two approaches are studied in detail in
\cite{Benatti5}; here, we briefly compare them in the case of two
two-mode indistinguishable bosons associated with orthogonal
single-particle pure states $\vert\psi_{1,2}\rangle\in\mathcal{H}$,
as, for instance, orthogonal polarization states, described by
creation and annihilation operators $a_i,a_i^\dag$, $i=,2$. The pure
state
\begin{equation}
\label{eq0}
\vert 1,1\rangle=a_1^\dag a_2^\dag\vert 0\rangle
\end{equation}
belongs the two-particle sector of the symmetric Fock
space and represents a pure state with one boson in each mode.
In the particle picture of first quantization, the same state corresponds to the symmetrized vector
\begin{equation}
\label{eq1}
\vert\psi_+\rangle=\frac{1}{\sqrt{2}}\Big(\vert\psi_1\rangle\otimes\vert\psi_2\rangle+\vert\psi_2\rangle\otimes\vert\psi_1\rangle\Big)\ ,
\end{equation}
which describes a balanced superposition of two states: one with the first particle in the state $\vert\psi_1\rangle$ and the second particle in the state $\vert\psi_2\rangle$, the other with the two states exchanged. The net result is that, given the state \eqref{eq1}, we can only say that one particle is surely in the state $\vert\psi_1\rangle$  and that the other one is surely in the state $\vert\psi_2\rangle$, but we cannot attribute a specific state to a specific particle.

As an example of the misunderstandings that can result from sticking
to the particle picture when dealing with identical particles,
state \eqref{eq1} is often referred to as entangled: actually, it is only
formally so in the particle picture, while, as we shall show below, it is
separable
in the mode picture.
Indeed,
\eqref{eq1} would clearly display particle entanglement  as the state
cannot be written as a tensor product of single particle states.
Instead, in \eqref{eq0}, the same state is expressed as the action on
the vacuum state of two independent creation operators and thus should
correspond to a mode-separable state under whichever meaningful
definition of mode-entanglement one should adopt.
We shall later give solid ground to this latter statement, but firstly we show that, while for distinguishable particles the state \eqref{eq1} is the prototype of an entangled pure state, it is nevertheless separable in the sense specified below.

\subsubsection{Particle entanglement}




For distinguishable particles,
bipartite observables are termed local if they are tensor products $O_{12}=O_1\otimes O_2$ of single particle observables $O_1\otimes\mathbb{I}$ pertaining to particle $1$ and $\mathbb{I}\otimes O_2$ pertaining to particle $2$, where $\mathbb{I}$ denotes the identity operator, namely if they address each particle independently. \colb{Consequently, in such a context, locality is associated to the addressability of single particles and referred to as particle-locality in the following.}
Then, a bipartite pure state $\vert\psi_{12}\rangle$ is separable if and only if the expectation values $\langle O_{12}\rangle_{12}=\langle\psi_{12}\vert O_1\otimes O_2\vert\psi_{12}\rangle$ of all local observables factorize:
\begin{equation}
\label{factor}
\langle O_{12}\rangle_{12}=\langle O_1\otimes \mathbb{I}\rangle_{12}\ \langle\mathbb{I}\otimes O_2\rangle_{12}\ .
\end{equation}
Indeed, if $\vert\psi_{12}\rangle=\vert\psi_1\rangle\otimes\vert\psi_2\rangle$ the above equality clearly holds. On the other hand, if the equality holds, by using the Schmidt decomposition of $\vert\psi_{12}\rangle=\sum_i\lambda_i\vert\phi^{(1)}_i\rangle\otimes\vert\phi_i^{(2)}\rangle$ and choosing $O_1=\vert\phi_i^{(1)}\rangle\langle\phi_i^{(1)}\vert$,
$O_2=\vert\phi_j^{(2)}\rangle\langle\phi_j^{(2)}\vert$ one finds that only one Schmidt coefficient can be different from zero. Instead, for distinguishable particles, the state $\vert\psi_+\rangle$ in \eqref{eq1} violates the above equality for $O_1=\vert\psi_1\rangle\langle\psi_1\vert$ and $O_2=\vert\psi_2\rangle\langle\psi_2\vert$ and is thus entangled.


If the two particles are identical, they cannot be addressed individually; one has thus to refer to observables that specify properties attributable to single particles without specifying to which one of them. \colb{It therefore follows that, in the case of two identical particles, appropriate single particle observables cannot be of the form $O\otimes \mathbb{I}$ or $\mathbb{I}\otimes O$, but of the symmetrized form
\begin{equation}
\label{singleobs}
O\otimes \mathbb{I}\,+\,\mathbb{I}\otimes O\ .
\end{equation}
}
Consider the single-particle property described by the one-dimensional projector $\vert\psi\rangle\langle\psi \vert$; for two indistinguishable particles the property "one particle is in the state $\vert\psi\rangle$", must then be represented by the symmetric projection~\cite{Ghirardi0,Ghirardi1}
\begin{eqnarray}
\nonumber
\mathcal{E}_{\psi} &=& \vert\psi\rangle\langle\psi\vert\otimes \Big(\mathbb{I}-\vert\psi\rangle\langle\psi\vert\Big)+ \Big(\mathbb{I}-\vert\psi\rangle\langle\psi\vert\Big)\otimes \vert\psi\rangle\langle\psi\vert\\
\label{new15}
&+& \vert\psi\rangle\langle\psi\vert \otimes \vert\psi\rangle\langle\psi\vert\ .
\end{eqnarray}
Similarly, the projector corresponding to the two qubits possessing two different properties $\vert\psi_{1,2}\rangle\langle\psi_{1,2} \vert$ must be
$P^{symm}_{12}=\vert\psi_{1}\rangle\langle\psi_{1} \vert\otimes
\vert\psi_{2}\rangle\langle\psi_{2} \vert   + \vert\psi_{2}\rangle\langle\psi_{2} \vert\otimes
\vert\psi_{1}\rangle\langle\psi_{1} \vert$.
If $\braket{\psi_1}{\psi_2}=0$, then
$P^{symm}_{12}=\mathcal{E}_{\psi_1}\mathcal{E}_{\psi_2}$.
It then follows that, despite the formal entanglement of $\vert\psi_+\rangle$, the factorization of mean-values as in~\eqref{factor} still holds; indeed,
\begin{equation}
\label{new1}
\langle\psi_+\vert\,\mathcal{E}_{\psi_1}
\vert\psi_+\rangle=\langle\psi_+\vert\,\mathcal{E}_{\psi_2}\,\vert\psi_+\rangle
=\langle\psi_+\vert\,P^{sym}_{12}\,\vert\psi_+\rangle=1\ .
\end{equation}
Therefore, from the particle point of view, namely of attributable properties, the formally entangled state $\vert\psi_+\rangle$ is indeed separable.

\subsubsection{Mode entanglement}


Basing on the attribution of properties to identical particles, the
previous discussion is developed in first quantization terms, namely
using symmetrized states and observables.
From the point of view of second quantization, separability and
entanglement are instead to be related to the algebraic structure of
Bose and Fermi systems rather than to the possibility of attributing
individual properties: this is the point of view
usually adopted in the analysis of many-body systems, {\redd for which} the primary object of investigation are the algebras of operators rather than their representations on particular Hilbert spaces
~\cite{Bratteli0,Emch0,Haag0,Strocchi0,Thirring0,Strocchi1,Strocchi2,Strocchi3}.

In order to appropriately formulate the notion of entanglement \colb{and non-locality} in systems made of identical particles, the leading intuition is that there is no a priori given tensor product structure, reminiscent of particle identification, either in the Hilbert space or in the algebra of observables. \colb{Therefore, questions about entanglement and separability are meaningful only with reference to specific classes of observables. Within this broader context, entanglement becomes a caption for non-local quantum correlations between observables exhibited by certain quantum states} \reddd{(the original discussion can be found in~\cite{Werner1,Werner2,Werner3}; further developments can be found 
in~\cite{Werner4,Werner5,Werner6,Halvorson0,Clifton0,Moriya1}}.

Polynomials in creation and annihilation operators can be used to generate bosonic and fermionic algebras $\cal A$ containing all physically relevant many-body observables.
Quite in general, physical states are given
by positive and normalized linear functionals $\omega:\mathcal{A}\to \mathbb{C}$
associating to any operator $A\in{\cal A}$ its mean value $\omega(A)$, such that
positive observables $A\geq 0$ are mapped to positive numbers $\omega(A)\geq 0$ and
$\omega(\mathbb{I})=1$. Typical instances of states are the standard expectations obtained by tracing with respect to a given density matrix $\rho$, $\omega(A)={\rm Tr}(\rho\,A)$; notice, however, that in presence of infinitely many degrees of freedom, not all physically meaningful states, like, for instance, the thermal ones, can be represented by density matrices \cite{Narnhofer0}.

Within such an algebraic approach, one can study the entanglement between observables of ${\cal A}$ with respect to a state $\omega$ by considering \textit{algebraic bipartitions}~\cite{Benatti1,Benatti6}, namely any couple of subalgebras $(\mathcal{A}_1,\mathcal{A}_2)$ of $\cal A$, having
only the identity in common, $\mathcal{A}_1\cap\mathcal{A}_2=\mathbb{I}$,
\reddd{such that the linear span of products of their operators generate the entire algebra $\mathcal{A}$} and
\begin{enumerate}
\item
$[A_1\,,\,A_2]=0$ for all $A_1\in\mathcal{A}_1$ and $A_2\in\mathcal{A}_2$ in the bosonic case;
\item
$[A^e_1\,,\,A_2]=0$ for all even elements $A^e_1\in\mathcal{A}^e_1$ and for all $A_2\in\mathcal{A}_2$, and, similarly, $[A_1\,,\,A^e_2]=0$ for all even elements $A^e_2\in\mathcal{A}^e_2$ and for all $A_1\in\mathcal{A}_1$, in the fermionic case,
\end{enumerate}
where the even elements $A_{1,2}^e\in\mathcal{A}_{1,2}$, are those remaining invariant under the transformation of creation and annihilation operators $\vartheta(a_i^\dag)=-a_i^\dag$ and similarly for $a_i$. They generate the so-called even subalgebras $\mathcal{A}_{1,2}^e\subset\mathcal{A}$ which contain the only fermionic operators accessible to experiments.

These 
desiderata embody the
notion of algebraic independence and generalize the tensor product structure typical of the particle picture.
\colb{Within this more general context, an operator $\cal O$ is called
$(\mathcal{A}_1,\mathcal{A}_2)$-local
if of the form ${\cal O}=A_1 A_2$, $A_1\in\mathcal{A}_1$ and
$A_2\in\mathcal{A}_2$.} Furthermore, a state $\omega$ is called $(\mathcal{A}_1,\mathcal{A}_2)$-separable if the expectation values of all $(\mathcal{A}_1,\mathcal{A}_2)$-local operators can be decomposed into convex combinations of expectations:
\begin{equation}
\label{def-sep-vec}
\omega(A_1 A_2)=
\sum_i\lambda_k\, \omega^{(1)}_k(A_1)\,\omega^{(2)}_k(A_2)\ ,
\end{equation}
in terms of other states $\omega^{(1)}_i$, $\omega^{(2)}_i$ with $\lambda_k>0$, $\sum_i\lambda_k=1$;
otherwise, $\omega$ is called $(\mathcal{A}_1,\mathcal{A}_2)$-entangled~\footnote{This generalized definition of separability
can be easily extended to the case
of more than two partitions;
for instance, in the case of an $n$-partition, Eq.(\ref{def-sep-vec})
would extend to
$\omega(A_1 A_2\cdots A_n)=\sum_k\lambda_k\, \omega_k^{(1)}(A_1)\,
\omega_k^{(2)}(A_2)\cdots\omega_k^{(n)}(A_n)$.
}.

In the standard case of bipartite entanglement
for pairs of distinguishable particles \cite{Horodecki09}, entangled states are all  density matrices $\rho$ which cannot be written in the form
\begin{equation}
\label{new2}
\rho=\sum_k\lambda_k\,\rho^{(1)}_k\otimes\rho^{(2)}_k\ ,\qquad \lambda_k\geq 0\ ,\quad \sum_k\lambda_k=1\ ,
\end{equation}
with $\rho^{(1)}_k$ and $\rho_k^{(2)}$ density matrices of the two parties.
That the algebraic definition reduces to the standard one becomes apparent in the case of two qubits by choosing the algebraic bipartition $\mathcal{A}_1=\mathcal{A}_2=\mathcal{M}_2$,
where $\mathcal{M}_2$
is the algebra of $2\times 2$ complex matrices and the expectation value
$\omega(A_1 A_2)={\rm Tr}\left(\rho\,A_1\otimes A_2\right)$.

Henceforth, we shall focus on many-body systems, whose elementary constituents
can be found in $\frak{M}$ different states or modes described by a discrete set of annihilation and creation operators $\{a_i,a^\dag_i\}_{i\in I}$; this is a very general framework, useful for the
description of physical systems in quantum optics, in atomic and condensed matter physics.
A bipartition of the $\frak{M}$-mode algebra ${\cal A}$ associated with the system
can then be easily obtained by considering two disjoint sets $\{a_i,\, a_i^\dagger \vert\, i=1,2,\ldots,m\}$
and $\{a_j,\, a_j^\dagger \vert\, j=m+1,m+2,\ldots,\frak{M}\}$, $\frak{M}$ being possibly infinite.
The two sets form subalgebras
${\cal A}_{1,2}$ that indeed constitute an algebraic bipartition of ${\cal A}$; in practice, it is determined by the integer $0<m< \frak{M}$.

As for distinguishable particles, the case of pure states states is
easier. Pure states in the algebraic context
~\footnote{In the algebraic descripiton, Hilbert spaces are a
  byproduct of the algebraic structure and of the expectation
  functional (state) defined on it \cite{Bratteli0}.} correspond to
those expectations on $\mathcal{A}$ that cannot be written as convex
combinations of other expectations.
For them the separability condition (\ref{def-sep-vec}) simplifies: one can indeed prove \cite{BenattiFloreanini2016} that
pure states $\omega$ on $\cal A$ are separable with respect to a given
bipartition $({\cal A}_1,{\cal A}_2)$ if and only if
\begin{equation}
\omega(A_1 A_2)=\omega(A_1)\, \omega(A_2)\ ,
\label{sep-pure}
\end{equation}
for all local operators $A_1 A_2$. In other terms, separable pure
states are just product states satisfying the factorization property~\eqref{factor} that holds for pure states of bipartite systems of distinguishable particles.

\subsection{Mode entanglement and metrology: bosons}\label{sec.bosons}

The differences between mode-entanglement and standard entanglement
are best appreciated in the case of $N$ bosons that can occupy $\frak{M}$
different modes; this is a very general situation encountered, for
instance, in ultracold gases consisting of bosonic atoms confined in a
multiple site optical lattices. These systems
turn out to be a unique laboratory for studying quantum effects in
many-body physics,
{\it e.g.} in quantum phase transition
and matter interference phenomena, and also for applications in quantum information
({\it e.g.} see
~\cite{Stringari0,Pethick0,Knight0,Haroche0,Leggett2,Kohl0,Inguscio0,Giorgini0,Cronin0,Yukalov0}, and references therein).

The algebra $\cal A$ is in this case generated by creation and annihilation operators $a^\dagger_i\,,\,a_i$, $i=1, 2,\ldots,\frak{M}$, obeying the commutation relations, $[a_i,\,a^\dagger_j]=\delta_{ij}$. The reference state $\omega$ is given by
the expectations with respect to the vacuum state $\vert0\rangle$, $a_i\vert0\rangle=0$ for all
$1\leq i\leq \frak{M}$, so that the natural states whose entanglement properties need be studied are vectors in the Fock Hilbert space $\mathcal{H}$ spanned by the many-body Fock states
\begin{eqnarray}
\nonumber
&&\hskip-.5cm
\vert n_1, n_2,\ldots,n_{\frak{M}}\rangle= \\
&&\hskip-.5cm
={1\over \sqrt{n_1!\, n_2!\cdots n_{\frak{M}}!}}
(a_1^\dagger)^{n_1}\, (a_2^\dagger)^{n_2}\, \cdots\, (a_M^\dagger)^{n_{\frak{M}}}\,\vert 0\rangle\ ,
\label{Fock-basis-bos}
\end{eqnarray}
or density matrices acting on it.

Given a bipartition $({\cal A}_1,{\cal A}_2)$ defined by two disjoint groups of creation and annihilation operators, any element $A_1\in {\cal A}_1$,
commutes with any element $A_2\in {\cal A}_2$, {\it i.e.} $[{\cal A}_1,\ {\cal A}_2]=\,0$.

In this framework, a necessary and sufficient condition for pure states $\vert\psi\rangle$ to be separable with respect to a given bipartition
$({\cal A}_1,\, {\cal A}_2)$ is that they be generated by acting on the vacuum state with $({\cal A}_1,\, {\cal A}_2)$-local operators~\cite{Benatti3},
\begin{equation}
\vert\psi\rangle={\cal P}(a^\dagger_1, \ldots, a^\dagger_m)\cdot
{\cal Q}(a^\dagger_{m+1},\ldots ,a^\dagger_{\frak{M}})\ \vert0\rangle\ ,
\label{sep-pure-state}
\end{equation}
where ${\cal P}$, ${\cal Q}$ are polynomials in the creation operators
relative to the first $m$ modes, the last $\frak{M}-m$ modes, respectively.
Pure states that can not be cast in the above form are thus $({\cal A}_1,\, {\cal A}_2)$-entangled.

When the state of the bosonic many-body system is not pure, it can be described by a density
matrix \colb{$\rho$, in general not diagonal with respect to the Fock basis (\ref{Fock-basis-bos}); since}
%
%
density matrices form a convex set \colb{whose
extremal points are projectors onto pure states, one deduces that generic mixed states
$\rho$ 
can be} $({\cal A}_1,\, {\cal A}_2)$-separable if and only if they are
convex combinations of $({\cal A}_1,\, {\cal A}_2)$-separable one-dimensional \colb{projections.}

An interesting application of these general considerations is given by a system
of $N$ bosons that can be found in just two modes, $\frak{M}=2$.
In the Bose-Hubbard approximation, $N$ ultracold bosonic atoms
confined in a double-well potential can be effectively described in
this way.
  The two creation operators $a_1^\dagger$ and $a_2^\dagger$
generate out of the vacuum bosons in the two wells, so that the Fock
basis (\ref{Fock-basis-bos})
can be conveniently relabeled in terms of the integer $k$ counting the number of bosons in the first well:
\begin{equation}
\label{NFock}
\vert k,N-k\rangle=\frac{(a_1^\dagger)^k (a_2^\dagger)^{N-k}}{\sqrt{k!(N-k)!}}\,
\vert0\rangle\ , \quad 0\leq k\leq N\ .
\end{equation}
Furthermore, $a_1,a_1^\dag$, respectively $a_2,a_2^\dag$,
generate two commuting subalgebras $\mathcal{A}_1$ and $\mathcal{A}_2$ that, together,
in turn generate the whole algebra $\mathcal{A}$; it is the
simplest bipartition of the system one can obtain by means of the operators $a_1,a_1^\dag$ and $a_2,a_2^\dag$.

Then,
the states $\vert k,N-k\rangle$ are separable: this agrees with the fact that they are created by the local operators $(a_1^\dagger)^k (a_2^\dagger)^{N-k}$.
Indeed, for any polynomial operator ${\cal P}_1\in {\cal A}_1$
and ${\cal P}_2\in {\cal A}_2$, the expectation value of the product ${\cal P}_1 {\cal P}_2$
is such that (compare with (\ref{sep-pure}))
\begin{eqnarray}
\nonumber
&&\langle {k,N-k}\vert{\cal P}_1 {\cal P}_2\vert{k,N-k}\rangle\\
\nonumber
&&={1\over k! (N-k)!}
\langle 0\vert a_1^k\, {\cal P}_1\, (a_1^\dag)^k \vert0\rangle\
\langle 0\vert a_2^{N-k}\, {\cal P}_2\, (a_2^\dag)^{N-k} \vert0\rangle\\
&&=\langle k\vert {\cal P}_1 \vert k\rangle\ \langle N-k\vert {\cal P}_2 \vert N-k\rangle\ ,
\label{new8}
\end{eqnarray}
where $\vert k\rangle$ and $\vert N-k\rangle$ are single-mode Fock states.
Consequently, mixed spearable states must be diagonal with respect to the Fock basis \eqref{NFock},
{\it i.e.} density matrices of the form \cite{Benatti1}:
\begin{equation}
\rho=\sum_{k=0}^N p_k\, \vert k,N-k\rangle\langle k, N-k \vert\ ,\quad
p_k\geq 0\ ,\quad \sum_{k=0}^N p_k=1\ .
\label{9}
\end{equation}

Unlike ${\cal P}_1{\cal P}_2$, most observables of physical interest are non-local with respect to the bipartition $({\cal A}_1, {\cal A}_2)$, \colb{{\it i.e.} they are not of the form ${\cal O}=A_1 A_2$, with $A_1\in\mathcal{A}_1$ and
$A_2\in\mathcal{A}_2$.} Prominent among them are those used in phase estimation protocols based on ultra-cold atoms trapped in double-well interferometers. \colb{Consider the generators of the rotations satisfying the $su(2)$ algebraic relations
$[J_i,J_j]=i\varepsilon_{ijk}\,J_k$, $i,j,k=x,y,z$. Among their possible representations in terms of two-mode creation and annihilation operators, let us focus upon the
following one}
\begin{eqnarray}
\label{15}
&&J_x={1\over2}\big(a_1^\dag a_2 + a_1 a_2^\dag\big)\ ,\\
\label{new9}
&&J_y={1\over2i}\big(a_1^\dag a_2 - a_1 a_2^\dag\big)\ ,\\
&&J_z={1\over2}\big(a_1^\dag a_1 - a_2^\dag a_2\big)\ .
\label{new10}
\end{eqnarray}
Notice that,
although the operators in \eqref{15}, as well as the exponentials
$e^{i\theta J_x}$ and $e^{i\theta J_y}$, are
non-local with respect to the bipartition $({\cal A}_1, {\cal A}_2)$,
 $\theta\in[0, 2\pi]$, the exponential of $J_z$ turns out to be local:
\begin{equation}
e^{i\theta J_z}=e^{i\theta a_1^\dag a_1/2}\quad e^{-i\theta a_2^\dag a_2/2}\ ,
\label{17}
\end{equation}
with $e^{i\theta a_1^\dag a_1/2}\in{\cal A}_1$ and $e^{-i\theta a_2^\dag a_2/2}\in{\cal A}_2$.
\noindent
By a linear transformation, one can always pass to new annihilation operators
\begin{equation}
b_1={a_1+a_2\over\sqrt{2}}\ ,\qquad
b_2={a_1-a_2\over\sqrt{2}}\ ,
\label{18}
\end{equation}
and corresponding creation operators $b^\dag_{1,2}$, and rewrite
the three operators in \eqref{15} as
\begin{eqnarray}
\label{new11}
&&J_x={1\over2}\big(b_1^\dag b_1 - b_2^\dag b_2\big)\ ,\\
\label{new12}
&&J_y={1\over2i}\big(b_1 b_2^\dag - b_1^\dag b_2\big)\ ,\\
&&J_z={1\over2}\big(b_1 b_2^\dag + b_1^\dag b_2\big)\ .
\label{new13}
\end{eqnarray}
To $b_i,b^\dag_i$, $i=1,2$ one associates the bipartition
$({\cal B}_1, {\cal B}_2)$ of $\cal A$ consisting of the subalgebras generated by $b_1,b^\dag_1$, respectively $b_2,b_2^\dag$: with respect to it, the exponential of $J_x$ becomes:
\begin{equation}
e^{i\theta J_x}=e^{i\theta b_1^\dag b_1/2}\quad e^{-i\theta b_2^\dag b_2/2}\ ,
\label{20}
\end{equation}
with $e^{i\theta b_1^\dag b_1/2}\in{\cal B}_1$ and $e^{-i\theta b_2^\dag b_2/2}\in{\cal B}_2$.
\noindent
This explicitly shows that an operator, local with respect to a given bipartition,
can 
become non-local if a different algebraic bipartition is chosen.

These considerations are relevant for metrological applications  \reddd{(see for instance
~\cite{caves_quantum-mechanical_1981,Yurke0,Klauder0,HollandPRL1993,Kitagawa0,Wineland0,Sanders0,
Bollinger0,Bouyer0,Dowling1,Sorensen0,Dowling2,Dunningham0,Wang0,GiovannettiS2004,
Korbicz0,Higgins07,Meystre0,Dowling3,Dorner0,Smerzi0,BoixoPRA2009,Briegel0,
Kacprowicz0,GiovannettiNPhot2011})}.
In fact, ultracold atoms trapped in a double-well optical potential realize
a very accurate interferometric device: state preparation and beam splitting can be precisely achieved by tuning the interatomic interaction and by acting on the height
of the potential barrier. The combination of standard Mach-Zehnder
type interferometric operations,
{\it i.e.} state preparation, beam splitting, phase shift and subsequent beam recombination,
can be effectively described as a suitable rotation of the initial state $\rho_{\rm in}$
by a unitary transformation \cite{Klauder0,Sanders0}:
\begin{equation}
\rho_{\rm in}\mapsto \rho_\theta=U_\theta\, \rho_{\rm in}\, U_\theta^\dag\ ,\qquad
U_\theta=e^{i\theta\, J_n}\ .
\label{22}
\end{equation}
The phase change is induced precisely by the operators in (\ref{15}) through the combination:
\begin{equation}
J_n\equiv n_x\, J_x + n_y\, J_y + n_z\, J_z\ ,\quad
n_x^2 + n_y^2 + n_z^2=1\ .
\label{23}
\end{equation}
In practice, the state transformation $\rho_{\rm in}\mapsto \rho_\theta$
inside the interferometer can be effectively modeled as a pseudo-spin rotation along the
unit vector $\vec{n}=(n_x, n_y, n_z)$, whose choice
depends on the specific realization of the interferometric apparatus
and of the adopted measurement procedure.

As discussed in
Section \ref{sec.Intro},
in the case of distinguishable particles,
for any separable state $\rho_{\rm sep}$
the quantum Fisher information is bounded by
$I_\theta[\rho_{\rm
  in},\ J_n]\leq c N$ where $c$ is a constant independent of $N$
that can be taken as the maximum quantum Fisher information of a single system in all
components of the mixed separable state (see eq.\eqref{eq:additivity_qfi})
\cite{fujiwara_additivity_2002,GiovannettiPRL2006,Smerzi0}.
Then, entangled initial states of distinguishable particles
are needed in order to obtain
sub-shot-noise accuracies.
A corresponding statement was proven in \cite{BenattiPRA2013} for a
system of $N$ indistinguishable bosons: The quantum Fisher information with respect to a
parameter $\theta$ imprinted onto a
$(\mathcal{A},\mathcal{B})$-separable state via a
$(\mathcal{A},\mathcal{B})$-local unitary operator
of the form $U(\theta)=\exp(i\theta(A(a,a^\dagger)+B(b,b^\dagger))$, where
$A$ (respectively $B$) are hermitian functions of $a,a^\dagger$
(respectively $b,b^\dagger$), strictly vanishes.  Hence, in order to
be able to estimate $\theta$ at all (and even more so to beat the
standard quantum limit in terms of the total number of bosons), the separability of the
input state or the locality of the unitary operator that imprints
$\theta$ on the state need to be broken.  Fortunately, mode
non-locality is easily achieved {\it e.g.}~in quantum optics with a simple
beam-splitter, without any particle interactions (see below).\\

In the case of the double-well system introduced above,
the
operator algebras defined by $a_1^\dagger$, $a_1$ and  $a_2^\dagger$,
$a_2$,
are the natural ones.
With respect to them, a balanced Fock number
state as in \eqref{NFock} of the
form $\rho_{N/2}=\vert\frac{N}{2}, \frac{N}{2}\rangle\langle \frac{N}{2},\frac{N}{2}\vert$ is separable.
By choosing  the vector $\vec n$ in the plane orthogonal to the $z$ direction one computes  $\displaystyle
I_\theta\big[\rho_{N/2}, J_n\big]=
{N^2\over2}+N$, thus approaching the
Heisenberg limit.

As a concrete example \cite{BenattiPRA2013}, consider the action of a beam-splitter described by the unitary $U_{BS}(\alpha)=\exp(\alpha a^\dag_1a_2-\alpha^* a_1a^\dag_2)$ involving two modes.
If the complex transparency parameter $\alpha=i\theta/2$, with $\theta$ real, then $U_{BS}(\alpha)=\exp(i\theta J_x)$.
In the case of $N$ distinguishable qubits, a state as the balanced Fock number state
$\vert\frac{N}{2},\frac{N}{2}\rangle$ has half qubits in the state $\vert0\rangle$ and half in the state $\vert 1\rangle$ such that $\sigma_z\vert i\rangle=(-)^i\vert i\rangle$, $i=0,1$. Then,
$\displaystyle J_{i}=\sum_{j=1}^N\frac{\sigma^{(j)}_{i}}{2}$ with
$i\in\{x,y,z\}$
have zero mean values, while the purity of
\colb{$\rho_{N/2}$ yields $I_\theta[\rho_{N/2},\ J_x]=4\,\langle J^2_x\rangle=N$}, since
\begin{equation}
\label{new3}
J_x^2=\frac{N}{4}+\frac{1}{4}\sum_{j\neq k=1}^N\sigma^{(j)}_x\sigma^{(k)}_x\ .
\end{equation}
Instead, in the case of indistinguishable bosonic qubits,
the mean values of $J_{i}$ for all $i\in\{x,y,z\}$
in \eqref{15}
vanish, while using~\eqref{15}, $\displaystyle I_\theta[\rho_N,\ J_x]=4\,\langle J^2_x\rangle={N^2\over 2}+N$.

Unlike in the case of distinguishable particles, the quantum Fisher information
can thus attain a value greater than $N$ even with initial states like
$\rho_{N/2}$ that are separable with
respect to the given bipartition.
As mentioned before,
the rotation operated by the beam-splitter is not around the $z$ axis
and is thus non-local with respect to the chosen bipartition.
From the point of view of mode-entanglement, it is thus not the entanglement
of the states fed into the beam-splitter that helps overcoming the
shot-noise-limit
in the transparency parameter $\theta$ estimation accuracy; rather,
the non-local character of the rotations operated by the apparatus
on initially separable states
allows $\sigma(\theta_\text{est})$ to be smaller 
than $1/\sqrt{N}$,
with the possibility of eventually reaching the Heisenberg $1/N$ limit
\cite{Benatti2,BenattiPRA2013}.\\
Notice that, if one does not take into account the identity of
particles, the beam-splitter action in~\eqref{20} is \colb{particle-local according to the discussion at the beginning of section A.1; indeed,}
\begin{equation}
\label{locop}
e^{i\theta J_x}=\bigotimes_{j=1}^N e^{i\theta\sigma^{(j)}_x/2}\ .
\end{equation}
Thus, a prior massive entanglement of the input state
\begin{equation}
\label{new4}
\vert0\rangle\otimes\cdots\vert0\rangle\otimes\vert 1\rangle\otimes\cdots\vert1\rangle
\end{equation}
with $k$ spins up, $\sigma_z\vert 0\rangle=\vert0\rangle$, and $N-k$ spins down, $\sigma_z\vert 1\rangle=-\vert1\rangle$, is needed.

\colb{Instead, if particle identity is considered, then the operator in~\eqref{locop},
is not particle-local since it cannot be written as a product of symmetrized single particle operators (see~\eqref{singleobs} for the case $N=2$).
Furthermore, in the formalism of first quantization, a Fock number state as in~\eqref{NFock} is represented in the symmetrized
form~\cite{BenattiPRA2013}}
\begin{equation}
\label{k1st}
\frac{1}{\mathcal{N}}\sum_\pi\vert0_{\pi(1)}\rangle\otimes\cdots\vert0_{\pi(k)}\rangle\otimes\vert1_{\pi(k+1)}\rangle\otimes\cdots\vert1_{\pi(N)}\rangle\ ,
\end{equation}
where the sum is over all possible permutations $\pi$ of the $N$ indices and $\mathcal{N}=\sqrt{N!k!(N-k)!}$.
\colb{Despite its formally entangled structure, such a state is the symmetrization of a tensor product state with the first $k$ particles in the state $\vert 0\rangle$ and the second $N-k$ particles in the state $\vert1\rangle$. Generalizing the argument briefly sketched in section A.1 in the case of two identical particles, individual properties can then be attributed to each of its constituents. Therefore, the state in~\eqref{k1st} carries no particle-entanglement, particle non-locality being instead provided by
the particle non-local operator in~\eqref{locop}.}

Of course, returning to the second quantization formalism, by changing bipartition from
$({\cal A}_1, {\cal A}_2)$ to
$({\cal B}_1, {\cal B}_2)$ via equations~\eqref{18}, the action of the beam-splitter, as outlined
in~\eqref{20}, is local.
In this bipartition, the non-locality
necessary for enhancing the sensitivity
completely resides in the state. Therefore, the mode-description, leaves the freedom to
locate the resources necessary to accuracy-enhancing either in the
entanglement of the state or in the non-locality of the operations
performed on it. \\

In \cite{BenattiPRA2013}, also the paradigmatic case of phase
estimation in a Mach-Zehnder interferometer was considered with
similar results: in the mode-bipartition corresponding to the two
modes after the first beam-splitter, the phase shift operation in one
arm is a local operator.  Hence, at that stage the state must be
mode-entangled to allow estimating the phase shift with an accuracy
better than the shot-noise limit. However, under
certain conditions
the first beam-splitter can generate enough mode-entanglement from a
mode-separable state fed into the two input ports of the interferometer to
beat the standard quantum limit.
For more general settings, the question of
what scaling of the quantum Fisher information can be
achieved with the number of indistinguishable bosons is still open.
\colb{Non-locality is partially attributed to operations like beamsplitting instead of entirely to states, even in cases when "fluffy-bunny" entanglement is turned into useful entanglement as discussed in~\cite{Plenio0}.}
 \\

For massive bosons one might think
that the tensor product of Fock states is a very natural state: As the
boson-number is conserved at the energies considered, one cannot make
coherent superpositions of different numbers of atoms.  However, in
typical experiments, one needs to average over many runs, and the real
difficulty consists in controlling the atom number with single-atom
precision from run to run \colb{\cite{Demkowicz0}}.  One has therefore effectively a mixed
state with a distribution of different atom numbers. So far the best
experimentally demonstrated approximations to Fock states with massive
bosons are number-squeezed
states with a few dB of squeezing
\cite{Oberthaler1,Treutlein10,EsteveN2008}.  One may hope that novel
measurement techniques such as the quantum gas microscope
\cite{bakr_quantum_2009}
may enable precise knowledge of boson numbers in the future and
thus the preparation of Fock states with a large number of
atoms at least in a post-selected fashion.

Recently, the authors of \cite{Oz16} used concentration of measure
techniques to investigate the usefulness of randomly sampled probe
states for unitary quantum metrology. They show that random pure
states drawn from the Hilbert space of distinguishable particles
typically do not lead to super-classical scaling of
precision. However, random states from the symmetric subspace,
i.e.~bosonic states, typically achieve the Heisenberg limit, even for very mixed
isospectral states. 
 Moreover, the quantum-enhancement is typically robust
against the loss of particles, in contrast to e.g.~GHZ-states.  It
remains to be seen how entangled these states are in the
sense of mode-entanglement, but independently of the outcome of such
an assessment, these results are in line with the finding that the
naturally symmetrized pure states of
bosons are  a useful resource for quantum metrology. \colb{The fact that certain bosonic states
can lead to the Heisenberg limit} while mode entanglement does not play any significant role has also been recently emphasized in \cite{Friis2015,Safranek2016}.

There are also metrological advantages achievable with bosons
that are beyond the context of ``standard quantum limit'' versus ``Heisenberg limit'' scaling: In
\cite{PhysRevA.95.012305} it
was shown that by using a grid-state \cite{gottesman_encoding_2001} of
a single bosonic mode, one can
determine both amplitude and phase of a Fourier-component of a small
driving field that adds at most $\pi/2$ photons, or equivalently,
both quadrature components of the
displacement operator of the state. Slightly biased estimators were
found whose sum of mean-square deviations from the true values scales
as $1/\sqrt{n}$    with the average number of photons $n$ in the
probe-state.  A ``compass state'' was proposed in
\cite{zurek_sub-planck_2001} that achieves similar sensitivity
for small displacements up to order $1/\sqrt{n}$.   These results
should be contrasted with the lower bound for any
single-mode Gaussian state as a probe state that is of order one,
independent of
$n$, and regardless the amount of squeezing.  With two-mode Gaussian states
one can beat this constant lower bound, but then the two modes must be
necessarily entangled \cite{PhysRevA.87.012107}. It is possible that the result
in \cite{PhysRevA.95.012305} may still be improved
upon with other states, as in \cite{PhysRevA.95.012305} also a lower
bound for all single-mode probe states was found that
scales as $1/(2n+1)$. This is the same lower bound as for arbitrary
(in particular: entangled) two-mode Gaussian states
\cite{PhysRevA.87.012107}.

\subsection{Mode entanglement and metrology: fermions}

As in the case of bosonic systems, we shall consider generic fermionic many-body systems made of $N$
elementary constituents that can occupy $\frak{M}$ different states or modes, $N<\frak{M}$. The creation $a_i^\dagger$ and
annihilation $a_i$ operators for mode $i$ obey now the anticommutation relations
$\{a_i\,,\,a^\dag_j\}=\delta_{ij}$ and generate the fermion algebra $\cal A$,
{\it i.e.} the norm closure of all polynomials in these operators.
As already specified before, a bipartition $({\cal A}_1,\ {\cal A}_2)$ of this algebra
is the splitting of
the collection of creation and annihilation operators into two disjoint sets.
The Hilbert space $\cal H$ of the system is again generated out of the vacuum state $\vert0\rangle$
by the action of the creation operators; it is spanned by the many-body Fock states,
\begin{equation}
\vert n_1, n_2,\ldots,n_{\frak{M}}\rangle=
(a_1^\dagger)^{n_1}\, (a_2^\dagger)^{n_2}\, \cdots\, (a_{\frak{M}}^\dagger)^{n_{\frak{M}}}\,\vert0\rangle\ ,
\label{Fock-basis-boson}
\end{equation}
where the integers $n_1, n_2, \ldots, n_{\frak{M}}$ are the occupation numbers of the different modes, with
$\sum_i n_i=N$; they can now take only the two values 0 or 1.

As already clear from the definition of fermionic algebraic bipartitions,
because of the anti-commutation relations, in dealing with fermions,
one must distinguish between even and odd operators.
While the even component $\mathcal{A}^e$ of $\mathcal{A}$ consists of elements $A^e\in\mathcal{A}$ such that $\vartheta(A^e)=A^e$, the odd component $\mathcal{A}^o$ of $\mathcal{A}$ consists
of those elements $A^o\in\mathcal{A}$ such that $\vartheta(A^o)=-A^o$. Even elements of $\cal A$
commute with all other elements, while odd elements commute only with even ones.

The anticommuting character of the fermion algebra $\cal A$ puts stringent constraints
on the form of the fermion states that can be represented as product of other states,
like the ones appearing in the decomposition \eqref{def-sep-vec} that defines separable states.
Specifically, as a consequence of the result in \cite{Araki0}, any product
$\omega_k^{(1)}(A_1)\, \omega_k^{(2)}(A_2)$ vanishes
whenever $A_1$ and $A_2$ both belong to the odd components of their respective subalgebras.
Then, given a mode bipartition $({\cal A}_1,{\cal A}_2)$ of the fermionic algebra $\cal A$,
{\it i.e.} a decomposition of $\cal A$ in the subalgebra ${\cal A}_1$ generated by the first
$m$ modes and the subalgebra ${\cal A}_2$, generated by the
remaining $\frak{M}-m$ ones, it follows that the decomposition \eqref{def-sep-vec} is meaningful only for local operators
$A_1 A_2$ for which $[A_1,\, A_2]=\,0$, so that the definition of separability it encodes
is completely equivalent to the one adopted for bosonic systems.

As a further consequence of the result in \cite{Araki0}, one derives that
if a state $\omega$ is non vanishing on a local operator $A_1^o A_2^o$, with the two components
$A_1^o \in {\cal A}_1^o$, $A_2^o \in {\cal A}_2^o$ both belonging to the odd part
of the two subalgebras, then $\omega$ is entangled with respect to the
bipartition $(\mathcal{A}_{1},\mathcal{A}_{2})$.
Indeed, if $\omega(A_1^o A_2^o)\neq0$, then $\omega$
cannot be written as in \eqref{def-sep-vec}, and therefore cannot be separable.

Using these results, as for the bosonic case, one shows that~\cite{Benatti6}, given a bipartition of the fermionic algebra $\cal A$ determined by the integer $m$, a pure state $\vert\psi\rangle$ results
separable if and only if it can be written in the form (\ref{sep-pure-state}).
Examples of pure separable states of $N$ fermions are the Fock states
in (\ref{Fock-basis-boson});
indeed, they can be recast in the form
\begin{eqnarray}
&&\vert k_1, \ldots, k_m; p_{m+1}, \ldots, p_{\frak{M}}\rangle\\
&&=\left[(\hat{a}_1^\dag)^{k_1}\cdots(\hat{a}_m^\dag)^{k_m}\right] \times
\left[(\hat{a}_{m+1}^\dag)^{p_{m+1}}\cdots(\hat{a}_{\frak{M}}^\dag)^{p_{\frak{M}}}\right]\vert0\rangle\ ,
\nonumber
\end{eqnarray}
where the $\cal P$ and $\cal Q$ appearing in (\ref{sep-pure-state})
are now monomials in the creation operators of the two partitions.

Concerning the metrological use of fermionic systems, the situation may appear more problematic than with bosons, for each mode can accommodate at most one fermion;
therefore, the scaling with $N$ of the sensitivity in the estimation
of physical parameters may worsen.
Indeed, while a two-mode bosonic apparatus, as a double-well interferometer,
filled with $N$ particles is sufficient to reach sub shot-noise sensitivities,
with fermions, a multimode interferometer is
needed in order to reach comparable sensitivities (for the use of
multi-mode interferometers see, for
instance~\cite{Dariano1,Dariano2,Soderholm0,Vourdas0,Cooper1}, and
\cite{Cooper2} in the fermionic case).
As an example, consider a system of $N$ fermions in $\frak{M}$ modes, with $\frak{M}$ even,
and let us fix the balanced bipartition $(\frak{M}/2, \frak{M}/2)$, in which each of the two parts
contain $m=\frak{M}/2$ modes, taking for simplicity $N\leq m$. As generator
of the unitary transformation $\rho\to\rho_\theta$ inside the measuring apparatus
let us take the following operator:
\begin{equation}
{\cal J}=\frac{1}{2}\sum_{k=1}^m \omega_k\ \Big(a_k^\dagger a_{m+k} + a_{m+k}^\dagger a_k\Big)\ ,
\label{4.40}
\end{equation}
where $\omega_k$ is a given spectral function, {\it e.g.} $\omega_k
\simeq k^p$, with $p$ positive.
\colb{The apparatus implementing the above state transformation is clearly non-local
with respect to the chosen bipartition: $e^{i\theta {\cal J}}$ can not be written
as the product $A_1 A_2$ of two components made of operators $A_1$ and $A_2$ belonging only
to the first, second partition, respectively.} It represents a generalized,
multimode beam splitter, and the whole measuring device behaves as
a multimode interferometer.

Let us feed the interferometer with a pure initial state, $\rho=\vert\psi\rangle\langle \psi\vert$,
\begin{equation}
\vert\psi\rangle=\vert \underbrace{1,\dots,1}_N,\underbrace{0,\ldots,0}_{m-N}\, ;\ \underbrace{0,\ldots,0}_m\, \rangle
=a_1^\dagger a_2^\dagger\cdots a_N^\dagger \vert0\rangle\ ,
\label{4.41}
\end{equation}
where the fermions occupy the first $N$ modes of the first partition;
$\vert\psi\rangle$ is a Fock state
and therefore it is separable, as already discussed. The quantum
Fisher information
can be easily computed \cite{Benatti6}
\begin{equation}
I_\theta\big[\rho, {\cal J}\big]= \sum_{k=1}^N \omega_k^2\ .
\label{4.42}
\end{equation}
Unless $\omega_k$ is $k$-independent, $I_\theta\big[\rho, {\cal
  J}\big]$ is larger than $N$ and therefore
the interferometric apparatus can beat the shot-noise limit in
$\theta$-estimation, even starting
with a separable state.
Actually, for $\omega_k \simeq k^p$, one gets: $I_\theta\big[\rho,
{\cal J}\big]\simeq N^{2p+1}$,
reaching sub-Heisenberg sensitivities with a linear device.
Note that this result
and the ability to go beyond the Heisenberg limit is not a
``geometrical'' phenomenon attributable to
a phase accumulation even on empty modes \cite{Dariano1}; rather, it
is a genuine quantum effect,
that scales as a function of the number of fermions, the resource
available in the measure.

Again, as in the bosonic case, it is not the entanglement
of the initial state that helps overcoming the shot-noise-limit
in the phase estimation;
rather, it is the non-local character of the
rotations operated by the apparatus on an initially separable state
that allows one 
beating the shot-noise limit.

%
%
%
%
%

\section{More general Hamiltonians}

\subsection{Non-linear Hamiltonians}

\newcommand{\mtext}[1]{{\color{magenta}#1}}
\newcommand{\btext}[1]{{\color{blue}#1}}
\newcommand{\gtext}[1]{{\color{green}#1}}
\newcommand{\Var}{\text{Var}}
\label{sec.genH}
Most discussions of quantum-enhanced measurements consider, implicitly or explicitly, evolution
under a Hamiltonian that is linear in a collective variable of the
system.  For illustration, consider Ramsey spectroscopy on a
collection of $N$ atoms with ground and excited states $|g\rangle$,
$|e\rangle$, respectively.  If these have energies $\pm \hbar
\omega/2$, then the Hamiltonian giving rise to the Ramsey oscillation
is
\begin{equation}
H =  \hbar \omega S_z \equiv \hbar \omega \sum_{i=1}^N s_z^{(i)}
\end{equation}
where $s_z \equiv \frac{1}{2} ( |e\rangle \langle e | - |g\rangle
\langle g | )$ is a pseudo spin-1/2 operator describing the transition
and the superscript $^{(i)}$ indicates the $i$-th atom.  This
Hamiltonian is manifestly linear in $S_z$, and can be trivially
decomposed into micro-Hamiltonians $H = \sum_i h^{(i)} \equiv \sum_i
\hbar \omega s_z^{(i)}$ that describe the uncoupled precession of each
atom in the ensemble.  For a single Ramsey sequence of duration $T$,
so that the unknown parameter is $\theta = \omega T$, the Standard Quantum Limit and Heisenberg-limit
sensitivities
are as described in Section \ref{sec:IntroI.1}, { $\Var
  (\theta_\text{est})_{\rm SQL} = N^{-1}$ and $\Var
(\theta_\text{est})_{\rm HL} = N^{-2}$.
The assumption of uncoupled particles is often physically reasonable, for example when describing photons in a linear interferometer or low-density atomic gases for which collisional interactions can be neglected.  In other systems, including  Bose-Einstein condensates \cite{GrossN2010,RiedelN2010}, laser interferometers at high power \cite{LIGONP2011,AasiNP2013others}, high-density atomic magnetometers \cite{DangAPL2010,KominisN2003,ShahPRL2010,VasilakisPRL2011} and high-density atomic clocks \cite{DeutschPRL2010}, the assumption of uncoupled particles is unwarranted.  This motivates the study of nonlinear Hamiltonians.

The unusual features of nonlinear Hamiltonians are well illustrated in the following example \cite{BoixoPRA2008}.
First, define the collective operator $S_0 \equiv \sum_i s_0^{(i)} \equiv \sum_i \mathds{1}^{(i)}$, where $\mathds{1}$ indicates the identity operator.  $S_0$ is clearly the total number of particles.  Now
consider the nonlinear Hamiltonian
\begin{equation}
\label{eq:HBoixo}
H =  \hbar \Omega  S_0 S_z =   \hbar \Omega \sum_{i=1}^N\sum_{j=1}^N s_0^{(i)} s_z^{(j)}.
\end{equation}
This is linear in the unknown $\Omega$, but of second order in the collective variables $S$.  At the microscopic level, the Hamiltonian describes a pair-wise interaction, with energy $\hbar \Omega s_0^{(i)} s_z^{(j)}$, between each pair of particles $(i,j)$.

\newcommand{\thetaNL}{\theta_{\rm NL}}
\renewcommand{\thetaNL}{\Theta}
For a system with a fixed number $N$ of particles, the consequence for the dynamics of the system is very simple: the operator $S_0$ can be replaced by its eigenvalue $N$, leading to an effective Hamiltonian
\begin{equation}
H_N =  \hbar N \Omega  S_z.
\end{equation}
Estimation of the product $N \Omega T = N \thetaNL$ now gives the same
uncertainties that we saw earlier in the estimation of $\theta$.  That
is, {$\Var(N \thetaNL_\text{est})_{\rm SQL} =  N^{-1}$ so that  $\Var(
\thetaNL_\text{est})_{\rm SQL} = N^{-3}$.}  Similarly,  $\Var(\thetaNL_\text{est})_{\rm
  HL} = N^{-4}$.  More generally,  a nonlinear Hamiltonian containing
$k$-order products of collective variables will contain $N^k$
microscopic interaction terms {that contribute to the Hamiltonian and thus to the
rate of change of an observable such as $S_z$ under time evolution}. In contrast, the
{variance of such a macroscopic observable,
e.g.~$\Var(S_z)$, scales as} $N^{1}$ (Standard Quantum Limit) or $N^0$
(Heisenberg-limit).  {This allows
signal-to-noise ratios scaling as $N^{2k-1}$ (Standard Quantum Limit) and as $N^{2k}$ (Heisenberg-limit) \cite{BoixoPRL2007}.}
These noise terms depend only on the size of the
system and nature of the initial state but not on the Hamiltonian
\cite{BoixoPRL2007}.

That nonlinearities lead to improved scaling of the sensitivity appears to have been independently discovered by A. Luis and J. Beltr\'{a}n \cite{LuisPL2004,BeltranPRA2005,LuisPR2007} and by S. Boixo and co-workers \cite{BoixoPRL2007,BoixoPRA2008,BoixoPRL2008,DattaMPLB2012}.  A related proposal using interactions to give a scaling  $\sigma(\theta_\text{est}) \propto 2^{-N}$  is described in \cite{RoyPRL2008}.

Prior to the appearance of these results, the term ``Heisenberg
limit,'' which was introduced into the literature by
\cite{HollandPRL1993} in the context of interferometric phase
estimation
with the definition $\sigma(\phi_\text{est}) =
1/N$, had been used, often indiscriminately, to describe 1) the
sensitivity $\sigma(\phi_\text{est}) = 1/N$, 2) the scaling
$\sigma(\phi_\text{est}) \propto 1/N$,
  3) the best possible sensitivity with $N$ particles, and 4) the best possible scaling with $N$ particles \cite{GiovannettiNPhot2011,GiovannettiPRL2006,GiovannettiS2004}.  Clearly these multiple definitions are not all compatible in a scenario with a nonlinear Hamiltonian.  Taking as a definition ``error \ldots bounded by the inverse
of the physical resources,'' and implicitly considering scaling,  \cite{ZwierzPRL2010} (see also  \cite{ZweirzPRL2010erratum} )
showed that an appropriate definition for ``physical resource'' is the query complexity of the system viewed as a quantum network.

For the simplest optical nonlinearity, $\theta \propto N^2$ and quadrature detection, it has been shown that quadrature squeezed states are near-optimal \cite{Maldonado-MundoPRA2009}.  Considering the same nonlinearity and limiting to classical inputs, i.e. coherent states and mixtures thereof, it is argued in \cite{RivasPRL2010} that non-linear strategies can out-perform linear ones by concentrating the available particles in a small number of high-intensity probes.  \cite{TilmaPRA2010} analyzed a variety of entangled coherent states for nonlinear interferometry of varying orders, and found that in most cases entanglement degraded the sensitivity for high-order nonlinearities.  \cite{BerradaPRA2013} considered the use of two-mode squeezed states as inputs to a non-linear interferometer, including the effects of loss, and showed a robust advantage for such states.

As already mentioned in Section \ref{sec:Fisherinfo}, the above results concern local measures implying some prior knowledge of the parameter being estimated. The situation for global measures without prior information is considered in \cite{HallPRX2012}.


%
%
%

\newcommand{\Szero}{{S_0}}
\newcommand{\Sx}{{S_x}}
\newcommand{\Sy}{{S_y}}
\newcommand{\Sz}{{S_z}}
\newcommand{\Jzero}{{J_0}}
\newcommand{\Jx}{{J_x}}
\newcommand{\Jy}{{J_y}}
\newcommand{\Jz}{{J_z}}
\newcommand{\fx}{{f_x}}
\newcommand{\fy}{{f_y}}
\newcommand{\fz}{{f_z}}

\subsection{Proposed experimental realizations}

A number of physical systems have been proposed for nonlinear quantum-enhanced measurements:  Propagation through nonlinear optical materials \cite{LuisPL2004,BeltranPRA2005,LuisPR2007}, collisional interactions in Bose-Einstein condensates \cite{BoixoPRA2009}, Duffing nonlinearity in nanomechanical resonators  \cite{WoolleyNJP2008}, and  nonlinear Faraday rotation probing of an atomic ensemble \cite{NapolitanoNJP2010}.

\subsubsection{Nonlinear optics} \label{sec.NLO}

The first proposals concerned nonlinear optics \cite{LuisPL2004,BeltranPRA2005,LuisPR2007}, in which a nonlinear optical susceptibility is directly responsible for a phase-shift $\theta \propto N^k$, where $k$ is the order of the nonlinear contribution to the refractive index.  In the simplest example, \cite{BeltranPRA2005} showed that an input coherent state $|\alpha\rangle$, experiencing a Kerr-type nonlinearity described by the unitary $\exp[i \Theta (a^\dagger a)^2]$, and detected in quadrature $X = \frac{1}{\sqrt{2}}(a + a^\dagger)$, gives an outcome distribution
\begin{equation}
P(X =x|\Theta) = | \langle x | e^{i \Theta (a^\dagger a)^2} | \alpha \rangle|^2.
\end{equation}
{If we consider the case of} small $\Theta$, imaginary $\alpha = i \sqrt{\langle N
  \rangle }$, and the 
estimator $\Theta_\text{est} = \bar{ X}
/|\partial \bar{X}/\partial \theta| $, where $\bar{X}\equiv
\sum_{i=1}^MX_i$ is the mean of the observed quadratures,
 {we find} the standard deviation
\begin{eqnarray}
\sigma(\Theta_\text{est}) &=& \frac{\sigma(X)}{\sqrt{M}|d \langle
  X\rangle/ d\Theta|} =
\frac{\sigma(X)}{ \sqrt{M}|\langle[X,(a^\dagger a)^2] \rangle|}\\
&=& \frac{1}{4M^{1/2} N^{-3/2}}\,.
\end{eqnarray}
Here we have used $\sigma(X)=1/\sqrt{2}$ for the quantum
mechanical uncertainty of $X$ in the initial state, and which up to
corrections of order $\Theta^2$  
holds also for the evolved state when $\Theta$ is small.
For large
$N$, this strategy saturates the quantum Cram\'er-Rao bound;  the quantum Fisher information
 is straightforwardly calculated to be $I_\Theta = 4N + 24 N^2 + 16 N^3$.


\subsubsection{Ultra-cold atoms}

Coherent interaction-based processes are well developed in Bose-Einstein condensates and have been used extensively for squeezing generation.  For example, a confined two-species Bose-Einstein condensate experiences collisional energy shifts described by an effective Hamiltonian
\begin{eqnarray}
H_{\rm eff} &\propto& a_{11} n_1(n_1-1) + 2 a_{12} n_1 n_2  + a_{22} n_2(n_2-1)  \\
& = & (a_{11} - 2 a_{12} + a_{22}) S_z^2  +  2 (a_{11} - a_{22}) S_z S_0  \nonumber \\ & & +
{\rm terms~in~} \Sz, \Szero
\end{eqnarray}
where $S_z \equiv \frac{1}{2} ( n_1 - n_2)$ and $S_0 \equiv \frac{1}{2} ( n_1 + n_2)$ are pseudo-spin operators, $n_{1}$ and $n_{2}$ are the number of atoms of species 1 and 2, respectively, and $a_{ij}$ are the collisional scattering lengths.  In  $^{87}$Rb, and with $|1\rangle \equiv |F=1,m=-1\rangle$, $|2\rangle \equiv |F=2,m=1\rangle$, the scattering lengths (near zero magnetic field) have the ratio $a_{11}:a_{12}:a_{22} = 1.03 : 1 : 0.97$.  A proven method to generate spin squeezing in this system is to increase $a_{12}$ using a Feshbach resonance to give the single-axis twisting Hamiltonian $H_{\rm eff} \propto S_z^2$ \cite{MuesselPRL2014}, plus terms proportional to $\Sz$ and $\Szero$, which induce a global rotation and a global phase shift, respectively.

It was observed  in \cite{BoixoPRL2008} that the zero-field scattering
lengths naturally give $a_{11} - 2 a_{12} + a_{22} \approx 0$, making
small the coefficient of $S_z^2$ and  leaving the $\Szero \Sz$ term as
the dominant nonlinear contribution.  Detailed analyses of the
Bose-Einstein condensate physics beyond the simplified single-mode treatment here are given in \cite{BoixoPRA2008,BoixoPRA2009, TaclaNJP2013}.  The strategy gives $N^{-3/2}$ scaling for estimation of  the relative scattering length $a_{11} - a_{22}$.  \cite{MahmudPRA2014} describe a strategy of dynamical decoupling to suppress the second-order terms in the Hamiltonian and thus make dominant three-body interactions, giving a sensitivity scaling of $N^{-5/2}$ for measurements of three-body collision strengths.

\subsubsection{Nano-mechanical oscillators}

 \cite{WoolleyNJP2008} propose a nonlinear interferometer using two modes of a nano-mechanical
oscillator, with amplitudes $x_a$ and $x_b$, experiencing the nonlinear Hamiltonian
\begin{equation}
H_{\rm eff} = H_{\rm SHO}^{(a)} + H_{\rm SHO}^{\reddd{(b)}} + \frac{1}{4} \chi_a m \omega^2 x_a^4 + \frac{1}{4} \chi_b m \omega^2 x_b^4 + C(t)
\end{equation}
where $H_{\rm SHO}$ indicates the simple harmonic oscillator Hamiltonian, $\chi$ is the Duffing nonlinearity coefficient, $\omega$ is the low-amplitude resonance frequency, and $C$ is an externally-controlled coupling between modes $a$ and $b$ that produces a beam-splitter interaction.  With an interferometric sequence resembling a Mach-Zehnder interferometer, the Duffing nonlinearity can be estimated with uncertainty scaling as $N^{-3/2}$, where $N$ is the number of excitations.

\subsubsection{Nonlinear Faraday rotation}\label{sec:Napo}

Whereas Luis and co-workers considered phenomenological models of optical nonlinearities, \cite{NapolitanoNJP2010} describes an {\em ab-initio} calculation of the optical nonlinearity produced on a particular atomic transition, using degenerate perturbation theory and a collective quantum variable description. This gives an effective Hamiltonian for the interaction of polarized light, described by the Stokes operators ${\bf S}$, with the collective orientation and alignment spin variables  ${\bf J}$  of an atomic ensemble:
\bea
H_{\rm eff} & = & H_{\rm eff}^{(2)}  + H_{\rm eff}^{(4)}  + O({\bf S}^3) \\
\label{eq:H2}
H_{\rm eff}^{(2)}  & = &  \alpha^{(1)}\Sz\Jz+\alpha^{(2)}\left(\Sx\Jx+\Sy\Jy\right)  \\
\label{eq:H4}
H_{\rm eff}^{(4)} &=& \beta_J^{(0)} \Sz^2 \Jzero + \beta_N^{(0)}
\Sz^2 N_A +  \beta^{(1)}  \Szero \Sz \Jz  \nonumber \\ & & +
\beta^{(2)}  S_0 (\Sx \Jx + \Sy \Jy).
\eea
where the $\alpha$ and $\beta$ coefficients are linear and non-linear polarizabilities that  depend on the detuning of the probe photons from the atomic resonance.  By proper choice of detuning and initial atomic polarization ${\bf J}$, the term $\beta^{(1)}  \Szero \Sz \Jz$ can be made dominant, making the Hamiltonian formally equivalent to that of Eq.~(\ref{eq:HBoixo}).  Note that  $\beta^{(1)}\Jz$, proportional to the atomic polarization $\Jz$, plays the role of the unknown interaction energy $\hbar \Omega$.  The photons are thus made to interact, mediated by and proportional to the atomic polarization $\Jz$.  For a different detuning, the term $ \alpha^{(1)}\Sz\Jz$ becomes dominant, allowing a linear measurement of the same quantity $\Jz$ with the same atomic system.\\

The experimental realization using a cold, optically-trapped $^{87}$Rb atomic ensemble is described in \cite{NapolitanoN2011}.  The experiment observed the predicted scaling of {$\Var(J_z) \propto N^{-3}$} over a range of photon numbers from $N=5 \times 10^5$ to $5 \times 10^7$.  For larger photon numbers the scaling worsened, i.e.  {$\Var(\Jz (N) )$} had a logarithmic derivative $> -3$.  Due to this limited range of the $N^{-3}$ scaling, and the difference in pre-factors $\beta^{(1)}$ versus $\alpha^{(1)}$, the nonlinear estimation never surpassed the sensitivity $\Var(\Jz)$ of the linear measurement for the same number of photons.  Nonetheless, due to a shorter measurement time $\tau$, the nonlinear measurement did surpass the linear measurement in spectral noise density {$\Var(\Jz) {\tau}$}, a common figure of merit for time- or frequency-resolved measurements.

\subsection{Observations and commentary}\label{obs}

Several differences between linear and nonlinear strategies, perhaps
surprising, deserve comment.

First, it should be obvious that there is no conflict with the
Heisenberg uncertainty principle.  $\theta$ and $\Theta$ are
parameters, not observables, and as such are not subject to
operator-based uncertainty relations, neither the Heisenberg
uncertainty principle nor
generalizations such as the Robertson uncertainty
relation \cite{RobertsonPR1929}. Moreover, the advantageous scaling in
$\delta \thetaNL$ is the result of a rapidly-growing {\em signal},
rather than a rapidly decreasing statistical noise. A nonlinear
Hamiltonian immediately leads to a strong change in the scaling of the
signal: even the simplest $k=2$ nonlinearity gives signal growing as $
\omega  \propto N^2$ and thus Standard Quantum Limit uncertainty $(\delta \thetaNL)_{\rm
  SQL} \propto N^{-3/2}$, which scales faster with $N$ than does
$(\delta \theta)_{\rm HL} \propto N^{-1}$.

Second, the estimated phases $\theta$ and $\thetaNL$ necessarily
reflect different physical quantities.  $\hbar \omega$ describes a
single-particle energy such as that due to an external field, whereas
$\hbar \Omega$ describes a pair-wise interaction energy.  As such, the
uncertainties $\delta \theta$ and $\delta \thetaNL$ are not {\em
  directly} comparable.  Any comparison of the efficacy of the
measurements must  introduce another element, a connection between a
third physical quantity, $\theta$, and $\Theta$.  This we have seen
in Section \ref{sec:Napo}, where the unknown $\Jz$ appears in both the
linear and nonlinear Hamiltonians.
In what follows, we describe an optimized linear/nonlinear comparison.

\subsection{Nonlinear measurement under number-optimized conditions }

A more extensive study of nonlinear spin measurements, using the same
system as \cite{NapolitanoN2011} is reported in \cite{SewellPRX2014}.
This work compared two estimation strategies, one linear and one
non-linear, for measuring the collective variable $J_y$, which
describes a component of the spin alignment tensor used in a style of
optical magnetometry known as alignment-to-orientation conversion
\cite{BudkerPRL2000,PustelnyJAP2008,SewellPRL2012}.   The linear
estimation used the term $\alpha^{(2)}  \Sy \Jy$, which   appears in
Eq. (\ref{eq:H2}) and produces a rotation from linearly polarized
light toward elliptically-polarized light.  The nonlinear estimation
in contrast used Eq.~(\ref{eq:H2}) in second order:  in the first
step, due to the $\alpha^{(2)}  \Sx \Jx$ term and the  input $\Sx$
optical polarization, an initial $\Jy$ atomic polarization is rotated
toward $\Jz$ by an angle $\phi \propto \langle\Sx\rangle$, 
and thus $\propto N$,
where $N$ is the number of photons.  In the second step, the term
$\alpha^{(1)} \Sz \Jz$ produces a Faraday rotation, i.e. from $\Sx$
toward $\Sy$, by an angle proportional to the $\Jz$ polarization
produced in the first step.  The resulting $\Sy$ polarization is $\Sy
\propto \Jy N^2$, while the statistical noise is $\sigma(\Sy) \propto
N^{1/2}$, giving sensitivity scaling $\sigma(\Jy) \propto
N^{-3/2}$.
Importantly, the two estimation strategies use the same
$\Sx$-polarized input, and thus have identical statistical noise and
cause identical damage in the form of spontaneous scattering, which
adds noise to the atomic polarization.

The experimentally-observed nonlinear sensitivity was compared against
the calculated ideal sensitivity of the linear measurement.  Owing to
its faster scaling, and more favorable pre-factors than in
\cite{NapolitanoN2011}, the nonlinear measurement's sensitivity
surpassed that of the ideal linear measurement at about $2 \times
10^7$ photons.  A comparison was also made when each measurement was
independently optimized for number $N$ and detuning, which affects
both the pre-factors $\alpha$ and the scattering.  The ability to
produce measurement-induced spin-squeezing was taken as the figure of
merit, and the fully-optimized nonlinear measurement gave more
squeezing than the fully-optimized linear measurement.   This shows
that for  some quantities of practical interest, a nonlinear
measurement can out-perform the best possible linear measurement.
\reddd{Similar conclusions have been drawn for the case of
number-optimized saturable spectroscopy \cite{MitchellQST2017}.}

\subsection{Signal amplification with nonlinear Hamiltonians}

The single-axis twisting Hamiltonian $H_{\rm twist} = \chi S_z^2$, in
addition to producing spin squeezed states, has been proposed as a
nonlinear amplifier to facilitate state readout in atom interferometry
\cite{DavisPRL2016}.  Starting from an $x$-polarized coherent spin
state $\ket{\bf x}$, and defining the unitary $U \equiv \exp[- i
H_{\rm twist} \tau/\hbar]$, the Wigner distribution of the squeezed
state $U\ket{\bf x}$ is  thin in the $z$ direction, and is thus
sensitive to rotations ${\cal R}_y(\phi)$ about the $y$-axis, so that
states of the form ${\cal R}_y(\phi)U\ket{\bf x}$ have large quantum
Fisher information with respect to $\phi$. Exploitation of this
in-principle sensitivity is challenging, however, because it requires
low-noise readout, detecting $S_z$ at the single-atom level if the
Heisenberg limit is to be approached. In contrast, a sequence that
applies  $H_{\rm twist}$, waits for rotation about the $y$-axis and
then applies $-H_{\rm twist}$ for an equal time generates the state
$U^\dagger{\cal R}_y(\phi)U\ket{\bf x}$.  Because $U^\dagger$ is
unitary, the  quantum Fisher information is unchanged, but the
perturbation implied by ${\cal R}_y(\phi)$ now manifests itself
at the scale of the original coherent spin state, which is to say it
is amplified from the Heisenberg-limit scale up to the Standard Quantum Limit scale, greatly
facilitating detection. \reddd{Implementations include a cold-atom
cavity QED system \cite{HostenS2016} and
Bose-Einstein condensates \cite{LinnemannPRL2016}.} While this strategy clearly uses
entanglement, it is nonetheless striking that un-doing the
entanglement-generation step provides an important benefit.

\subsection{Other modifications of the Hamiltonian}

The assumption of a Hamiltonian $H=\sum_{k=1}^N h_k$ considered for
the derivation of
eq.\eqref{dxminUs}, where $\Lambda$ and $\lambda$ are the largest and
smallest eigenvalues of $h_k$, respectively,
not only implies distinguishable subsystems, it is also restrictive in
two other important regards: a.) The existence of such bounds on the
spectrum of $h_k$ may not be warranted, and b.) interactions between
the subsystems are excluded.  In this section we explore the
consequences of lifting these restrictions.

\subsubsection{Lifting spectral limitations}
A large portion of the work on quantum-enhanced measurements stems from quantum optics, where
the basic dynamical objects are modes of the e.m.~field, corresponding
to simple harmonic oscillators, $h_k=\hbar\omega_k(n_k+1/2)$.  A phase
shift in mode $k$ can be implemented by $U=\exp(i n_k \theta)$. Clearly,
for the relevant Hamiltonian $h_k=n_k$ acting as generator of the phase shift,
$\Lambda=\infty$ and
$\lambda=0$.  Hence,  eq.\eqref{dxminUs} would imply
a minimal uncertainty $\text{Var}(\theta_{\rm est})=0$. Of course, one
may argue that in reality
one can never use states of infinite energy, such that there is
effectively a maximum energy.  However, it need not be that the
maximum energy sustainable by the system must be distributed over $N$
modes.  Indeed, what is typically counted in quantum optics in terms
of resources is not the number of modes $N$, but the total number of photons
$n$, directly linked to the total energy.  It turns out, that the
total number of modes (or subsystems, in general) is completely
irrelevant for achieving optimal
sensitivity, even if the parameter is coded in several modes or subsystems,
e.g.~with a general unitary transformation of the form $U=\exp(i \theta
\sum_{k=1}^N h_k )$, if one can stock the same amount of energy in a single
system as in the total system.  Note that this is often the case in
quantum optics, where different modes can be spatially confined or
parametrically influenced by the
same optical elements such as mirrors, beam-splitters, and phase
shifters whose material properties ultimately determine the maximum
amount of energy that can be used. \\

To see the liberating effect of unbound spectra, recall that for any
initial pure state $\ket{\psi}$ propagated by a Hamiltonian of the form
$H=\theta G$ with a hermitian generator $G$ for a time $T$ the quantum Fisher information is
given by \cite{Braunstein94,braunstein_generalized_1996}
\begin{equation}
  \label{eq:Ic}
  I_{\theta}=4 \text{Var}(G) T^2 \equiv 4(\langle G^2\rangle-\langle
  G\rangle^2)T^2 \,.
\end{equation}
Let $G=\sum_i e_i \ketbra{i}{i}$ be the spectral decomposition of $G$,
and $\ket{\psi}=\sum_{i=1}^L c_i\ket{i}$, where we assume that
$\ket{1}$ ($\ket{L}$) are the states of lowest (largest) energy
available. Then $
\text{Var}(G)=\sum_{i=1}^Lp_i e_i^2-(\sum_{i=1}^Lp_i
e_i)^2$ with $p_i=|c_i|^2$ and $\sum_{i=1}^Lp_i=1$.
The Popoviciu inequality \cite{Popoviciu35} states $
\text{Var}(G)\le (e_L-e_1)^2/4$. It is saturated for $p_1=p_L=1/2,\,\,p_i=0$
else.
The state $\ket{\psi}=(|1\rangle+e^{i\varphi}|L\rangle)/\sqrt{2}$ with an arbitrary phase
$\varphi$ saturates the inequality and thus   maximizes $I_\theta$.
 If $e_L$ or $e_1$ is
degenerate, only the total probability for the degenerate energy levels
is fixed to 1/2, and arbitrary linear combinations
in the degenerate subspace are allowed.  But the value of $\text{Var}(G)$
remains unchanged under such redistributions, and we may still choose
just two non-vanishing probabilities $p_1=p_L=1/2$.
The derivation did not make use of a multi-mode structure
of the energy eigenstates. Hence, exactly the same minimal
uncertainty of $\theta_\text{est}$ can
be obtained by
superposing the ground state of a single mode with a Fock state of
given maximum allowed energy as with an arbitrarily entangled multi-mode
state containing components of up to the same maximum energy. \\
For a specific example, consider phase estimation in a Mach-Zehnder
interferometer.  It has $N=2$ modes, and a phase shift just in one of
them, i.e.~the relevant Hamiltonian is $H=\theta n_1$.  Adding energy
conservation of the two modes at the beam-splitters (i.e.~the fact
that the accessible states are two-mode Fock states of the form
$\ket{n-n_2,n_2}$, where $n_2$ with $0\le n_2\le n$ is the number of photons
in the second mode), one immediately finds that the optimal two mode state is
$(\ket{n,0}+\ket{0,n})/\sqrt{2}$, i.e.~the highly entangled N00N state
\cite{Boto00}.
However, we can achieve exactly the same variance of $G$ and hence
sensitivity with the single-mode state
$(\ket{n}+\ket{0})/\sqrt{2}\otimes \rho_2$, i.e.~a product state
where we keep the second mode in {\em any} state $\rho_2$.  In both
cases the maximum energy of the first mode is $n\hbar \omega$
(assuming $\omega_1=\omega_2=\omega$), and the average energy in
the interferometer $n\hbar\omega/2$ (neglecting the vacuum energy
$\hbar\omega/2$).  Hence, also from the perspective of maximum energy
deposit in the optical components, there is no advantage in using two
entangled modes. If the Mach-Zehnder interferometer is realized
abstractly via Ramsey-pulses on $N$ two-level systems (states
$\ket{0},\ket{1}$) for the
beam-splitters, and a phase shift $\exp(i \theta J_z)$,
$J_z=\sum_{i=1}^N \sigma_z^{(i)}/2$, the state that
maximizes  ${\rm Var}(J_z)$ is the (maximally entangled) GHZ state
$(\ket{0\ldots   0}+\ket{1\ldots 1})/\sqrt{2}$.  But exactly the same
uncertainty can be obtained with a single spin-j ($j=N/2$) in the state
$(\ket{j,j}+\ket{j,-j})/\sqrt{2}$ (in the usual $\ket{j,m}$ notation,
where $j$ is the total angular momentum and $m$ its
$z$-component). Clearly, allowing as large a
spectrum for a single system as for the combined systems makes
entanglement entirely unnecessary here.
\\

These considerations teach us that the relevant quantity to be
maximized is the quantum uncertainty of the generator $G$.  This can
be understood in terms of generalized Heisenberg uncertainty
relations, {\redd in which} the generator $G$ plays the quantity
complementary to
$\theta$, as was found early on \cite{braunstein_generalized_1996}.  In a
multi-component system maximizing $\text{Var}(G)$ may be achieved with
highly entangled states, but if the spectral range of a single system
admits the same $\text{Var}(G)$, there is no need for
entanglement.
If unbound spectra are permitted, one can in fact do much better than
the Heisenberg-limit:  In \cite{PhysRevX.5.031018} the single-mode state
$\frac{\sqrt{3}}{2}\sum_{n=0}^\infty 2^{-n}|2^n\rangle$ was pointed
out that has diverging
$\text{Var}(n)$ with, at the same time finite $\bar n$. It
therefore  allows, at least  in principle and in an ideal setting,
arbitrarily precise phase measurements while using finite
energy. \\

In \cite{Braun11.2,braun_ultimate_2012} a state of the form
$(\ket{0}+\ket{2n})/\sqrt{2}$ was found to be optimal for mass
measurements with a nano-mechanical oscillator given a maximum
allowed number of excitation quanta $2n$ and times much larger than
the oscillation period (for shorter times there are contributions also
from the dependence on frequency of the energy eigenfunctions). The
same state of a single
mode of the e.m.~field is optimal for measuring the speed of light
\cite{braun_how_2015}. In both cases the quantum uncertainties scale as
$1/n$ (quantum Fisher information
proportional to $n^2$), and obviously no entanglement is needed.
Of
course, a state of the form $(\ket{0}+\ket{2n})/\sqrt{2}$ (called
``half a N00N'' state in \cite{Braun11.2}) is still highly
non-classical (see also \cite{PhysRevA.92.042115}).  In fact, already
a single Fock state $\ket{n}$ is
highly non-classical, as is witnessed by its highly oscillatory Wigner
function \cite{Schleich01} with substantial negative parts. The superposition
$(\ket{0}+\ket{2n})/\sqrt{2}$ leads in addition to $2n$ lobes in
azimuthal direction that explain the sensitivity of phase measurements
$\propto n^{-1}$. Alternatively, one can use superpositions of coherent states
\cite{braginsky_quantum_1995,lund_conditional_2004,suzuki_practical_2006,neergaard-nielsen_generation_2006,wakui_photon_2007,bimbard_quantum-optical_2010,yukawa_generating_2013}, i.e.~''Schr\"odinger-cat'' type states of the form
$(\ket{\alpha}+\ket{-\alpha})/\sqrt{2}$. They have been created in
quantum optics with values of $\alpha=0.79$ in
\cite{ourjoumtsev_generating_2006}.  In \cite{lund_conditional_2004} a
``breeding method'' based on  weak squeezing, beam
mixing with an auxiliary coherent field, and photon detecting
with threshold detectors was proposed to achieve values up to
$\alpha\le 2.5$, but the success
probability has been found to be too low for a realistic iterated
protocol. An alternative based on homodyning was proposed in
\cite{laghaout_amplification_2013,etesse_proposal_2014} and
implemented in \cite{etesse_experimental_2015}, leading to
$\alpha\simeq 1.63$.     The current record
in the optical domain for
``large'' $\alpha$ appears to be $\alpha\simeq \sqrt{3}$, achieved
from two-mode squeezed vacuum and
$n$-photon detection on one of the modes \cite{huang_optical_2015}.
In the microwave regime, superpositions of coherent states with
$\alpha=\sqrt{7}$ have been
generated, as well as superposition of coherent states with smaller
phase differences with up to 111 photons
\cite{vlastakis_deterministically_2013}.  In \cite{Monroe96}
superpositions of coherent  states of the
vibrational motion of a $^9\text{Be}^+$ ion in a one-dimensional trap with
$\alpha\simeq 3$ were reported. Almost arbitrary
superpositions with a small number of photons can  be generated by
using couplings of a mode with two-level systems that can be tuned in
and out of resonance, and a plethora of methods for generating
superpositions of coherent
states were proposed,
but reviewing the entire literature of non-classical states in general
and even all the proposals for generating superpositions of coherent
states
is beyond the scope of the present review (see
e.g.~\cite{gottesman_encoding_2001,deleglise_reconstruction_2008,hofheinz_synthesizing_2009}
and the Nobel lectures of Serge Haroche and David J.~Wineland
\cite{haroche_nobel_2013,wineland_nobel_2013} for
historical accounts of the development of these fields and many more
references, as well as the literature citing \cite{PhysRevA.58.3472}
where superpositions of coherent states in a
Mach-Zehnder interferometer were studied with respect to the
limitations arising from imperfect photodetectors).

The use of superpositions of coherent  states for metrology was examined in
\cite{1464-4266-6-8-032,PhysRevA.65.042313} and it was found that the
Heisenberg limit
can be reached. In \cite{PhysRevA.58.3472} .
superpositions of coherent states in atom interferometers may even
exhibit quantum-enhanced sensitivity to parameters that have
have no classical analog.  For example, in \cite{PhysRevA.92.010101}
it was shown
that monitoring the decoherence rate of an superposition of
atomic coherent  states may
uncover clues about so-far undetected particles that couple softly
(i.e.~via weak momentum transfer, but not weakly) to the atoms. This
is reminiscent of previous ideas of using decoherence as a sensitive
probe \cite{Braun11}.  The
decoherence rate can be detected with
sensitivity that is limited only by the spatial size of the
superposition, and the situation is quite similar to
the estimation of boson loss discussed in \ref{sec.BL}.

\subsubsection{Decoherence-enhanced measurements}
Decoherence is arguably the most fundamental issue that plagues
quantum enhancements of all kinds, and quantum enhanced measurements
are no exception.  However, decoherence has interesting physical
properties which imply that it can also be useful for precision
measurements.  This goes beyond the benefits of decoherence and open system
dynamics found as early as the late 1990s and the early 2000s, when it was
realized that entanglement can be created through decay processes or
more generally through coupling to common environments
\cite{Plenio99,Braun02,benatti_environment_2003,Braun05,benatti_environment_2008,benatti_environment-induced_2009,benatti_entangling_2010},
and, paradoxically, that decoherence of quantum computations can be reduced  by
rather strong dissipation that confines the computation to a
decoherence-free subspace through a
Zeno-type effect due to the rapid decay of states outside
the decoherence-free subspace \cite{Beige00,Beige00b}.
Recently such ideas have found renewed interest, and meanwhile
techniques have been proposed to create steady state
entanglement in driven open quantum systems, such as cold Rydberg
gases in the Rydberg-blockade regime \cite{lee_emergence_2015}. It
remains to be seen
whether such stabilized entangled states are useful for precision
measurements.

Here, on the contrary, we focus on the dynamics created by decoherence
processes themselves. Decoherence arises from an interaction with an environment
described by a non-trivial interaction Hamiltonian
$H_{\rm int}=\sum_{i}S_iB_i$ with a similar structure as
the non-linear Hamiltonians considered above, where {\redd in $H_{\rm
    int}$}, however, one
distinguishes operators $S_i$ pertaining to the system and others ($B_i$)
pertaining to the environment.  The environment is typically
considered as heat bath
with a large number of degrees of freedom that
may not be entirely accessible \cite{Weiss99,Breuer02,Benatti05}.
In addition, system and heat bath have their own Hamiltonian, such
that the total Hamiltonian reads $H=H_S+H_B+H_{\rm int}$.
From simple model systems it is known that decoherence tends to become
extremely fast for quantum superpositions of states that
differ macroscopically in terms of the eigenvalues of one of the
$S_i$. For example it was shown that superposing two Gaussian
wave packets of a free particle in one dimenson that is coupled
through its position to a
heat bath of harmonic oscillators leads to
decoherence times that scale as powers of $\hbar$ that depend on how
the wave packets are localized in phase space: The shortest
decoherence time, scaling as $\hbar/|q_1-q_2|$ results from wave packets
distinguished only by positions $q_1,q_2$, the longest  one $\sim
\hbar^{1/2}/\sqrt{|p_1-p_2|}$ from wave packets that differ only in
their mean momenta $p_1,p_2$, and an intermediate one scaling as
$\hbar^{2/3}/|(q_1-q_2)(p_1-p_2)|^{1/3}$. For systems of finite
Hilbert-space dimensions such as angular momenta, one can often
identify an effective
$\hbar$ that scales like the inverse Hilbert-space dimension which
suggests that monitoring the decoherence process can lead to highly
sensitive measurements, possibly surpassing the $1/\sqrt{N}$ scaling
of the Standard Quantum Limit.

That this intuition is correct was shown in \cite{Braun11},
where a method war proposed for measuring the length of a cavity by
monitoring the decoherence process of $N$ atoms inside the cavity.
The atoms are initially prepared in a highly excited dark state,
{\redd in which}
destructive interference prevents them from transferring their energy
to a mode of the cavity with which they are resonant. For example, if
one has two atoms coupling via an interaction
$(g_1\sigma_-^{(1)}+g_2\sigma_-^{(2)})^\dagger+h.c.$ to a mode of the cavity
with annihilation operator $a$, a state $\propto
(1/g_1)|10\rangle-(1/g_2)|01\rangle$ of the atoms (where $|0\rangle$
and $|1\rangle$ are the ground and excited states of the atoms) is a
dark state, also known as decoherence-free state: the amplitudes of
photon transfer from the two atoms to the cavity cancel. However, the
couplings $g_i$ depend on the position of the atoms relative to the
cavity due to the envelope of the e.m.~field.  If the cavity changes
its length $L$ with the atoms at fixed positions, the $g_i$ change, such that
the original state becomes slightly bright.  This manifests itself
through the transfer of atomic excitations to the mode of the cavity,
from where photons can escape and be detected outside. In
\cite{Braun11} it was shown that through this procedure the
minimal uncertainty with which $L$ can be estimated (according to the quantum Cram\'er-Rao bound) scales as $1/N$ even when using an initial product
state of $N/2$ pairs of atoms.  This scaling applies both for a
perfect cavity, and in the bad cavity limit {\redd in which}  superradiance
arises.  Thus, at least in principle, Heisenberg-limited scaling
can arise here without the need for an entangled state, and in spite
of the inherently decoherent nature of superradiance.  However, the
prefactor matters also here: Superradiance leads to a rapid decay of
all states that are not dark, such that the available signal and with
it the prefactor of the $1/N$ scaling law deteriorate rapidly with
time. \\

Given the delocalized nature of the cavity mode in this example, it is
possible in principle to make the number of atoms that interact with the mode
arbitrarily large, in contrast to the non-linear schemes above, {\redd
  for which}
the interactions have to decrease if the total energy is to remain an
extensive quantity.  But if the volume is kept fixed the atoms will start to interact
so that the simple independent atom model of superradiance breaks
down.  For even larger $N$ one has to increase the size of the cavity
in order to accommodate all atoms. When the largest possible density is
reached, the volume will have to
grow $\propto N$, which implies coupling constants of the atoms to the
cavity that decay as $1/\sqrt{N}$ and leads back to Standard Quantum Limit scaling.  In
addition, the number of atoms has to be macroscopic in order to compete
with the best classical sensitivities reached  with
interferometers such as LIGO:  Assuming that the prefactor in the $1/N$-scaling
is of order one, one needs
$\sim 10^{21}$ atoms for a minimal uncertainty of $10^{-21}$. A cubic optical
lattice with one atom every $\mu$m, the lattice would have to span
10m in $x,y,z$-direction in order to accommodate that many atoms. When
using a
diamond with regularly arranged NV-centers every
$10\,$nm (such dense samples have been fabricated
\cite{acosta_diamonds_2009}), one would still require
a cubic diamond of edge length 10cm.
These examples show that competing with the best classical techniques
is very challenging even if one can achieve Heisenberg-limit scaling, as in
classical protocols it is relatively easy
to scale up the number of photons, compensating thus for a less
favorable scaling with $N$.

\subsubsection{Coherent averaging} \label{sec.cohav}
There is nothing inherently quantum about the $1/\sqrt{N}$ scaling of
the Standard Quantum Limit.  Rather, this behavior is a simple
consequence of the central
limit theorem applied to $N$ independently acquired measurement
results that are averaged as part of a classical noise-reduction
procedure. The idea of coherent averaging is to replace the averaging
by a coherent procedure, {\redd in which} the $N$ probes all interact
with a
central quantum system (a ``quantum bus'').  In the end one measures
the quantum bus or the entire system. \\

A simple example shows how this can lead to Heisenberg-limit
sensitivity without needing any entanglement: Consider $N$ spins-1/2
interacting with a central spin-1/2 with the Ising-interaction $H_{\rm
  int}=
\sum_{i=1}^Ng_i\sigma_z^{(0)}\otimes\sigma_z^{(i)}$, where the index
zero indicates the central spin.  The interaction commutes with the
free Hamiltonian of all spins
$H_s=\hbar\sum_{i=0}^N\omega_i\sigma_z^{(i)}$,
and we can solve the time-evolution exactly, starting from the initial
product state
$|\psi(0)\rangle=\frac{1}{\sqrt{2}}(|0\rangle_0+|1\rangle_0)\otimes_{i=1}^N|0\rangle_i$.
At time $t$, the state has evolved to
$|\psi(t)\rangle=\frac{1}{\sqrt{2}}(e^{i(\omega_0/2-N\bar{g})t}|0\rangle_0
+e^{-i(\omega_0/2-N\bar{g})t}|1\rangle_0)\otimes_{i=1}^N|0\rangle_i$,
up to an unimportant global phase factor. In particular, the state
remains a product state at all times. If we measure
$\sigma_x^{(0)}$ of the central spin, we find $\langle
\sigma_x^{(0)}(t)\rangle=\cos((\omega_0-2N \bar{g})t)$, i.e.~the
oscillation frequency increases for large $N$ proportional to $N$.  As the
quantum fluctuations of the central spin are independent of $N$, this
implies a standard deviation in the estimation  of the average coupling
$\bar{g}$ that scales as $1/N$, which can be confirmed by calculating
the quantum Fisher information. Clearly, this is not an effect of
entanglement, but simply of a phase accumulation.  In this respect the
approach is reminiscent to the sequential phase accumulation protocols,
{\redd in which} the precision of a phase shift $\varphi$ measurement
is enhanced by
sending the light several times through the same phase shifter
\cite{Higgins07}.  However, there the losses increase exponentially
with the number of passes, and the sequential nature of the
interaction leads to a bandwidth penalty that is absent for the
simultaneous interaction described by $H_{\rm int}$. In
\cite{birchall_quantum-classical_2016} the
exponential loss of photons with the number of passes was taken into
account, and quantum Fisher information per scattered photon optimized. It was found that
when both probe and reference beam are
subject to photon loss, the
reduction of  $\sigma(\varphi_\text{est})$ by a non-classical state
compared to an optimal classical multi-pass strategy is only at most
$\sim$19.5\%, and an optimal number of passes independent of the
initial photon number was found, resulting in a loss of
about 80\% of all input photons. For a single-mode lossy phase the
possible sensitivity gain is even smaller. Multi-pass microscopy was
proposed in \cite{juffmann_multi-pass_2016} and it was experimentally
demonstrated that at a
constant number of photon sample interactions retardance and
transmission measurements with a sensitivity beyond the
single-pass shot-noise limit could be achieved. Similar ideas are
being developed for electron-microscopy \cite{juffmann_multi-pass_2016e}.
\\

The fact that an interaction with $N$ probes and a central quantum bus
can lead to Heisenberg-limit scaling of the sensitivity was  first found in
\cite{Braun11}, where a more
general model was studied. It will be noted that the
total Hamiltonian has exactly the same structure as for a decoherence
model, with the $N$ probes playing the role of the original system,
and the bus the role of an environment. However, in contrast to the
standard scenario in decoherence,
it is assumed here that at least part of the
``environment'' is accessible.  The above example shows that the
``environment'' can be as simple as a single spin-1/2.
One can also extend the model to include additional decoherence processes. In
\cite{Braun11} a phase-flip channel with rate $\Gamma$ on all spins was
considered. It was found that phase flips on any of the $N$ probes has
no effect, whereas phase flip of the central spin introduces a
prefactor that decays exponentially as $\exp(-2\Gamma t)$ with time in $\langle
\sigma_x^{(0)}(t)\rangle$. Since the prefactor is independent of $N$
the power-law scaling of the sensitivity with $N$ is unchanged, but it
is clear that the prefactor matters and leads to a sensitivity that
deteriorates exponentially with time. \\

The fact that the parameter estimated in the above example is the interaction
strength between the bus and the $N$ probes prevents a comparison with
other schemes that do not use interactions.  In
\cite{fraisse_coherent_2015} a more comprehensive study of two
different spin-systems was
undertaken where also parameters describing the probes themselves and the
bus were examined.  Regimes of strong and weak interaction
were analyzed, and different initial states considered.  The two
models, called ZZZZ and ZZXX models, are respectively given by the Hamiltonians
\begin{eqnarray}
  \label{eq:models}
  H_1&=&\frac{\omega_0}{2}\sigma_z^{(0)}+\frac{\omega_1}{2}\sum_{i=1}^N\sigma_z^{(i)}+
  \frac{     g}{2}\sum_{i=1}^N\sigma_z^{(0)}\sigma_z^{(i)}\,,\nonumber\\
  H_2&=&\frac{\omega_0}{2}\sigma_z^{(0)}+\frac{\omega_1}{2}\sum_{i=1}^N\sigma_z^{(i)}+
  \frac{     g}{2}\sum_{i=1}^N\sigma_x^{(0)}\sigma_x^{(i)}\,,
\end{eqnarray}
where $\hbar=1$.
The ZZZZ model is an exactly solvable dephasing model, the ZZXX
can be analyzed numerically and with perturbation theory.  The
analysis is simplified when starting with a product state that is symmetric
under exchange of the probes, in which case the probes can be
considered as a single spin with total spin quantum number
$j=N/2$. It was found that for $\omega_1$, Heisenberg-limit scaling can be achieved
in the ZZXX model when measuring the entire system, but not when only
measuring the quantum bus, and not for the ZZZZ model.  Heisenberg-limit scaling for
the uncertainty of $\omega_0$ is not possible in either model.  For
the interaction strength $g$, Heisenberg-limit scaling is found for a large set of initial
states and range of interaction strengths when measuring
the entire system, but only for a small set of initial states in the
vicinity of the state considered in the simple example above, when measuring
only the quantum bus. Interestingly, in the ZZZZ model Heisenberg-limit scaling of
the sensitivity for measuring $g$ also arises with the probes in a
thermal state at any finite temperature, as long as the quantum bus
can be brought into an initially pure state.  This is reminiscent of
the ``power of one-qubit'' \cite{knill_power_1998}: With a set of
qubits of which only a single one is initially in a pure state, a
quantum enhancement is already possible in quantum computation, providing
evidence for an important role of quantum discord
\cite{datta_entanglement_2005,lanyon_experimental_2008} (see
Sec.~\ref{sec:discordmodi}).
As detailed in Sec.\ref{sec:discordmodi}, the
DQC1-scheme can provide better-than-Standard Quantum Limit sensitivity as
soon as the ancillas have a finite purity.  The control qubit
plays the role of the quantum bus, and the
dipole-dipole interaction between the Rydberg atoms implements the
$XX$-interaction considered in \cite{fraisse_coherent_2015}.

An important limitation of such schemes was proven in
\cite{fraisse_Hamiltonian_2016,BoixoPRL2007}, where it was
shown very generally that with a Hamiltonian extension to an ancilla
system the sensitivity of a phase shift measurement cannot be improved
beyond the best sensitivity achievable with the original system
itself.  Coherent averaging is nevertheless interesting, as it allows
one to achieve without injecting
entanglement better-than-Standard Quantum Limit sensitivity for which the non-interacting phase
shift measurement would need a highly entangled state.
\red{In \cite{PhysRevA.95.062342} it was shown that for general
parameter dependent Hamiltonians $H(\theta)$ the largest sensitivity
is achieved if the eigenvectors of $(d/d\theta)H(\theta)$ to the
largest and smallest eigenvalue are also eigenvectors of
$H(\theta)$. If these eigenvalues are non-degenerate, the condition is
also necessary. For a phase shift-Hamiltonian the condition is
obviously satisfied.  This insight opens the way to Hamiltonian engineering
techniques by adding parameter-independent parts to the Hamiltonian
that remove or overwhelm parts that spoil the commutativity of
$(d/d\theta)H(\theta)$ and $H(\theta)$ in the subspace of the largest
and smallest eigenvalues of $(d/d\theta)H(\theta)$.  These techniques
were called ``Hamiltonian subtraction'' and ``signal flooding'',
respectively, and proposed to improve magnetometry with NV-centers. \\}

\red{Another opportunity for Hamiltonian engineering arises
if the eigenvalues of $H(\theta)$ do not depend on the parameter.  It
was shown in \cite{PhysRevA.90.022117,PhysRevA.93.059901} that in such
a case the quantum Fisher information becomes periodic in time. This is particularly
pernicious for quantum-enhanced measurement schemes that allow long measurement times, as
under the condition of quantum coherence the quantum Fisher information typically increases
quadratically with time if the eigenvalues of $H(\theta)$ depend on
$\theta$.   Adding another
parameter-independent Hamiltonian might lead to parameter-dependent
eigenvalues and hence unlock unbound increase of the quantum Fisher information with time. }
\\

The existence of Heisenberg-limit scaling of sensitivities for product states,
suggests that ``coherent averaging'' might even be possible
classically. This question was investigated in \cite{braun_coherently_2014} in
a purely classical model of harmonic oscillators, {\redd in which} $N$
``probe''
oscillators interact with a central oscillator.  It was found that
indeed for weak interaction strengths a regime of Heisenberg-limit scaling of the
sensitivity exists, even though the scaling crosses over to Standard Quantum Limit
scaling for sufficiently large $N$. Nevertheless, it was proposed that
this could be useful for measuring very weak interactions, and notably
improve measurements of the gravitational constant.

\subsubsection{Quantum feedback schemes}\label{sec.qfeed}
In the context of having probes interact with additional (ancilla)
systems, quantum feedback schemes should be mentioned.  This is a
whole field by itself (see
e.g.~\cite{wiseman_quantum_2009,dalessandro_introduction_2007,serafini_feedback_2012}
for reviews). Quantum feedback generalizes classical
feedback loops to the quantum world: one tries to stabilize, or more
generally dynamically
control, a desired state of or an operation on a quantum system by
obtaining information about its actual state or operation, and feeding
back corrective actions into the controls of the system that bring it
back to that desired state or operation if any deviation occurs.
As a consequence, the field
can be broadly classified according to two different categories:
Firstly, the
object to be controlled maybe a quantum state or an entire operation.
And secondly, the type of
information fed back can be classical or quantum. With classical
information is meant information that is obtained from a measurement,
and which is then typically processed on a classical computer and used
to re-adjust the classical control knobs of the experiment. Such
schemes are called ``measurement-based feedback''. In
contrast to this, ``coherent feedback'' schemes directly use quantum
systems that are then manipulated unitarily and fed back to the
system.

Measurement-based feedback schemes in metrology are also known
as ``adaptive measurements'' (see also Sec.\ref{sec.adaptive} for
adaptive measurements in the context of
phase transitions). An adaptive scheme was proposed as
early as 1988 by Nagaoka for mending the problem that the optimal POVM
obtained in
standard quantum parameter estimation depends on the a priori unknown parameter
\cite{Nagaoka88}. One starts with a random
estimate, uses its value to determine the corresponding optimal POVM,
measures the POVM, updates the estimate, uses that new value to determine
a new optimal POVM, and so on.  The scheme was shown to be strongly
consistent (meaning unbiased for the number of
iterations going to infinity), and asymptotically efficient
(i.e.~saturates the quantum Cram\'er-Rao bound in that limit) by Fujiwara
\cite{fujiwara_strong_2006}.  It was experimentally implemented in
\cite{okamoto_experimental_2012} as an adaptive quantum state
estimation scheme for
measuring polarization of a light beam, but is in principle a general
purpose estimation scheme
applicable to any quantum statistical model using identical
copies of an unknown quantum state.

 In the context of quantum optics,  an
adaptive homodyne   scheme was proposed  by Wiseman for   measuring the phase of
an optical mode, {\redd in which}
the reference phase of the local oscillator $\Phi(t)$ is adjusted in
real-time to
$\Phi(t)\simeq \pi/2+\varphi(t)$, where $\varphi(t)$ is the latest
estimate of the phase carried by a continuous-wave, phase-squeezed light signal
\cite{wiseman_adaptive_1995}. This reference phase corresponds to the highest
sensitivity in a homodyne scheme.
Keeping the local oscillator phase through feedback close to this
optimal operating point can beat non-adpative heterodyning  in single shot phase
decoding, as experimentally demonstrated in
\cite{armen_adaptive_2002}. In \cite{PhysRevA.70.043812} it was shown
that adaptive measurements have a finite factor
advantage even in the limit of arbitrarily weak coherent states.
Phase estimation
using feedback was also
studied in
\cite{berry_optimal_2000,PhysRevA.65.043803,PhysRevA.73.063824,Higgins07}.
In
\cite{PhysRevA.65.043803} it was investigated how well a
stochastically varying, white
noise-correlated  phase  can be estimated.  The theoretical analysis
showed that
the  variance of the phase estimation could be
reduced by a factor $\sqrt{2}$ by a simple adaptive scheme compared to
a non-adaptive heterodyne scheme, resulting in a value of
$n^{-1/2}/\sqrt{2}$, where $n$ is the number of photons per coherence
time. With a squeezed beam and a more accurate feedback, the scaling
can be improved to  $n^{-2/3}$. The latter result was also found
for a narrow-band squeezed beam
\cite{PhysRevA.73.063824,PhysRevA.87.019901}.
In \cite{yonezawa_quantum-enhanced_2012} $15\pm 4$\% reduction of mean
square error of the phase
below the coherent-state limit was reported with this scheme in
optical-phase tracking, i.e.~in a case without any a priori information
about the value of the signal phase.  The broad support of the signal phase
implies that there is an optimal amount of squeezing, and the
sensitivity enhancement is directly given by the squeezing.  The
scheme can therefore be seen as a generalization of
Caves' idea  of reducing
the uncertainty with which a (fixed)
phase shift in one arm of an interferometer can be measured
\cite{caves_quantum-mechanical_1981}. Instead of having a fixed phase
reference by the beam in the other arm of the interferometer, the
feedback allows one to continuously adjust the phase of the reference
in the homodyning scheme to the optimal operating point.
\cite{PhysRevA.94.023840} proposed a feedback scheme based on measured temporal
correlations ($g^{(2)}$ correlation function) for estimating the phase
of a coherent state inside a cavity and find that the uncertainty
scales better than the Standard Quantum Limit, namely as $n^{-0.65}$, where $n$ is the
mean photon number of the coherent state.
 In
\cite{PhysRevLett.104.093601} an ``adaptive quantum smoothing method''
was used experimentally for estimating a stochastically fluctuating
phase on a coherent beam. ``Smoothing'' refers to the fact that
estimates are obtained not only from data measured up to the time
{\redd when} one wants to estimate the phase, but also on later data.  This
implies, of course, that these smoothed estimates can only be
calculated
after a sufficient delay or at the end of the experiment, whereas
feedback itself at time $t$ can only use data from times $t'\le t$ (or
even $t'<t$ when considering finite propagation times).  Theory
predicts a reduction of
the mean square error by a factor 2$\sqrt{2}$ compared to the Standard Quantum Limit
(achievable by non-adaptive filtering, i.e.~without feedback and using
only previous data at any time), and an experimental
reduction of about $2.24\pm 0.14$ was achieved.
\\

Quantum error
correction  for quantum-enhanced measurements, recently
introduced in
\cite{kessler_quantum_2014,dur_improved_2014}, can also be seen in the
context of quantum feedback schemes \cite{PhysRevA.65.042301}.
Quantum error correction is one of the
most important ingredients of
quantum computing \cite{Shor95,Steane96,Gottesman96}.  The general idea,
both for quantum computing and
quantum-enhanced measurements, is that one would like to apply
recovery operations $\mathcal{R}$ to a state that after encoding the
desired information through an operation $\mathcal M$ has been
corrupted by a noise process $\mathcal{E}$, such that
$\mathcal{R}\circ\mathcal{E}\circ\mathcal{M}(\rho)\propto\mathcal{M}\rho$.
In \cite{kessler_quantum_2014} it was shown that this can be achieved
for the sensing
of a single qubit subject to dephasing noise if it is coupled to a pure
ancilla bit.  Syndrome measurements (i.e.~measurements of collective
observables which do not destroy the relevant phase information, a
concept developed in quantum error correction) of both qubits at a rate faster than
the dephasing rate allow one to detect whether a phase flip has
occurred and to correct it, extending in this way the coherence time
available for Ramsey interferometry to much longer times and thus better
maximum sensitivities.  When using $N$ qubits in parallel, an ancilla is not
necessary.  The method then operates directly on an initially
entangled state, such as the GHZ-state, and measures error-syndromes on
pairs of spins. Thus, the idea is here not so much to avoid
entanglement but rather to stabilize through rapid multi-spin
error-syndrome measurements the correct imprinting of the
information on the quantum state against unwanted decoherence
processes. For phase estimation on $N$ qubits evolving in parallel
under individual and identical Pauli rank-one noisy channels, fast
control schemes based on quantum error correction allow one to restore
the Heisenberg-limit by completely eliminating the noise, at the cost of slowing
down the unitary evolution by a constant factor, unless the noise is
dephasing noise 
that couples to the same Pauli-operator as the Hamiltonian generating
the phase shift
\cite{Sekatski2016}.

More generally, one can prove that sequential metrological schemes
involving an initial probe entangled with an ancilla, with the probe
undergoing $N$ passes of a transformation encoding the parameter of
interest, interspersed by arbitrary feedback control operations acting
on probe and ancilla, and followed by a joint measurement on the two
particles at the output, can outperform any parallel metrological
scheme relying on an initial $N$-particle entangled state
\cite{Maccone2014,Sekatski2016,Maccone2016,Yuan2015,Yuan2016}. In
particular, in \cite{Yuan2016} it was shown that a sequential
feedback scheme allows one to realize a joint quantum-enhanced
measurement of all the three
components of a magnetic field on a single-qubit probe. As remarked in
Sec.~\ref{sec:parvseq}, the use of sequential schemes assisted by
suitable control reduces the input demand from multipartite to
bipartite entanglement, resulting in a notable technological
advantage.

\section{Thermodynamical and non-equilibrium steady states}

In this section, we discuss precision parameter estimations when probes are
thermal states or non-equilibrium steady states of dissipative
dynamics. \red{These states have the advantage to be stationary, and describe
mesoscopic systems. From measurements on these probes, intensive parameters, like temperature, chemical potential, or couplings of Hamiltonians or of dissipators, are infered with a sensitivity given by the inverse of the quantum Fisher Information.} Thermal probes are crucial for both fundamental
issues and technological applications
\cite{Benedict1984,Childs2001,Giazotto2006}. Estimations with
dissipative dynamics \cite{Bellomo2009,Bellomo2010-1,Bellomo2010-2,Alipour2014,Zhang2015}
are also instances of process tomography
\cite{Mohseni2008,Merkel2013,Bendersky2013,Baldwin2014} with partial
prior knowledge.
\red{The identification of the quantum Fisher Information with the Bures metric clarifies the role
of criticality as a resource for estimation sensitivity. With
extensive, i.e. linear in the system size, interactions and away from
critical behaviors, the Bures distance between the probe state and its
infinitesimal perturbation is at most extensive. Critical behaviors,
e.g.~separation between different states of matter or long-range
correlations, are thus characterised by superextensive Bures metric
and the quantum Fisher Information. We thus focus on superextensivity of the quantum Fisher Information as a signature
of enhancements in precision measurements on thermodynamical and
non-equilibrium steady states.}
We
lay emphasis on highly sensitive probes that do not need to be
entangled, and in certain cases not even quantum.


\subsection{Thermodynamical states and thermal phase transitions}

Thermodynamic states at equilibrium are derived by the maximization of
the information-theoretic Shannon's entropy
\cite{Jaynes1957-1,Jaynes1957-2}, equivalent to the maximization of
the number of microscopic configurations compatible with physical
constraints. Given the probability distribution $\{p_j\}_j$ of a set
of configurations $\{j\}_j$, the constraints are the normalization
$\sum_jp_j=1$ and the averages of 
certain quantities $\langle F^{(k)}\rangle=\sum_jp_j f_j^{(k)}$,
$f_j^{(k)}$ being the values of the quantity $F^{(k)}$ corresponding
to the $j$-th configuration.
The solution of the maximization is the well-known Boltzmann-Gibbs distribution

\begin{equation}
p_j=\frac{e^{-\sum_k\theta_k f_j^{(k)}}}{Z}, \qquad Z=\sum_j e^{-\sum_k\theta_k f_j^{(k)}},
\end{equation}
where $Z$ is the partition function, and $\theta_k$ is the Lagrange
multiplier corresponding to the 
quantity $F^{(k)}$ fixed on average.

This formalism is equally adequate for both classical and quantum
thermodynamic systems. In the quantum case all the 
quantities $F^{(k)}$ are commuting operators, the configurations are
labeled by the set of eigenvalues of these operators
and possibly additional quantum numbers in the case of degeneracy,
and
the thermal state is the density matrix $\rho$ diagonal in the common
eigenbasis of the 
$F^{(k)}$ with eigenvalues $p_j$:

\begin{equation} \label{Gibbs.state}
\rho=\sum_j p_j|j\rangle\langle j|, \qquad F^{(k)}|j\rangle=f_j^{(k)}|j\rangle.
\end{equation}

Lagrange multipliers are the thermodynamic parameters to be
estimated. Since the density matrix in
\eqref{Gibbs.state} depends on them only through its eigenvalues $p_j$, the quantum Fisher matrix $I=[I_{\theta_k,\theta_{k'}}]_{k,k'}$ coincides with the classical Fisher matrix of the probability distribution $\{p_j\}_j$. A straightforward computation shows

\begin{equation} \label{thermal.Fisher.matrix}
I_{\theta_k,\theta_{k'}}=\frac{\partial^2\ln Z}{\partial\theta_k\partial\theta_{k'}}=\text{Cov}\Big(F^{(k)},F^{(k')}\Big).
\end{equation}
See also \cite{Jiang2014} for the computation of the quantum Fisher Information with density
matrices in exponential form. The diagonal element
$I_{\theta_k,\theta_k}$ is the largest inverse sensitivity for a
single estimation of the parameter $\theta_k$, while the Fisher matrix
$I$ bounds the inverse covariance matrix of the multiparameter
estimation, see eq.\eqref{eq:QCRBm}. The Cram\'er-Rao bound hence reads
\begin{equation}
\label{thermal.CRB}
\big[\text{Cov}(\theta_{k,\text{est}},\theta_{k',\text{est}})\big]\ \big[\text{Cov}\big(F^{(k)},F^{(k')}\big)\big]\geq\frac{1}{M},
\end{equation}
which is the uncertainty relation for conjugate variables in statistical mechanics, \reddd{see e.g.} \cite{Gilmore1985,Davis2012}.

The computation of the Fisher matrix \eqref{thermal.Fisher.matrix}, together with the Cram\'er-Rao bound \eqref{thermal.CRB}, tells us that the best sensitivity of Lagrange multipliers $\{\theta_k\}_k$ is inversely proportional to squared thermal fluctuations, and thus susceptibilities, \reddd{see e.g.} \cite{Reichl1998}. For connections among metric of thermal states, Fisher information, and susceptibilities see \cite{Weinhold1974,Ruppeiner1979,Ruppeiner1981,Salamon1984,Diosi1984,Mrugala1984,
Nulton1985,Janyszek1986,Janyszek1989,Janyszek1990,Ruppeiner1991,Ruppeiner1995,
Brody1995,Dolan1998,Janke2002,Janke2003,Brody2003,Crooks2007,Prokopenko2011,
Davis2012} for classical systems and \cite{Janyszek1986-2,Janyszek1990-2,Zanardi2007-3,You2007,Paunkovic2008-2,
Zanardi2008,Quan2009,marzolino_precision_2013,Marzolino2015} for quantum systems.

Due to the pairwise commutativity of the $F^{(k)}$, 
 the estimations of the parameters $\{\theta_k\}_k$ are reduced to
 parameter estimations with the classical probability distribution
 $\{p_j\}_j$. Thus, the maximum likelihood estimator is asymptotically
 unbiased and optimal, in the sense of achieving the Cram\'er-Rao
 bound, in the limit of infinitely many measurements
 \cite{Helstrom1976,Holevo1982}. This estimator consists in measuring
 each 
$F^{(k)}$ with outcomes $\{f_j^{(k)}\}_{j=1,\dots,M}$, and in maximizing the average logarithmic likelihood $\ell=\frac{1}{M}\ln\prod_j p_j$ with respect to the parameters $\{\theta_k\}_k$.

Among the most common statistical ensembles for equilibrium systems,
there are the canonical ensemble and the grandcanonical ensemble. The
canonical ensemble describes systems that only exchange energy with
their surrounding: the only 
quantity $F_1=H$ fixed on average is the
Hamiltonian, the Lagrange multiplier is
$\theta_1=\beta=\frac{1}{k_BT}$ where $k_B$ is the Boltzmann constant
and $T$ is the absolute temperature, and the Fisher information is
proportional to the heat capacity $C_V=\Big(\frac{\partial\langle
  H\rangle}{\partial T}\Big)_V$,
\begin{equation}
I_{\beta,\beta}=\text{Var}(H)=k_B T^2 C_V.
\end{equation}
The grandcanonical ensemble describes systems that exchange energy and
particles with the surrounding: the 
quantities fixed on average are the
Hamiltonian $F_1=H$ and the particle number $F_2=N$, with Lagrange
multipliers being the inverse temperature
$\theta_1=\beta=\frac{1}{k_BT}$ and $\theta_2=-\beta\mu$ where $\mu$
is the chemical potential. The Fisher information of temperature and
chemical potential are linked to thermal fluctuations, i.e.~heat
capacity $C_V=\Big(\frac{\partial\langle H\rangle}{\partial T}\Big)_V$
and isothermal compressibility
$\kappa_T=-\frac{1}{V}\Big(\frac{\partial V}{\partial P}\Big)_T$
respectively, where $V$ is the volume and $P$ is the pressure
\cite{marzolino_precision_2013,Marzolino2015}:

\begin{align}
I_{\beta,\beta}= & \text{Var}(\mu N-H)=\frac{\partial \, \mu\langle N\rangle}{\partial\beta}+k_B T^2 C_V \\
I_{\mu,\mu}= & \beta^2 \, \text{Var}(N)=\frac{\langle N\rangle^2}{\beta V} \, \kappa_T.
\end{align}

Another parameter that can be estimated within this framework is the
magnetic field. For certain classical magnetic or spin systems, the
interaction with a magnetic field $\bB$ is $\bB\cdot \bM$,  with $\bM$
being the total
magnetization. This interaction term can represent a contribution to
the Hamiltonian as well as additional ``
fixed-on-average quantities'' $\bB$ with
Lagrange multipliers $\beta \bM$. The magnetic field is also linked to
magnetic susceptibility $\chi=\frac{\partial\langle M\rangle}{\partial
  B}$ (where we consider for simplicity only a single component of
$\bB$ and $\bM$, $M\equiv M_z$, $\equiv Bz$):
\begin{equation}
I_{B,B}=\beta^2 \, \text{Var}(M)=\beta\chi,
\end{equation}
This picture of the magnetic field as a Lagrange multiplier is valid
also for coupling constants whenever the Hamiltonian is
$H=\sum_j\lambda_j H_j$, where $\beta\lambda_j$ is the Lagrange
multiplier of $H_j$. For general quantum systems, the
non-commutativity of magnetization or other Hamiltonian contributions
$H_j$ with the rest of the Hamiltonian gives rise to quantum phase
transitions that occur
also at zero temperature without thermal fluctuations.
The above considerations also apply to the so-called \emph{generalized
  Gibbs ensembles}, i.e.~with arbitrary 
fixed-on-average quantities
$F^{(k)}$, {\redd for which} estimations of parameters $\theta_k$ are under
experimental \cite{Langen2015} and theoretical \cite{Foini2017}
study.

Thermal susceptibilities are typically extensive except in the presence of phase transitions. Thus, their connection with Fisher information suggests that thermal states at critical points \cite{Reichl1998,Baxter,Diosi1984,Janyszek1989,Janyszek1990,Ruppeiner1991,
Ruppeiner1995,Brody1995,Dolan1998,Janke2002,Janke2003,Brody2003,Prokopenko2011} with divergent susceptibilities are probes for enhanced measurements. Thermal susceptibility divergences occur also in classical systems, proving precision measurements without entanglement.

\subsubsection{Role of quantum statistics}

We now discuss the estimation of Lagrange multipliers of quantum gases
in the grandcanonical ensemble and the role of quantum statistics
therein \cite{marzolino_precision_2013,Marzolino2015}. Consider ideal
gases in a  homogeneous or harmonic trap. $I_{\beta,\beta}$ is always
extensive in the average particle number $\langle N\rangle$.
The corresponding relative error found in \cite{marzolino_precision_2013,Marzolino2015}
for temperature estimation is still one order of magnitude smaller than
the standard deviations obtained experimentally via density
measurements of Bosons \cite{Leanhardt2003}
and Fermions \cite{Muller2010,Sanner2010}.

Estimations of chemical potentials are more sensitive to quantum
statistics than estimations of temperature,
 because the chemical potential is the conjugate Lagrange multiplier
 of the particle number which in turn reveals clear signatures of
 quantum statistics, such as bunching and antibunching. Effects of
 quantum statistics are more evident in quantum degenerate gases,
 i.e.~at low temperatures.

In Fermion gases, $I_{\mu,\mu}$ is extensive but diverges at zero temperature.
A change in the chemical potential corresponds to the addition or the
subtraction of particles, thus achieving a state orthogonal to the
previous one. This sudden state change makes the chemical potential
estimation very sensitive.
A generalization of the Cram\'er-Rao bound,
called Hammerseley-Chapman-Robbins-Kshirsagar bound that is
suitable for
non-differentiable statistical models \cite{Tsuda2005} must be
applied.
This may lead to superextensive $I_{\mu,\mu}$, depending on
the degree of rotational symmetry breaking or confinement anisotropy
and dimension (see appendix B in
\cite{marzolino_precision_2013}).

Bose gases undergo a phase transition to a Bose-Einstein condensate
in three
dimensions for homogeneous confinement, and in three or two dimensions
in a harmonic trap. Approaching
from above
the critical temperature, or zero
temperature with large density when there is no phase transition, $I_{\mu,\mu}$ is
superextensive: for homogeneous and harmonic trap respectively

\begin{align}
\label{idgas1}
I_{\mu,\mu}\lesssim &
\begin{cases}
\mathcal{O}\left(\beta^2\langle N\rangle^{\frac{4}{3}}\right) & \text{in three dimensions} \\
\mathcal{O}\left(\beta^2\frac{\langle N\rangle^2}{\log\langle N\rangle}\right) & \text{in two dimensions} \\
\mathcal{O}\left(\beta^2\langle N\rangle^2\right) & \text{in one dimension}
\end{cases}, \\
\label{idgas2}
I_{\mu,\mu}\lesssim &
\begin{cases}
\mathcal{O}\left(\beta^2\langle N\rangle\right) & \text{in three dimensions} \\
\mathcal{O}\left(\beta^2\langle N\rangle\log\langle N\rangle\right) & \text{in two dimensions} \\
\mathcal{O}\left(\beta^2\frac{\langle N\rangle^2}{\log\langle N\rangle}\right) & \text{in one dimension}
\end{cases}.
\end{align}

Below the critical temperature, the Bose-Einstein condensate phase depends on the
anisotropy of the external potential. If the gas is much less confined
along certain directions, the Bose-Einstein condensate is an effective low dimensional gas
with excitations restricted to directions along the less confined dimensions. A hierarchy of condensations to subsequent lower-dimensional gases is possible. These Bose-Einstein condensates have been studied both at finite size and in the thermodynamic limit focusing on mathematical structures and general properties \cite{Girardeau1960,Girardeau1965,Casimir1968,Krueger1968,Rehr1970,vandenBerg1982,
vandenBerg1983,vandenBerg1986-1,vandenBerg1986-2,Ketterle1996,vanDruten1997,Mullin1997,
Zobay2004,Beau2010,Mullin2012},
in connection with liquid helium in thin films \cite{Osborne1949,Mills1964,Douglass1964,Khorana1965,Goble67,Goble1965,Goble1966}, magnetic flux of superconducting rings \cite{Sonin1969}, and gravito-optical traps \cite{Wallis1996}. Experimental realizations of effective low dimensional Bose-Einstein condensates with trapped atoms have been reported in \cite{Gorlitz2001,Greiner2001,Esteve2006,vanAmerongen2008,vanAmerongen2008-2,
Bouchoule2011,Armijo2011}.
$I_{\mu,\mu}$ in these Bose-Einstein condensate phases is
superextensive and interpolates between the scaling above the critical
temperature and $I_{\mu,\mu}=\mathcal{O}\big(\beta^2\langle
N\rangle^2\big)$ for a standard Bose-Einstein condensate only
consisting of the ground state. The advantage of Bose-Einstein
condensate probes for precision estimations is that $I_{\mu,\mu}$ is
superextensive in the entire Bose-Einstein condensate phase and not
only at critical points as for susceptibilities of other thermal phase
transition.

In a mean field model with
interactions treated perturbatively, 
if the ideal system exhibits a superlinear scaling of the
Fisher information, the interaction strength $\lambda$ has to go to
zero for $N\to\infty$ for the perturbation theory to remain valid. In this
limit, the superlinear scaling disappears
for any non-zero value of $\lambda$, but at finite $N$ there are values of
$\lambda$ which do not destroy the sub-shot-noise.

Moreover, superextensive quantum Fisher Information
in one dimension at fixed volume $L_x$ and small contact interaction coupling $\frac{c}{L_x}$ results
\begin{align}
I_{\mu,\mu}\simeq & \frac{\beta^2\lambda_T^2\langle N\rangle^3}{2\pi L_x^2}+\beta^2\langle N\rangle-\frac{\beta^2\lambda_T^4\langle N\rangle^4}{8\pi^2 L_x^4}\left(1-e^{-\frac{4\pi}{\lambda_T^2\rho^2}\langle N\rangle}\right) \nonumber \\
& +c\Bigg(\frac{3\beta^3\lambda_T^6\langle N\rangle^7}{16\pi^3
  L_x^7}\left(1-e^{-\frac{4\pi}{\lambda_T^2\rho^2}\langle
  N\rangle}\right)\nonumber\\
&-\frac{\beta^3\lambda_T^4\langle N\rangle^6}{4\pi^2 L_x^4}\left(2+e^{-\frac{4\pi}{\lambda_T^2\rho^2}\langle N\rangle}\right)\Bigg),
\end{align}
where $\lambda_T$ is the thermal wavelength,
in agreement with experimental measurements on atom chips using $^{87}$Rb atoms \cite{Armijo2011}.
Superextensive grandcanonical fluctuations of particle
number, and thus superextensive $I_{\mu,\mu}$, have been observed in
a photon Bose-Einstein condensate, \reddd{see e.g.} \cite{Schmitt2014}, which can be realized at room temperature
\cite{Klaers2014}.

\subsubsection{Interferometric thermometry}

We now discuss a protocol for precision thermometry proposed in \cite{Stace2010}, using a Mach-Zehnder interferometer coupled to an ideal gas consisting of $N$ two-level atoms in the canonical ensemble.
The gas Hamiltonian is $H_0=\sum_{j=1}^N\epsilon \,
|\epsilon\rangle_j\langle\epsilon|$, where $|\epsilon\rangle_j$ is the
$j$-th particle excited state with single-particle energy $\epsilon$,
while the single-particle lowest energy is zero.
We skip the label $j$ at the bra-vector in the
projector for brevity.

The interferometer is injected with $K$ two-level atoms 
that interact with the gas in one arm of the interferometer with the interaction Hamiltonian
$H_I=\alpha\sum_{j=1}^K\sum_{l=1}^N|\epsilon\rangle_j\langle\epsilon|\otimes|\epsilon\rangle_l\langle\epsilon|$, where the index $j$ labels the atoms in the interferometer and $l$ refers to the atoms in the gas.
\red{In order for the interaction not to sensitively perturb the gas}, the
interferometer should be much smaller than the gas, thus $K\ll N$.
Each atom in the interferometers acquires a relative phase $\phi=\alpha m\tau$ between the arms, where $\tau$ is the interaction time and $m$ is the number of excited atoms in the gas whose expectation $\langle m\rangle=N/(1+e^{\beta\epsilon})$ depends on the temperature.

The inverse temperature can be estimated from the interferometric phase measurement with the sensitivity

\begin{equation}
\label{sensit.interf} \sigma(\beta)=\frac{\delta\phi}{\left|\alpha\tau\frac{d\langle m\rangle}{d\beta}\right|}=\frac{\big(1+e^{\beta\epsilon}\big)^2}{\epsilon \, e^{\beta\epsilon}N} \cdot \frac{\delta\phi}{\alpha\tau}\,,
\end{equation}
resulting from error propagation,
where $\delta\phi$ is the best sensitivity of the phase estimation
according to the quantum Cram\'er-Rao
bound.
The scaling with the number $K$ of probes comes from $\delta\phi$. For distinguishable atoms, separable states imply shot-noise $\delta\phi=1/K^{1/2}$, while sub-shot-noise can be achieved with separable states of
identical atoms as discussed in section \ref{sec.IdPar}, leading to $
\delta\beta\propto 1/K$ for ideal non-interacting bosons, or $
\delta\beta\propto 1/K^{p+1/2}$ for ideal fermions with a dispersion
relation of the probe atoms as discussed after \eqref{4.42}.
Note also that the scaling of the sensitivity \eqref{sensit.interf}
with respect to the particle number $N$ in the gas looks like a
Heisenberg scaling
irrespectively of
the probe state, and it is not possible to achieve this scaling
by direct measurement of the gas (i.e.~without any probe) because the
Fisher information is extensive except
at critical points.  What is conventionally considered is however the
scaling with the number of probes, in this case $K$, which can be
controlled.

Optimal thermometry with a single quantum probe (and hence
by definition without entanglement) was discussed in
\cite{Correa15}. It was found that the quantum Cram\'er-Rao
bound for $T$ of a thermalizing
probe reproduces the well-known
relation of temperature fluctuations to specific heat $C_V(T)$,
$(T/\text{Var}(T_\text{est}))^2\le C_V(T)$ (with Boltzmann constant
$k_B=1$, see also \cite{PhysRevE.83.011109}). The level structure of
the probe was then optimized to obtain maximum heat capacity and it
was found that the probe should have only two different energy levels,
with the highest one maximally degenerate and a non-trivial dependence
of the optimal energy gap on temperature (see also
\cite{reeb_tight_2015}).  The sensitivity of the probe increases with
the number of levels, but the role of the  quantumness of the initial
state of the probe on the sensitivity of thermometry is not fully
understood yet. Interferometric thermometry with a single probe was realized
experimentally in an NMR setup in \cite{Raitz2015} and the role of
quantum coherence emphasized. Experimental simulations in quantum
optical setups and
investigations of the role of quantum coherence
were reported in \cite{tham_simulating_2016,mancino_quantum_2016}. The
theoretical study in
\cite{PhysRevA.91.012331}  examined thermometry with two qubits.
Numerical evidence suggested that while initial quantum coherences can
improve the sensitivity, the optimal initial state is not maximally
entangled.

\subsection{Thermodynamical states and quantum phase transitions} \label{QPT}

Quantum phase transitions are sudden changes of the ground state for varying Hamiltonian parameters. If the Hamiltonian $H(\bm{\theta})=\sum_{j\geq 0} E_j(\bm{\theta})|E_j(\bm{\theta})\rangle\langle E_j(\bm{\theta})|$, with $E_j\leq E_{j+1}$, has a unique pure ground state $|E_0(\bm{\theta})\rangle$, the Fisher matrix \cite{You2007,Zanardi2007-2,CamposVenuti2007,Gu2009} is

\begin{equation} \label{QFI.ground}
I_{\theta_k,\theta_{k'}}=4 \, \text{Re}\sum_{j>0}\frac{\langle E_0|\partial_{\theta_k}H|E_j\rangle\langle E_j|\partial_{\theta_{k'}}H|E_0\rangle}{\left(E_j-E_0\right)^2}
\end{equation}
which follows from the differentiation of the eigenvalue equation $H|E_j\rangle=E_j|E_j\rangle$ or from the standard time-independent perturbation theory with respect to small variations $d\bm{\theta}$. The quantum Fisher Information is also expressed in terms of the imaginary time correlation function or dynamical response function \cite{You2007,CamposVenuti2007,Gu2009,You2015}, and has been used to count avoided crossings \cite{Wimberger2016}.

Equation \eqref{QFI.ground} tells us that the Fisher matrix can diverge only for divergent Hamiltonian derivatives or for gapless systems $E_1-E_0\to0$ in the thermodynamic limit. Thus, the divergence or the superextensivity of the quantum Fisher Information reveals a quantum phase transition but the converse does not hold: see \cite{Gu2010} for a review. Finite size scaling of the quantum Fisher Information \cite{CamposVenuti2007,Zanardi2008,Gu2009} can be derived using finite size scaling at criticality \cite{BrankovDanchevTonchev,Continentino}.

Superextensive quantum Fisher Information was observed at low order symmetry breaking quantum phase transitions,
topological quantum phase transitions, and gapless phases, but this criterion may fail at
high order symmetry breaking quantum phase transitions \cite{Tzeng2008-2} and
Berezinskii-Kosterlitz-Thouless (BKT) quantum phase transitions \cite{Chen2008,Sun2015}. We
now review quantum critical systems at zero temperature exhibiting
superextensive quantum Fisher Information of Hamiltonian parameters without entanglement.
We adapt and unify the notation of existing literature, \red{by writing
  down a general parameterized
  Hamiltonian whose different special cases are studied in the
  literature.}

\subsubsection{Quasi-free Fermion models}

\red{
We start with non-interacting many-body Hamiltonians, i.e.
\begin{equation} \label{free.ham}
H(\bm{\theta})=\sum_j\omega_j(\bm{\theta})a_j^\dag(\bm{\theta})a_j(\bm{\theta}),
\end{equation}
where $a_j^\dag$ ($a_j$) creates (annihilates) a Fermion in the $j$-th eigenmode.
The dependence of the eigenmodes $a_j^\dag,a_j$ on the parameters corresponds in second quantization to the dependence of the Hamiltonian eigenvectors on $\bm{\theta}$, as required for the ground state to be sensitive to variations of $\bm{\theta}$.
The eigenstates of \eqref{free.ham}, and thus thermal probes including the ground state, are not entangled in the eigenmodes. Therefore, measurements on the ground state provide parameter estimation without entanglement and with enhanced precision at phase transitions, as shown by the following examples.
}

\red{
Under Bogoliubov transformations the Hamiltonian \eqref{free.ham} can
be mapped into many models studied in literature. Bogoliubov
transformations preserve neither mode entanglement nor operation
locality, as discussed in Section \ref{sec.IdPar}. The relativity of
entanglement with respect to the basis of modes provides complementary
pictures of the Hamiltonian and of estimation protocols of its
parameters but does not change the physics: either the probe state is
entangled in one basis, or it is not in another and the enhanced
estimation precision is achieved by non-local measurements as a
consequence of long-range correlations at criticality. This situation
is reminiscent of the case of interferometry with identical particles,
discussed in Section \ref{sec.IdPar}, {\redd for which} rotations of
modes redistributes quantum resources between initial entanglement and
non-local interferometers. Moreover, Bogolibov transformations and
corresponding rotated modes are experimentally addressed in several
physical systems
\cite{Segovia1999,Vogels2002,Moritz2003,Davis2006,Robillard2008,
Sattler2011,InguscioFallani,Hu2014,Yan2016}, showing that quasi-particles represent legitimate and experimentally relevant subsystems.
}

\red{
Hamiltonians equivalent to \eqref{free.ham} under Bogoliubov transformations are quasi-free Fermion models \cite{Cozzini2007}, i.e. quadratic Hamiltonians in the creation $c_j^\dag$ and annihilation $c_j$ operators of $L$ Fermionic modes:
}

\begin{equation} \label{ham.quasifree.ferm1}
H_{\text{quasi-free}}=\sum_{j,l=1}^L\left(c_j^\dag A_{j,l}c_l+\frac{1}{2}\left(c_j^\dag B_{j,l}c_l^\dag+\text{h.c.}\right)\right),
\end{equation}
where $L$ is a measure of the system volume.
Translationally 
 invariant Hamiltonians \eqref{ham.quasifree.ferm1} with
periodic boundary conditions and tunneling in the modes
$\{c_j,c_j^\dag\}_j$ of range $r$ have

\begin{align} \label{Ham.var.range}
A_{j,l}= & \, (J-\mu)\delta_{j,l}-J\theta(r-|j-l|)-J\theta(|j-l|-L+r), \nonumber \\
B_{j,l}= & \, J\gamma \, \text{sign}(j-l)\big(\theta(r-|j-l|)-\theta(|j-l|-L+r)\big),
\end{align}
where $J>0$, $\text{sign}(0)=0$, and $\theta(\cdot)$ is the unit step
function with $\theta(0)=1$. Here $J$ corresponds to a tunneling
energy between different sites, and $J\gamma$ to an effective
interaction; $\mu$ multiplies the total particle number and hence
corresponds to a chemical potential.
This Hamiltonian can be analytically diagonalized
\cite{Cozzini2007}. For large $L$ and a fully connected system,
$r=\big\lfloor\frac{L}{2}\big\rfloor$ with periodic boundary conditions,
the Fisher matrix with respect
to $(\mu,\gamma)$ is

\begin{equation}
I_{\mu,\mu}=\gamma^2 S, \quad I_{\gamma,\gamma}=(\mu-J)^2 S, \quad I_{\mu,\gamma}=-(\mu-J)\gamma S,
\end{equation}
with

\begin{equation} \label{scaling.fully.connected}
S\simeq
\begin{cases}
\displaystyle \frac{L}{2} \left|\frac{J}{(\mu-J)\gamma}\right|
\frac{1}{\big(|\mu-J|+|J\gamma|\big)^2} & \text{if } (\mu\neq
J,\gamma\neq0) \\
\displaystyle \frac{L^2}{3J^2\gamma^4} & \text{if } (\mu=J,\gamma\neq0) \\
\displaystyle \frac{L^2 J^2}{(\mu-J)^4} & \text{if } (\mu\neq
J,\gamma=0)
\end{cases}
\end{equation}
The lines $(\mu=J,\gamma\neq0)$ and $(\mu\neq J,\gamma=0)$ reveal second-order quantum phase transitions, as well as superextensivity of the Fisher matrix in the volume $L$. At the critical line $\gamma=0$, the ground state is unentangled not only in the eigenmodes but also in the original modes $\{c_j,c_j^\dag\}_j$.

The fully connected translationally invariant Hamiltonian with open boundary conditions reads

\begin{equation}
A_{j,l}=(\mu-J)\delta_{j,l}+J, \qquad B_{j,l}=J\gamma \, \text{sign}(l-j),
\end{equation}
with second-order quantum phase transitions at lines $\mu=J$ and $\gamma=0$, and
superextensive Fisher matrix as for periodic boundary conditions with
different prefactors \cite{Cozzini2007,Zanardi2007}.
For instance, at $(\mu>J,\gamma=0)$

\begin{equation}
I_{\mu,\mu}=0, \qquad I_{\gamma,\gamma}=\frac{L^2 J^2}{3(\mu-J)^2}, \qquad I_{\mu,\gamma}=0.
\end{equation}
Another interesting case is the Hamiltonian with nearest neighbor tunneling in the modes $\{c_j,c_j^\dag\}_j$, periodic boundary conditions, and $J>0$:

\begin{align} \label{Ham.nn.ferm}
A_{j,l}= & \, (J-\mu) \, \delta_{j,l}-J \, \theta(1-|j-l|), \nonumber \\
B_{j,l}= & \, J\gamma \, \text{sign}(l-j) \, \theta(1-|j-l|),
\end{align}
Such Hamiltonian \cite{Zanardi2008} is also equal to

\begin{equation} \label{quasifree.ferm.spin.ham}
H_{\text{quasi-spin}}=\sum_{n=1}^{\left\lfloor\frac{L}{2}\right\rfloor}\big(-\epsilon_n\sigma_n^z+\Delta_n\sigma_n^y\big),
\end{equation}
with $\epsilon_n=-J\cos\left(\frac{2\pi n}{L}\right)-\frac{\mu}{2}$,
 $\Delta_n=-J\gamma\sin\left(\frac{2\pi n}{L}\right)$, and $\{\sigma_n^{y,z}\}_n$ are Pauli matrices on $n$ orthogonal $\mathbbm{C}^2$ subspaces. Eq.~\eqref{quasifree.ferm.spin.ham}
provides an alternative representation of the Hamiltonian in terms of
non-interacting quasi-spins, and its eigenstates are separable with
respect to both the eigenmodes and the quasi-spins. The same quantum Fisher Information
scaling holds in both representations as theoretical independent
models.

The quantum Fisher Information is linear in $L$ away from the critical points, but superextensive at criticality in the leading order for large $L$ \cite{Zanardi2006,Zanardi2008}:

\begin{align}
\label{I_J.nn.ferm} & I_{J,J}\big(|\mu|=2J,\gamma\big)=\mathcal{O}\left(\frac{L^2}{J^2\gamma^2}\right), \\
\label{I_gamma.nn.ferm} & I_{\gamma,\gamma}\big(|\mu|\leq 2J,\gamma=0\big)=\mathcal{O}\left(L^2\right).
\end{align}
The origin of the superextensivity of $I_{J,J}$ stems from the fact
that the symmetric logarithmic derivative, and thus the optimal
estimation of $J$, is close to a single particle operator in the
Fermion representation away from the critical point, but 
is a
genuine multi-particle operator close to the critical point. In the
canonical ensemble,
at $\left|\frac{\mu}{2}-J\right|\lesssim\frac{1}{\beta}$, the quantum Fisher Information has a
divergence around zero temperature,
$I_{J,J}=\mathcal{O}\big(\frac{L\beta}{|J\gamma|}\big)$.
The superextensivity of $I_{J,J}$, together with the divergence of the
derivative of the geometric phase, is also a universal feature
depending only on the slope for the closing of the energy gap between
one and zero Fermion occupations of certain eigenmodes
\cite{Cheng2017}.

Superextensive quantum Fisher Information is observed also for tight-binding electrons on the triangular lattice with magnetic flux $\frac{\phi}{2}$ within each triangle, hopping constants $t_a$ ($t_b$) at edge along the $x$ ($y$) direction and $t_c$ at the third edge \cite{Gong2008}. Assuming zero momentum in the $y$ direction \cite{Ino2006}, the Hamiltonian can be transformed into \eqref{ham.quasifree.ferm1}, with $B_{j,l}=0$ and

\begin{align}
A_{j,l}= & -2t_b\cos\left(2\pi\phi j\right)\delta_{l,j}-\left(t_a+t_c e^{-2\pi i\phi\left(j-\frac{1}{2}\right)}\right)\delta_{l,j-1} \nonumber \\
& -\left(t_a+t_c e^{2\pi
    i\phi\left(j+\frac{1}{2}\right)}\right)\delta_{l,j+1},
\end{align}
with $j$ and $l$ labeling the sites in the $x$ direction. Consider
$L=F_m$, the $m$-th Fibonacci number, and
$\phi=\frac{F_{m-1}}{F_m}\xrightarrow[m\to\infty]{}\frac{\sqrt{5}-1}{2}$,
the inverse of the golden ratio in the thermodynamic limit. At the
critical line $t_c=t_a$, numerical computations result in
$I_{t_c/t_a}=\mathcal{O}\big(L^{4.9371}\big)$ if $m=3L+1$, 
otherwise
$I_{t_c/t_a}=\mathcal{O}\big(L^{2.0}\big)$. $I_{t_b/t_a}$ has the same
size dependence at the critical line $2t_b=t_a$.
A more complicated quasi-free Fermion Hamiltonian describes a
superconductor with a magnetic impurity \cite{Paunkovic2008}. A sudden
increase of the Bures distance between two reduced ground states of
few modes around the impurity at very close exchange interaction with
the impurity was numerically observed, but it is unclear whether the
quantum Fisher Information with respect to exchange interaction is superextensive.

\subsubsection{Hubbard models}

Another class of Fermionic systems is described by $L$-mode Hubbard Hamiltonians

\begin{align} \label{Hubbard}
H_{\text{HM}}= & -\sum_{\substack{j=1,\dots,L \\ \sigma=\uparrow,\downarrow}}t_{\sigma}\left(c_{j,\sigma}^\dag c_{j+1,\sigma}+\text{h.c.}\right) \nonumber \\
& +U\sum_{j=1}^L n_{j,\uparrow}n_{j,\downarrow}-\mu\sum_{j,\sigma}n_{j,\sigma},
\end{align}
with $n_{j,\sigma}=c_{j,\sigma}^\dag c_{j,\sigma}$, $t_{\sigma}$ the
hopping constants, $U$ the interaction strength, and $\mu$ the
external potential. When $t_{\uparrow}=t_{\downarrow}$, the system
undergoes a Berezinskii-Kosterlitz-Thouless (BKT) quantum phase transition at $U=0$ and
half filling $n=\frac{1}{L}\sum_{j,\sigma}\langle
n_{j,\sigma}\rangle=1$. $I_U$ is extensive, but $I_U/L$ diverges as
$1/n$ for $n\to0$ and $U=0$, and as $1/U^4$ around the BKT critical
point only if the system size is much larger than the correlation
length
\cite{CamposVenuti2008}. At zero interaction \eqref{Hubbard} is a free Fermion Hamiltonian, and thus the ground state is not entangled in the eigenmodes.

In the large $U$ limit and at $n=\frac{2}{3}$, a quantum phase transition occurs with control parameter $t_{\downarrow}/t_{\uparrow}$. The quantum Fisher Information $I_{t_{\downarrow}/t_{\uparrow}}$ at critical points is superextensive $\mathcal{O}\big(L^\alpha\big)$, where the exponent was numerically computed $\alpha\simeq 5.3$ \cite{Gu2008}. In the large $U$ limit, the eigenstates of the Hamiltonian are perturbations of those of the interaction term, that are Fock states, thus with vanishingly small entanglement with respect to the modes $\{c_{j,\sigma},c_{j,\sigma}^\dag\}_{j,\sigma}$.

\subsubsection{Spin-$1/2$ systems}

We now discuss systems of $N$ spins-$1/2$, which provide alternative
representations of quasi-free models. Consider the following complete
Hamiltonian, to be specialised later on,
\begin{align} \label{spin.ham}
H_{\text{spin}}= &
                   -J\sum_{j=1}^{N-1}\Big(\frac{1+\gamma}{2}\sigma_j^x\sigma_{j+1}^x+\frac{1-\gamma}{2}\sigma_j^y\sigma_{j+1}^y\nonumber\\
& +\Delta\sigma_j^z\sigma_{j+1}^z\Big) \nonumber \\
& +d\sum_{j=1}^{N-1}\big(\vec\sigma_j\wedge\vec\sigma_{j+1}\big)^z-\sum_{j=1}^N\big(h-gj\big)\sigma_j^z,
\end{align}
with anisotropies $\gamma$ and $\Delta$, Dzyaloshinskii-Moriya coupling
$d$, magnetic field with uniform value $h$ and gradient $g$,
and $\vec\sigma_j=\big(\sigma_j^x,\sigma_j^y,\sigma_j^z\big)$.

For the moment, we focus on the XY model with transverse field,
i.e.~$\Delta=d=g=0$, with $J>0$ and periodic boundary conditions. This
Hamiltonian can be transformed into \eqref{ham.quasifree.ferm1} with
\eqref{Ham.nn.ferm}, $\mu=2h$, and $L=N$ using the Jordan-Wigner
transformation, \reddd{see e.g.} \cite{Giamarchi}. Thus, the XY Hamiltonian provides an
alternative physical setting to implement precision metrology without
entanglement in the Fermionic eigenmodes.

At zero temperature, $I_{h,h}=\big(\frac{J^2}{h^2}\big)I_{J,J}$, since the ground state depends on $h$ and $J$ only via the ratio $\frac{h}{J}$. The Fisher matrix is extensive with divergent prefactors around critical regions \cite{Zanardi2007-2,Cheng2017}, and is superextensive at criticality as in (\ref{I_J.nn.ferm},\ref{I_gamma.nn.ferm}).

In the Ising model, i.e.~$\gamma=1$,
corrections to the quantum Fisher Information around the critical points $|h|=J$ \cite{Chen2008,Zhou2008,Zhou2008-2,You2015} and at small temperature \cite{Zanardi2007-3,Invernizzi2008} are also linear in $N$ with divergent prefactors. $I_{h,h}$ was also explicitly computed in \cite{Damski2013,Damski2014-2,You2015}. A good estimator of the parameter $J$ is the value inferred by measurements of the total magnetization, at least at small system size \cite{Invernizzi2008}.

The divergence of $I_{h,h}/N$ was numerically observed also in the XX model, i.e. $\gamma=0$, approaching the critical field $|h|=J$, with the reduced state of spin blocks \cite{Sacramento2011}. This implies precision estimations of the magnetic field looking only at a part of the system.

Divergent $I_{h,h}/N$ at critical field $|h|=J$ was numerically computed also when an alternating magnetic field $\sum_{j=1}^N(-)^{j+1}\delta\sigma_j^z$ is added to the XY Hamiltonian \cite{You2015}. The corresponding quasi-free Fermion Hamiltonian has the additional term $\sum_{j=1}^N(-)^{j+1}\delta c_j^\dag c_j$.

The XY and the Ising models are also interesting because the
superextensivity of the Fisher matrix
$I=[I_{\theta,\theta'}]_{\theta,\theta'=h,\gamma}$ is robust against
disorder, i.e.~with Hamiltonian parameters being Gaussian random
variables \cite{Garnerone2009-2}. In the presence of disorder, quantum phase transitions
are broadened to
Griffiths phases \cite{Griffiths1969,Fisher1992,Fisher1995,Sachdev}. Thus, the exponent of $N$ in average $I_{h,h}$ and $I_{\gamma,\gamma}$ is slightly reduced but the superextensivity is broadened in the parameter range.
Superextensivity of $I_{h,h}$ at the critical Ising point $|h|=J$
becomes a broad peak, and the superextensive $I_{\gamma,\gamma}$ at
the $\gamma=0$ critical line is split into two symmetric broad peaks
around $\gamma=0$, where at $\gamma=0$
a local minimum with extensive $I_{\gamma,\gamma}$ arises.
Furthermore, the scaling of $I_{h,h}$ of the XY model at the critical point $|h|=J$ is preserved when a periodic time-oscillating transverse magnetic field is considered, and the state is the Floquet time-evolution of the ground state at $t=0$ up to times that scale linearly with the system size for small time-dependent driving \cite{Lorenzo2017}.
These robustness features are suitable for practical implementations of precision measurements.

Another interesting system is the XXZ model, i.e.~$\gamma=0$ and $J<0$, whose low-energy spectrum for $|\Delta|<\frac{1}{2}$ is equivalent to a quasi-free Boson Hamiltonian known as Luttinger liquid, \reddd{see e.g.} \cite{Giamarchi}. Thus, superextensive ground state quantum Fisher Information implies precision measurements without entanglement in the Boson eigenmodes.
Given the Luttinger liquid parameter
$K=\frac{\pi}{2\arccos(-\frac{2\Delta}{\sqrt{1+2d/J}})}$,
superextensive quantum Fisher Information $I_d$ is observed with periodic boundary conditions at $d=h=g=0$,

\begin{equation}
I_d=
\begin{cases}
\mathcal{O}\left(\frac{N\ln N}{J^2}\right) & \textnormal{if } \Delta=\frac{1+\sqrt{5}}{8} \\
\mathcal{O}\left(\frac{N^{6-8K}}{J^2}\right) & \textnormal{ if }
\frac{1+\sqrt{5}}{8}<\Delta< \frac{1}{2} \\
\mathcal{O}\left(\frac{N}{J\ln N}\right)^2 & \textnormal{if } \Delta=\frac{1}{2}
\end{cases},
\end{equation}
while, with open boundary conditions, $I_d=\mathcal{O}\big(\frac{J^2 K
  N^2}{(J^2+4d^2)^2}\big)$ at $h=g=0$
and
$I_g=\mathcal{O}\big(\frac{KN^4}{J^2 u^2}\big)$,
with $u=\frac{\pi\sqrt{1-4\Delta^2}}{2\arccos(2\Delta)}$, at $d=h=g=0$
\cite{Greschner2013}. Here, the quantum Fisher Information is superextensive mainly because the system
is gapless, while the quantum phase transition only affects subleading in $N$, even though superextensive, orders of the quantum Fisher Information.
$I_d$ is also equal to the quantum Fisher Information $I_\phi$ with respect to a twist phase
$\phi$ of spin operators
$\sigma_j^+\sigma_{j+1}^-\to\sigma_j^+\sigma_{j+1}^-e^{i\phi}$
\cite{Thesberg2011}. Via the Jordan-Wigner transformation, the
$\Delta=\gamma=0$ case in \eqref{spin.ham}
is equivalent to a quasi-free Fermion Hamiltonian with matrix elements
given by \eqref{Ham.nn.ferm} with $\gamma=\mu=0$
and with the
Dzyaloshinskii-Moriya and the gradient field terms
$-\frac{id}{2}\sum_{j=1}^N c_j^\dag c_{j+1}+\textnormal{h.c.}$ and
$\sum_{j=1}^N gj c_j^\dag c_j$, respectively. $I_d$ and $I_g$ have the
same scaling in a generalization of the latter model with spin-$1/2$
Fermions \cite{Greschner2013}, \colb{providing a further} example of precision
measurements without Fermion entanglement. \\

Another model, mapped to a quasi-free Fermionic Hamiltonian and
exhibiting superextensive quantum Fisher Information, is the quantum compass chain with
periodic boundary conditions \cite{Motamedifar2013}
 \begin{equation}
H_{\text{QCC}}=\sum_{j=1}^{\frac{N}{2}}\left(\sum_{\alpha=x,y}J_\alpha\sigma_{2j-1}^\alpha\sigma_{2j}^\alpha+
J_z\sigma_{2j}^z\sigma_{2j+1}^z\right)-h\sum_{j=1}^N\sigma_j^y.
\end{equation}
Quantum phase transitions occur at critical fields $h_{1,2}=\sqrt{J_x(J_y\pm J_z)}/2$ when these values are real. Numerical computations show that $I_h=\mathcal{O}\big(N^\alpha\big)$ is superextensive with

\begin{equation}
\alpha\simeq
\begin{cases}
1.80\pm 0.02 & \text{for } \frac{J_x}{J_z}>0, \, \frac{J_y}{J_z}>1 \text{ and } h=h_1 \\
1.98\pm 0.02 & \text{for } \frac{J_x}{J_z}>0, \, \frac{J_y}{J_z}>1 \text{ and } h=h_2 \\
1.94\pm 0.02 & \text{for } \frac{J_x}{J_z}<0, \, \frac{J_y}{J_z}<1 \text{ and } h=h_2 \\
2.02\pm 0.02 & \text{for } \frac{J_x}{J_z}>0, \, \frac{J_y}{J_z}<1 \text{ and } h=h_1 \\
\end{cases}.
\end{equation}

\subsubsection{Topological quantum phase transitions}

The quantum Fisher Information is superextensive also in topological quantum phase transition which have non-local order parameters \cite{Zeng2015}. Mosaic models defined on two-dimensional lattices with trivalent vertices, i.e.~each vertex is the border among three polygons, and with three-body interaction were numerically studied. The $N$-particle Hamiltonian is

\begin{equation}
H_{\textnormal{mosaic}}=-\sum_{\substack{\alpha=x,y,z\\(j,l)\in S(\alpha)}}J_{j,l}^\alpha \, \sigma_j^\alpha\sigma_l^\alpha-K\sum_{j,l,k}\sigma_j^x\sigma_l^y\sigma_k^z,
\end{equation}
where $S(\alpha)$ is the set of edges in the $\alpha\in\{x,y,z\}$ direction. $H_{\textnormal{mosaic}}$ can be mapped onto a free Majorana Fermion Hamiltonian, thus without entanglement in Fermion eigenmodes. When the edge numbers of the three polygons are $(4,8,8)$, $J_{j,l}^{x,y}=J$, $J_{j,l}^z=J_z$, and $K=0$, the quantum Fisher Information is $I_{J_z}=\mathcal{O}\big(N^{1.07615\pm0.00005}\big)$ at the critical point $J_z=\sqrt{2} \, J>0$ \cite{Garnerone2009}. When the edge numbers are $(3,12,12)$, $J_{j,l}^{x,y,z}=J$ for edges within triangular elementary subcells \cite{Yao2007}, $J_{j,l}^{x,y,z}=J'$ for other links, and $K=0$, the quantum Fisher Information is $I_{J'}=\mathcal{O}\big(N^{1.078\pm0.005}\big)$ at the critical point $J'=\sqrt{3} \, J>0$ \cite{Garnerone2009}. $H_{\textnormal{mosaic}}$ on the honeycomb lattice with $J_{j,l}^\alpha=J_\alpha$ has topological quantum phase transition at the boundaries $|J_x|=|J_y|+|J_z|$, $|J_y|=|J_z|+|J_x|$, and $|J_z|=|J_x|+|J_y|$ with superextensive quantum Fisher Information, e.g. $I_{J_x}=\mathcal{O}(N\ln N)$ for $J_y=J_z=\frac{J_x}{2}$ and $K=0$ \cite{Zhao2009}, and $I_{J_z}=\mathcal{O}\big(N^{1.08675\pm0.00005}\big)$ for $J_z=\frac{J}{2}$, $J_x=J_y=\frac{J-J_z}{2}$ and $K=\frac{1}{15}$ \cite{Garnerone2009}.

Topological quantum phase transition are particularly relevant because of superextensive quantum Fisher Information in gapless phases and not only at critical points. In the honeycomb lattice with $K=0$, $J_x+J_y+J_z=J$ and $J_x=J_y$, the quantum Fisher Information is $I_{J_z}=\mathcal{O}\big(N^{1.2535\pm0.00005}\big)$ at the critical point $J_z=\frac{J}{2}$ \cite{Yang2008}, and $I_{J_z}=\mathcal{O}(N\ln N)$ in the gapless phase $J_z<\frac{J}{2}$ \cite{Gu2009}. Superextensivity in the gapless phase is due to the algebraic decay of the correlation function of the $z$-edge, in contrast to exponential decay in the gapped phase $J_z>\frac{J}{2}$ leading to extensive quantum Fisher Information.

\subsection{Non-equilibrium steady states}

Consider now parameter estimation with probes in non-equilibrium steady states of spin-$1/2$ chains with boundary noise \cite{Zunkovic2010,Prosen2015,Marzolino2016} described by the Markovian master equation \cite{Breuer02,Benatti05}

\begin{align}
\frac{\partial\rho_t}{\partial t}= & -i[H_{\text{XYZ}},\rho_t] \nonumber \\
& +\lambda\sum_{\substack{\alpha=1,2 \\ \j=1,N}}\left(\mathcal{L}_{\alpha,j}\rho_t\mathcal{L}_{\alpha,j}^\dag-\frac{1}{2}\left\{\mathcal{L}_{\alpha,j}^\dag\mathcal{L}_{\alpha,j},\rho_t\right\}\right),
\end{align}
with XYZ Hamiltonian, i.e.~\eqref{spin.ham} with $d=g=0$,
and Lindblad operators $\mathcal{L}_{1(2),j}=\sqrt{(1\pm\mu)/2} \, \sigma_j^\pm$.

Superextensive quantum Fisher Information of the non-equilibrium steady states $\rho_\infty$ is observed at
non-equilibrium phase transition and in phases with long-range correlations. In the
XY model, $\Delta=0$, superextensive quantum Fisher Information was computed at the critical
lines $h=0$ ($I_{h,h}=\mathcal{O}(N^6)$) and $\gamma=0$ with
$|h|<J|1-\gamma^2|$ ($I_{\gamma,\gamma}=\mathcal{O}(N^2)$), at the
critical points $|h|=J|1-\gamma^2|$
($[I_{\theta,\theta'}]_{\theta,\theta'=h,\gamma}=\mathcal{O}(N^6)$),
and in the phase with long-range correlations $|h|<J|1-\gamma^2|$
($[I_{\theta,\theta'}]_{\theta,\theta'=h,\gamma}=\mathcal{O}(N^3)$)
\cite{Banchi2014}.  

In the XXZ model $\gamma=0$, the quantum Fisher Information $I_\Delta$ is superextensive in the limit of small $\frac{\lambda}{J}$ and for $|\Delta|\leq\frac{1}{2}$ at irrational $\frac{\arccos\Delta}{\pi}$ \cite{Marzolino2014,Marzolino2016-2,Marzolino2017}.
If $\frac{\arccos\Delta}{\pi}$ is rational
$I_\Delta=\frac{\lambda^2\mu^2}{J^2}(\tilde\xi N^2+\xi N)$ where $\xi$
and $\tilde\xi$ are constants in $N$ and for
$\frac{\lambda}{J}<\frac{1}{\sqrt{N}}$, thus the quantum Fisher Information is not
superextensive. Nevertheless, after inserting the value of $\Delta$ in
$\frac{\arccos\Delta}{\pi}$ and reducing the fraction to lowest term
$\frac{\arccos\Delta}{\pi}=\frac{q}{p}$ with coprime integers $q,p$,
one realizes that the coefficient $\xi$ is unbounded when the
denominator $p$ grows.
Therefore, $I_\Delta$, as a function of $\Delta$, exhibits a fractal-like structure with a different size scaling at irrational $\frac{\arccos\Delta}{\pi}$.
Moreover, the limit of $\frac{\arccos\Delta}{\pi}$ approaching irrational numbers from rationals and the thermodynamic limit do not commute. If the thermodynamic limit is first performed then
$I_\Delta=\mathcal{O}\left(\frac{\lambda^2\mu^2}{J^2}N^{\sim 5}\right)$ with $\frac{\lambda}{J}<\frac{1}{\sqrt{N}}$. If the thermodynamic limit is postponed after the limit to irrational $\frac{\arccos\Delta}{\pi}$ then the
quantum Fisher Information is still fitted by a superextensive power law with exponent depending on the value of
$\Delta$,
e.g.~$I_\Delta=\mathcal{O}\left(\frac{\lambda^2\mu^2}{J^2}N^{2.32788\pm0.0009}\right)$
with $\frac{\lambda}{J}<\frac{1}{\sqrt{N}}$ and
$\frac{\arccos\Delta}{\pi}$ being the golden ratio.
When $\Delta=\frac{1}{2}$,
$I_\Delta=\mathcal{O}\left(\frac{\lambda^2\mu^2}{J^2}N^4\right)$ with
$\frac{\lambda}{J}<\frac{1}{N}$ in both cases. 

In addition, the quantum Fisher Information of the reduced state of a single spin at position $k$ scales superextensively also for arbitrary dissipation strength $\lambda$ but only at the critical points $|\Delta|=1$,
and with a power law depending on the position $k$ of the spin \cite{Marzolino2017}:
e.g.~at $\lambda=1$, $I_\Delta=\mathcal{O}(N^{\sim 2})$ for $k=1$,
$k=\lfloor\frac{N}{2}\rfloor$, or $k=N$, and
$I_\Delta=\mathcal{O}(N^{\sim 4})$ for $k=\lfloor\frac{N}{4}\rfloor$ or
$k=\lfloor\frac{3N}{4}\rfloor)$.
This proves that the anisotropy at $|\Delta|=1$ can be
precisely estimated measuring single spin magnetisations along the $z$ axis, or measuring the magnetisations $\sum_{j\in\mathcal{P}}\langle\sigma_j^z\rangle$ for any non-centrosymmetric portion $\mathcal{P}$ of the chain, or $\sum_{j\in\mathcal{P}}f\left(\langle\sigma_j^z\rangle\right)$ with even functions $f(\cdot)$ for any set $\mathcal{P}$, even centrosymmetric ones.

For either small $\lambda$ or small $\mu$ the steady states $\rho_\infty$ of the above models are perturbations of the completely mixed state and thus not entangled, \reddd{see e.g.} \cite{Bengtsson06}. In the XX model $\gamma=\Delta=0$, there is no nearest neighbor spin entanglement for a wide range of parameters \cite{Znidaric2012}. Non-equilibrium steady state probes are favorable because the quantum Fisher Information is superextensive in a whole phase and not only at exceptional parameters. Moreover, the distinguishability of non-equilibrium steady states via Fisher information, thus detectability of non-equilibrium criticality and metrological performances, are enhanced compared to thermal equilibrium systems: e.g. the BKT quantum phase transition at $\Delta=\frac{1}{2}$ in the ground state of the XXZ model does not correspond to superextensive quantum Fisher Information \cite{Chen2008,Sun2015}.

\subsection{Adaptive measurements}\label{sec.adaptive}

Since critical points without critical phases are isolated values, they should be known in advance in order to set the system at criticality and benefit of superextensive quantum Fisher Information. A partial solution is an adaptive approach \cite{Mehboudi2016} that we generalize to phase transitions under reasonable conditions.
The idea is to perform several estimates changing the thermal state or the non-equilibrium steady state, in particular the critical point, at each step according to previous estimates, in order to approach the phase transition ensuring enhanced sensitivity of the control parameter estimation.

Consider any phase transition with control parameter $\theta$
to be estimated, and
assume the quantum Fisher Information $I_{\theta-\theta_c}=\frac{\xi N}{|\theta-\theta_c|^\alpha}$ close
to a critical point $\theta_c$ with $\alpha>0$ and a prefactor
$\xi$. We use the notation $I_{\theta-\theta_c}$ to make clear that
the quantum Fisher Information depends on the difference $\theta-\theta_c$, rather than on
$\theta$ alone.  Sub-shot-noise sensitivity is shown only at
$\theta=\theta_c$. Assume also that $\theta$ is initially known within a fixed
interval
$\theta\in[\theta_{\textnormal{min}},\theta_{\textnormal{max}}]$
enclosing the value $\theta_c$ which can be controlled by other system
parameters. First, set the critical point to
$\theta_c^{(1)}=\theta_{\textnormal{max}}$,
ensuring that $\theta<\theta_c^{(1)}$, i.e.~one is on a well-defined
side of the phase transition.
Find a first estimate $\theta_\text{est}^{(1)}$
for the parameter $\theta$, with an uncertainty that saturates the quantum Cram\'er-Rao
bound.   Use therefore a number
of measurements $M$ that ensures a small error compared to the
original confidence interval,
$\sigma\big(\theta_\text{est}^{(1)}\big)=\left(\frac{1}{M
I_{\theta-\theta_c^{(1)}}}\right)^{1/2}\ll (\theta_\text{max}-\theta_\text{min})$,
where once more
$\sigma\big(\theta_\text{est}^{(1)}\big)=\text{Var}\big(\theta_\text{est}^{(1)}\big)^{1/2}$
is the standard deviation of the estimate.
Then, update the critical parameter to
$\theta_c^{(2)}=\theta_\text{est}^{(1)}+\sigma\big(\theta_\text{est}^{(1)}\big)$. Since
$\sigma\big(\theta_\text{est}^{(1)}\big)\ll
(\theta_\text{max}-\theta_\text{min})$, the new critical point
$\theta_c^{(2)}$ is now
much closer to the true value of $\theta$ than $\theta_c^{(1)}$, assuming that
the obtained
estimate $\theta_\text{est}^{(1)}$ (which is random and only on
average agrees for an unbiased estimator with $\theta$) is indeed
within an interval of order $\sigma(\theta_\text{est}^{(1)})$ of the
true $\theta$.  Hence, in the next round, the quantum Fisher Information should be
substantially larger.
Perform then again sufficiently
many times a POVM that allows saturating the quantum Cram\'er-Rao
bound for a new estimate
$\theta_\text{est}^{(2)}$,
$\sigma^2\big(\theta_\text{est}^{(2)}\big)=\frac{1}{M
  I_{\theta-\theta_c^{(2)}}}\simeq\big(\frac{(\sigma(\theta_\text{est}^{(1)})^\alpha}{\xi
  MN}\big)\propto(MN)^{-\frac{2+\alpha}{2}}$. After $k$
iterations, the sensitivity of the estimate $\theta^{(k)}$ saturating the
quantum Cram\'er-Rao
bound is
\begin{equation} \label{adapt}
\sigma^2\big(\theta^{(k)}_\text{est}\big)=\frac{1}{M I_{\theta-\theta_c^{(k)}}}\propto(MN)^{-\sum_{j=0}^{k-1}\frac{\alpha^j}{2^j}}=(MN)^{\frac{1-(\alpha/2)^k}{\alpha/2-1}},
\end{equation}
achieving sub-shot-noise for $\alpha<2$ with the limiting scaling
$\sigma^2\big(\theta_\text{est}\big)\propto(MN)^{\frac{2}{\alpha-2}}$
for $k\to\infty$.

Within the above adaptive scheme, control of Hamiltonian parameters, estimations at each step, and re-materialization or re-stabilization of states at each step are assumed to be efficiently implementable.
This adaptive measurement was proposed to estimate the critical
magnetic field $h=J$ of the XX model at small non-zero temperature
when the quantum phase transition is smoothened to a phase crossover \cite{Mehboudi2016},
achieving the limiting scaling
$\sigma^2\big(\theta_\text{est}\big)\propto(MN)^{-\frac{4}{3}}$. The
optimal estimate is derived from measurements of the magnetization
along the $z$ direction $\sum_j\sigma_j^z$, since the magnetic field
commutes with the spin interaction restoring the classical picture of
Lagrange multipliers and 
quantities fixed on average. The same estimate and
that derived from measurements of the variance of $\sum_j\sigma_j^x$
are nearly optimal for the XY model. It is remarkable that, when the
adaptive measurement is applied to the model \eqref{Ham.var.range}
approaching the critical lines in \eqref{scaling.fully.connected} from
the non-critical region, the critical scalings are consistently recovered.

\section{Outlook}
Most work on quantum-enhanced measurements has investigated the
benefits of using quantum entanglement (see \cite{Paris2009,GiovannettiNPhot2011,SmerziR,toth_quantum_2014,pezze_non-classical_2016} for
recent reviews). Indeed, under certain
restrictive assumptions (see Introduction), entanglement can be shown
to be necessary if
one wants to improve over classical sensitivity. However, going beyond
these restrictive assumptions opens up a host of new possibilities of
which we have explored a large number in this review.  Given the
difficulty of producing and maintaining entangled states of a large
number of subsystems, some of these may open up new roads to better
sensitivities than classically possible with a comparable number of
resources in actual experiments.  \\

We conclude this review by challenging yet another common mind-set in
the field (which we could not quite escape in this review either),
namely the hunt for faster {\em scaling} of the
sensitivity, in particular the quest for a scaling faster than $1/\sqrt{N}$
with the number of subsystems $N$, and the goal of reaching
``Heisenberg-limited'' scaling $1/N$:  It should be clear (see also
Sec.\ref{obs} and  the linear/nonlinear comparison in
\cite{NapolitanoN2011})  that {\em
  scaling} of the sensitivity is not
{\em per se} a desideratum.  Any given instrument or measurement is
judged by its sensitivity, not the scaling thereof.  When the
sensitivity is $\sigma(\theta_\text{est}) = \alpha N^d$, it is sometimes argued
that the pre-factor $\alpha$ is irrelevant, because a more rapid
scaling necessarily leads to better sensitivity for
sufficiently large $N$.  While mathematically impeccable, this
argument assumes that the scaling persists to sufficiently large $N$
where the possibly small prefactor can be compensated --- an assumption
that may not be valid in practice.  
Typically, at some point the
model breaks down, and systematic errors arise that scale with a
positive power of $N$ and at some point become comparable
to the stochastic error quantified by the quantum Cram\'er-Rao bound.
And finally, some large-$N$ catastrophe destroys the
instrument and its measurement capability.
Such concerns are of course very relevant for real-life experiments,
{\redd for which} material properties have to be taken into account.  But
they may also determine fundamental bounds to achievable precision for
various physical
quantities that are ultimately linked to the fabric of space-time at extremely
small length- and time-scales.
Such ideas were advanced early
on notably by Wigner, who estimated the ultimate achievable precision
of atomic clocks: increasing the energy of the used clock states more
and more for improving its precision leads ultimately to the formation
of a black hole and hence renders reading off the clock impossible
\cite{wigner_relativistic_1957}.
Similar limitations of this kind exist for measurements of lengths
\cite{amelino-camelia_gravity-wave_1999,Ng00}, and have been recently
explored in more detail for the speed of light in
vacuum \cite{braun_how_2015}. \\   

While current technology is still far from probing such extreme
conditions, we nevertheless arrive to the conclusion that the
importance of scaling is to  give the functional form for an {\em
  extrapolation} of the sensitivity, a prediction of how well one
could measure if one had a given large $N$.  The validity of this
extrapolation will be limited by the range over which the scaling
persists, an additional datum not described by the scaling nor by the
prefactor.  Moreover, there is the possibility of an {\em optimum}
$N$,  beyond which the sensitivity worsens~\cite{Nichols2016}.  In such a case, the
interesting questions are ``is the optimum $N_{\rm opt}$ achievable
given available resources'' and ``what is the actual sensitivity at
this optimum?''  At such an optimum the local scaling is flat,
i.e.~the smallest possible uncertainty of an unbiased estimate of the
parameter 
is independent of $N$.  Ironically, our
discussion of advantageous scaling leads us to the conclusion that the
best scaling may be no scaling at all.\\

The
relevance of the actually achievable smallest uncertainty rather than
its scaling
with the number of resources makes it particularly important that
alternatives to the use of massive entanglement be investigated, as so
far the number of subsystems that could be entangled experimentally has
remained relatively small. We hope that the present review will
stimulate further research in this direction.

\begin{acknowledgments}
DB thanks Julien Fra\"isse for useful discussions, a careful
reading of the manuscript, and pointing out
ref.~\cite{Popoviciu35}. GA acknowledges fruitful discussions with
Kavan Modi. SP would like to thank S. L. Braunstein, Cosmo Lupo
and Leonardo Banchi for comments on Section II. This work was
supported in part by the Deutsche Forschungsgemeinschaft through
SFB TRR21 and Grant BR 5221/1-1, by the European Research Council
(ERC) through the StG GQCOP (Grant Agreement No.~637352) and
AQUMET (Grant Agreement No.~280169) and the  PoC ERIDIAN (Grant
Agreement No.~713682), by the Royal Society through the
International Exchanges Programme (Grant No.~IE150570), by the
Foundational Questions Institute (fqxi.org) through the Physics of
the Observer Programme (Grant No.~FQXi-RFP-1601), the Spanish
MINECO/FEDER, the MINECO projects MAQRO (Ref.~FIS2015-68039-P),
XPLICA (Ref.~FIS2014-62181-EXP) and Severo Ochoa grant
SEV-2015-0522, Catalan 2014-SGR-1295, by the European Union
Project QUIC (Grant Agreement No.~641122), by Fundaci\'{o} Privada
CELLEX, by the Engineering and Physical Sciences Research Council
(EPSRC) through the UK Quantum Communications Hub (EP/M013472/1),
by the grants J1-5439 and N1-0025 of Slovenian Research Agency,
and by H2020 CSA Twinning project No.~692194, ``RBI-T-WINNING''.
\end{acknowledgments}

\bibliographystyle{apsrmp4-1}
\bibliography{QEMnoE_resub2_171211}

\begin{thebibliography}{523}%
\makeatletter
\providecommand \@ifxundefined [1]{%
 \@ifx{#1\undefined}
}%
\providecommand \@ifnum [1]{%
 \ifnum #1\expandafter \@firstoftwo
 \else \expandafter \@secondoftwo
 \fi
}%
\providecommand \@ifx [1]{%
 \ifx #1\expandafter \@firstoftwo
 \else \expandafter \@secondoftwo
 \fi
}%
\providecommand \natexlab [1]{#1}%
\providecommand \enquote  [1]{``#1''}%
\providecommand \bibnamefont  [1]{#1}%
\providecommand \bibfnamefont [1]{#1}%
\providecommand \citenamefont [1]{#1}%
\providecommand \href@noop [0]{\@secondoftwo}%
\providecommand \href [0]{\begingroup \@sanitize@url \@href}%
\providecommand \@href[1]{\@@startlink{#1}\@@href}%
\providecommand \@@href[1]{\endgroup#1\@@endlink}%
\providecommand \@sanitize@url [0]{\catcode `\\12\catcode `\$12\catcode
  `\&12\catcode `\#12\catcode `\^12\catcode `\_12\catcode `\%12\relax}%
\providecommand \@@startlink[1]{}%
\providecommand \@@endlink[0]{}%
\providecommand \url  [0]{\begingroup\@sanitize@url \@url }%
\providecommand \@url [1]{\endgroup\@href {#1}{\urlprefix }}%
\providecommand \urlprefix  [0]{URL }%
\providecommand \Eprint [0]{\href }%
\providecommand \doibase [0]{http://dx.doi.org/}%
\providecommand \selectlanguage [0]{\@gobble}%
\providecommand \bibinfo  [0]{\@secondoftwo}%
\providecommand \bibfield  [0]{\@secondoftwo}%
\providecommand \translation [1]{[#1]}%
\providecommand \BibitemOpen [0]{}%
\providecommand \bibitemStop [0]{}%
\providecommand \bibitemNoStop [0]{.\EOS\space}%
\providecommand \EOS [0]{\spacefactor3000\relax}%
\providecommand \BibitemShut  [1]{\csname bibitem#1\endcsname}%
\let\auto@bib@innerbib\@empty
\bibitem [{\citenamefont {Aasi}\ \emph {et~al.}(2013)\citenamefont {Aasi} \emph
  {et~al.}}]{AasiNP2013others}%
  \BibitemOpen
  \bibfield  {author} {\bibinfo {author} {\bibnamefont {Aasi}, \bibfnamefont
  {J.}},  \emph {et~al.}} (\bibinfo {year} {2013}),\ \href
  {http://dx.doi.org/10.1038/nphoton.2013.177} {\bibfield  {journal} {\bibinfo
  {journal} {Nat Photon}\ }\textbf {\bibinfo {volume} {7}}~(\bibinfo {number}
  {8}),\ \bibinfo {pages} {613}}\BibitemShut {NoStop}%
\bibitem [{\citenamefont {Acin}(2001)}]{Acin01}%
  \BibitemOpen
  \bibfield  {author} {\bibinfo {author} {\bibnamefont {Acin}, \bibfnamefont
  {A.}}} (\bibinfo {year} {2001}),\ \href@noop {} {\bibfield  {journal}
  {\bibinfo  {journal} {Phys. Rev. Lett.}\ }\textbf {\bibinfo {volume} {87}},\
  \bibinfo {pages} {177901}}\BibitemShut {NoStop}%
\bibitem [{\citenamefont {Acosta}\ \emph {et~al.}(2009)\citenamefont {Acosta},
  \citenamefont {Bauch}, \citenamefont {Ledbetter}, \citenamefont {Santori},
  \citenamefont {Fu}, \citenamefont {Barclay}, \citenamefont {Beausoleil},
  \citenamefont {Linget}, \citenamefont {Roch}, \citenamefont {Treussart},
  \citenamefont {Chemerisov}, \citenamefont {Gawlik},\ and\ \citenamefont
  {Budker}}]{acosta_diamonds_2009}%
  \BibitemOpen
  \bibfield  {author} {\bibinfo {author} {\bibnamefont {Acosta}, \bibfnamefont
  {V.~M.}}, \bibinfo {author} {\bibfnamefont {E.}~\bibnamefont {Bauch}},
  \bibinfo {author} {\bibfnamefont {M.~P.}\ \bibnamefont {Ledbetter}}, \bibinfo
  {author} {\bibfnamefont {C.}~\bibnamefont {Santori}}, \bibinfo {author}
  {\bibfnamefont {K.-M.~C.}\ \bibnamefont {Fu}}, \bibinfo {author}
  {\bibfnamefont {P.~E.}\ \bibnamefont {Barclay}}, \bibinfo {author}
  {\bibfnamefont {R.~G.}\ \bibnamefont {Beausoleil}}, \bibinfo {author}
  {\bibfnamefont {H.}~\bibnamefont {Linget}}, \bibinfo {author} {\bibfnamefont
  {J.~F.}\ \bibnamefont {Roch}}, \bibinfo {author} {\bibfnamefont
  {F.}~\bibnamefont {Treussart}}, \bibinfo {author} {\bibfnamefont
  {S.}~\bibnamefont {Chemerisov}}, \bibinfo {author} {\bibfnamefont
  {W.}~\bibnamefont {Gawlik}}, \ and\ \bibinfo {author} {\bibfnamefont
  {D.}~\bibnamefont {Budker}}} (\bibinfo {year} {2009}),\ \href@noop {}
  {\bibfield  {journal} {\bibinfo  {journal} {Phys. Rev. B}\ }\textbf {\bibinfo
  {volume} {80}}~(\bibinfo {number} {11}),\ \bibinfo {pages}
  {115202}}\BibitemShut {NoStop}%
\bibitem [{\citenamefont {Adesso}(2014)}]{Adesso2014}%
  \BibitemOpen
  \bibfield  {author} {\bibinfo {author} {\bibnamefont {Adesso}, \bibfnamefont
  {G.}}} (\bibinfo {year} {2014}),\ \href {\doibase 10.1103/PhysRevA.90.022321}
  {\bibfield  {journal} {\bibinfo  {journal} {Phys. Rev. A}\ }\textbf {\bibinfo
  {volume} {90}}~(\bibinfo {number} {2}),\ \bibinfo {pages}
  {022321}}\BibitemShut {NoStop}%
\bibitem [{\citenamefont {Adesso}\ \emph {et~al.}(2016)\citenamefont {Adesso},
  \citenamefont {Bromley},\ and\ \citenamefont {Cianciaruso}}]{ABC}%
  \BibitemOpen
  \bibfield  {author} {\bibinfo {author} {\bibnamefont {Adesso}, \bibfnamefont
  {G.}}, \bibinfo {author} {\bibfnamefont {T.~R.}\ \bibnamefont {Bromley}}, \
  and\ \bibinfo {author} {\bibfnamefont {M.}~\bibnamefont {Cianciaruso}}}
  (\bibinfo {year} {2016}),\ \href@noop {} {\bibfield  {journal} {\bibinfo
  {journal} {J. Phys. A: Math. Theor.}\ }\textbf {\bibinfo {volume} {49}},\
  \bibinfo {pages} {473001}}\BibitemShut {NoStop}%
\bibitem [{\citenamefont {Adesso}\ \emph {et~al.}(2009)\citenamefont {Adesso},
  \citenamefont {Dell'Anno}, \citenamefont {Siena}, \citenamefont
  {Illuminati},\ and\ \citenamefont {Souza}}]{AdessoAnno09}%
  \BibitemOpen
  \bibfield  {author} {\bibinfo {author} {\bibnamefont {Adesso}, \bibfnamefont
  {G.}}, \bibinfo {author} {\bibfnamefont {F.}~\bibnamefont {Dell'Anno}},
  \bibinfo {author} {\bibfnamefont {S.~D.}\ \bibnamefont {Siena}}, \bibinfo
  {author} {\bibfnamefont {F.}~\bibnamefont {Illuminati}}, \ and\ \bibinfo
  {author} {\bibfnamefont {L.~A.~M.}\ \bibnamefont {Souza}}} (\bibinfo {year}
  {2009}),\ \href@noop {} {\bibfield  {journal} {\bibinfo  {journal} {Phys.
  Rev. A}\ }\textbf {\bibinfo {volume} {79}},\ \bibinfo {pages}
  {040305(R)}}\BibitemShut {NoStop}%
\bibitem [{\citenamefont {Ahn}\ \emph {et~al.}(2002)\citenamefont {Ahn},
  \citenamefont {Doherty},\ and\ \citenamefont {Landahl}}]{PhysRevA.65.042301}%
  \BibitemOpen
  \bibfield  {author} {\bibinfo {author} {\bibnamefont {Ahn}, \bibfnamefont
  {C.}}, \bibinfo {author} {\bibfnamefont {A.~C.}\ \bibnamefont {Doherty}}, \
  and\ \bibinfo {author} {\bibfnamefont {A.~J.}\ \bibnamefont {Landahl}}}
  (\bibinfo {year} {2002}),\ \href {\doibase 10.1103/PhysRevA.65.042301}
  {\bibfield  {journal} {\bibinfo  {journal} {Phys. Rev. A}\ }\textbf {\bibinfo
  {volume} {65}},\ \bibinfo {pages} {042301}}\BibitemShut {NoStop}%
\bibitem [{\citenamefont {Alipour}\ \emph {et~al.}(2014)\citenamefont
  {Alipour}, \citenamefont {Mehboudi},\ and\ \citenamefont
  {Rezakhani}}]{Alipour2014}%
  \BibitemOpen
  \bibfield  {author} {\bibinfo {author} {\bibnamefont {Alipour}, \bibfnamefont
  {S.}}, \bibinfo {author} {\bibfnamefont {M.}~\bibnamefont {Mehboudi}}, \ and\
  \bibinfo {author} {\bibfnamefont {A.~T.}\ \bibnamefont {Rezakhani}}}
  (\bibinfo {year} {2014}),\ \href {\doibase 10.1103/PhysRevLett.112.120405}
  {\bibfield  {journal} {\bibinfo  {journal} {Phys. Rev. Lett.}\ }\textbf
  {\bibinfo {volume} {112}},\ \bibinfo {pages} {120405}}\BibitemShut {NoStop}%
\bibitem [{\citenamefont
  {Amelino-Camelia}(1999)}]{amelino-camelia_gravity-wave_1999}%
  \BibitemOpen
  \bibfield  {author} {\bibinfo {author} {\bibnamefont {Amelino-Camelia},
  \bibfnamefont {G.}}} (\bibinfo {year} {1999}),\ \href {\doibase
  10.1038/18377} {\bibfield  {journal} {\bibinfo  {journal} {Nature}\ }\textbf
  {\bibinfo {volume} {398}}~(\bibinfo {number} {6724}),\ \bibinfo {pages}
  {216}}\BibitemShut {NoStop}%
\bibitem [{\citenamefont {van Amerongen}(2008)}]{vanAmerongen2008-2}%
  \BibitemOpen
  \bibfield  {author} {\bibinfo {author} {\bibnamefont {van Amerongen},
  \bibfnamefont {A.}}} (\bibinfo {year} {2008}),\ \href@noop {} {\bibfield
  {journal} {\bibinfo  {journal} {Ann. Phys. Fr.}\ }\textbf {\bibinfo {volume}
  {22}},\ \bibinfo {pages} {1}}\BibitemShut {NoStop}%
\bibitem [{\citenamefont {van Amerongen}\ \emph {et~al.}(2008)\citenamefont
  {van Amerongen}, \citenamefont {van Es}, \citenamefont {Wicke}, \citenamefont
  {Kheruntsyan},\ and\ \citenamefont {van Druten}}]{vanAmerongen2008}%
  \BibitemOpen
  \bibfield  {author} {\bibinfo {author} {\bibnamefont {van Amerongen},
  \bibfnamefont {A.}}, \bibinfo {author} {\bibfnamefont {J.}~\bibnamefont {van
  Es}}, \bibinfo {author} {\bibfnamefont {P.}~\bibnamefont {Wicke}}, \bibinfo
  {author} {\bibfnamefont {K.}~\bibnamefont {Kheruntsyan}}, \ and\ \bibinfo
  {author} {\bibfnamefont {N.}~\bibnamefont {van Druten}}} (\bibinfo {year}
  {2008}),\ \href@noop {} {\bibfield  {journal} {\bibinfo  {journal} {Phys.
  Rev. Lett.}\ }\textbf {\bibinfo {volume} {100}},\ \bibinfo {pages}
  {090402}}\BibitemShut {NoStop}%
\bibitem [{\citenamefont {Amico}\ \emph {et~al.}(2008)\citenamefont {Amico},
  \citenamefont {Fazio}, \citenamefont {Osterloh},\ and\ \citenamefont
  {Vedral}}]{Amico0}%
  \BibitemOpen
  \bibfield  {author} {\bibinfo {author} {\bibnamefont {Amico}, \bibfnamefont
  {L.}}, \bibinfo {author} {\bibfnamefont {R.}~\bibnamefont {Fazio}}, \bibinfo
  {author} {\bibfnamefont {A.}~\bibnamefont {Osterloh}}, \ and\ \bibinfo
  {author} {\bibfnamefont {V.}~\bibnamefont {Vedral}}} (\bibinfo {year}
  {2008}),\ \href@noop {} {\bibfield  {journal} {\bibinfo  {journal} {Rev. Mod.
  Phys.}\ }\textbf {\bibinfo {volume} {80}},\ \bibinfo {pages}
  {517}}\BibitemShut {NoStop}%
\bibitem [{\citenamefont {Araki}\ and\ \citenamefont {Moriya}(2003)}]{Araki0}%
  \BibitemOpen
  \bibfield  {author} {\bibinfo {author} {\bibnamefont {Araki}, \bibfnamefont
  {H.}}, \ and\ \bibinfo {author} {\bibfnamefont {H.}~\bibnamefont {Moriya}}}
  (\bibinfo {year} {2003}),\ \href@noop {} {\bibfield  {journal} {\bibinfo
  {journal} {Commun. Math. Phys.}\ }\textbf {\bibinfo {volume} {237}},\
  \bibinfo {pages} {105}}\BibitemShut {NoStop}%
\bibitem [{\citenamefont {Argentieri}\ \emph {et~al.}(2011)\citenamefont
  {Argentieri}, \citenamefont {Benatti}, \citenamefont {Floreanini},\ and\
  \citenamefont {Marzolino}}]{Argentieri0}%
  \BibitemOpen
  \bibfield  {author} {\bibinfo {author} {\bibnamefont {Argentieri},
  \bibfnamefont {G.}}, \bibinfo {author} {\bibfnamefont {F.}~\bibnamefont
  {Benatti}}, \bibinfo {author} {\bibfnamefont {R.}~\bibnamefont {Floreanini}},
  \ and\ \bibinfo {author} {\bibfnamefont {U.}~\bibnamefont {Marzolino}}}
  (\bibinfo {year} {2011}),\ \href@noop {} {\bibfield  {journal} {\bibinfo
  {journal} {Int. J. Quant. Inf.}\ }\textbf {\bibinfo {volume} {9}},\ \bibinfo
  {pages} {1745}}\BibitemShut {NoStop}%
\bibitem [{\citenamefont {Armen}\ \emph {et~al.}(2002)\citenamefont {Armen},
  \citenamefont {Au}, \citenamefont {Stockton}, \citenamefont {Doherty},\ and\
  \citenamefont {Mabuchi}}]{armen_adaptive_2002}%
  \BibitemOpen
  \bibfield  {author} {\bibinfo {author} {\bibnamefont {Armen}, \bibfnamefont
  {M.~A.}}, \bibinfo {author} {\bibfnamefont {J.~K.}\ \bibnamefont {Au}},
  \bibinfo {author} {\bibfnamefont {J.~K.}\ \bibnamefont {Stockton}}, \bibinfo
  {author} {\bibfnamefont {A.~C.}\ \bibnamefont {Doherty}}, \ and\ \bibinfo
  {author} {\bibfnamefont {H.}~\bibnamefont {Mabuchi}}} (\bibinfo {year}
  {2002}),\ \href {\doibase 10.1103/PhysRevLett.89.133602} {\bibfield
  {journal} {\bibinfo  {journal} {Phys. Rev. Lett.}\ }\textbf {\bibinfo
  {volume} {89}}~(\bibinfo {number} {13}),\ \bibinfo {pages}
  {133602}}\BibitemShut {NoStop}%
\bibitem [{\citenamefont {Armijo}\ \emph {et~al.}(2011)\citenamefont {Armijo},
  \citenamefont {Jacqmin}, \citenamefont {Kheruntsyan},\ and\ \citenamefont
  {Bouchoule}}]{Armijo2011}%
  \BibitemOpen
  \bibfield  {author} {\bibinfo {author} {\bibnamefont {Armijo}, \bibfnamefont
  {J.}}, \bibinfo {author} {\bibfnamefont {T.}~\bibnamefont {Jacqmin}},
  \bibinfo {author} {\bibfnamefont {K.}~\bibnamefont {Kheruntsyan}}, \ and\
  \bibinfo {author} {\bibfnamefont {I.}~\bibnamefont {Bouchoule}}} (\bibinfo
  {year} {2011}),\ \href@noop {} {\bibfield  {journal} {\bibinfo  {journal}
  {Phys. Rev. A}\ }\textbf {\bibinfo {volume} {83}},\ \bibinfo {pages}
  {021605}}\BibitemShut {NoStop}%
\bibitem [{\citenamefont {Asb\'oth}\ \emph {et~al.}(2005)\citenamefont
  {Asb\'oth}, \citenamefont {Calsamiglia},\ and\ \citenamefont
  {Ritsch}}]{Asboth2005}%
  \BibitemOpen
  \bibfield  {author} {\bibinfo {author} {\bibnamefont {Asb\'oth},
  \bibfnamefont {J.~K.}}, \bibinfo {author} {\bibfnamefont {J.}~\bibnamefont
  {Calsamiglia}}, \ and\ \bibinfo {author} {\bibfnamefont {H.}~\bibnamefont
  {Ritsch}}} (\bibinfo {year} {2005}),\ \href {\doibase
  10.1103/PhysRevLett.94.173602} {\bibfield  {journal} {\bibinfo  {journal}
  {Phys. Rev. Lett.}\ }\textbf {\bibinfo {volume} {94}},\ \bibinfo {pages}
  {173602}}\BibitemShut {NoStop}%
\bibitem [{\citenamefont {Audenaert}\ \emph {et~al.}(2007)\citenamefont
  {Audenaert}, \citenamefont {Calsamiglia}, \citenamefont {Mu\~noz Tapia},
  \citenamefont {Bagan}, \citenamefont {Masanes}, \citenamefont {Acin},\ and\
  \citenamefont {Verstraete}}]{Audenaert2007}%
  \BibitemOpen
  \bibfield  {author} {\bibinfo {author} {\bibnamefont {Audenaert},
  \bibfnamefont {K.~M.~R.}}, \bibinfo {author} {\bibfnamefont {J.}~\bibnamefont
  {Calsamiglia}}, \bibinfo {author} {\bibfnamefont {R.}~\bibnamefont {Mu\~noz
  Tapia}}, \bibinfo {author} {\bibfnamefont {E.}~\bibnamefont {Bagan}},
  \bibinfo {author} {\bibfnamefont {L.}~\bibnamefont {Masanes}}, \bibinfo
  {author} {\bibfnamefont {A.}~\bibnamefont {Acin}}, \ and\ \bibinfo {author}
  {\bibfnamefont {F.}~\bibnamefont {Verstraete}}} (\bibinfo {year} {2007}),\
  \href {\doibase 10.1103/PhysRevLett.98.160501} {\bibfield  {journal}
  {\bibinfo  {journal} {Phys. Rev. Lett.}\ }\textbf {\bibinfo {volume} {98}},\
  \bibinfo {pages} {160501}}\BibitemShut {NoStop}%
\bibitem [{\citenamefont {Augusiak}\ \emph {et~al.}(2016)\citenamefont
  {Augusiak}, \citenamefont {Ko\l{}ody\'{n}ski}, \citenamefont {Streltsov},
  \citenamefont {Bera}, \citenamefont {Ac\'{\i}n},\ and\ \citenamefont
  {Lewenstein}}]{Augusiak15}%
  \BibitemOpen
  \bibfield  {author} {\bibinfo {author} {\bibnamefont {Augusiak},
  \bibfnamefont {R.}}, \bibinfo {author} {\bibfnamefont {J.}~\bibnamefont
  {Ko\l{}ody\'{n}ski}}, \bibinfo {author} {\bibfnamefont {A.}~\bibnamefont
  {Streltsov}}, \bibinfo {author} {\bibfnamefont {M.~N.}\ \bibnamefont {Bera}},
  \bibinfo {author} {\bibfnamefont {A.}~\bibnamefont {Ac\'{\i}n}}, \ and\
  \bibinfo {author} {\bibfnamefont {M.}~\bibnamefont {Lewenstein}}} (\bibinfo
  {year} {2016}),\ \href {\doibase 10.1103/PhysRevA.94.012339} {\bibfield
  {journal} {\bibinfo  {journal} {Phys. Rev. A}\ }\textbf {\bibinfo {volume}
  {94}},\ \bibinfo {pages} {012339}}\BibitemShut {NoStop}%
\bibitem [{\citenamefont {Bakr}\ \emph {et~al.}(2009)\citenamefont {Bakr},
  \citenamefont {Gillen}, \citenamefont {Peng}, \citenamefont {F{\"o}lling},\
  and\ \citenamefont {Greiner}}]{bakr_quantum_2009}%
  \BibitemOpen
  \bibfield  {author} {\bibinfo {author} {\bibnamefont {Bakr}, \bibfnamefont
  {W.~S.}}, \bibinfo {author} {\bibfnamefont {J.~I.}\ \bibnamefont {Gillen}},
  \bibinfo {author} {\bibfnamefont {A.}~\bibnamefont {Peng}}, \bibinfo {author}
  {\bibfnamefont {S.}~\bibnamefont {F{\"o}lling}}, \ and\ \bibinfo {author}
  {\bibfnamefont {M.}~\bibnamefont {Greiner}}} (\bibinfo {year} {2009}),\ \href
  {\doibase 10.1038/nature08482} {\bibfield  {journal} {\bibinfo  {journal}
  {Nature}\ }\textbf {\bibinfo {volume} {462}}~(\bibinfo {number} {7269}),\
  \bibinfo {pages} {74}}\BibitemShut {NoStop}%
\bibitem [{\citenamefont {Balachandran}\ \emph {et~al.}(2013)\citenamefont
  {Balachandran}, \citenamefont {Govindarajan}, \citenamefont {de~Queiroz},\
  and\ \citenamefont {Reyes-Lega}}]{Balachandran0}%
  \BibitemOpen
  \bibfield  {author} {\bibinfo {author} {\bibnamefont {Balachandran},
  \bibfnamefont {A.}}, \bibinfo {author} {\bibfnamefont {T.}~\bibnamefont
  {Govindarajan}}, \bibinfo {author} {\bibfnamefont {A.}~\bibnamefont
  {de~Queiroz}}, \ and\ \bibinfo {author} {\bibfnamefont {A.}~\bibnamefont
  {Reyes-Lega}}} (\bibinfo {year} {2013}),\ \href@noop {} {\bibfield  {journal}
  {\bibinfo  {journal} {Phys. Rev. Lett.}\ }\textbf {\bibinfo {volume} {110}},\
  \bibinfo {pages} {080503}}\BibitemShut {NoStop}%
\bibitem [{\citenamefont {Baldwin}\ \emph {et~al.}(2014)\citenamefont
  {Baldwin}, \citenamefont {Kalev},\ and\ \citenamefont
  {Deutsch}}]{Baldwin2014}%
  \BibitemOpen
  \bibfield  {author} {\bibinfo {author} {\bibnamefont {Baldwin}, \bibfnamefont
  {C.~H.}}, \bibinfo {author} {\bibfnamefont {A.}~\bibnamefont {Kalev}}, \ and\
  \bibinfo {author} {\bibfnamefont {I.~H.}\ \bibnamefont {Deutsch}}} (\bibinfo
  {year} {2014}),\ \href {\doibase 10.1103/PhysRevA.90.012110} {\bibfield
  {journal} {\bibinfo  {journal} {Phys. Rev. A}\ }\textbf {\bibinfo {volume}
  {90}},\ \bibinfo {pages} {012110}}\BibitemShut {NoStop}%
\bibitem [{\citenamefont {Banchi}\ \emph {et~al.}(2015)\citenamefont {Banchi},
  \citenamefont {Braunstein},\ and\ \citenamefont {Pirandola}}]{Banchi15}%
  \BibitemOpen
  \bibfield  {author} {\bibinfo {author} {\bibnamefont {Banchi}, \bibfnamefont
  {L.}}, \bibinfo {author} {\bibfnamefont {S.~L.}\ \bibnamefont {Braunstein}},
  \ and\ \bibinfo {author} {\bibfnamefont {S.}~\bibnamefont {Pirandola}}}
  (\bibinfo {year} {2015}),\ \href@noop {} {\bibfield  {journal} {\bibinfo
  {journal} {Phys. Rev. Lett.}\ }\textbf {\bibinfo {volume} {115}},\ \bibinfo
  {pages} {260501}}\BibitemShut {NoStop}%
\bibitem [{\citenamefont {Banchi}\ \emph {et~al.}(2014)\citenamefont {Banchi},
  \citenamefont {Giorda},\ and\ \citenamefont {Zanardi}}]{Banchi2014}%
  \BibitemOpen
  \bibfield  {author} {\bibinfo {author} {\bibnamefont {Banchi}, \bibfnamefont
  {L.}}, \bibinfo {author} {\bibfnamefont {P.}~\bibnamefont {Giorda}}, \ and\
  \bibinfo {author} {\bibfnamefont {P.}~\bibnamefont {Zanardi}}} (\bibinfo
  {year} {2014}),\ \href {\doibase 10.1103/PhysRevE.89.022102} {\bibfield
  {journal} {\bibinfo  {journal} {Phys. Rev. E}\ }\textbf {\bibinfo {volume}
  {89}},\ \bibinfo {pages} {022102}}\BibitemShut {NoStop}%
\bibitem [{\citenamefont {Banuls}\ \emph {et~al.}(2006)\citenamefont {Banuls},
  \citenamefont {Cirac},\ and\ \citenamefont {Wolf}}]{Cirac0}%
  \BibitemOpen
  \bibfield  {author} {\bibinfo {author} {\bibnamefont {Banuls}, \bibfnamefont
  {M.-C.}}, \bibinfo {author} {\bibfnamefont {J.}~\bibnamefont {Cirac}}, \ and\
  \bibinfo {author} {\bibfnamefont {M.}~\bibnamefont {Wolf}}} (\bibinfo {year}
  {2006}),\ \href@noop {} {\bibfield  {journal} {\bibinfo  {journal} {Phys.
  Rev. A}\ }\textbf {\bibinfo {volume} {76}},\ \bibinfo {pages}
  {022311}}\BibitemShut {NoStop}%
\bibitem [{\citenamefont {Barnett}\ \emph {et~al.}(2003)\citenamefont
  {Barnett}, \citenamefont {Fabre},\ and\ \citenamefont
  {Ma{\^\i}tre}}]{refId0}%
  \BibitemOpen
  \bibfield  {author} {\bibinfo {author} {\bibnamefont {Barnett}, \bibfnamefont
  {S.~M.}}, \bibinfo {author} {\bibfnamefont {C.}~\bibnamefont {Fabre}}, \ and\
  \bibinfo {author} {\bibfnamefont {A.}~\bibnamefont {Ma{\^\i}tre}}} (\bibinfo
  {year} {2003}),\ \href {\doibase 10.1140/epjd/e2003-00003-3} {\bibfield
  {journal} {\bibinfo  {journal} {Eur. Phys. J. D}\ }\textbf {\bibinfo {volume}
  {22}}~(\bibinfo {number} {3}),\ \bibinfo {pages} {513}}\BibitemShut {NoStop}%
\bibitem [{\citenamefont {Barnum}\ \emph {et~al.}(2004)\citenamefont {Barnum},
  \citenamefont {Knill}, \citenamefont {Ortiz}, \citenamefont {Somma},\ and\
  \citenamefont {Viola}}]{Viola1}%
  \BibitemOpen
  \bibfield  {author} {\bibinfo {author} {\bibnamefont {Barnum}, \bibfnamefont
  {H.}}, \bibinfo {author} {\bibfnamefont {E.}~\bibnamefont {Knill}}, \bibinfo
  {author} {\bibfnamefont {G.}~\bibnamefont {Ortiz}}, \bibinfo {author}
  {\bibfnamefont {R.}~\bibnamefont {Somma}}, \ and\ \bibinfo {author}
  {\bibfnamefont {L.}~\bibnamefont {Viola}}} (\bibinfo {year} {2004}),\
  \href@noop {} {\bibfield  {journal} {\bibinfo  {journal} {Phys. Rev. Lett.}\
  }\textbf {\bibinfo {volume} {92}},\ \bibinfo {pages} {107902}}\BibitemShut
  {NoStop}%
\bibitem [{\citenamefont {Barnum}\ \emph {et~al.}(2005)\citenamefont {Barnum},
  \citenamefont {Ortiz}, \citenamefont {Somma},\ and\ \citenamefont
  {Viola}}]{Viola2}%
  \BibitemOpen
  \bibfield  {author} {\bibinfo {author} {\bibnamefont {Barnum}, \bibfnamefont
  {H.}}, \bibinfo {author} {\bibfnamefont {G.}~\bibnamefont {Ortiz}}, \bibinfo
  {author} {\bibfnamefont {R.}~\bibnamefont {Somma}}, \ and\ \bibinfo {author}
  {\bibfnamefont {L.}~\bibnamefont {Viola}}} (\bibinfo {year} {2005}),\
  \href@noop {} {\bibfield  {journal} {\bibinfo  {journal} {Int. J. Theor.
  Phys.}\ }\textbf {\bibinfo {volume} {44}},\ \bibinfo {pages}
  {2127}}\BibitemShut {NoStop}%
\bibitem [{\citenamefont {Barzanjeh}\ \emph {et~al.}(2015)\citenamefont
  {Barzanjeh}, \citenamefont {Guha}, \citenamefont {Weedbrook}, \citenamefont
  {Vitali}, \citenamefont {Shapiro},\ and\ \citenamefont
  {Pirandola}}]{Barzanjeh15}%
  \BibitemOpen
  \bibfield  {author} {\bibinfo {author} {\bibnamefont {Barzanjeh},
  \bibfnamefont {S.}}, \bibinfo {author} {\bibfnamefont {S.}~\bibnamefont
  {Guha}}, \bibinfo {author} {\bibfnamefont {C.}~\bibnamefont {Weedbrook}},
  \bibinfo {author} {\bibfnamefont {D.}~\bibnamefont {Vitali}}, \bibinfo
  {author} {\bibfnamefont {J.~H.}\ \bibnamefont {Shapiro}}, \ and\ \bibinfo
  {author} {\bibfnamefont {S.}~\bibnamefont {Pirandola}}} (\bibinfo {year}
  {2015}),\ \href@noop {} {\bibfield  {journal} {\bibinfo  {journal} {Phys.
  Rev. Lett.}\ }\textbf {\bibinfo {volume} {114}},\ \bibinfo {pages}
  {080503}}\BibitemShut {NoStop}%
\bibitem [{\citenamefont {Baumgratz}\ \emph {et~al.}(2014)\citenamefont
  {Baumgratz}, \citenamefont {Cramer},\ and\ \citenamefont
  {Plenio}}]{Baumgratz2014}%
  \BibitemOpen
  \bibfield  {author} {\bibinfo {author} {\bibnamefont {Baumgratz},
  \bibfnamefont {T.}}, \bibinfo {author} {\bibfnamefont {M.}~\bibnamefont
  {Cramer}}, \ and\ \bibinfo {author} {\bibfnamefont {M.~B.}\ \bibnamefont
  {Plenio}}} (\bibinfo {year} {2014}),\ \href {\doibase
  10.1103/PhysRevLett.113.140401} {\bibfield  {journal} {\bibinfo  {journal}
  {Phys. Rev. Lett.}\ }\textbf {\bibinfo {volume} {113}},\ \bibinfo {pages}
  {140401}}\BibitemShut {NoStop}%
\bibitem [{\citenamefont {Baxter}(1982)}]{Baxter}%
  \BibitemOpen
  \bibfield  {author} {\bibinfo {author} {\bibnamefont {Baxter}, \bibfnamefont
  {R.~J.}}} (\bibinfo {year} {1982}),\ \href@noop {} {\emph {\bibinfo {title}
  {Exactly Solved Models in Statistical Mechanics}}}\ (\bibinfo  {publisher}
  {Academic, New York})\BibitemShut {NoStop}%
\bibitem [{\citenamefont {Beau}\ and\ \citenamefont
  {Zagrebnov}(2010)}]{Beau2010}%
  \BibitemOpen
  \bibfield  {author} {\bibinfo {author} {\bibnamefont {Beau}, \bibfnamefont
  {M.}}, \ and\ \bibinfo {author} {\bibfnamefont {V.~A.}\ \bibnamefont
  {Zagrebnov}}} (\bibinfo {year} {2010}),\ \href@noop {} {\bibfield  {journal}
  {\bibinfo  {journal} {Condens. Matter Phys.}\ }\textbf {\bibinfo {volume}
  {13}},\ \bibinfo {pages} {23003}}\BibitemShut {NoStop}%
\bibitem [{\citenamefont {Beenakker}(2006)}]{Beenakker06}%
  \BibitemOpen
  \bibfield  {author} {\bibinfo {author} {\bibnamefont {Beenakker},
  \bibfnamefont {C.~W.~J.}}} (\bibinfo {year} {2006}),\ in\ \href@noop {}
  {\emph {\bibinfo {booktitle} {Quantum Computers, Algorithms and Chaos}}},\
  \bibinfo {series and number} {Proceedings of the International School of
  Physics "Enrico Fermi"}\ (\bibinfo  {publisher} {Societa Italiana di
  fisica},\ \bibinfo {address} {Bologna - Italy})\BibitemShut {NoStop}%
\bibitem [{\citenamefont {Beige}\ \emph
  {et~al.}(2000{\natexlab{a}})\citenamefont {Beige}, \citenamefont {Braun},\
  and\ \citenamefont {Knight}}]{Beige00b}%
  \BibitemOpen
  \bibfield  {author} {\bibinfo {author} {\bibnamefont {Beige}, \bibfnamefont
  {A.}}, \bibinfo {author} {\bibfnamefont {D.}~\bibnamefont {Braun}}, \ and\
  \bibinfo {author} {\bibfnamefont {P.~L.}\ \bibnamefont {Knight}}} (\bibinfo
  {year} {2000}{\natexlab{a}}),\ \href@noop {} {\bibfield  {journal} {\bibinfo
  {journal} {New Journal of Physics}\ }\textbf {\bibinfo {volume} {2}},\
  \bibinfo {pages} {22.1}}\BibitemShut {NoStop}%
\bibitem [{\citenamefont {Beige}\ \emph
  {et~al.}(2000{\natexlab{b}})\citenamefont {Beige}, \citenamefont {Braun},
  \citenamefont {Tregenna},\ and\ \citenamefont {Knight}}]{Beige00}%
  \BibitemOpen
  \bibfield  {author} {\bibinfo {author} {\bibnamefont {Beige}, \bibfnamefont
  {A.}}, \bibinfo {author} {\bibfnamefont {D.}~\bibnamefont {Braun}}, \bibinfo
  {author} {\bibfnamefont {B.}~\bibnamefont {Tregenna}}, \ and\ \bibinfo
  {author} {\bibfnamefont {P.~L.}\ \bibnamefont {Knight}}} (\bibinfo {year}
  {2000}{\natexlab{b}}),\ \href@noop {} {\bibfield  {journal} {\bibinfo
  {journal} {Phys. Rev. Lett.}\ }\textbf {\bibinfo {volume} {85}},\ \bibinfo
  {pages} {1762}}\BibitemShut {NoStop}%
\bibitem [{\citenamefont {Bellomo}\ \emph {et~al.}(2009)\citenamefont
  {Bellomo}, \citenamefont {De~Pasquale}, \citenamefont {Gualdi},\ and\
  \citenamefont {Marzolino}}]{Bellomo2009}%
  \BibitemOpen
  \bibfield  {author} {\bibinfo {author} {\bibnamefont {Bellomo}, \bibfnamefont
  {B.}}, \bibinfo {author} {\bibfnamefont {A.}~\bibnamefont {De~Pasquale}},
  \bibinfo {author} {\bibfnamefont {G.}~\bibnamefont {Gualdi}}, \ and\ \bibinfo
  {author} {\bibfnamefont {U.}~\bibnamefont {Marzolino}}} (\bibinfo {year}
  {2009}),\ \href {\doibase 10.1103/PhysRevA.80.052108} {\bibfield  {journal}
  {\bibinfo  {journal} {Phys. Rev. A}\ }\textbf {\bibinfo {volume} {80}},\
  \bibinfo {pages} {052108}}\BibitemShut {NoStop}%
\bibitem [{\citenamefont {Bellomo}\ \emph
  {et~al.}(2010{\natexlab{a}})\citenamefont {Bellomo}, \citenamefont
  {De~Pasquale}, \citenamefont {Gualdi},\ and\ \citenamefont
  {Marzolino}}]{Bellomo2010-2}%
  \BibitemOpen
  \bibfield  {author} {\bibinfo {author} {\bibnamefont {Bellomo}, \bibfnamefont
  {B.}}, \bibinfo {author} {\bibfnamefont {A.}~\bibnamefont {De~Pasquale}},
  \bibinfo {author} {\bibfnamefont {G.}~\bibnamefont {Gualdi}}, \ and\ \bibinfo
  {author} {\bibfnamefont {U.}~\bibnamefont {Marzolino}}} (\bibinfo {year}
  {2010}{\natexlab{a}}),\ \href {\doibase 10.1103/PhysRevA.82.062104}
  {\bibfield  {journal} {\bibinfo  {journal} {Phys. Rev. A}\ }\textbf {\bibinfo
  {volume} {82}},\ \bibinfo {pages} {062104}}\BibitemShut {NoStop}%
\bibitem [{\citenamefont {Bellomo}\ \emph
  {et~al.}(2010{\natexlab{b}})\citenamefont {Bellomo}, \citenamefont
  {De~Pasquale}, \citenamefont {Gualdi},\ and\ \citenamefont
  {Marzolino}}]{Bellomo2010-1}%
  \BibitemOpen
  \bibfield  {author} {\bibinfo {author} {\bibnamefont {Bellomo}, \bibfnamefont
  {B.}}, \bibinfo {author} {\bibfnamefont {A.}~\bibnamefont {De~Pasquale}},
  \bibinfo {author} {\bibfnamefont {G.}~\bibnamefont {Gualdi}}, \ and\ \bibinfo
  {author} {\bibfnamefont {U.}~\bibnamefont {Marzolino}}} (\bibinfo {year}
  {2010}{\natexlab{b}}),\ \href@noop {} {\bibfield  {journal} {\bibinfo
  {journal} {J. Phys. A}\ }\textbf {\bibinfo {volume} {43}},\ \bibinfo {pages}
  {395303}}\BibitemShut {NoStop}%
\bibitem [{\citenamefont {Beltr\'an}\ and\ \citenamefont
  {Luis}(2005)}]{BeltranPRA2005}%
  \BibitemOpen
  \bibfield  {author} {\bibinfo {author} {\bibnamefont {Beltr\'an},
  \bibfnamefont {J.}}, \ and\ \bibinfo {author} {\bibfnamefont
  {A.}~\bibnamefont {Luis}}} (\bibinfo {year} {2005}),\ \href {\doibase
  10.1103/PhysRevA.72.045801} {\bibfield  {journal} {\bibinfo  {journal} {Phys.
  Rev. A}\ }\textbf {\bibinfo {volume} {72}}~(\bibinfo {number} {4}),\ \bibinfo
  {pages} {045801}}\BibitemShut {NoStop}%
\bibitem [{\citenamefont {Benatti}\ and\ \citenamefont
  {Braun}(2013)}]{BenattiPRA2013}%
  \BibitemOpen
  \bibfield  {author} {\bibinfo {author} {\bibnamefont {Benatti}, \bibfnamefont
  {F.}}, \ and\ \bibinfo {author} {\bibfnamefont {D.}~\bibnamefont {Braun}}}
  (\bibinfo {year} {2013}),\ \href {\doibase 10.1103/PhysRevA.87.012340}
  {\bibfield  {journal} {\bibinfo  {journal} {Phys. Rev. A}\ }\textbf {\bibinfo
  {volume} {87}},\ \bibinfo {pages} {012340}}\BibitemShut {NoStop}%
\bibitem [{\citenamefont {Benatti}\ and\ \citenamefont
  {Floreanini}(2005)}]{Benatti05}%
  \BibitemOpen
  \bibfield  {author} {\bibinfo {author} {\bibnamefont {Benatti}, \bibfnamefont
  {F.}}, \ and\ \bibinfo {author} {\bibfnamefont {R.}~\bibnamefont
  {Floreanini}}} (\bibinfo {year} {2005}),\ \href@noop {} {\bibfield  {journal}
  {\bibinfo  {journal} {Int. J. Mod. Phys. B}\ }\textbf {\bibinfo {volume}
  {19}},\ \bibinfo {pages} {3063}}\BibitemShut {NoStop}%
\bibitem [{\citenamefont {Benatti}\ and\ \citenamefont
  {Floreanini}(2014)}]{Benatti7}%
  \BibitemOpen
  \bibfield  {author} {\bibinfo {author} {\bibnamefont {Benatti}, \bibfnamefont
  {F.}}, \ and\ \bibinfo {author} {\bibfnamefont {R.}~\bibnamefont
  {Floreanini}}} (\bibinfo {year} {2014}),\ \href@noop {} {\bibfield  {journal}
  {\bibinfo  {journal} {Int. J. Quant. Inf.}\ }\textbf {\bibinfo {volume}
  {12}},\ \bibinfo {pages} {1461002}}\BibitemShut {NoStop}%
\bibitem [{\citenamefont {Benatti}\ and\ \citenamefont
  {Floreanini}(2016)}]{BenattiFloreanini2016}%
  \BibitemOpen
  \bibfield  {author} {\bibinfo {author} {\bibnamefont {Benatti}, \bibfnamefont
  {F.}}, \ and\ \bibinfo {author} {\bibfnamefont {R.}~\bibnamefont
  {Floreanini}}} (\bibinfo {year} {2016}),\ \href@noop {} {\bibfield  {journal}
  {\bibinfo  {journal} {J. Phys. A}\ }\textbf {\bibinfo {volume} {49}},\
  \bibinfo {pages} {305303}}\BibitemShut {NoStop}%
\bibitem [{\citenamefont {Benatti}\ \emph {et~al.}(2009)\citenamefont
  {Benatti}, \citenamefont {Floreanini},\ and\ \citenamefont
  {Marzolino}}]{benatti_environment-induced_2009}%
  \BibitemOpen
  \bibfield  {author} {\bibinfo {author} {\bibnamefont {Benatti}, \bibfnamefont
  {F.}}, \bibinfo {author} {\bibfnamefont {R.}~\bibnamefont {Floreanini}}, \
  and\ \bibinfo {author} {\bibfnamefont {U.}~\bibnamefont {Marzolino}}}
  (\bibinfo {year} {2009}),\ \href {\doibase 10.1209/0295-5075/88/20011}
  {\bibfield  {journal} {\bibinfo  {journal} {EPL}\ }\textbf {\bibinfo {volume}
  {88}}~(\bibinfo {number} {2}),\ \bibinfo {pages} {20011}}\BibitemShut
  {NoStop}%
\bibitem [{\citenamefont {Benatti}\ \emph
  {et~al.}(2010{\natexlab{a}})\citenamefont {Benatti}, \citenamefont
  {Floreanini},\ and\ \citenamefont {Marzolino}}]{Benatti1}%
  \BibitemOpen
  \bibfield  {author} {\bibinfo {author} {\bibnamefont {Benatti}, \bibfnamefont
  {F.}}, \bibinfo {author} {\bibfnamefont {R.}~\bibnamefont {Floreanini}}, \
  and\ \bibinfo {author} {\bibfnamefont {U.}~\bibnamefont {Marzolino}}}
  (\bibinfo {year} {2010}{\natexlab{a}}),\ \href@noop {} {\bibfield  {journal}
  {\bibinfo  {journal} {Ann. Phys.}\ }\textbf {\bibinfo {volume} {325}},\
  \bibinfo {pages} {924}}\BibitemShut {NoStop}%
\bibitem [{\citenamefont {Benatti}\ \emph
  {et~al.}(2010{\natexlab{b}})\citenamefont {Benatti}, \citenamefont
  {Floreanini},\ and\ \citenamefont {Marzolino}}]{benatti_entangling_2010}%
  \BibitemOpen
  \bibfield  {author} {\bibinfo {author} {\bibnamefont {Benatti}, \bibfnamefont
  {F.}}, \bibinfo {author} {\bibfnamefont {R.}~\bibnamefont {Floreanini}}, \
  and\ \bibinfo {author} {\bibfnamefont {U.}~\bibnamefont {Marzolino}}}
  (\bibinfo {year} {2010}{\natexlab{b}}),\ \href {\doibase
  10.1103/PhysRevA.81.012105} {\bibfield  {journal} {\bibinfo  {journal} {Phys.
  Rev. A}\ }\textbf {\bibinfo {volume} {81}}~(\bibinfo {number} {1}),\ \bibinfo
  {pages} {012105}}\BibitemShut {NoStop}%
\bibitem [{\citenamefont {Benatti}\ \emph {et~al.}(2011)\citenamefont
  {Benatti}, \citenamefont {Floreanini},\ and\ \citenamefont
  {Marzolino}}]{Benatti2}%
  \BibitemOpen
  \bibfield  {author} {\bibinfo {author} {\bibnamefont {Benatti}, \bibfnamefont
  {F.}}, \bibinfo {author} {\bibfnamefont {R.}~\bibnamefont {Floreanini}}, \
  and\ \bibinfo {author} {\bibfnamefont {U.}~\bibnamefont {Marzolino}}}
  (\bibinfo {year} {2011}),\ \href@noop {} {\bibfield  {journal} {\bibinfo
  {journal} {J. Phys. B}\ }\textbf {\bibinfo {volume} {44}}~(\bibinfo {number}
  {091001}),\ \bibinfo {pages} {924}}\BibitemShut {NoStop}%
\bibitem [{\citenamefont {Benatti}\ \emph
  {et~al.}(2012{\natexlab{a}})\citenamefont {Benatti}, \citenamefont
  {Floreanini},\ and\ \citenamefont {Marzolino}}]{Benatti3}%
  \BibitemOpen
  \bibfield  {author} {\bibinfo {author} {\bibnamefont {Benatti}, \bibfnamefont
  {F.}}, \bibinfo {author} {\bibfnamefont {R.}~\bibnamefont {Floreanini}}, \
  and\ \bibinfo {author} {\bibfnamefont {U.}~\bibnamefont {Marzolino}}}
  (\bibinfo {year} {2012}{\natexlab{a}}),\ \href@noop {} {\bibfield  {journal}
  {\bibinfo  {journal} {Ann. Phys.}\ }\textbf {\bibinfo {volume} {327}},\
  \bibinfo {pages} {1304}}\BibitemShut {NoStop}%
\bibitem [{\citenamefont {Benatti}\ \emph
  {et~al.}(2012{\natexlab{b}})\citenamefont {Benatti}, \citenamefont
  {Floreanini},\ and\ \citenamefont {Marzolino}}]{Benatti4}%
  \BibitemOpen
  \bibfield  {author} {\bibinfo {author} {\bibnamefont {Benatti}, \bibfnamefont
  {F.}}, \bibinfo {author} {\bibfnamefont {R.}~\bibnamefont {Floreanini}}, \
  and\ \bibinfo {author} {\bibfnamefont {U.}~\bibnamefont {Marzolino}}}
  (\bibinfo {year} {2012}{\natexlab{b}}),\ \href@noop {} {\bibfield  {journal}
  {\bibinfo  {journal} {Phys. Rev. A}\ }\textbf {\bibinfo {volume} {85}},\
  \bibinfo {pages} {042329}}\BibitemShut {NoStop}%
\bibitem [{\citenamefont {Benatti}\ \emph
  {et~al.}(2014{\natexlab{a}})\citenamefont {Benatti}, \citenamefont
  {Floreanini},\ and\ \citenamefont {Marzolino}}]{Benatti6}%
  \BibitemOpen
  \bibfield  {author} {\bibinfo {author} {\bibnamefont {Benatti}, \bibfnamefont
  {F.}}, \bibinfo {author} {\bibfnamefont {R.}~\bibnamefont {Floreanini}}, \
  and\ \bibinfo {author} {\bibfnamefont {U.}~\bibnamefont {Marzolino}}}
  (\bibinfo {year} {2014}{\natexlab{a}}),\ \href@noop {} {\bibfield  {journal}
  {\bibinfo  {journal} {Phys. Rev. A}\ }\textbf {\bibinfo {volume} {89}},\
  \bibinfo {pages} {032326}}\BibitemShut {NoStop}%
\bibitem [{\citenamefont {Benatti}\ \emph {et~al.}(2003)\citenamefont
  {Benatti}, \citenamefont {Floreanini},\ and\ \citenamefont
  {Piani}}]{benatti_environment_2003}%
  \BibitemOpen
  \bibfield  {author} {\bibinfo {author} {\bibnamefont {Benatti}, \bibfnamefont
  {F.}}, \bibinfo {author} {\bibfnamefont {R.}~\bibnamefont {Floreanini}}, \
  and\ \bibinfo {author} {\bibfnamefont {M.}~\bibnamefont {Piani}}} (\bibinfo
  {year} {2003}),\ \href {\doibase 10.1103/PhysRevLett.91.070402} {\bibfield
  {journal} {\bibinfo  {journal} {Phys. Rev. Lett.}\ }\textbf {\bibinfo
  {volume} {91}}~(\bibinfo {number} {7}),\ \bibinfo {pages}
  {070402}}\BibitemShut {NoStop}%
\bibitem [{\citenamefont {Benatti}\ \emph
  {et~al.}(2014{\natexlab{b}})\citenamefont {Benatti}, \citenamefont
  {Floreanini},\ and\ \citenamefont {Titimbo}}]{Benatti5}%
  \BibitemOpen
  \bibfield  {author} {\bibinfo {author} {\bibnamefont {Benatti}, \bibfnamefont
  {F.}}, \bibinfo {author} {\bibfnamefont {R.}~\bibnamefont {Floreanini}}, \
  and\ \bibinfo {author} {\bibfnamefont {K.}~\bibnamefont {Titimbo}}} (\bibinfo
  {year} {2014}{\natexlab{b}}),\ \href@noop {} {\bibfield  {journal} {\bibinfo
  {journal} {Open. Sys. Inf. Dyn.}\ }\textbf {\bibinfo {volume} {21}},\
  \bibinfo {pages} {1440003}}\BibitemShut {NoStop}%
\bibitem [{\citenamefont {Benatti}\ \emph {et~al.}(2017)\citenamefont
  {Benatti}, \citenamefont {Franchini}, \citenamefont {Floreanini},\ and\
  \citenamefont {Marzolino}}]{Benatti8}%
  \BibitemOpen
  \bibfield  {author} {\bibinfo {author} {\bibnamefont {Benatti}, \bibfnamefont
  {F.}}, \bibinfo {author} {\bibfnamefont {F.}~\bibnamefont {Franchini}},
  \bibinfo {author} {\bibfnamefont {R.}~\bibnamefont {Floreanini}}, \ and\
  \bibinfo {author} {\bibfnamefont {U.}~\bibnamefont {Marzolino}}} (\bibinfo
  {year} {2017}),\ \href@noop {} {\bibfield  {journal} {\bibinfo  {journal}
  {Open Syst. Inf. Dyn.}\ }\textbf {\bibinfo {volume} {24}},\ \bibinfo {pages}
  {1740004}}\BibitemShut {NoStop}%
\bibitem [{\citenamefont {Benatti}\ \emph {et~al.}(2008)\citenamefont
  {Benatti}, \citenamefont {Liguori},\ and\ \citenamefont
  {Nagy}}]{benatti_environment_2008}%
  \BibitemOpen
  \bibfield  {author} {\bibinfo {author} {\bibnamefont {Benatti}, \bibfnamefont
  {F.}}, \bibinfo {author} {\bibfnamefont {A.~M.}\ \bibnamefont {Liguori}}, \
  and\ \bibinfo {author} {\bibfnamefont {A.}~\bibnamefont {Nagy}}} (\bibinfo
  {year} {2008}),\ \href {\doibase 10.1063/1.2889716} {\bibfield  {journal}
  {\bibinfo  {journal} {J. Math. Phys.}\ }\textbf {\bibinfo {volume}
  {49}}~(\bibinfo {number} {4}),\ \bibinfo {pages} {042103}}\BibitemShut
  {NoStop}%
\bibitem [{\citenamefont {Bendersky}\ and\ \citenamefont
  {Paz}(2013)}]{Bendersky2013}%
  \BibitemOpen
  \bibfield  {author} {\bibinfo {author} {\bibnamefont {Bendersky},
  \bibfnamefont {A.}}, \ and\ \bibinfo {author} {\bibfnamefont {J.~P.}\
  \bibnamefont {Paz}}} (\bibinfo {year} {2013}),\ \href {\doibase
  10.1103/PhysRevA.87.012122} {\bibfield  {journal} {\bibinfo  {journal} {Phys.
  Rev. A}\ }\textbf {\bibinfo {volume} {87}},\ \bibinfo {pages}
  {012122}}\BibitemShut {NoStop}%
\bibitem [{\citenamefont {Benedict}(1984)}]{Benedict1984}%
  \BibitemOpen
  \bibfield  {author} {\bibinfo {author} {\bibnamefont {Benedict},
  \bibfnamefont {R.~P.}}} (\bibinfo {year} {1984}),\ \href@noop {} {\emph
  {\bibinfo {title} {{Fundamentals of temperature, pressure and flow
  measurements}}}}\ (\bibinfo  {publisher} {Hogn Wiley \& sons})\BibitemShut
  {NoStop}%
\bibitem [{\citenamefont {Bengtsson}\ and\ \citenamefont
  {\.{Z}yczkowski}(2006)}]{Bengtsson06}%
  \BibitemOpen
  \bibfield  {author} {\bibinfo {author} {\bibnamefont {Bengtsson},
  \bibfnamefont {I.}}, \ and\ \bibinfo {author} {\bibfnamefont
  {K.}~\bibnamefont {\.{Z}yczkowski}}} (\bibinfo {year} {2006}),\ \href@noop {}
  {\emph {\bibinfo {title} {Geometry of quantum states: an introduction to
  quantum entanglement}}}\ (\bibinfo  {publisher} {Cambride University
  Press})\BibitemShut {NoStop}%
\bibitem [{\citenamefont {Bennett}\ \emph {et~al.}(1999)\citenamefont
  {Bennett}, \citenamefont {DiVincenzo}, \citenamefont {Fuchs}, \citenamefont
  {Mor}, \citenamefont {Rains}, \citenamefont {Shor}, \citenamefont {Smolin},\
  and\ \citenamefont {Wootters}}]{Bennett99}%
  \BibitemOpen
  \bibfield  {author} {\bibinfo {author} {\bibnamefont {Bennett}, \bibfnamefont
  {C.~H.}}, \bibinfo {author} {\bibfnamefont {D.~P.}\ \bibnamefont
  {DiVincenzo}}, \bibinfo {author} {\bibfnamefont {C.~A.}\ \bibnamefont
  {Fuchs}}, \bibinfo {author} {\bibfnamefont {T.}~\bibnamefont {Mor}}, \bibinfo
  {author} {\bibfnamefont {E.}~\bibnamefont {Rains}}, \bibinfo {author}
  {\bibfnamefont {P.~W.}\ \bibnamefont {Shor}}, \bibinfo {author}
  {\bibfnamefont {J.~A.}\ \bibnamefont {Smolin}}, \ and\ \bibinfo {author}
  {\bibfnamefont {W.~K.}\ \bibnamefont {Wootters}}} (\bibinfo {year} {1999}),\
  \href@noop {} {\bibfield  {journal} {\bibinfo  {journal} {Phys. Rev. A}\
  }\textbf {\bibinfo {volume} {59}}~(\bibinfo {number} {2}),\ \bibinfo {pages}
  {1070}}\BibitemShut {NoStop}%
\bibitem [{\citenamefont {van~den Berg}(1983)}]{vandenBerg1983}%
  \BibitemOpen
  \bibfield  {author} {\bibinfo {author} {\bibnamefont {van~den Berg},
  \bibfnamefont {M.}}} (\bibinfo {year} {1983}),\ \href@noop {} {\bibfield
  {journal} {\bibinfo  {journal} {J. Stat. Phys.}\ }\textbf {\bibinfo {volume}
  {31}},\ \bibinfo {pages} {623}}\BibitemShut {NoStop}%
\bibitem [{\citenamefont {van~den Berg}\ and\ \citenamefont
  {Lewis}(1982)}]{vandenBerg1982}%
  \BibitemOpen
  \bibfield  {author} {\bibinfo {author} {\bibnamefont {van~den Berg},
  \bibfnamefont {M.}}, \ and\ \bibinfo {author} {\bibfnamefont {J.~T.}\
  \bibnamefont {Lewis}}} (\bibinfo {year} {1982}),\ \href@noop {} {\bibfield
  {journal} {\bibinfo  {journal} {Physica A}\ }\textbf {\bibinfo {volume}
  {110}},\ \bibinfo {pages} {550}}\BibitemShut {NoStop}%
\bibitem [{\citenamefont {van~den Berg}\ \emph
  {et~al.}(1986{\natexlab{a}})\citenamefont {van~den Berg}, \citenamefont
  {Lewis},\ and\ \citenamefont {Munn}}]{vandenBerg1986-2}%
  \BibitemOpen
  \bibfield  {author} {\bibinfo {author} {\bibnamefont {van~den Berg},
  \bibfnamefont {M.}}, \bibinfo {author} {\bibfnamefont {J.~T.}\ \bibnamefont
  {Lewis}}, \ and\ \bibinfo {author} {\bibfnamefont {M.}~\bibnamefont {Munn}}}
  (\bibinfo {year} {1986}{\natexlab{a}}),\ \href@noop {} {\bibfield  {journal}
  {\bibinfo  {journal} {Helv. Phys. Acta}\ }\textbf {\bibinfo {volume} {59}},\
  \bibinfo {pages} {1289}}\BibitemShut {NoStop}%
\bibitem [{\citenamefont {van~den Berg}\ \emph
  {et~al.}(1986{\natexlab{b}})\citenamefont {van~den Berg}, \citenamefont
  {Lewis},\ and\ \citenamefont {Pul\'e}}]{vandenBerg1986-1}%
  \BibitemOpen
  \bibfield  {author} {\bibinfo {author} {\bibnamefont {van~den Berg},
  \bibfnamefont {M.}}, \bibinfo {author} {\bibfnamefont {J.~T.}\ \bibnamefont
  {Lewis}}, \ and\ \bibinfo {author} {\bibfnamefont {J.~V.}\ \bibnamefont
  {Pul\'e}}} (\bibinfo {year} {1986}{\natexlab{b}}),\ \href@noop {} {\bibfield
  {journal} {\bibinfo  {journal} {Helv. Phys. Acta}\ }\textbf {\bibinfo
  {volume} {59}},\ \bibinfo {pages} {1271}}\BibitemShut {NoStop}%
\bibitem [{\citenamefont {Berrada}(2013)}]{BerradaPRA2013}%
  \BibitemOpen
  \bibfield  {author} {\bibinfo {author} {\bibnamefont {Berrada}, \bibfnamefont
  {K.}}} (\bibinfo {year} {2013}),\ \href {\doibase 10.1103/PhysRevA.88.013817}
  {\bibfield  {journal} {\bibinfo  {journal} {Phys. Rev. A}\ }\textbf {\bibinfo
  {volume} {88}},\ \bibinfo {pages} {013817}}\BibitemShut {NoStop}%
\bibitem [{\citenamefont {Berry}\ \emph {et~al.}(2015)\citenamefont {Berry},
  \citenamefont {Tsang}, \citenamefont {Hall},\ and\ \citenamefont
  {Wiseman}}]{PhysRevX.5.031018}%
  \BibitemOpen
  \bibfield  {author} {\bibinfo {author} {\bibnamefont {Berry}, \bibfnamefont
  {D.~W.}}, \bibinfo {author} {\bibfnamefont {M.}~\bibnamefont {Tsang}},
  \bibinfo {author} {\bibfnamefont {M.~J.~W.}\ \bibnamefont {Hall}}, \ and\
  \bibinfo {author} {\bibfnamefont {H.~M.}\ \bibnamefont {Wiseman}}} (\bibinfo
  {year} {2015}),\ \href {\doibase 10.1103/PhysRevX.5.031018} {\bibfield
  {journal} {\bibinfo  {journal} {Phys. Rev. X}\ }\textbf {\bibinfo {volume}
  {5}},\ \bibinfo {pages} {031018}}\BibitemShut {NoStop}%
\bibitem [{\citenamefont {Berry}\ and\ \citenamefont
  {Wiseman}(2000)}]{berry_optimal_2000}%
  \BibitemOpen
  \bibfield  {author} {\bibinfo {author} {\bibnamefont {Berry}, \bibfnamefont
  {D.~W.}}, \ and\ \bibinfo {author} {\bibfnamefont {H.~M.}\ \bibnamefont
  {Wiseman}}} (\bibinfo {year} {2000}),\ \href {\doibase
  10.1103/PhysRevLett.85.5098} {\bibfield  {journal} {\bibinfo  {journal}
  {Phys. Rev. Lett.}\ }\textbf {\bibinfo {volume} {85}}~(\bibinfo {number}
  {24}),\ \bibinfo {pages} {5098}}\BibitemShut {NoStop}%
\bibitem [{\citenamefont {Berry}\ and\ \citenamefont
  {Wiseman}(2002)}]{PhysRevA.65.043803}%
  \BibitemOpen
  \bibfield  {author} {\bibinfo {author} {\bibnamefont {Berry}, \bibfnamefont
  {D.~W.}}, \ and\ \bibinfo {author} {\bibfnamefont {H.~M.}\ \bibnamefont
  {Wiseman}}} (\bibinfo {year} {2002}),\ \href {\doibase
  10.1103/PhysRevA.65.043803} {\bibfield  {journal} {\bibinfo  {journal} {Phys.
  Rev. A}\ }\textbf {\bibinfo {volume} {65}},\ \bibinfo {pages}
  {043803}}\BibitemShut {NoStop}%
\bibitem [{\citenamefont {Berry}\ and\ \citenamefont
  {Wiseman}(2006)}]{PhysRevA.73.063824}%
  \BibitemOpen
  \bibfield  {author} {\bibinfo {author} {\bibnamefont {Berry}, \bibfnamefont
  {D.~W.}}, \ and\ \bibinfo {author} {\bibfnamefont {H.~M.}\ \bibnamefont
  {Wiseman}}} (\bibinfo {year} {2006}),\ \href {\doibase
  10.1103/PhysRevA.73.063824} {\bibfield  {journal} {\bibinfo  {journal} {Phys.
  Rev. A}\ }\textbf {\bibinfo {volume} {73}},\ \bibinfo {pages}
  {063824}}\BibitemShut {NoStop}%
\bibitem [{\citenamefont {Berry}\ and\ \citenamefont
  {Wiseman}(2013)}]{PhysRevA.87.019901}%
  \BibitemOpen
  \bibfield  {author} {\bibinfo {author} {\bibnamefont {Berry}, \bibfnamefont
  {D.~W.}}, \ and\ \bibinfo {author} {\bibfnamefont {H.~M.}\ \bibnamefont
  {Wiseman}}} (\bibinfo {year} {2013}),\ \href {\doibase
  10.1103/PhysRevA.87.019901} {\bibfield  {journal} {\bibinfo  {journal} {Phys.
  Rev. A}\ }\textbf {\bibinfo {volume} {87}},\ \bibinfo {pages}
  {019901}}\BibitemShut {NoStop}%
\bibitem [{\citenamefont {Bimbard}\ \emph {et~al.}(2010)\citenamefont
  {Bimbard}, \citenamefont {Jain}, \citenamefont {{MacRae}},\ and\
  \citenamefont {Lvovsky}}]{bimbard_quantum-optical_2010}%
  \BibitemOpen
  \bibfield  {author} {\bibinfo {author} {\bibnamefont {Bimbard}, \bibfnamefont
  {E.}}, \bibinfo {author} {\bibfnamefont {N.}~\bibnamefont {Jain}}, \bibinfo
  {author} {\bibfnamefont {A.}~\bibnamefont {{MacRae}}}, \ and\ \bibinfo
  {author} {\bibfnamefont {A.~I.}\ \bibnamefont {Lvovsky}}} (\bibinfo {year}
  {2010}),\ \href@noop {} {\bibfield  {journal} {\bibinfo  {journal} {Nat
  Photon}\ }\textbf {\bibinfo {volume} {4}}~(\bibinfo {number} {4}),\ \bibinfo
  {pages} {243}}\BibitemShut {NoStop}%
\bibitem [{\citenamefont {Birchall}\ \emph {et~al.}(2017)\citenamefont
  {Birchall}, \citenamefont {O'Brien}, \citenamefont {Matthews},\ and\
  \citenamefont {Cable}}]{birchall_quantum-classical_2016}%
  \BibitemOpen
  \bibfield  {author} {\bibinfo {author} {\bibnamefont {Birchall},
  \bibfnamefont {P.~M.}}, \bibinfo {author} {\bibfnamefont {J.~L.}\
  \bibnamefont {O'Brien}}, \bibinfo {author} {\bibfnamefont {J.~C.~F.}\
  \bibnamefont {Matthews}}, \ and\ \bibinfo {author} {\bibfnamefont
  {H.}~\bibnamefont {Cable}}} (\bibinfo {year} {2017}),\ \href {\doibase
  10.1103/PhysRevA.96.062109} {\bibfield  {journal} {\bibinfo  {journal} {Phys.
  Rev. A}\ }\textbf {\bibinfo {volume} {96}},\ \bibinfo {pages}
  {062109}}\BibitemShut {NoStop}%
\bibitem [{\citenamefont {Bloch}\ \emph {et~al.}(2008)\citenamefont {Bloch},
  \citenamefont {Dalibard},\ and\ \citenamefont {Zwerger}}]{Bloch0}%
  \BibitemOpen
  \bibfield  {author} {\bibinfo {author} {\bibnamefont {Bloch}, \bibfnamefont
  {I.}}, \bibinfo {author} {\bibfnamefont {J.}~\bibnamefont {Dalibard}}, \ and\
  \bibinfo {author} {\bibfnamefont {W.}~\bibnamefont {Zwerger}}} (\bibinfo
  {year} {2008}),\ \href@noop {} {\bibfield  {journal} {\bibinfo  {journal}
  {Rev. Mod. Phys.}\ }\textbf {\bibinfo {volume} {80}},\ \bibinfo {pages}
  {885}}\BibitemShut {NoStop}%
\bibitem [{\citenamefont {Boixo}\ \emph
  {et~al.}(2008{\natexlab{a}})\citenamefont {Boixo}, \citenamefont {Datta},
  \citenamefont {Davis}, \citenamefont {Flammia}, \citenamefont {Shaji},\ and\
  \citenamefont {Caves}}]{BoixoPRL2008}%
  \BibitemOpen
  \bibfield  {author} {\bibinfo {author} {\bibnamefont {Boixo}, \bibfnamefont
  {S.}}, \bibinfo {author} {\bibfnamefont {A.}~\bibnamefont {Datta}}, \bibinfo
  {author} {\bibfnamefont {M.~J.}\ \bibnamefont {Davis}}, \bibinfo {author}
  {\bibfnamefont {S.~T.}\ \bibnamefont {Flammia}}, \bibinfo {author}
  {\bibfnamefont {A.}~\bibnamefont {Shaji}}, \ and\ \bibinfo {author}
  {\bibfnamefont {C.~M.}\ \bibnamefont {Caves}}} (\bibinfo {year}
  {2008}{\natexlab{a}}),\ \href@noop {} {\bibfield  {journal} {\bibinfo
  {journal} {Phys. Rev. Lett.}\ }\textbf {\bibinfo {volume} {101}}~(\bibinfo
  {number} {4}),\ \bibinfo {pages} {040403}}\BibitemShut {NoStop}%
\bibitem [{\citenamefont {Boixo}\ \emph {et~al.}(2009)\citenamefont {Boixo},
  \citenamefont {Datta}, \citenamefont {Davis}, \citenamefont {Shaji},
  \citenamefont {Tacla},\ and\ \citenamefont {Caves}}]{BoixoPRA2009}%
  \BibitemOpen
  \bibfield  {author} {\bibinfo {author} {\bibnamefont {Boixo}, \bibfnamefont
  {S.}}, \bibinfo {author} {\bibfnamefont {A.}~\bibnamefont {Datta}}, \bibinfo
  {author} {\bibfnamefont {M.~J.}\ \bibnamefont {Davis}}, \bibinfo {author}
  {\bibfnamefont {A.}~\bibnamefont {Shaji}}, \bibinfo {author} {\bibfnamefont
  {A.~B.}\ \bibnamefont {Tacla}}, \ and\ \bibinfo {author} {\bibfnamefont
  {C.~M.}\ \bibnamefont {Caves}}} (\bibinfo {year} {2009}),\ \href {\doibase
  10.1103/PhysRevA.80.032103} {\bibfield  {journal} {\bibinfo  {journal} {Phys.
  Rev. A}\ }\textbf {\bibinfo {volume} {80}},\ \bibinfo {pages}
  {032103}}\BibitemShut {NoStop}%
\bibitem [{\citenamefont {Boixo}\ \emph
  {et~al.}(2008{\natexlab{b}})\citenamefont {Boixo}, \citenamefont {Datta},
  \citenamefont {Flammia}, \citenamefont {Shaji}, \citenamefont {Bagan},\ and\
  \citenamefont {Caves}}]{BoixoPRA2008}%
  \BibitemOpen
  \bibfield  {author} {\bibinfo {author} {\bibnamefont {Boixo}, \bibfnamefont
  {S.}}, \bibinfo {author} {\bibfnamefont {A.}~\bibnamefont {Datta}}, \bibinfo
  {author} {\bibfnamefont {S.~T.}\ \bibnamefont {Flammia}}, \bibinfo {author}
  {\bibfnamefont {A.}~\bibnamefont {Shaji}}, \bibinfo {author} {\bibfnamefont
  {E.}~\bibnamefont {Bagan}}, \ and\ \bibinfo {author} {\bibfnamefont {C.~M.}\
  \bibnamefont {Caves}}} (\bibinfo {year} {2008}{\natexlab{b}}),\ \href@noop {}
  {\bibfield  {journal} {\bibinfo  {journal} {Phys. Rev. A}\ }\textbf {\bibinfo
  {volume} {77}}~(\bibinfo {number} {1}),\ \bibinfo {pages}
  {{012317}}}\BibitemShut {NoStop}%
\bibitem [{\citenamefont {Boixo}\ \emph {et~al.}(2007)\citenamefont {Boixo},
  \citenamefont {Flammia}, \citenamefont {Caves},\ and\ \citenamefont
  {Geremia}}]{BoixoPRL2007}%
  \BibitemOpen
  \bibfield  {author} {\bibinfo {author} {\bibnamefont {Boixo}, \bibfnamefont
  {S.}}, \bibinfo {author} {\bibfnamefont {S.~T.}\ \bibnamefont {Flammia}},
  \bibinfo {author} {\bibfnamefont {C.~M.}\ \bibnamefont {Caves}}, \ and\
  \bibinfo {author} {\bibfnamefont {J.~M.}\ \bibnamefont {Geremia}}} (\bibinfo
  {year} {2007}),\ \href {\doibase 10.1103/PhysRevLett.98.090401} {\bibfield
  {journal} {\bibinfo  {journal} {Phys. Rev. Lett.}\ }\textbf {\bibinfo
  {volume} {98}}~(\bibinfo {number} {9}),\ \bibinfo {pages}
  {{090401}}}\BibitemShut {NoStop}%
\bibitem [{\citenamefont {Boixo}\ and\ \citenamefont
  {Heunen}(2012)}]{Boixo2012}%
  \BibitemOpen
  \bibfield  {author} {\bibinfo {author} {\bibnamefont {Boixo}, \bibfnamefont
  {S.}}, \ and\ \bibinfo {author} {\bibfnamefont {C.}~\bibnamefont {Heunen}}}
  (\bibinfo {year} {2012}),\ \href {\doibase 10.1103/PhysRevLett.108.120402}
  {\bibfield  {journal} {\bibinfo  {journal} {Phys. Rev. Lett.}\ }\textbf
  {\bibinfo {volume} {108}},\ \bibinfo {pages} {120402}}\BibitemShut {NoStop}%
\bibitem [{\citenamefont {Boixo}\ and\ \citenamefont
  {Somma}(2008)}]{BoixoSomma}%
  \BibitemOpen
  \bibfield  {author} {\bibinfo {author} {\bibnamefont {Boixo}, \bibfnamefont
  {S.}}, \ and\ \bibinfo {author} {\bibfnamefont {R.~D.}\ \bibnamefont
  {Somma}}} (\bibinfo {year} {2008}),\ \href {\doibase
  10.1103/PhysRevA.77.052320} {\bibfield  {journal} {\bibinfo  {journal} {Phys.
  Rev. A}\ }\textbf {\bibinfo {volume} {77}},\ \bibinfo {pages}
  {052320}}\BibitemShut {NoStop}%
\bibitem [{\citenamefont {Bollinger}\ \emph {et~al.}(1996)\citenamefont
  {Bollinger}, \citenamefont {Itano}, \citenamefont {Wineland},\ and\
  \citenamefont {Heinzen}}]{Bollinger0}%
  \BibitemOpen
  \bibfield  {author} {\bibinfo {author} {\bibnamefont {Bollinger},
  \bibfnamefont {J.~J.}}, \bibinfo {author} {\bibfnamefont {W.~M.}\
  \bibnamefont {Itano}}, \bibinfo {author} {\bibfnamefont {D.~J.}\ \bibnamefont
  {Wineland}}, \ and\ \bibinfo {author} {\bibfnamefont {D.~J.}\ \bibnamefont
  {Heinzen}}} (\bibinfo {year} {1996}),\ \href {\doibase
  10.1103/PhysRevA.54.R4649} {\bibfield  {journal} {\bibinfo  {journal} {Phys.
  Rev. A}\ }\textbf {\bibinfo {volume} {54}},\ \bibinfo {pages}
  {R4649}}\BibitemShut {NoStop}%
\bibitem [{\citenamefont {Boto}\ \emph {et~al.}(2000)\citenamefont {Boto},
  \citenamefont {Kok}, \citenamefont {Abrams}, \citenamefont {Braunstein},
  \citenamefont {Williams},\ and\ \citenamefont {Dowling}}]{Boto00}%
  \BibitemOpen
  \bibfield  {author} {\bibinfo {author} {\bibnamefont {Boto}, \bibfnamefont
  {A.~N.}}, \bibinfo {author} {\bibfnamefont {P.}~\bibnamefont {Kok}}, \bibinfo
  {author} {\bibfnamefont {D.~S.}\ \bibnamefont {Abrams}}, \bibinfo {author}
  {\bibfnamefont {S.~L.}\ \bibnamefont {Braunstein}}, \bibinfo {author}
  {\bibfnamefont {C.~P.}\ \bibnamefont {Williams}}, \ and\ \bibinfo {author}
  {\bibfnamefont {J.~P.}\ \bibnamefont {Dowling}}} (\bibinfo {year} {2000}),\
  \href {\doibase 10.1103/PhysRevLett.85.2733} {\bibfield  {journal} {\bibinfo
  {journal} {Phys. Rev. Lett.}\ }\textbf {\bibinfo {volume} {85}}~(\bibinfo
  {number} {13}),\ \bibinfo {pages} {2733}}\BibitemShut {NoStop}%
\bibitem [{\citenamefont {Bouchoule}\ \emph {et~al.}(2011)\citenamefont
  {Bouchoule}, \citenamefont {van Druten},\ and\ \citenamefont
  {Westbrook}}]{Bouchoule2011}%
  \BibitemOpen
  \bibfield  {author} {\bibinfo {author} {\bibnamefont {Bouchoule},
  \bibfnamefont {I.}}, \bibinfo {author} {\bibfnamefont {N.~J.}\ \bibnamefont
  {van Druten}}, \ and\ \bibinfo {author} {\bibfnamefont {C.~I.}\ \bibnamefont
  {Westbrook}}} (\bibinfo {year} {2011}),\ in\ \href@noop {} {\emph {\bibinfo
  {booktitle} {{Atom Chips}}}},\ \bibinfo {editor} {edited by\ \bibinfo
  {editor} {\bibfnamefont {J.}~\bibnamefont {Reichel}}\ and\ \bibinfo {editor}
  {\bibfnamefont {V.}~\bibnamefont {Vuleti\'c}}}\ (\bibinfo  {publisher}
  {Wiley-VCH Verlag GmbH \& Co. KGaA, Weinheim, Germany})\BibitemShut {NoStop}%
\bibitem [{\citenamefont {Bouyer}\ and\ \citenamefont
  {Kasevich}(1997)}]{Bouyer0}%
  \BibitemOpen
  \bibfield  {author} {\bibinfo {author} {\bibnamefont {Bouyer}, \bibfnamefont
  {P.}}, \ and\ \bibinfo {author} {\bibfnamefont {M.}~\bibnamefont {Kasevich}}}
  (\bibinfo {year} {1997}),\ \href@noop {} {\bibfield  {journal} {\bibinfo
  {journal} {Phys. Rev. A}\ }\textbf {\bibinfo {volume} {56}},\ \bibinfo
  {pages} {R1083}}\BibitemShut {NoStop}%
\bibitem [{\citenamefont {Bradshaw}\ \emph {et~al.}(2016)\citenamefont
  {Bradshaw}, \citenamefont {Assad}, \citenamefont {Haw}, \citenamefont {Tan},
  \citenamefont {Lam},\ and\ \citenamefont {Gu}}]{Bradshaw2016}%
  \BibitemOpen
  \bibfield  {author} {\bibinfo {author} {\bibnamefont {Bradshaw},
  \bibfnamefont {M.}}, \bibinfo {author} {\bibfnamefont {S.~M.}\ \bibnamefont
  {Assad}}, \bibinfo {author} {\bibfnamefont {J.~Y.}\ \bibnamefont {Haw}},
  \bibinfo {author} {\bibfnamefont {S.-H.}\ \bibnamefont {Tan}}, \bibinfo
  {author} {\bibfnamefont {P.~K.}\ \bibnamefont {Lam}}, \ and\ \bibinfo
  {author} {\bibfnamefont {M.}~\bibnamefont {Gu}}} (\bibinfo {year} {2016}),\
  \href@noop {} {\enquote {\bibinfo {title} {{T}he overarching framework
  between {G}aussian quantum discord and {G}aussian quantum illumination},}\
  }\bibinfo {note} {ArXiv:1611.10020v1}\BibitemShut {NoStop}%
\bibitem [{\citenamefont {Braginsky}\ \emph {et~al.}(1995)\citenamefont
  {Braginsky}, \citenamefont {Khalili},\ and\ \citenamefont
  {Thorne}}]{braginsky_quantum_1995}%
  \BibitemOpen
  \bibfield  {author} {\bibinfo {author} {\bibnamefont {Braginsky},
  \bibfnamefont {V.~B.}}, \bibinfo {author} {\bibfnamefont {F.~Y.}\
  \bibnamefont {Khalili}}, \ and\ \bibinfo {author} {\bibfnamefont {K.~S.}\
  \bibnamefont {Thorne}}} (\bibinfo {year} {1995}),\ \href@noop {} {\emph
  {\bibinfo {title} {Quantum {Measurement}}}},\ \bibinfo {edition} {1st}\ ed.\
  (\bibinfo  {publisher} {Cambridge University Press},\ \bibinfo {address}
  {Cambridge})\BibitemShut {NoStop}%
\bibitem [{\citenamefont {Brand\~ao}\ and\ \citenamefont
  {Gour}(2015)}]{PhysRevLett.115.070503}%
  \BibitemOpen
  \bibfield  {author} {\bibinfo {author} {\bibnamefont {Brand\~ao},
  \bibfnamefont {F.~G. S.~L.}}, \ and\ \bibinfo {author} {\bibfnamefont
  {G.}~\bibnamefont {Gour}}} (\bibinfo {year} {2015}),\ \href {\doibase
  10.1103/PhysRevLett.115.070503} {\bibfield  {journal} {\bibinfo  {journal}
  {Phys. Rev. Lett.}\ }\textbf {\bibinfo {volume} {115}},\ \bibinfo {pages}
  {070503}}\BibitemShut {NoStop}%
\bibitem [{\citenamefont {Brankov}\ \emph {et~al.}(2000)\citenamefont
  {Brankov}, \citenamefont {Danchev},\ and\ \citenamefont
  {Tonchev}}]{BrankovDanchevTonchev}%
  \BibitemOpen
  \bibfield  {author} {\bibinfo {author} {\bibnamefont {Brankov}, \bibfnamefont
  {J.~G.}}, \bibinfo {author} {\bibfnamefont {D.~M.}\ \bibnamefont {Danchev}},
  \ and\ \bibinfo {author} {\bibfnamefont {N.~S.}\ \bibnamefont {Tonchev}}}
  (\bibinfo {year} {2000}),\ \href@noop {} {\emph {\bibinfo {title} {Theory of
  Critical Phenomena in Finite-Size Systems}}}\ (\bibinfo  {publisher} {World
  Scientific, Singapore})\BibitemShut {NoStop}%
\bibitem [{\citenamefont {Bratteli}\ and\ \citenamefont
  {Robinson}(1987)}]{Bratteli0}%
  \BibitemOpen
  \bibfield  {author} {\bibinfo {author} {\bibnamefont {Bratteli},
  \bibfnamefont {O.}}, \ and\ \bibinfo {author} {\bibfnamefont
  {D.}~\bibnamefont {Robinson}}} (\bibinfo {year} {1987}),\ \href@noop {}
  {\emph {\bibinfo {title} {Operator Algebras and Quantum Statistical
  Mechanics}}}\ (\bibinfo  {publisher} {Springer},\ \bibinfo {address}
  {Heidelberg})\BibitemShut {NoStop}%
\bibitem [{\citenamefont {Braun}(2002)}]{Braun02}%
  \BibitemOpen
  \bibfield  {author} {\bibinfo {author} {\bibnamefont {Braun}, \bibfnamefont
  {D.}}} (\bibinfo {year} {2002}),\ \href@noop {} {\bibfield  {journal}
  {\bibinfo  {journal} {Phys. Rev. Lett.}\ }\textbf {\bibinfo {volume} {89}},\
  \bibinfo {pages} {277901}}\BibitemShut {NoStop}%
\bibitem [{\citenamefont {Braun}(2005)}]{Braun05}%
  \BibitemOpen
  \bibfield  {author} {\bibinfo {author} {\bibnamefont {Braun}, \bibfnamefont
  {D.}}} (\bibinfo {year} {2005}),\ \href@noop {} {\bibfield  {journal}
  {\bibinfo  {journal} {Phys. Rev. A}\ }\textbf {\bibinfo {volume} {72}},\
  \bibinfo {pages} {062324}}\BibitemShut {NoStop}%
\bibitem [{\citenamefont {Braun}(2010)}]{Braun10}%
  \BibitemOpen
  \bibfield  {author} {\bibinfo {author} {\bibnamefont {Braun}, \bibfnamefont
  {D.}}} (\bibinfo {year} {2010}),\ \href@noop {} {\bibfield  {journal}
  {\bibinfo  {journal} {Eur. Phys. J. D}\ }\textbf {\bibinfo {volume} {59}},\
  \bibinfo {pages} {521}}\BibitemShut {NoStop}%
\bibitem [{\citenamefont {Braun}(2011)}]{Braun11.2}%
  \BibitemOpen
  \bibfield  {author} {\bibinfo {author} {\bibnamefont {Braun}, \bibfnamefont
  {D.}}} (\bibinfo {year} {2011}),\ \href@noop {} {\bibfield  {journal}
  {\bibinfo  {journal} {Europhys. Lett.}\ }\textbf {\bibinfo {volume}
  {94}}~(\bibinfo {number} {6}),\ \bibinfo {pages} {68007}}\BibitemShut
  {NoStop}%
\bibitem [{\citenamefont {Braun}(2012)}]{braun_ultimate_2012}%
  \BibitemOpen
  \bibfield  {author} {\bibinfo {author} {\bibnamefont {Braun}, \bibfnamefont
  {D.}}} (\bibinfo {year} {2012}),\ \href {\doibase 10.1209/0295-5075/99/49901}
  {\bibfield  {journal} {\bibinfo  {journal} {{EPL}}\ }\textbf {\bibinfo
  {volume} {99}}~(\bibinfo {number} {4}),\ \bibinfo {pages}
  {49901}}\BibitemShut {NoStop}%
\bibitem [{\citenamefont {Braun}\ and\ \citenamefont {Martin}(2011)}]{Braun11}%
  \BibitemOpen
  \bibfield  {author} {\bibinfo {author} {\bibnamefont {Braun}, \bibfnamefont
  {D.}}, \ and\ \bibinfo {author} {\bibfnamefont {J.}~\bibnamefont {Martin}}}
  (\bibinfo {year} {2011}),\ \href {\doibase 10.1038/ncomms1220} {\bibfield
  {journal} {\bibinfo  {journal} {Nat. Commun.}\ }\textbf {\bibinfo {volume}
  {2}},\ \bibinfo {pages} {223}}\BibitemShut {NoStop}%
\bibitem [{\citenamefont {Braun}\ and\ \citenamefont
  {Popescu}(2014)}]{braun_coherently_2014}%
  \BibitemOpen
  \bibfield  {author} {\bibinfo {author} {\bibnamefont {Braun}, \bibfnamefont
  {D.}}, \ and\ \bibinfo {author} {\bibfnamefont {S.}~\bibnamefont {Popescu}}}
  (\bibinfo {year} {2014}),\ \href
  {http://www.degruyter.com/view/j/qmetro.2014.2.issue-1/qmetro-2014-0003/qmetro-2014-0003.xml}
  {\bibfield  {journal} {\bibinfo  {journal} {Quantum Measurements and Quantum
  Metrology}\ }\textbf {\bibinfo {volume} {2}}~(\bibinfo {number}
  {1})}\BibitemShut {NoStop}%
\bibitem [{\citenamefont {Braun}\ \emph {et~al.}(2017)\citenamefont {Braun},
  \citenamefont {Schneiter},\ and\ \citenamefont {Fischer}}]{braun_how_2015}%
  \BibitemOpen
  \bibfield  {author} {\bibinfo {author} {\bibnamefont {Braun}, \bibfnamefont
  {D.}}, \bibinfo {author} {\bibfnamefont {F.}~\bibnamefont {Schneiter}}, \
  and\ \bibinfo {author} {\bibfnamefont {U.~R.}\ \bibnamefont {Fischer}}}
  (\bibinfo {year} {2017}),\ \href@noop {} {\bibfield  {journal} {\bibinfo
  {journal} {Classical and Quantum Gravity}\ }\textbf {\bibinfo {volume}
  {34}}~(\bibinfo {number} {17}),\ \bibinfo {pages} {175009}}\BibitemShut
  {NoStop}%
\bibitem [{\citenamefont {Braunstein}\ and\ \citenamefont
  {Caves}(1994)}]{Braunstein94}%
  \BibitemOpen
  \bibfield  {author} {\bibinfo {author} {\bibnamefont {Braunstein},
  \bibfnamefont {S.~L.}}, \ and\ \bibinfo {author} {\bibfnamefont {C.~M.}\
  \bibnamefont {Caves}}} (\bibinfo {year} {1994}),\ \href {\doibase
  10.1103/PhysRevLett.72.3439} {\bibfield  {journal} {\bibinfo  {journal}
  {Phys. Rev. Lett.}\ }\textbf {\bibinfo {volume} {72}},\ \bibinfo {pages}
  {3439}}\BibitemShut {NoStop}%
\bibitem [{\citenamefont {Braunstein}\ \emph {et~al.}(1996)\citenamefont
  {Braunstein}, \citenamefont {Caves},\ and\ \citenamefont
  {Milburn}}]{braunstein_generalized_1996}%
  \BibitemOpen
  \bibfield  {author} {\bibinfo {author} {\bibnamefont {Braunstein},
  \bibfnamefont {S.~L.}}, \bibinfo {author} {\bibfnamefont {C.~M.}\
  \bibnamefont {Caves}}, \ and\ \bibinfo {author} {\bibfnamefont {G.~J.}\
  \bibnamefont {Milburn}}} (\bibinfo {year} {1996}),\ \href@noop {} {\bibfield
  {journal} {\bibinfo  {journal} {Annals of Physics}\ }\textbf {\bibinfo
  {volume} {247}}~(\bibinfo {number} {1}),\ \bibinfo {pages} {135}}\BibitemShut
  {NoStop}%
\bibitem [{\citenamefont {Breuer}\ and\ \citenamefont
  {Petruccione}(2002)}]{Breuer02}%
  \BibitemOpen
  \bibfield  {author} {\bibinfo {author} {\bibnamefont {Breuer}, \bibfnamefont
  {H.-P.}}, \ and\ \bibinfo {author} {\bibfnamefont {F.}~\bibnamefont
  {Petruccione}}} (\bibinfo {year} {2002}),\ \href@noop {} {\emph {\bibinfo
  {title} {The Theory of Open Quantum Systems}}}\ (\bibinfo  {publisher}
  {Oxford University Press})\BibitemShut {NoStop}%
\bibitem [{\citenamefont {Brody}\ and\ \citenamefont
  {Rivier}(1995)}]{Brody1995}%
  \BibitemOpen
  \bibfield  {author} {\bibinfo {author} {\bibnamefont {Brody}, \bibfnamefont
  {D.}}, \ and\ \bibinfo {author} {\bibfnamefont {N.}~\bibnamefont {Rivier}}}
  (\bibinfo {year} {1995}),\ \href {\doibase 10.1103/PhysRevE.51.1006}
  {\bibfield  {journal} {\bibinfo  {journal} {Phys. Rev. E}\ }\textbf {\bibinfo
  {volume} {51}},\ \bibinfo {pages} {1006}}\BibitemShut {NoStop}%
\bibitem [{\citenamefont {Brody}\ and\ \citenamefont {Ritz}(2003)}]{Brody2003}%
  \BibitemOpen
  \bibfield  {author} {\bibinfo {author} {\bibnamefont {Brody}, \bibfnamefont
  {D.~C.}}, \ and\ \bibinfo {author} {\bibfnamefont {A.}~\bibnamefont {Ritz}}}
  (\bibinfo {year} {2003}),\ \href@noop {} {\bibfield  {journal} {\bibinfo
  {journal} {Journal of Geometry and Physics}\ }\textbf {\bibinfo {volume}
  {47}},\ \bibinfo {pages} {207}}\BibitemShut {NoStop}%
\bibitem [{\citenamefont {Bromley}\ \emph {et~al.}(2017)\citenamefont
  {Bromley}, \citenamefont {Silva}, \citenamefont {Oncebay-Segura},
  \citenamefont {Soares-Pinto}, \citenamefont {deAzevedo}, \citenamefont
  {Tufarelli},\ and\ \citenamefont {Adesso}}]{thereismore}%
  \BibitemOpen
  \bibfield  {author} {\bibinfo {author} {\bibnamefont {Bromley}, \bibfnamefont
  {T.~R.}}, \bibinfo {author} {\bibfnamefont {I.~A.}\ \bibnamefont {Silva}},
  \bibinfo {author} {\bibfnamefont {C.~O.}\ \bibnamefont {Oncebay-Segura}},
  \bibinfo {author} {\bibfnamefont {D.~O.}\ \bibnamefont {Soares-Pinto}},
  \bibinfo {author} {\bibfnamefont {E.~R.}\ \bibnamefont {deAzevedo}}, \bibinfo
  {author} {\bibfnamefont {T.}~\bibnamefont {Tufarelli}}, \ and\ \bibinfo
  {author} {\bibfnamefont {G.}~\bibnamefont {Adesso}}} (\bibinfo {year}
  {2017}),\ \href {\doibase 10.1103/PhysRevA.95.052313} {\bibfield  {journal}
  {\bibinfo  {journal} {Phys. Rev. A}\ }\textbf {\bibinfo {volume} {95}},\
  \bibinfo {pages} {052313}}\BibitemShut {NoStop}%
\bibitem [{\citenamefont {Budker}\ \emph {et~al.}(2000)\citenamefont {Budker},
  \citenamefont {Kimball}, \citenamefont {Rochester},\ and\ \citenamefont
  {Yashchuk}}]{BudkerPRL2000}%
  \BibitemOpen
  \bibfield  {author} {\bibinfo {author} {\bibnamefont {Budker}, \bibfnamefont
  {D.}}, \bibinfo {author} {\bibfnamefont {D.~F.}\ \bibnamefont {Kimball}},
  \bibinfo {author} {\bibfnamefont {S.~M.}\ \bibnamefont {Rochester}}, \ and\
  \bibinfo {author} {\bibfnamefont {V.~V.}\ \bibnamefont {Yashchuk}}} (\bibinfo
  {year} {2000}),\ \href {\doibase 10.1103/PhysRevLett.85.2088} {\bibfield
  {journal} {\bibinfo  {journal} {Phys. Rev. Lett.}\ }\textbf {\bibinfo
  {volume} {85}},\ \bibinfo {pages} {2088}}\BibitemShut {NoStop}%
\bibitem [{\citenamefont {Cable}\ \emph {et~al.}(2016)\citenamefont {Cable},
  \citenamefont {Gu},\ and\ \citenamefont {Modi}}]{Cable2015}%
  \BibitemOpen
  \bibfield  {author} {\bibinfo {author} {\bibnamefont {Cable}, \bibfnamefont
  {H.}}, \bibinfo {author} {\bibfnamefont {M.}~\bibnamefont {Gu}}, \ and\
  \bibinfo {author} {\bibfnamefont {K.}~\bibnamefont {Modi}}} (\bibinfo {year}
  {2016}),\ \href {\doibase 10.1103/PhysRevA.93.040304} {\bibfield  {journal}
  {\bibinfo  {journal} {Phys. Rev. A}\ }\textbf {\bibinfo {volume} {93}},\
  \bibinfo {pages} {040304}}\BibitemShut {NoStop}%
\bibitem [{\citenamefont {Calabrese}\ \emph {et~al.}(2012)\citenamefont
  {Calabrese}, \citenamefont {Mintchev},\ and\ \citenamefont
  {Vicari}}]{Calabrese0}%
  \BibitemOpen
  \bibfield  {author} {\bibinfo {author} {\bibnamefont {Calabrese},
  \bibfnamefont {P.}}, \bibinfo {author} {\bibfnamefont {M.}~\bibnamefont
  {Mintchev}}, \ and\ \bibinfo {author} {\bibfnamefont {E.}~\bibnamefont
  {Vicari}}} (\bibinfo {year} {2012}),\ \href@noop {} {\bibfield  {journal}
  {\bibinfo  {journal} {Europhys. Lett.}\ }\textbf {\bibinfo {volume} {98}},\
  \bibinfo {pages} {20003}}\BibitemShut {NoStop}%
\bibitem [{\citenamefont {Campos~Venuti}\ \emph {et~al.}(2008)\citenamefont
  {Campos~Venuti}, \citenamefont {Cozzini}, \citenamefont {Buonsante},
  \citenamefont {Massel}, \citenamefont {Bray-Ali},\ and\ \citenamefont
  {Zanardi}}]{CamposVenuti2008}%
  \BibitemOpen
  \bibfield  {author} {\bibinfo {author} {\bibnamefont {Campos~Venuti},
  \bibfnamefont {L.}}, \bibinfo {author} {\bibfnamefont {M.}~\bibnamefont
  {Cozzini}}, \bibinfo {author} {\bibfnamefont {P.}~\bibnamefont {Buonsante}},
  \bibinfo {author} {\bibfnamefont {F.}~\bibnamefont {Massel}}, \bibinfo
  {author} {\bibfnamefont {N.}~\bibnamefont {Bray-Ali}}, \ and\ \bibinfo
  {author} {\bibfnamefont {P.}~\bibnamefont {Zanardi}}} (\bibinfo {year}
  {2008}),\ \href {\doibase 10.1103/PhysRevB.78.115410} {\bibfield  {journal}
  {\bibinfo  {journal} {Phys. Rev. B}\ }\textbf {\bibinfo {volume} {78}},\
  \bibinfo {pages} {115410}}\BibitemShut {NoStop}%
\bibitem [{\citenamefont {Campos~Venuti}\ and\ \citenamefont
  {Zanardi}(2007)}]{CamposVenuti2007}%
  \BibitemOpen
  \bibfield  {author} {\bibinfo {author} {\bibnamefont {Campos~Venuti},
  \bibfnamefont {L.}}, \ and\ \bibinfo {author} {\bibfnamefont
  {P.}~\bibnamefont {Zanardi}}} (\bibinfo {year} {2007}),\ \href {\doibase
  10.1103/PhysRevLett.99.095701} {\bibfield  {journal} {\bibinfo  {journal}
  {Phys. Rev. Lett.}\ }\textbf {\bibinfo {volume} {99}},\ \bibinfo {pages}
  {095701}}\BibitemShut {NoStop}%
\bibitem [{\citenamefont {Casimir}(1968)}]{Casimir1968}%
  \BibitemOpen
  \bibfield  {author} {\bibinfo {author} {\bibnamefont {Casimir}, \bibfnamefont
  {H.~B.~G.}}} (\bibinfo {year} {1968}),\ in\ \href@noop {} {\emph {\bibinfo
  {booktitle} {{Fundamental problems in statistical mechanics II}}}},\ \bibinfo
  {editor} {edited by\ \bibinfo {editor} {\bibfnamefont {E.~G.~D.}\
  \bibnamefont {Cohen}}},\ p.\ \bibinfo {pages} {188}\BibitemShut {NoStop}%
\bibitem [{\citenamefont {Caves}(1980)}]{caves_measurement_1980}%
  \BibitemOpen
  \bibfield  {author} {\bibinfo {author} {\bibnamefont {Caves}, \bibfnamefont
  {C.~M.}}} (\bibinfo {year} {1980}),\ \href {\doibase
  10.1103/RevModPhys.52.341} {\bibfield  {journal} {\bibinfo  {journal} {Rev.
  Mod. Phys.}\ }\textbf {\bibinfo {volume} {52}}~(\bibinfo {number} {2}),\
  \bibinfo {pages} {341}}\BibitemShut {NoStop}%
\bibitem [{\citenamefont {Caves}(1981)}]{caves_quantum-mechanical_1981}%
  \BibitemOpen
  \bibfield  {author} {\bibinfo {author} {\bibnamefont {Caves}, \bibfnamefont
  {C.~M.}}} (\bibinfo {year} {1981}),\ \href@noop {} {\bibfield  {journal}
  {\bibinfo  {journal} {Phys. Rev. D}\ }\textbf {\bibinfo {volume}
  {23}}~(\bibinfo {number} {8}),\ \bibinfo {pages} {1693}}\BibitemShut
  {NoStop}%
\bibitem [{\citenamefont {Paunkovi\ifmmode~\acute{c}\else \'{c}\fi{}}\ \emph
  {et~al.}(2008)\citenamefont {Paunkovi\ifmmode~\acute{c}\else \'{c}\fi{}},
  \citenamefont {Sacramento}, \citenamefont {Nogueira}, \citenamefont
  {Vieira},\ and\ \citenamefont {Dugaev}}]{Paunkovic2008}%
  \BibitemOpen
  \bibfield  {author} {\bibinfo {author} {\bibnamefont
  {Paunkovi\ifmmode~\acute{c}\else \'{c}\fi{}}, \bibfnamefont {N.}}, \bibinfo
  {author} {\bibfnamefont {P.~D.}\ \bibnamefont {Sacramento}}, \bibinfo
  {author} {\bibfnamefont {P.}~\bibnamefont {Nogueira}}, \bibinfo {author}
  {\bibfnamefont {V.~R.}\ \bibnamefont {Vieira}}, \ and\ \bibinfo {author}
  {\bibfnamefont {V.~K.}\ \bibnamefont {Dugaev}}} (\bibinfo {year} {2008}),\
  \href {\doibase 10.1103/PhysRevA.77.052302} {\bibfield  {journal} {\bibinfo
  {journal} {Phys. Rev. A}\ }\textbf {\bibinfo {volume} {77}},\ \bibinfo
  {pages} {052302}}\BibitemShut {NoStop}%
\bibitem [{\citenamefont {Paunkovi\ifmmode~\acute{c}\else \'{c}\fi{}}\ and\
  \citenamefont {Vieira}(2008)}]{Paunkovic2008-2}%
  \BibitemOpen
  \bibfield  {author} {\bibinfo {author} {\bibnamefont
  {Paunkovi\ifmmode~\acute{c}\else \'{c}\fi{}}, \bibfnamefont {N.}}, \ and\
  \bibinfo {author} {\bibfnamefont {V.~R.}\ \bibnamefont {Vieira}}} (\bibinfo
  {year} {2008}),\ \href {\doibase 10.1103/PhysRevE.77.011129} {\bibfield
  {journal} {\bibinfo  {journal} {Phys. Rev. E}\ }\textbf {\bibinfo {volume}
  {77}},\ \bibinfo {pages} {011129}}\BibitemShut {NoStop}%
\bibitem [{\citenamefont
  {Chapeau-Blondeau}(2015)}]{chapeau-blondeau_optimized_2015}%
  \BibitemOpen
  \bibfield  {author} {\bibinfo {author} {\bibnamefont {Chapeau-Blondeau},
  \bibfnamefont {F.}}} (\bibinfo {year} {2015}),\ \href {\doibase
  10.1103/PhysRevA.91.052310} {\bibfield  {journal} {\bibinfo  {journal} {Phys.
  Rev. A}\ }\textbf {\bibinfo {volume} {91}},\ \bibinfo {pages}
  {052310}}\BibitemShut {NoStop}%
\bibitem [{\citenamefont {Chen}\ \emph {et~al.}(2008)\citenamefont {Chen},
  \citenamefont {Wang}, \citenamefont {Hao},\ and\ \citenamefont
  {Wang}}]{Chen2008}%
  \BibitemOpen
  \bibfield  {author} {\bibinfo {author} {\bibnamefont {Chen}, \bibfnamefont
  {S.}}, \bibinfo {author} {\bibfnamefont {L.}~\bibnamefont {Wang}}, \bibinfo
  {author} {\bibfnamefont {Y.}~\bibnamefont {Hao}}, \ and\ \bibinfo {author}
  {\bibfnamefont {Y.}~\bibnamefont {Wang}}} (\bibinfo {year} {2008}),\ \href
  {\doibase 10.1103/PhysRevA.77.032111} {\bibfield  {journal} {\bibinfo
  {journal} {Phys. Rev. A}\ }\textbf {\bibinfo {volume} {77}},\ \bibinfo
  {pages} {032111}}\BibitemShut {NoStop}%
\bibitem [{\citenamefont {Cheng}\ \emph {et~al.}(2017)\citenamefont {Cheng},
  \citenamefont {Gong}, \citenamefont {Guo},\ and\ \citenamefont
  {Zhou}}]{Cheng2017}%
  \BibitemOpen
  \bibfield  {author} {\bibinfo {author} {\bibnamefont {Cheng}, \bibfnamefont
  {J.-M.}}, \bibinfo {author} {\bibfnamefont {M.}~\bibnamefont {Gong}},
  \bibinfo {author} {\bibfnamefont {G.-C.}\ \bibnamefont {Guo}}, \ and\
  \bibinfo {author} {\bibfnamefont {Z.-W.}\ \bibnamefont {Zhou}}} (\bibinfo
  {year} {2017}),\ \href@noop {} {\bibfield  {journal} {\bibinfo  {journal}
  {Phys. Rev. A}\ }\textbf {\bibinfo {volume} {95}},\ \bibinfo {pages}
  {062117}}\BibitemShut {NoStop}%
\bibitem [{\citenamefont {Childs}\ \emph {et~al.}(2000)\citenamefont {Childs},
  \citenamefont {Preskill},\ and\ \citenamefont {Renes}}]{Childs00}%
  \BibitemOpen
  \bibfield  {author} {\bibinfo {author} {\bibnamefont {Childs}, \bibfnamefont
  {A.}}, \bibinfo {author} {\bibfnamefont {J.}~\bibnamefont {Preskill}}, \ and\
  \bibinfo {author} {\bibfnamefont {J.}~\bibnamefont {Renes}}} (\bibinfo {year}
  {2000}),\ \href@noop {} {\bibfield  {journal} {\bibinfo  {journal} {J. Mod.
  Opt.}\ }\textbf {\bibinfo {volume} {47}},\ \bibinfo {pages}
  {155}}\BibitemShut {NoStop}%
\bibitem [{\citenamefont {Childs}(2001)}]{Childs2001}%
  \BibitemOpen
  \bibfield  {author} {\bibinfo {author} {\bibnamefont {Childs}, \bibfnamefont
  {P.~R.~N.}}} (\bibinfo {year} {2001}),\ \href@noop {} {\emph {\bibinfo
  {title} {{Practical Temperature Measurement}}}}\ (\bibinfo  {publisher}
  {Butterworth-Heinemann})\BibitemShut {NoStop}%
\bibitem [{\citenamefont {Chin}\ \emph {et~al.}(2012)\citenamefont {Chin},
  \citenamefont {Huelga},\ and\ \citenamefont
  {Plenio}}]{PhysRevLett.109.233601}%
  \BibitemOpen
  \bibfield  {author} {\bibinfo {author} {\bibnamefont {Chin}, \bibfnamefont
  {A.~W.}}, \bibinfo {author} {\bibfnamefont {S.~F.}\ \bibnamefont {Huelga}}, \
  and\ \bibinfo {author} {\bibfnamefont {M.~B.}\ \bibnamefont {Plenio}}}
  (\bibinfo {year} {2012}),\ \href {\doibase 10.1103/PhysRevLett.109.233601}
  {\bibfield  {journal} {\bibinfo  {journal} {Phys. Rev. Lett.}\ }\textbf
  {\bibinfo {volume} {109}},\ \bibinfo {pages} {233601}}\BibitemShut {NoStop}%
\bibitem [{\citenamefont {Chua}(2015)}]{chua_quantum_2015}%
  \BibitemOpen
  \bibfield  {author} {\bibinfo {author} {\bibnamefont {Chua}, \bibfnamefont
  {S.~S.~Y.}}} (\bibinfo {year} {2015}),\ \href@noop {} {\emph {\bibinfo
  {title} {Quantum {Enhancement} of a 4 km {Laser} {Interferometer}
  {Gravitational}-{Wave} {Detector}}}}\ (\bibinfo  {publisher}
  {Springer})\BibitemShut {NoStop}%
\bibitem [{\citenamefont {Clark}\ \emph {et~al.}(2016)\citenamefont {Clark},
  \citenamefont {Stokes},\ and\ \citenamefont {Beige}}]{PhysRevA.94.023840}%
  \BibitemOpen
  \bibfield  {author} {\bibinfo {author} {\bibnamefont {Clark}, \bibfnamefont
  {L.~A.}}, \bibinfo {author} {\bibfnamefont {A.}~\bibnamefont {Stokes}}, \
  and\ \bibinfo {author} {\bibfnamefont {A.}~\bibnamefont {Beige}}} (\bibinfo
  {year} {2016}),\ \href@noop {} {\bibfield  {journal} {\bibinfo  {journal}
  {Phys. Rev. A}\ }\textbf {\bibinfo {volume} {94}},\ \bibinfo {pages}
  {023840}}\BibitemShut {NoStop}%
\bibitem [{\citenamefont {Clifton}\ and\ \citenamefont
  {Halvorson}(2001)}]{Clifton0}%
  \BibitemOpen
  \bibfield  {author} {\bibinfo {author} {\bibnamefont {Clifton}, \bibfnamefont
  {R.}}, \ and\ \bibinfo {author} {\bibfnamefont {H.}~\bibnamefont
  {Halvorson}}} (\bibinfo {year} {2001}),\ \href@noop {} {\bibfield  {journal}
  {\bibinfo  {journal} {Stud. Hist. Phil. Mod. Phys.}\ }\textbf {\bibinfo
  {volume} {32}},\ \bibinfo {pages} {31}}\BibitemShut {NoStop}%
\bibitem [{\citenamefont {Continentino}(2001)}]{Continentino}%
  \BibitemOpen
  \bibfield  {author} {\bibinfo {author} {\bibnamefont {Continentino},
  \bibfnamefont {M.~A.}}} (\bibinfo {year} {2001}),\ \href@noop {} {\emph
  {\bibinfo {title} {Quantum Scaling in Many-Body Systems}}}\ (\bibinfo
  {publisher} {World Scientific Publishing Co. Pte. Ltd.})\BibitemShut
  {NoStop}%
\bibitem [{\citenamefont {Cooper}\ \emph {et~al.}(2009)\citenamefont {Cooper},
  \citenamefont {Hallwood},\ and\ \citenamefont {Dunningham}}]{Cooper1}%
  \BibitemOpen
  \bibfield  {author} {\bibinfo {author} {\bibnamefont {Cooper}, \bibfnamefont
  {J.~J.}}, \bibinfo {author} {\bibfnamefont {D.~W.}\ \bibnamefont {Hallwood}},
  \ and\ \bibinfo {author} {\bibfnamefont {J.~A.}\ \bibnamefont {Dunningham}}}
  (\bibinfo {year} {2009}),\ \href@noop {} {\bibfield  {journal} {\bibinfo
  {journal} {J. Phys. B}\ }\textbf {\bibinfo {volume} {42}},\ \bibinfo {pages}
  {105301}}\BibitemShut {NoStop}%
\bibitem [{\citenamefont {Cooper}\ \emph {et~al.}(2012)\citenamefont {Cooper},
  \citenamefont {Hallwood}, \citenamefont {Dunningham},\ and\ \citenamefont
  {Brand}}]{Cooper2}%
  \BibitemOpen
  \bibfield  {author} {\bibinfo {author} {\bibnamefont {Cooper}, \bibfnamefont
  {J.~J.}}, \bibinfo {author} {\bibfnamefont {D.~W.}\ \bibnamefont {Hallwood}},
  \bibinfo {author} {\bibfnamefont {J.~A.}\ \bibnamefont {Dunningham}}, \ and\
  \bibinfo {author} {\bibfnamefont {J.}~\bibnamefont {Brand}}} (\bibinfo {year}
  {2012}),\ \href {\doibase 10.1103/PhysRevLett.108.130402} {\bibfield
  {journal} {\bibinfo  {journal} {Phys. Rev. Lett.}\ }\textbf {\bibinfo
  {volume} {108}},\ \bibinfo {pages} {130402}}\BibitemShut {NoStop}%
\bibitem [{\citenamefont {Correa}\ \emph {et~al.}(2015)\citenamefont {Correa},
  \citenamefont {Mehboudi}, \citenamefont {Adesso},\ and\ \citenamefont
  {Sanpera}}]{Correa15}%
  \BibitemOpen
  \bibfield  {author} {\bibinfo {author} {\bibnamefont {Correa}, \bibfnamefont
  {L.~A.}}, \bibinfo {author} {\bibfnamefont {M.}~\bibnamefont {Mehboudi}},
  \bibinfo {author} {\bibfnamefont {G.}~\bibnamefont {Adesso}}, \ and\ \bibinfo
  {author} {\bibfnamefont {A.}~\bibnamefont {Sanpera}}} (\bibinfo {year}
  {2015}),\ \href {\doibase 10.1103/PhysRevLett.114.220405} {\bibfield
  {journal} {\bibinfo  {journal} {Phys. Rev. Lett.}\ }\textbf {\bibinfo
  {volume} {114}},\ \bibinfo {pages} {220405}}\BibitemShut {NoStop}%
\bibitem [{\citenamefont {Cozzini}\ \emph {et~al.}(2007)\citenamefont
  {Cozzini}, \citenamefont {Giorda},\ and\ \citenamefont
  {Zanardi}}]{Cozzini2007}%
  \BibitemOpen
  \bibfield  {author} {\bibinfo {author} {\bibnamefont {Cozzini}, \bibfnamefont
  {M.}}, \bibinfo {author} {\bibfnamefont {P.}~\bibnamefont {Giorda}}, \ and\
  \bibinfo {author} {\bibfnamefont {P.}~\bibnamefont {Zanardi}}} (\bibinfo
  {year} {2007}),\ \href {\doibase 10.1103/PhysRevB.75.014439} {\bibfield
  {journal} {\bibinfo  {journal} {Phys. Rev. B}\ }\textbf {\bibinfo {volume}
  {75}},\ \bibinfo {pages} {014439}}\BibitemShut {NoStop}%
\bibitem [{\citenamefont {Cronin}\ \emph {et~al.}(2009)\citenamefont {Cronin},
  \citenamefont {Schmiedmayer},\ and\ \citenamefont {Pritchard}}]{Cronin0}%
  \BibitemOpen
  \bibfield  {author} {\bibinfo {author} {\bibnamefont {Cronin}, \bibfnamefont
  {A.}}, \bibinfo {author} {\bibfnamefont {J.}~\bibnamefont {Schmiedmayer}}, \
  and\ \bibinfo {author} {\bibfnamefont {D.}~\bibnamefont {Pritchard}}}
  (\bibinfo {year} {2009}),\ \href@noop {} {\bibfield  {journal} {\bibinfo
  {journal} {Rev. Mod. Phys.}\ }\textbf {\bibinfo {volume} {81}},\ \bibinfo
  {pages} {1051}}\BibitemShut {NoStop}%
\bibitem [{\citenamefont {Crooks}(2007)}]{Crooks2007}%
  \BibitemOpen
  \bibfield  {author} {\bibinfo {author} {\bibnamefont {Crooks}, \bibfnamefont
  {G.~E.}}} (\bibinfo {year} {2007}),\ \href {\doibase
  10.1103/PhysRevLett.99.100602} {\bibfield  {journal} {\bibinfo  {journal}
  {Phys. Rev. Lett.}\ }\textbf {\bibinfo {volume} {99}},\ \bibinfo {pages}
  {100602}}\BibitemShut {NoStop}%
\bibitem [{\citenamefont {D'Alessandro}(2007)}]{dalessandro_introduction_2007}%
  \BibitemOpen
  \bibfield  {author} {\bibinfo {author} {\bibnamefont {D'Alessandro},
  \bibfnamefont {D.}}} (\bibinfo {year} {2007}),\ \href@noop {} {\emph
  {\bibinfo {title} {Introduction to {Quantum} {Control} and {Dynamics}}}},\
  \bibinfo {edition} {1st}\ ed.\ (\bibinfo  {publisher} {Chapman and
  Hall/CRC},\ \bibinfo {address} {Boca Raton})\BibitemShut {NoStop}%
\bibitem [{\citenamefont {Damanet}\ \emph {et~al.}(2016)\citenamefont
  {Damanet}, \citenamefont {Braun},\ and\ \citenamefont
  {Martin}}]{PhysRevA.94.033838}%
  \BibitemOpen
  \bibfield  {author} {\bibinfo {author} {\bibnamefont {Damanet}, \bibfnamefont
  {F.}}, \bibinfo {author} {\bibfnamefont {D.}~\bibnamefont {Braun}}, \ and\
  \bibinfo {author} {\bibfnamefont {J.}~\bibnamefont {Martin}}} (\bibinfo
  {year} {2016}),\ \href {\doibase 10.1103/PhysRevA.94.033838} {\bibfield
  {journal} {\bibinfo  {journal} {Phys. Rev. A}\ }\textbf {\bibinfo {volume}
  {94}},\ \bibinfo {pages} {033838}}\BibitemShut {NoStop}%
\bibitem [{\citenamefont {Damski}(2013)}]{Damski2013}%
  \BibitemOpen
  \bibfield  {author} {\bibinfo {author} {\bibnamefont {Damski}, \bibfnamefont
  {B.}}} (\bibinfo {year} {2013}),\ \href {\doibase 10.1103/PhysRevE.87.052131}
  {\bibfield  {journal} {\bibinfo  {journal} {Phys. Rev. E}\ }\textbf {\bibinfo
  {volume} {87}},\ \bibinfo {pages} {052131}}\BibitemShut {NoStop}%
\bibitem [{\citenamefont {Damski}\ and\ \citenamefont
  {Rams}(2014)}]{Damski2014-2}%
  \BibitemOpen
  \bibfield  {author} {\bibinfo {author} {\bibnamefont {Damski}, \bibfnamefont
  {B.}}, \ and\ \bibinfo {author} {\bibfnamefont {M.}~\bibnamefont {Rams}}}
  (\bibinfo {year} {2014}),\ \href@noop {} {\bibfield  {journal} {\bibinfo
  {journal} {J. Phys. A}\ }\textbf {\bibinfo {volume} {47}},\ \bibinfo {pages}
  {025303}}\BibitemShut {NoStop}%
\bibitem [{\citenamefont {Dang}\ \emph {et~al.}(2010)\citenamefont {Dang},
  \citenamefont {Maloof},\ and\ \citenamefont {Romalis}}]{DangAPL2010}%
  \BibitemOpen
  \bibfield  {author} {\bibinfo {author} {\bibnamefont {Dang}, \bibfnamefont
  {H.~B.}}, \bibinfo {author} {\bibfnamefont {A.~C.}\ \bibnamefont {Maloof}}, \
  and\ \bibinfo {author} {\bibfnamefont {M.~V.}\ \bibnamefont {Romalis}}}
  (\bibinfo {year} {2010}),\ \href {\doibase 10.1063/1.3491215} {\bibfield
  {journal} {\bibinfo  {journal} {Applied Physics Letters}\ }\textbf {\bibinfo
  {volume} {97}}~(\bibinfo {number} {15}),\ \bibinfo {eid}
  {151110}}\BibitemShut {NoStop}%
\bibitem [{\citenamefont {D'Ariano}\ \emph {et~al.}(1998)\citenamefont
  {D'Ariano}, \citenamefont {Macchiavello},\ and\ \citenamefont
  {Sacchi}}]{Dariano2}%
  \BibitemOpen
  \bibfield  {author} {\bibinfo {author} {\bibnamefont {D'Ariano},
  \bibfnamefont {G.}}, \bibinfo {author} {\bibfnamefont {C.}~\bibnamefont
  {Macchiavello}}, \ and\ \bibinfo {author} {\bibfnamefont {M.}~\bibnamefont
  {Sacchi}}} (\bibinfo {year} {1998}),\ \href@noop {} {\bibfield  {journal}
  {\bibinfo  {journal} {Phys. Lett.}\ }\textbf {\bibinfo {volume} {A248}},\
  \bibinfo {pages} {103}}\BibitemShut {NoStop}%
\bibitem [{\citenamefont {D'Ariano}\ and\ \citenamefont
  {Paris}(1997)}]{Dariano1}%
  \BibitemOpen
  \bibfield  {author} {\bibinfo {author} {\bibnamefont {D'Ariano},
  \bibfnamefont {G.}}, \ and\ \bibinfo {author} {\bibfnamefont
  {M.}~\bibnamefont {Paris}}} (\bibinfo {year} {1997}),\ \href@noop {}
  {\bibfield  {journal} {\bibinfo  {journal} {Phys. Rev. A}\ }\textbf {\bibinfo
  {volume} {55}},\ \bibinfo {pages} {2267}}\BibitemShut {NoStop}%
\bibitem [{\citenamefont {Datta}\ \emph {et~al.}(2005)\citenamefont {Datta},
  \citenamefont {Flammia},\ and\ \citenamefont
  {Caves}}]{datta_entanglement_2005}%
  \BibitemOpen
  \bibfield  {author} {\bibinfo {author} {\bibnamefont {Datta}, \bibfnamefont
  {A.}}, \bibinfo {author} {\bibfnamefont {S.~T.}\ \bibnamefont {Flammia}}, \
  and\ \bibinfo {author} {\bibfnamefont {C.~M.}\ \bibnamefont {Caves}}}
  (\bibinfo {year} {2005}),\ \href {\doibase 10.1103/PhysRevA.72.042316}
  {\bibfield  {journal} {\bibinfo  {journal} {Phys. Rev. A}\ }\textbf {\bibinfo
  {volume} {72}}~(\bibinfo {number} {4}),\
  10.1103/PhysRevA.72.042316}\BibitemShut {NoStop}%
\bibitem [{\citenamefont {Datta}\ and\ \citenamefont
  {Shaji}(2012)}]{DattaMPLB2012}%
  \BibitemOpen
  \bibfield  {author} {\bibinfo {author} {\bibnamefont {Datta}, \bibfnamefont
  {A.}}, \ and\ \bibinfo {author} {\bibfnamefont {A.}~\bibnamefont {Shaji}}}
  (\bibinfo {year} {2012}),\ \href {\doibase 10.1142/S0217984912300104}
  {\bibfield  {journal} {\bibinfo  {journal} {Modern Physics Letters B}\
  }\textbf {\bibinfo {volume} {26}}~(\bibinfo {number} {18}),\ \bibinfo {pages}
  {1230010}}\BibitemShut {NoStop}%
\bibitem [{\citenamefont {Davis}\ \emph {et~al.}(2016)\citenamefont {Davis},
  \citenamefont {Bentsen},\ and\ \citenamefont
  {Schleier-Smith}}]{DavisPRL2016}%
  \BibitemOpen
  \bibfield  {author} {\bibinfo {author} {\bibnamefont {Davis}, \bibfnamefont
  {E.}}, \bibinfo {author} {\bibfnamefont {G.}~\bibnamefont {Bentsen}}, \ and\
  \bibinfo {author} {\bibfnamefont {M.}~\bibnamefont {Schleier-Smith}}}
  (\bibinfo {year} {2016}),\ \href {\doibase 10.1103/PhysRevLett.116.053601}
  {\bibfield  {journal} {\bibinfo  {journal} {Phys. Rev. Lett.}\ }\textbf
  {\bibinfo {volume} {116}},\ \bibinfo {pages} {053601}}\BibitemShut {NoStop}%
\bibitem [{\citenamefont {Davis}\ \emph {et~al.}(2006)\citenamefont {Davis},
  \citenamefont {Choi}, \citenamefont {Pollanen},\ and\ \citenamefont
  {Halperin}}]{Davis2006}%
  \BibitemOpen
  \bibfield  {author} {\bibinfo {author} {\bibnamefont {Davis}, \bibfnamefont
  {J.~P.}}, \bibinfo {author} {\bibfnamefont {H.}~\bibnamefont {Choi}},
  \bibinfo {author} {\bibfnamefont {J.}~\bibnamefont {Pollanen}}, \ and\
  \bibinfo {author} {\bibfnamefont {W.~P.}\ \bibnamefont {Halperin}}} (\bibinfo
  {year} {2006}),\ \href {\doibase 10.1103/PhysRevLett.97.115301} {\bibfield
  {journal} {\bibinfo  {journal} {Phys. Rev. Lett.}\ }\textbf {\bibinfo
  {volume} {97}},\ \bibinfo {pages} {115301}}\BibitemShut {NoStop}%
\bibitem [{\citenamefont {Davis}\ and\ \citenamefont
  {Guti\'errez}(2012)}]{Davis2012}%
  \BibitemOpen
  \bibfield  {author} {\bibinfo {author} {\bibnamefont {Davis}, \bibfnamefont
  {S.}}, \ and\ \bibinfo {author} {\bibfnamefont {G.}~\bibnamefont
  {Guti\'errez}}} (\bibinfo {year} {2012}),\ \href {\doibase
  10.1103/PhysRevE.86.051136} {\bibfield  {journal} {\bibinfo  {journal} {Phys.
  Rev. E}\ }\textbf {\bibinfo {volume} {86}},\ \bibinfo {pages}
  {051136}}\BibitemShut {NoStop}%
\bibitem [{\citenamefont {De~Pasquale}\ \emph {et~al.}(2015)\citenamefont
  {De~Pasquale}, \citenamefont {Facchi}, \citenamefont {Florio}, \citenamefont
  {Giovannetti}, \citenamefont {Matsuoka},\ and\ \citenamefont
  {Yuasa}}]{PhysRevA.92.042115}%
  \BibitemOpen
  \bibfield  {author} {\bibinfo {author} {\bibnamefont {De~Pasquale},
  \bibfnamefont {A.}}, \bibinfo {author} {\bibfnamefont {P.}~\bibnamefont
  {Facchi}}, \bibinfo {author} {\bibfnamefont {G.}~\bibnamefont {Florio}},
  \bibinfo {author} {\bibfnamefont {V.}~\bibnamefont {Giovannetti}}, \bibinfo
  {author} {\bibfnamefont {K.}~\bibnamefont {Matsuoka}}, \ and\ \bibinfo
  {author} {\bibfnamefont {K.}~\bibnamefont {Yuasa}}} (\bibinfo {year}
  {2015}),\ \href {\doibase 10.1103/PhysRevA.92.042115} {\bibfield  {journal}
  {\bibinfo  {journal} {Phys. Rev. A}\ }\textbf {\bibinfo {volume} {92}},\
  \bibinfo {pages} {042115}}\BibitemShut {NoStop}%
\bibitem [{\citenamefont {Degen}\ \emph {et~al.}(2016)\citenamefont {Degen},
  \citenamefont {Reinhard},\ and\ \citenamefont
  {Cappellaro}}]{degen_quantum_2016}%
  \BibitemOpen
  \bibfield  {author} {\bibinfo {author} {\bibnamefont {Degen}, \bibfnamefont
  {C.~L.}}, \bibinfo {author} {\bibfnamefont {F.}~\bibnamefont {Reinhard}}, \
  and\ \bibinfo {author} {\bibfnamefont {P.}~\bibnamefont {Cappellaro}}}
  (\bibinfo {year} {2016}),\ \href {http://arxiv.org/abs/1611.02427} {\enquote
  {\bibinfo {title} {Quantum sensing},}\ }\bibinfo {note}
  {ArXiv:1611.02427}\BibitemShut {NoStop}%
\bibitem [{\citenamefont {Del\'eglise}\ \emph {et~al.}(2008)\citenamefont
  {Del\'eglise}, \citenamefont {Dotsenko}, \citenamefont {Sayrin},
  \citenamefont {Bernu}, \citenamefont {Brune}, \citenamefont {Raimond},\ and\
  \citenamefont {Haroche}}]{deleglise_reconstruction_2008}%
  \BibitemOpen
  \bibfield  {author} {\bibinfo {author} {\bibnamefont {Del\'eglise},
  \bibfnamefont {S.}}, \bibinfo {author} {\bibfnamefont {I.}~\bibnamefont
  {Dotsenko}}, \bibinfo {author} {\bibfnamefont {C.}~\bibnamefont {Sayrin}},
  \bibinfo {author} {\bibfnamefont {J.}~\bibnamefont {Bernu}}, \bibinfo
  {author} {\bibfnamefont {M.}~\bibnamefont {Brune}}, \bibinfo {author}
  {\bibfnamefont {J.-M.}\ \bibnamefont {Raimond}}, \ and\ \bibinfo {author}
  {\bibfnamefont {S.}~\bibnamefont {Haroche}}} (\bibinfo {year} {2008}),\ \href
  {\doibase 10.1038/nature07288} {\bibfield  {journal} {\bibinfo  {journal}
  {Nature}\ }\textbf {\bibinfo {volume} {455}}~(\bibinfo {number} {7212}),\
  \bibinfo {pages} {510}}\BibitemShut {NoStop}%
\bibitem [{\citenamefont {Demkowicz-Dobrza{\'n}ski}\ \emph
  {et~al.}(2014)\citenamefont {Demkowicz-Dobrza{\'n}ski}, \citenamefont
  {Jarzyna},\ and\ \citenamefont {Ko{\l}odi{\'n}ski}}]{Demkowicz0}%
  \BibitemOpen
  \bibfield  {author} {\bibinfo {author} {\bibnamefont
  {Demkowicz-Dobrza{\'n}ski}, \bibfnamefont {R.}}, \bibinfo {author}
  {\bibfnamefont {M.}~\bibnamefont {Jarzyna}}, \ and\ \bibinfo {author}
  {\bibfnamefont {J.}~\bibnamefont {Ko{\l}odi{\'n}ski}}} (\bibinfo {year}
  {2014}),\ \href@noop {} {\bibfield  {journal} {\bibinfo  {journal} {Progress
  in Optics}\ }\textbf {\bibinfo {volume} {60}},\ \bibinfo {pages}
  {345}}\BibitemShut {NoStop}%
\bibitem [{\citenamefont {Demkowicz-Dobrza{\'n}ski}\ and\ \citenamefont
  {Maccone}(2014)}]{Maccone2014}%
  \BibitemOpen
  \bibfield  {author} {\bibinfo {author} {\bibnamefont
  {Demkowicz-Dobrza{\'n}ski}, \bibfnamefont {R.}}, \ and\ \bibinfo {author}
  {\bibfnamefont {L.}~\bibnamefont {Maccone}}} (\bibinfo {year} {2014}),\ \href
  {\doibase 10.1103/PhysRevLett.113.250801} {\bibfield  {journal} {\bibinfo
  {journal} {Phys. Rev. Lett.}\ }\textbf {\bibinfo {volume} {113}},\ \bibinfo
  {pages} {250801}}\BibitemShut {NoStop}%
\bibitem [{\citenamefont {Deutsch}\ \emph {et~al.}(2010)\citenamefont
  {Deutsch}, \citenamefont {Ramirez-Martinez}, \citenamefont {Lacro\^ute},
  \citenamefont {Reinhard}, \citenamefont {Schneider}, \citenamefont {Fuchs},
  \citenamefont {Pi\'echon}, \citenamefont {Lalo\"e}, \citenamefont {Reichel},\
  and\ \citenamefont {Rosenbusch}}]{DeutschPRL2010}%
  \BibitemOpen
  \bibfield  {author} {\bibinfo {author} {\bibnamefont {Deutsch}, \bibfnamefont
  {C.}}, \bibinfo {author} {\bibfnamefont {F.}~\bibnamefont
  {Ramirez-Martinez}}, \bibinfo {author} {\bibfnamefont {C.}~\bibnamefont
  {Lacro\^ute}}, \bibinfo {author} {\bibfnamefont {F.}~\bibnamefont
  {Reinhard}}, \bibinfo {author} {\bibfnamefont {T.}~\bibnamefont {Schneider}},
  \bibinfo {author} {\bibfnamefont {J.~N.}\ \bibnamefont {Fuchs}}, \bibinfo
  {author} {\bibfnamefont {F.}~\bibnamefont {Pi\'echon}}, \bibinfo {author}
  {\bibfnamefont {F.}~\bibnamefont {Lalo\"e}}, \bibinfo {author} {\bibfnamefont
  {J.}~\bibnamefont {Reichel}}, \ and\ \bibinfo {author} {\bibfnamefont
  {P.}~\bibnamefont {Rosenbusch}}} (\bibinfo {year} {2010}),\ \href {\doibase
  10.1103/PhysRevLett.105.020401} {\bibfield  {journal} {\bibinfo  {journal}
  {Phys. Rev. Lett.}\ }\textbf {\bibinfo {volume} {105}},\ \bibinfo {pages}
  {020401}}\BibitemShut {NoStop}%
\bibitem [{\citenamefont {Di\'osi}\ \emph {et~al.}(1984)\citenamefont
  {Di\'osi}, \citenamefont {Forg\'acs}, \citenamefont {Luk\'acs},\ and\
  \citenamefont {Frisch}}]{Diosi1984}%
  \BibitemOpen
  \bibfield  {author} {\bibinfo {author} {\bibnamefont {Di\'osi}, \bibfnamefont
  {L.}}, \bibinfo {author} {\bibfnamefont {G.}~\bibnamefont {Forg\'acs}},
  \bibinfo {author} {\bibfnamefont {B.}~\bibnamefont {Luk\'acs}}, \ and\
  \bibinfo {author} {\bibfnamefont {H.~L.}\ \bibnamefont {Frisch}}} (\bibinfo
  {year} {1984}),\ \href {\doibase 10.1103/PhysRevA.29.3343} {\bibfield
  {journal} {\bibinfo  {journal} {Phys. Rev. A}\ }\textbf {\bibinfo {volume}
  {29}},\ \bibinfo {pages} {3343}}\BibitemShut {NoStop}%
\bibitem [{\citenamefont {Dolan}(1998)}]{Dolan1998}%
  \BibitemOpen
  \bibfield  {author} {\bibinfo {author} {\bibnamefont {Dolan}, \bibfnamefont
  {B.~P.}}} (\bibinfo {year} {1998}),\ \href@noop {} {\bibfield  {journal}
  {\bibinfo  {journal} {Proc. Roy. Soc. A}\ }\textbf {\bibinfo {volume}
  {454}},\ \bibinfo {pages} {2655}}\BibitemShut {NoStop}%
\bibitem [{\citenamefont {Dorner}\ \emph {et~al.}(2009)\citenamefont {Dorner},
  \citenamefont {Demkowicz-Dobrza{\'n}ski}, \citenamefont {Smith},
  \citenamefont {Lundeen}, \citenamefont {Wasilewski}, \citenamefont
  {Banaszek},\ and\ \citenamefont {Walmsley}}]{Dorner0}%
  \BibitemOpen
  \bibfield  {author} {\bibinfo {author} {\bibnamefont {Dorner}, \bibfnamefont
  {U.}}, \bibinfo {author} {\bibfnamefont {R.}~\bibnamefont
  {Demkowicz-Dobrza{\'n}ski}}, \bibinfo {author} {\bibfnamefont {B.~J.}\
  \bibnamefont {Smith}}, \bibinfo {author} {\bibfnamefont {J.~S.}\ \bibnamefont
  {Lundeen}}, \bibinfo {author} {\bibfnamefont {W.}~\bibnamefont {Wasilewski}},
  \bibinfo {author} {\bibfnamefont {K.}~\bibnamefont {Banaszek}}, \ and\
  \bibinfo {author} {\bibfnamefont {I.~A.}\ \bibnamefont {Walmsley}}} (\bibinfo
  {year} {2009}),\ \href {\doibase 10.1103/PhysRevLett.102.040403} {\bibfield
  {journal} {\bibinfo  {journal} {Phys. Rev. Lett.}\ }\textbf {\bibinfo
  {volume} {102}},\ \bibinfo {pages} {040403}}\BibitemShut {NoStop}%
\bibitem [{\citenamefont {Douglass}\ \emph {et~al.}(1964)\citenamefont
  {Douglass}, \citenamefont {Khorana},\ and\ \citenamefont
  {Brij}}]{Douglass1964}%
  \BibitemOpen
  \bibfield  {author} {\bibinfo {author} {\bibnamefont {Douglass},
  \bibfnamefont {D.~H.~J.}}, \bibinfo {author} {\bibfnamefont {B.~M.}\
  \bibnamefont {Khorana}}, \ and\ \bibinfo {author} {\bibfnamefont
  {M.}~\bibnamefont {Brij}}} (\bibinfo {year} {1964}),\ \href@noop {}
  {\bibfield  {journal} {\bibinfo  {journal} {Phys. Rev.}\ }\textbf {\bibinfo
  {volume} {138}},\ \bibinfo {pages} {A35}}\BibitemShut {NoStop}%
\bibitem [{\citenamefont {Dowling}(1998)}]{Dowling1}%
  \BibitemOpen
  \bibfield  {author} {\bibinfo {author} {\bibnamefont {Dowling}, \bibfnamefont
  {J.}}} (\bibinfo {year} {1998}),\ \href@noop {} {\bibfield  {journal}
  {\bibinfo  {journal} {Phys. Rev. A}\ }\textbf {\bibinfo {volume} {57}},\
  \bibinfo {pages} {4736}}\BibitemShut {NoStop}%
\bibitem [{\citenamefont {Dowling}(2008)}]{Dowling3}%
  \BibitemOpen
  \bibfield  {author} {\bibinfo {author} {\bibnamefont {Dowling}, \bibfnamefont
  {J.}}} (\bibinfo {year} {2008}),\ \href@noop {} {\bibfield  {journal}
  {\bibinfo  {journal} {Contemp. Phys.}\ }\textbf {\bibinfo {volume} {49}},\
  \bibinfo {pages} {125}}\BibitemShut {NoStop}%
\bibitem [{\citenamefont {Dowling}\ \emph {et~al.}(2006)\citenamefont
  {Dowling}, \citenamefont {Doherty},\ and\ \citenamefont
  {Wiseman}}]{Dowling0}%
  \BibitemOpen
  \bibfield  {author} {\bibinfo {author} {\bibnamefont {Dowling}, \bibfnamefont
  {M.}}, \bibinfo {author} {\bibfnamefont {A.}~\bibnamefont {Doherty}}, \ and\
  \bibinfo {author} {\bibfnamefont {H.}~\bibnamefont {Wiseman}}} (\bibinfo
  {year} {2006}),\ \href@noop {} {\bibfield  {journal} {\bibinfo  {journal}
  {Phys. Rev. A}\ }\textbf {\bibinfo {volume} {73}},\ \bibinfo {pages}
  {052323}}\BibitemShut {NoStop}%
\bibitem [{\citenamefont {van Druten}\ and\ \citenamefont
  {Ketterle}(1982)}]{vanDruten1997}%
  \BibitemOpen
  \bibfield  {author} {\bibinfo {author} {\bibnamefont {van Druten},
  \bibfnamefont {N.~J.}}, \ and\ \bibinfo {author} {\bibfnamefont
  {W.}~\bibnamefont {Ketterle}}} (\bibinfo {year} {1982}),\ \href@noop {}
  {\bibfield  {journal} {\bibinfo  {journal} {Phys. Rev. Lett.}\ }\textbf
  {\bibinfo {volume} {79}},\ \bibinfo {pages} {549}}\BibitemShut {NoStop}%
\bibitem [{\citenamefont {Duivenvoorden}\ \emph {et~al.}(2017)\citenamefont
  {Duivenvoorden}, \citenamefont {Terhal},\ and\ \citenamefont
  {Weigand}}]{PhysRevA.95.012305}%
  \BibitemOpen
  \bibfield  {author} {\bibinfo {author} {\bibnamefont {Duivenvoorden},
  \bibfnamefont {K.}}, \bibinfo {author} {\bibfnamefont {B.~M.}\ \bibnamefont
  {Terhal}}, \ and\ \bibinfo {author} {\bibfnamefont {D.}~\bibnamefont
  {Weigand}}} (\bibinfo {year} {2017}),\ \href {\doibase
  10.1103/PhysRevA.95.012305} {\bibfield  {journal} {\bibinfo  {journal} {Phys.
  Rev. A}\ }\textbf {\bibinfo {volume} {95}},\ \bibinfo {pages}
  {012305}}\BibitemShut {NoStop}%
\bibitem [{\citenamefont {Dunningham}\ \emph {et~al.}(2002)\citenamefont
  {Dunningham}, \citenamefont {Buenett},\ and\ \citenamefont
  {Barnett}}]{Dunningham0}%
  \BibitemOpen
  \bibfield  {author} {\bibinfo {author} {\bibnamefont {Dunningham},
  \bibfnamefont {J.}}, \bibinfo {author} {\bibfnamefont {K.}~\bibnamefont
  {Buenett}}, \ and\ \bibinfo {author} {\bibfnamefont {S.}~\bibnamefont
  {Barnett}}} (\bibinfo {year} {2002}),\ \href@noop {} {\bibfield  {journal}
  {\bibinfo  {journal} {Phys. Rev. Lett.}\ }\textbf {\bibinfo {volume} {89}},\
  \bibinfo {pages} {150401}}\BibitemShut {NoStop}%
\bibitem [{\citenamefont {D\"ur}\ \emph {et~al.}(2014)\citenamefont {D\"ur},
  \citenamefont {Skotiniotis}, \citenamefont {Fr\"owis},\ and\ \citenamefont
  {Kraus}}]{dur_improved_2014}%
  \BibitemOpen
  \bibfield  {author} {\bibinfo {author} {\bibnamefont {D\"ur}, \bibfnamefont
  {W.}}, \bibinfo {author} {\bibfnamefont {M.}~\bibnamefont {Skotiniotis}},
  \bibinfo {author} {\bibfnamefont {F.}~\bibnamefont {Fr\"owis}}, \ and\
  \bibinfo {author} {\bibfnamefont {B.}~\bibnamefont {Kraus}}} (\bibinfo {year}
  {2014}),\ \href {\doibase 10.1103/PhysRevLett.112.080801} {\bibfield
  {journal} {\bibinfo  {journal} {Phys. Rev. Lett.}\ }\textbf {\bibinfo
  {volume} {112}}~(\bibinfo {number} {8}),\ \bibinfo {pages}
  {080801}}\BibitemShut {NoStop}%
\bibitem [{\citenamefont {Eckert}\ \emph {et~al.}(2002)\citenamefont {Eckert},
  \citenamefont {Schliemann}, \citenamefont {Bru\ss},\ and\ \citenamefont
  {Lewenstein}}]{Eckert0}%
  \BibitemOpen
  \bibfield  {author} {\bibinfo {author} {\bibnamefont {Eckert}, \bibfnamefont
  {K.}}, \bibinfo {author} {\bibfnamefont {J.}~\bibnamefont {Schliemann}},
  \bibinfo {author} {\bibfnamefont {D.}~\bibnamefont {Bru\ss}}, \ and\ \bibinfo
  {author} {\bibfnamefont {M.}~\bibnamefont {Lewenstein}}} (\bibinfo {year}
  {2002}),\ \href@noop {} {\bibfield  {journal} {\bibinfo  {journal} {Ann. of
  Phys.}\ }\textbf {\bibinfo {volume} {299}},\ \bibinfo {pages}
  {88}}\BibitemShut {NoStop}%
\bibitem [{\citenamefont {Emch}(1972)}]{Emch0}%
  \BibitemOpen
  \bibfield  {author} {\bibinfo {author} {\bibnamefont {Emch}, \bibfnamefont
  {G.}}} (\bibinfo {year} {1972}),\ \href@noop {} {\emph {\bibinfo {title}
  {Algebraic Methods in Statistical Mechanics and Quantum Field Theory}}}\
  (\bibinfo  {publisher} {Wiley},\ \bibinfo {address} {New York})\BibitemShut
  {NoStop}%
\bibitem [{\citenamefont {Escher}\ \emph {et~al.}(2011)\citenamefont {Escher},
  \citenamefont {de~Matos~Filho},\ and\ \citenamefont
  {Davidovich}}]{escher_general_2011}%
  \BibitemOpen
  \bibfield  {author} {\bibinfo {author} {\bibnamefont {Escher}, \bibfnamefont
  {B.~M.}}, \bibinfo {author} {\bibfnamefont {R.~L.}\ \bibnamefont
  {de~Matos~Filho}}, \ and\ \bibinfo {author} {\bibfnamefont {L.}~\bibnamefont
  {Davidovich}}} (\bibinfo {year} {2011}),\ \href@noop {} {\bibfield  {journal}
  {\bibinfo  {journal} {Nat. Phys.}\ }\textbf {\bibinfo {volume} {7}}~(\bibinfo
  {number} {5}),\ \bibinfo {pages} {406}}\BibitemShut {NoStop}%
\bibitem [{\citenamefont {Esteve}\ \emph {et~al.}(2008)\citenamefont {Esteve},
  \citenamefont {Gross}, \citenamefont {Weller}, \citenamefont {Giovanazzi},\
  and\ \citenamefont {Oberthaler}}]{EsteveN2008}%
  \BibitemOpen
  \bibfield  {author} {\bibinfo {author} {\bibnamefont {Esteve}, \bibfnamefont
  {J.}}, \bibinfo {author} {\bibfnamefont {C.}~\bibnamefont {Gross}}, \bibinfo
  {author} {\bibfnamefont {A.}~\bibnamefont {Weller}}, \bibinfo {author}
  {\bibfnamefont {S.}~\bibnamefont {Giovanazzi}}, \ and\ \bibinfo {author}
  {\bibfnamefont {M.~K.}\ \bibnamefont {Oberthaler}}} (\bibinfo {year}
  {2008}),\ \href {http://dx.doi.org/10.1038/nature07332} {\bibfield  {journal}
  {\bibinfo  {journal} {Nature}\ }\textbf {\bibinfo {volume} {455}}~(\bibinfo
  {number} {7217}),\ \bibinfo {pages} {1216}}\BibitemShut {NoStop}%
\bibitem [{\citenamefont {Esteve}\ \emph {et~al.}(2006)\citenamefont {Esteve},
  \citenamefont {Trebbia}, \citenamefont {Schumm}, \citenamefont {Aspect},
  \citenamefont {Westbrook},\ and\ \citenamefont {Bouchoule}}]{Esteve2006}%
  \BibitemOpen
  \bibfield  {author} {\bibinfo {author} {\bibnamefont {Esteve}, \bibfnamefont
  {J.}}, \bibinfo {author} {\bibfnamefont {J.-B.}\ \bibnamefont {Trebbia}},
  \bibinfo {author} {\bibfnamefont {T.}~\bibnamefont {Schumm}}, \bibinfo
  {author} {\bibfnamefont {A.}~\bibnamefont {Aspect}}, \bibinfo {author}
  {\bibfnamefont {C.}~\bibnamefont {Westbrook}}, \ and\ \bibinfo {author}
  {\bibfnamefont {I.}~\bibnamefont {Bouchoule}}} (\bibinfo {year} {2006}),\
  \href@noop {} {\bibfield  {journal} {\bibinfo  {journal} {Phys. Rev. Lett.}\
  }\textbf {\bibinfo {volume} {96}},\ \bibinfo {pages} {130403}}\BibitemShut
  {NoStop}%
\bibitem [{\citenamefont {Etesse}\ \emph {et~al.}(2014)\citenamefont {Etesse},
  \citenamefont {Blandino}, \citenamefont {Kanseri},\ and\ \citenamefont
  {Tualle-Brouri}}]{etesse_proposal_2014}%
  \BibitemOpen
  \bibfield  {author} {\bibinfo {author} {\bibnamefont {Etesse}, \bibfnamefont
  {J.}}, \bibinfo {author} {\bibfnamefont {R.}~\bibnamefont {Blandino}},
  \bibinfo {author} {\bibfnamefont {B.}~\bibnamefont {Kanseri}}, \ and\
  \bibinfo {author} {\bibfnamefont {R.}~\bibnamefont {Tualle-Brouri}}}
  (\bibinfo {year} {2014}),\ \href {\doibase 10.1088/1367-2630/16/5/053001}
  {\bibfield  {journal} {\bibinfo  {journal} {New J. Phys.}\ }\textbf {\bibinfo
  {volume} {16}}~(\bibinfo {number} {5}),\ \bibinfo {pages}
  {053001}}\BibitemShut {NoStop}%
\bibitem [{\citenamefont {Etesse}\ \emph {et~al.}(2015)\citenamefont {Etesse},
  \citenamefont {Bouillard}, \citenamefont {Kanseri},\ and\ \citenamefont
  {Tualle-Brouri}}]{etesse_experimental_2015}%
  \BibitemOpen
  \bibfield  {author} {\bibinfo {author} {\bibnamefont {Etesse}, \bibfnamefont
  {J.}}, \bibinfo {author} {\bibfnamefont {M.}~\bibnamefont {Bouillard}},
  \bibinfo {author} {\bibfnamefont {B.}~\bibnamefont {Kanseri}}, \ and\
  \bibinfo {author} {\bibfnamefont {R.}~\bibnamefont {Tualle-Brouri}}}
  (\bibinfo {year} {2015}),\ \href {\doibase 10.1103/PhysRevLett.114.193602}
  {\bibfield  {journal} {\bibinfo  {journal} {Phys. Rev. Lett.}\ }\textbf
  {\bibinfo {volume} {114}}~(\bibinfo {number} {19}),\ \bibinfo {pages}
  {193602}}\BibitemShut {NoStop}%
\bibitem [{\citenamefont {Farace}\ \emph {et~al.}(2014)\citenamefont {Farace},
  \citenamefont {{De Pasquale}}, \citenamefont {Rigovacca},\ and\ \citenamefont
  {Giovannetti}}]{Farace2014}%
  \BibitemOpen
  \bibfield  {author} {\bibinfo {author} {\bibnamefont {Farace}, \bibfnamefont
  {A.}}, \bibinfo {author} {\bibfnamefont {A.}~\bibnamefont {{De Pasquale}}},
  \bibinfo {author} {\bibfnamefont {L.}~\bibnamefont {Rigovacca}}, \ and\
  \bibinfo {author} {\bibfnamefont {V.}~\bibnamefont {Giovannetti}}} (\bibinfo
  {year} {2014}),\ \href {\doibase 10.1088/1367-2630/16/7/073010} {\bibfield
  {journal} {\bibinfo  {journal} {New Journal of Physics}\ }\textbf {\bibinfo
  {volume} {16}}~(\bibinfo {number} {7}),\ \bibinfo {pages}
  {073010}}\BibitemShut {NoStop}%
\bibitem [{\citenamefont {Farace}\ \emph {et~al.}(2016)\citenamefont {Farace},
  \citenamefont {Pasquale}, \citenamefont {Adesso},\ and\ \citenamefont
  {Giovannetti}}]{Farace2016}%
  \BibitemOpen
  \bibfield  {author} {\bibinfo {author} {\bibnamefont {Farace}, \bibfnamefont
  {A.}}, \bibinfo {author} {\bibfnamefont {A.~D.}\ \bibnamefont {Pasquale}},
  \bibinfo {author} {\bibfnamefont {G.}~\bibnamefont {Adesso}}, \ and\ \bibinfo
  {author} {\bibfnamefont {V.}~\bibnamefont {Giovannetti}}} (\bibinfo {year}
  {2016}),\ \href@noop {} {\bibfield  {journal} {\bibinfo  {journal} {New J.
  Phys.}\ }\textbf {\bibinfo {volume} {18}},\ \bibinfo {pages}
  {013049}}\BibitemShut {NoStop}%
\bibitem [{\citenamefont {Ferraro}\ \emph {et~al.}(2010)\citenamefont
  {Ferraro}, \citenamefont {Aolita}, \citenamefont {Cavalcanti}, \citenamefont
  {Cucchietti},\ and\ \citenamefont {Ac\'in}}]{FerraroAcin2010}%
  \BibitemOpen
  \bibfield  {author} {\bibinfo {author} {\bibnamefont {Ferraro}, \bibfnamefont
  {A.}}, \bibinfo {author} {\bibfnamefont {L.}~\bibnamefont {Aolita}}, \bibinfo
  {author} {\bibfnamefont {D.}~\bibnamefont {Cavalcanti}}, \bibinfo {author}
  {\bibfnamefont {F.}~\bibnamefont {Cucchietti}}, \ and\ \bibinfo {author}
  {\bibfnamefont {A.}~\bibnamefont {Ac\'in}}} (\bibinfo {year} {2010}),\
  \href@noop {} {\bibfield  {journal} {\bibinfo  {journal} {Phys. Rev. A}\
  }\textbf {\bibinfo {volume} {81}}~(\bibinfo {number} {5}),\ \bibinfo {pages}
  {052318}}\BibitemShut {NoStop}%
\bibitem [{\citenamefont {Feynman}(1994)}]{Feynman0}%
  \BibitemOpen
  \bibfield  {author} {\bibinfo {author} {\bibnamefont {Feynman}, \bibfnamefont
  {R.}}} (\bibinfo {year} {1994}),\ \href@noop {} {\emph {\bibinfo {title}
  {Statistical Mechanics}}}\ (\bibinfo  {publisher} {Benjamin},\ \bibinfo
  {address} {Reading (MA)})\BibitemShut {NoStop}%
\bibitem [{\citenamefont {Fisher}(1992)}]{Fisher1992}%
  \BibitemOpen
  \bibfield  {author} {\bibinfo {author} {\bibnamefont {Fisher}, \bibfnamefont
  {D.~S.}}} (\bibinfo {year} {1992}),\ \href {\doibase
  10.1103/PhysRevLett.69.534} {\bibfield  {journal} {\bibinfo  {journal} {Phys.
  Rev. Lett.}\ }\textbf {\bibinfo {volume} {69}},\ \bibinfo {pages}
  {534}}\BibitemShut {NoStop}%
\bibitem [{\citenamefont {Fisher}(1995)}]{Fisher1995}%
  \BibitemOpen
  \bibfield  {author} {\bibinfo {author} {\bibnamefont {Fisher}, \bibfnamefont
  {D.~S.}}} (\bibinfo {year} {1995}),\ \href {\doibase
  10.1103/PhysRevB.51.6411} {\bibfield  {journal} {\bibinfo  {journal} {Phys.
  Rev. B}\ }\textbf {\bibinfo {volume} {51}},\ \bibinfo {pages}
  {6411}}\BibitemShut {NoStop}%
\bibitem [{\citenamefont {Foini}\ \emph {et~al.}(2017)\citenamefont {Foini},
  \citenamefont {Gambassi}, \citenamefont {Konik},\ and\ \citenamefont
  {Cugliandolo}}]{Foini2017}%
  \BibitemOpen
  \bibfield  {author} {\bibinfo {author} {\bibnamefont {Foini}, \bibfnamefont
  {L.}}, \bibinfo {author} {\bibfnamefont {A.}~\bibnamefont {Gambassi}},
  \bibinfo {author} {\bibfnamefont {R.}~\bibnamefont {Konik}}, \ and\ \bibinfo
  {author} {\bibfnamefont {L.~F.}\ \bibnamefont {Cugliandolo}}} (\bibinfo
  {year} {2017}),\ \href@noop {} {\bibfield  {journal} {\bibinfo  {journal}
  {Phys. Rev. E}\ }\textbf {\bibinfo {volume} {95}},\ \bibinfo {pages}
  {052116}}\BibitemShut {NoStop}%
\bibitem [{\citenamefont {Fra\"isse}(2017)}]{JulienThesis2017}%
  \BibitemOpen
  \bibfield  {author} {\bibinfo {author} {\bibnamefont {Fra\"isse},
  \bibfnamefont {J.~M.~E.}}} (\bibinfo {year} {2017}),\ \href@noop {} {\enquote
  {\bibinfo {title} {Ph.d. thesis, university t\"ubingen (2017)},}\
  }\BibitemShut {NoStop}%
\bibitem [{\citenamefont {Fra{\"\i}sse}\ and\ \citenamefont
  {Braun}(2015)}]{fraisse_coherent_2015}%
  \BibitemOpen
  \bibfield  {author} {\bibinfo {author} {\bibnamefont {Fra{\"\i}sse},
  \bibfnamefont {J.~M.~E.}}, \ and\ \bibinfo {author} {\bibfnamefont
  {D.}~\bibnamefont {Braun}}} (\bibinfo {year} {2015}),\ \href@noop {}
  {\bibfield  {journal} {\bibinfo  {journal} {Annalen der Physik}\ }\textbf
  {\bibinfo {volume} {527}},\ \bibinfo {pages} {701}}\BibitemShut {NoStop}%
\bibitem [{\citenamefont {Fra{\"\i}sse}\ and\ \citenamefont
  {Braun}(2016)}]{fraisse_Hamiltonian_2016}%
  \BibitemOpen
  \bibfield  {author} {\bibinfo {author} {\bibnamefont {Fra{\"\i}sse},
  \bibfnamefont {J.~M.~E.}}, \ and\ \bibinfo {author} {\bibfnamefont
  {D.}~\bibnamefont {Braun}}} (\bibinfo {year} {2016}),\ \href@noop {}
  {\enquote {\bibinfo {title} {Hamiltonian extensions in quantum metrology},}\
  }\bibinfo {note} {ArXiv:1610.05974}\BibitemShut {NoStop}%
\bibitem [{\citenamefont {Fra\"{\i}sse}\ and\ \citenamefont
  {Braun}(2017)}]{PhysRevA.95.062342}%
  \BibitemOpen
  \bibfield  {author} {\bibinfo {author} {\bibnamefont {Fra\"{\i}sse},
  \bibfnamefont {J.~M.~E.}}, \ and\ \bibinfo {author} {\bibfnamefont
  {D.}~\bibnamefont {Braun}}} (\bibinfo {year} {2017}),\ \href {\doibase
  10.1103/PhysRevA.95.062342} {\bibfield  {journal} {\bibinfo  {journal} {Phys.
  Rev. A}\ }\textbf {\bibinfo {volume} {95}},\ \bibinfo {pages}
  {062342}}\BibitemShut {NoStop}%
\bibitem [{\citenamefont {Friis}\ \emph {et~al.}(2015)\citenamefont {Friis},
  \citenamefont {Skotiniotis}, \citenamefont {Fuentes},\ and\ \citenamefont
  {D\"ur}}]{Friis2015}%
  \BibitemOpen
  \bibfield  {author} {\bibinfo {author} {\bibnamefont {Friis}, \bibfnamefont
  {N.}}, \bibinfo {author} {\bibfnamefont {M.}~\bibnamefont {Skotiniotis}},
  \bibinfo {author} {\bibfnamefont {I.}~\bibnamefont {Fuentes}}, \ and\
  \bibinfo {author} {\bibfnamefont {W.}~\bibnamefont {D\"ur}}} (\bibinfo {year}
  {2015}),\ \href {\doibase 10.1103/PhysRevA.92.022106} {\bibfield  {journal}
  {\bibinfo  {journal} {Phys. Rev. A}\ }\textbf {\bibinfo {volume} {92}},\
  \bibinfo {pages} {022106}}\BibitemShut {NoStop}%
\bibitem [{\citenamefont {Fuchs}\ and\ \citenamefont
  {de~Graaf}(1999)}]{Fuchs99}%
  \BibitemOpen
  \bibfield  {author} {\bibinfo {author} {\bibnamefont {Fuchs}, \bibfnamefont
  {C.~A.}}, \ and\ \bibinfo {author} {\bibfnamefont {J.~V.}\ \bibnamefont
  {de~Graaf}}} (\bibinfo {year} {1999}),\ \href@noop {} {\bibfield  {journal}
  {\bibinfo  {journal} {IEEE Trans. Inf. Theory}\ }\textbf {\bibinfo {volume}
  {45}},\ \bibinfo {pages} {1216}}\BibitemShut {NoStop}%
\bibitem [{\citenamefont
  {Fujiwara}(2001{\natexlab{a}})}]{fujiwara_quantum_2001}%
  \BibitemOpen
  \bibfield  {author} {\bibinfo {author} {\bibnamefont {Fujiwara},
  \bibfnamefont {A.}}} (\bibinfo {year} {2001}{\natexlab{a}}),\ \href
  {http://link.aps.org/doi/10.1103/PhysRevA.63.042304} {\bibfield  {journal}
  {\bibinfo  {journal} {Physical Review A}\ }\textbf {\bibinfo {volume}
  {63}}~(\bibinfo {number} {4})}\BibitemShut {NoStop}%
\bibitem [{\citenamefont {Fujiwara}(2001{\natexlab{b}})}]{Fujiwara01}%
  \BibitemOpen
  \bibfield  {author} {\bibinfo {author} {\bibnamefont {Fujiwara},
  \bibfnamefont {A.}}} (\bibinfo {year} {2001}{\natexlab{b}}),\ \href@noop {}
  {\bibfield  {journal} {\bibinfo  {journal} {Phys. Rev. A}\ }\textbf {\bibinfo
  {volume} {63}}~(\bibinfo {number} {4}),\ \bibinfo {pages}
  {042304}}\BibitemShut {NoStop}%
\bibitem [{\citenamefont {Fujiwara}(2006)}]{fujiwara_strong_2006}%
  \BibitemOpen
  \bibfield  {author} {\bibinfo {author} {\bibnamefont {Fujiwara},
  \bibfnamefont {A.}}} (\bibinfo {year} {2006}),\ \href {\doibase
  10.1088/0305-4470/39/40/014} {\bibfield  {journal} {\bibinfo  {journal} {J.
  Phys. A: Math. Gen.}\ }\textbf {\bibinfo {volume} {39}}~(\bibinfo {number}
  {40}),\ \bibinfo {pages} {12489}}\BibitemShut {NoStop}%
\bibitem [{\citenamefont {Fujiwara}\ and\ \citenamefont
  {Hashizum\'e}(2002)}]{fujiwara_additivity_2002}%
  \BibitemOpen
  \bibfield  {author} {\bibinfo {author} {\bibnamefont {Fujiwara},
  \bibfnamefont {A.}}, \ and\ \bibinfo {author} {\bibfnamefont
  {T.}~\bibnamefont {Hashizum\'e}}} (\bibinfo {year} {2002}),\ \href {\doibase
  10.1016/S0375-9601(02)00735-1} {\bibfield  {journal} {\bibinfo  {journal}
  {Physics Letters A}\ }\textbf {\bibinfo {volume} {299}},\ \bibinfo {pages}
  {469}}\BibitemShut {NoStop}%
\bibitem [{\citenamefont {Fujiwara}\ and\ \citenamefont
  {Imai}(2003)}]{fujiwara_quantum_2003}%
  \BibitemOpen
  \bibfield  {author} {\bibinfo {author} {\bibnamefont {Fujiwara},
  \bibfnamefont {A.}}, \ and\ \bibinfo {author} {\bibfnamefont
  {H.}~\bibnamefont {Imai}}} (\bibinfo {year} {2003}),\ \href
  {http://iopscience.iop.org/0305-4470/36/29/314} {\bibfield  {journal}
  {\bibinfo  {journal} {Journal of Physics A: Mathematical and General}\
  }\textbf {\bibinfo {volume} {36}}~(\bibinfo {number} {29}),\ \bibinfo {pages}
  {8093}}\BibitemShut {NoStop}%
\bibitem [{\citenamefont {Gagatsos}\ \emph {et~al.}(2016)\citenamefont
  {Gagatsos}, \citenamefont {Branford},\ and\ \citenamefont
  {Datta}}]{Gagatsos2016}%
  \BibitemOpen
  \bibfield  {author} {\bibinfo {author} {\bibnamefont {Gagatsos},
  \bibfnamefont {C.~N.}}, \bibinfo {author} {\bibfnamefont {D.}~\bibnamefont
  {Branford}}, \ and\ \bibinfo {author} {\bibfnamefont {A.}~\bibnamefont
  {Datta}}} (\bibinfo {year} {2016}),\ \href {\doibase
  10.1103/PhysRevA.94.042342} {\bibfield  {journal} {\bibinfo  {journal} {Phys.
  Rev. A}\ }\textbf {\bibinfo {volume} {94}},\ \bibinfo {pages}
  {042342}}\BibitemShut {NoStop}%
\bibitem [{\citenamefont {Gaiba}\ and\ \citenamefont {Paris}(2009)}]{Gaiba09}%
  \BibitemOpen
  \bibfield  {author} {\bibinfo {author} {\bibnamefont {Gaiba}, \bibfnamefont
  {R.}}, \ and\ \bibinfo {author} {\bibfnamefont {M.~G.~A.}\ \bibnamefont
  {Paris}}} (\bibinfo {year} {2009}),\ \href@noop {} {\bibfield  {journal}
  {\bibinfo  {journal} {Phys. Lett. A}\ }\textbf {\bibinfo {volume} {373}},\
  \bibinfo {pages} {934}}\BibitemShut {NoStop}%
\bibitem [{\citenamefont {Gao}\ and\ \citenamefont {Lee}(2014)}]{Gao2014}%
  \BibitemOpen
  \bibfield  {author} {\bibinfo {author} {\bibnamefont {Gao}, \bibfnamefont
  {Y.}}, \ and\ \bibinfo {author} {\bibfnamefont {H.}~\bibnamefont {Lee}}}
  (\bibinfo {year} {2014}),\ \href {\doibase 10.1140/epjd/e2014-50560-1}
  {\bibfield  {journal} {\bibinfo  {journal} {The European Physical Journal D}\
  }\textbf {\bibinfo {volume} {68}}~(\bibinfo {number} {11}),\ \bibinfo {pages}
  {347}}\BibitemShut {NoStop}%
\bibitem [{\citenamefont {Garnerone}\ \emph
  {et~al.}(2009{\natexlab{a}})\citenamefont {Garnerone}, \citenamefont
  {Abasto}, \citenamefont {Haas},\ and\ \citenamefont
  {Zanardi}}]{Garnerone2009}%
  \BibitemOpen
  \bibfield  {author} {\bibinfo {author} {\bibnamefont {Garnerone},
  \bibfnamefont {S.}}, \bibinfo {author} {\bibfnamefont {D.}~\bibnamefont
  {Abasto}}, \bibinfo {author} {\bibfnamefont {S.}~\bibnamefont {Haas}}, \ and\
  \bibinfo {author} {\bibfnamefont {P.}~\bibnamefont {Zanardi}}} (\bibinfo
  {year} {2009}{\natexlab{a}}),\ \href {\doibase 10.1103/PhysRevA.79.032302}
  {\bibfield  {journal} {\bibinfo  {journal} {Phys. Rev. A}\ }\textbf {\bibinfo
  {volume} {79}},\ \bibinfo {pages} {032302}}\BibitemShut {NoStop}%
\bibitem [{\citenamefont {Garnerone}\ \emph
  {et~al.}(2009{\natexlab{b}})\citenamefont {Garnerone}, \citenamefont
  {Jacobson}, \citenamefont {Haas},\ and\ \citenamefont
  {Zanardi}}]{Garnerone2009-2}%
  \BibitemOpen
  \bibfield  {author} {\bibinfo {author} {\bibnamefont {Garnerone},
  \bibfnamefont {S.}}, \bibinfo {author} {\bibfnamefont {N.~T.}\ \bibnamefont
  {Jacobson}}, \bibinfo {author} {\bibfnamefont {S.}~\bibnamefont {Haas}}, \
  and\ \bibinfo {author} {\bibfnamefont {P.}~\bibnamefont {Zanardi}}} (\bibinfo
  {year} {2009}{\natexlab{b}}),\ \href {\doibase
  10.1103/PhysRevLett.102.057205} {\bibfield  {journal} {\bibinfo  {journal}
  {Phys. Rev. Lett.}\ }\textbf {\bibinfo {volume} {102}},\ \bibinfo {pages}
  {057205}}\BibitemShut {NoStop}%
\bibitem [{\citenamefont {Genoni}\ \emph {et~al.}(2013)\citenamefont {Genoni},
  \citenamefont {Paris}, \citenamefont {Adesso}, \citenamefont {Nha},
  \citenamefont {Knight},\ and\ \citenamefont {Kim}}]{PhysRevA.87.012107}%
  \BibitemOpen
  \bibfield  {author} {\bibinfo {author} {\bibnamefont {Genoni}, \bibfnamefont
  {M.~G.}}, \bibinfo {author} {\bibfnamefont {M.~G.~A.}\ \bibnamefont {Paris}},
  \bibinfo {author} {\bibfnamefont {G.}~\bibnamefont {Adesso}}, \bibinfo
  {author} {\bibfnamefont {H.}~\bibnamefont {Nha}}, \bibinfo {author}
  {\bibfnamefont {P.~L.}\ \bibnamefont {Knight}}, \ and\ \bibinfo {author}
  {\bibfnamefont {M.~S.}\ \bibnamefont {Kim}}} (\bibinfo {year} {2013}),\ \href
  {\doibase 10.1103/PhysRevA.87.012107} {\bibfield  {journal} {\bibinfo
  {journal} {Phys. Rev. A}\ }\textbf {\bibinfo {volume} {87}},\ \bibinfo
  {pages} {012107}}\BibitemShut {NoStop}%
\bibitem [{\citenamefont {Gerry}\ and\ \citenamefont {Knight}(2005)}]{Knight0}%
  \BibitemOpen
  \bibfield  {author} {\bibinfo {author} {\bibnamefont {Gerry}, \bibfnamefont
  {C.}}, \ and\ \bibinfo {author} {\bibfnamefont {P.}~\bibnamefont {Knight}}}
  (\bibinfo {year} {2005}),\ \href@noop {} {\emph {\bibinfo {title}
  {Introductory Quantum Optics}}}\ (\bibinfo  {publisher} {Cambridge University
  Press},\ \bibinfo {address} {Cambridge})\BibitemShut {NoStop}%
\bibitem [{\citenamefont {Ghirardi}\ \emph {et~al.}(2002)\citenamefont
  {Ghirardi}, \citenamefont {Marinatto},\ and\ \citenamefont
  {Weber}}]{Ghirardi0}%
  \BibitemOpen
  \bibfield  {author} {\bibinfo {author} {\bibnamefont {Ghirardi},
  \bibfnamefont {G.}}, \bibinfo {author} {\bibfnamefont {L.}~\bibnamefont
  {Marinatto}}, \ and\ \bibinfo {author} {\bibfnamefont {T.}~\bibnamefont
  {Weber}}} (\bibinfo {year} {2002}),\ \href@noop {} {\bibfield  {journal}
  {\bibinfo  {journal} {J. Stat Phys.}\ }\textbf {\bibinfo {volume} {108}},\
  \bibinfo {pages} {49}}\BibitemShut {NoStop}%
\bibitem [{\citenamefont {Ghirardi}\ \emph {et~al.}(2004)\citenamefont
  {Ghirardi}, \citenamefont {Marinatto},\ and\ \citenamefont
  {Weber}}]{Ghirardi1}%
  \BibitemOpen
  \bibfield  {author} {\bibinfo {author} {\bibnamefont {Ghirardi},
  \bibfnamefont {G.}}, \bibinfo {author} {\bibfnamefont {L.}~\bibnamefont
  {Marinatto}}, \ and\ \bibinfo {author} {\bibfnamefont {T.}~\bibnamefont
  {Weber}}} (\bibinfo {year} {2004}),\ \href@noop {} {\bibfield  {journal}
  {\bibinfo  {journal} {Phys. Rev. A}\ }\textbf {\bibinfo {volume} {70}},\
  \bibinfo {pages} {012109}}\BibitemShut {NoStop}%
\bibitem [{\citenamefont {Giamarchi}(2003)}]{Giamarchi}%
  \BibitemOpen
  \bibfield  {author} {\bibinfo {author} {\bibnamefont {Giamarchi},
  \bibfnamefont {T.}}} (\bibinfo {year} {2003}),\ \href@noop {} {\emph
  {\bibinfo {title} {{Quantum Physics in One Dimension}}}}\ (\bibinfo
  {publisher} {Oxford University Press})\BibitemShut {NoStop}%
\bibitem [{\citenamefont {Giazotto}\ \emph {et~al.}(2006)\citenamefont
  {Giazotto}, \citenamefont {Heikkil\"a}, \citenamefont {Luukanen},
  \citenamefont {Savin},\ and\ \citenamefont {Pekola}}]{Giazotto2006}%
  \BibitemOpen
  \bibfield  {author} {\bibinfo {author} {\bibnamefont {Giazotto},
  \bibfnamefont {F.}}, \bibinfo {author} {\bibfnamefont {T.~T.}\ \bibnamefont
  {Heikkil\"a}}, \bibinfo {author} {\bibfnamefont {A.}~\bibnamefont
  {Luukanen}}, \bibinfo {author} {\bibfnamefont {A.~M.}\ \bibnamefont {Savin}},
  \ and\ \bibinfo {author} {\bibfnamefont {J.~P.}\ \bibnamefont {Pekola}}}
  (\bibinfo {year} {2006}),\ \href {\doibase 10.1103/RevModPhys.78.217}
  {\bibfield  {journal} {\bibinfo  {journal} {Rev. Mod. Phys.}\ }\textbf
  {\bibinfo {volume} {78}},\ \bibinfo {pages} {217}}\BibitemShut {NoStop}%
\bibitem [{\citenamefont {Gilchrist}\ \emph {et~al.}(2004)\citenamefont
  {Gilchrist}, \citenamefont {Nemoto}, \citenamefont {Munro}, \citenamefont
  {Ralph}, \citenamefont {Glancy}, \citenamefont {Braunstein},\ and\
  \citenamefont {Milburn}}]{1464-4266-6-8-032}%
  \BibitemOpen
  \bibfield  {author} {\bibinfo {author} {\bibnamefont {Gilchrist},
  \bibfnamefont {A.}}, \bibinfo {author} {\bibfnamefont {K.}~\bibnamefont
  {Nemoto}}, \bibinfo {author} {\bibfnamefont {W.~J.}\ \bibnamefont {Munro}},
  \bibinfo {author} {\bibfnamefont {T.~C.}\ \bibnamefont {Ralph}}, \bibinfo
  {author} {\bibfnamefont {S.}~\bibnamefont {Glancy}}, \bibinfo {author}
  {\bibfnamefont {S.~L.}\ \bibnamefont {Braunstein}}, \ and\ \bibinfo {author}
  {\bibfnamefont {G.~J.}\ \bibnamefont {Milburn}}} (\bibinfo {year} {2004}),\
  \href {http://stacks.iop.org/1464-4266/6/i=8/a=032} {\bibfield  {journal}
  {\bibinfo  {journal} {Journal of Optics B: Quantum and Semiclassical Optics}\
  }\textbf {\bibinfo {volume} {6}}~(\bibinfo {number} {8}),\ \bibinfo {pages}
  {S828}}\BibitemShut {NoStop}%
\bibitem [{\citenamefont {Gill}\ and\ \citenamefont
  {Levit}(1995)}]{gill_applications_1995}%
  \BibitemOpen
  \bibfield  {author} {\bibinfo {author} {\bibnamefont {Gill}, \bibfnamefont
  {R.~D.}}, \ and\ \bibinfo {author} {\bibfnamefont {B.~Y.}\ \bibnamefont
  {Levit}}} (\bibinfo {year} {1995}),\ \href@noop {} {\bibfield  {journal}
  {\bibinfo  {journal} {Bernoulli}\ }\textbf {\bibinfo {volume} {1}}~(\bibinfo
  {number} {1-2}),\ \bibinfo {pages} {59}}\BibitemShut {NoStop}%
\bibitem [{\citenamefont {Gilmore}(1985)}]{Gilmore1985}%
  \BibitemOpen
  \bibfield  {author} {\bibinfo {author} {\bibnamefont {Gilmore}, \bibfnamefont
  {R.}}} (\bibinfo {year} {1985}),\ \href {\doibase 10.1103/PhysRevA.31.3237}
  {\bibfield  {journal} {\bibinfo  {journal} {Phys. Rev. A}\ }\textbf {\bibinfo
  {volume} {31}},\ \bibinfo {pages} {3237}}\BibitemShut {NoStop}%
\bibitem [{\citenamefont {Giorgini}\ \emph {et~al.}(2008)\citenamefont
  {Giorgini}, \citenamefont {Pitaevskii},\ and\ \citenamefont
  {Stringari}}]{Giorgini0}%
  \BibitemOpen
  \bibfield  {author} {\bibinfo {author} {\bibnamefont {Giorgini},
  \bibfnamefont {S.}}, \bibinfo {author} {\bibfnamefont {L.}~\bibnamefont
  {Pitaevskii}}, \ and\ \bibinfo {author} {\bibfnamefont {S.}~\bibnamefont
  {Stringari}}} (\bibinfo {year} {2008}),\ \href@noop {} {\bibfield  {journal}
  {\bibinfo  {journal} {Rev. Mod. Phys.}\ }\textbf {\bibinfo {volume} {80}},\
  \bibinfo {pages} {1215}}\BibitemShut {NoStop}%
\bibitem [{\citenamefont {Giovannetti}\ \emph {et~al.}(2004)\citenamefont
  {Giovannetti}, \citenamefont {Lloyd},\ and\ \citenamefont
  {Maccone}}]{GiovannettiS2004}%
  \BibitemOpen
  \bibfield  {author} {\bibinfo {author} {\bibnamefont {Giovannetti},
  \bibfnamefont {V.}}, \bibinfo {author} {\bibfnamefont {S.}~\bibnamefont
  {Lloyd}}, \ and\ \bibinfo {author} {\bibfnamefont {L.}~\bibnamefont
  {Maccone}}} (\bibinfo {year} {2004}),\ \href {\doibase
  10.1126/science.1104149} {\bibfield  {journal} {\bibinfo  {journal}
  {Science}\ }\textbf {\bibinfo {volume} {306}}~(\bibinfo {number} {5700}),\
  \bibinfo {pages} {1330}}\BibitemShut {NoStop}%
\bibitem [{\citenamefont {Giovannetti}\ \emph {et~al.}(2006)\citenamefont
  {Giovannetti}, \citenamefont {Lloyd},\ and\ \citenamefont
  {Maccone}}]{GiovannettiPRL2006}%
  \BibitemOpen
  \bibfield  {author} {\bibinfo {author} {\bibnamefont {Giovannetti},
  \bibfnamefont {V.}}, \bibinfo {author} {\bibfnamefont {S.}~\bibnamefont
  {Lloyd}}, \ and\ \bibinfo {author} {\bibfnamefont {L.}~\bibnamefont
  {Maccone}}} (\bibinfo {year} {{2006}}),\ \href {\doibase
  {10.1103/PhysRevLett.96.010401}} {\bibfield  {journal} {\bibinfo  {journal}
  {{Phys. Rev. Lett.}}\ }\textbf {\bibinfo {volume} {{96}}}~(\bibinfo {number}
  {{1}}),\ \bibinfo {pages} {{010401}}}\BibitemShut {NoStop}%
\bibitem [{\citenamefont {Giovannetti}\ \emph {et~al.}(2011)\citenamefont
  {Giovannetti}, \citenamefont {Lloyd},\ and\ \citenamefont
  {Maccone}}]{GiovannettiNPhot2011}%
  \BibitemOpen
  \bibfield  {author} {\bibinfo {author} {\bibnamefont {Giovannetti},
  \bibfnamefont {V.}}, \bibinfo {author} {\bibfnamefont {S.}~\bibnamefont
  {Lloyd}}, \ and\ \bibinfo {author} {\bibfnamefont {L.}~\bibnamefont
  {Maccone}}} (\bibinfo {year} {2011}),\ \href
  {http://dx.doi.org/10.1038/nphoton.2011.35} {\bibfield  {journal} {\bibinfo
  {journal} {Nat Photon}\ }\textbf {\bibinfo {volume} {5}}~(\bibinfo {number}
  {4}),\ \bibinfo {pages} {222}}\BibitemShut {NoStop}%
\bibitem [{\citenamefont {Girardeau}(1960)}]{Girardeau1960}%
  \BibitemOpen
  \bibfield  {author} {\bibinfo {author} {\bibnamefont {Girardeau},
  \bibfnamefont {M.}}} (\bibinfo {year} {1960}),\ \href@noop {} {\bibfield
  {journal} {\bibinfo  {journal} {J. Math. Phys.}\ }\textbf {\bibinfo {volume}
  {1}},\ \bibinfo {pages} {516}}\BibitemShut {NoStop}%
\bibitem [{\citenamefont {Girardeau}(1965)}]{Girardeau1965}%
  \BibitemOpen
  \bibfield  {author} {\bibinfo {author} {\bibnamefont {Girardeau},
  \bibfnamefont {M.}}} (\bibinfo {year} {1965}),\ \href@noop {} {\bibfield
  {journal} {\bibinfo  {journal} {J. Math. Phys.}\ }\textbf {\bibinfo {volume}
  {6}},\ \bibinfo {pages} {1083}}\BibitemShut {NoStop}%
\bibitem [{\citenamefont {Girolami}\ \emph {et~al.}(2014)\citenamefont
  {Girolami}, \citenamefont {Souza}, \citenamefont {Giovannetti}, \citenamefont
  {Tufarelli}, \citenamefont {Filgueiras}, \citenamefont {Sarthour},
  \citenamefont {Soares-Pinto}, \citenamefont {Oliveira},\ and\ \citenamefont
  {Adesso}}]{Girolami2014}%
  \BibitemOpen
  \bibfield  {author} {\bibinfo {author} {\bibnamefont {Girolami},
  \bibfnamefont {D.}}, \bibinfo {author} {\bibfnamefont {A.~M.}\ \bibnamefont
  {Souza}}, \bibinfo {author} {\bibfnamefont {V.}~\bibnamefont {Giovannetti}},
  \bibinfo {author} {\bibfnamefont {T.}~\bibnamefont {Tufarelli}}, \bibinfo
  {author} {\bibfnamefont {J.~G.}\ \bibnamefont {Filgueiras}}, \bibinfo
  {author} {\bibfnamefont {R.~S.}\ \bibnamefont {Sarthour}}, \bibinfo {author}
  {\bibfnamefont {D.~O.}\ \bibnamefont {Soares-Pinto}}, \bibinfo {author}
  {\bibfnamefont {I.~S.}\ \bibnamefont {Oliveira}}, \ and\ \bibinfo {author}
  {\bibfnamefont {G.}~\bibnamefont {Adesso}}} (\bibinfo {year} {2014}),\ \href
  {\doibase 10.1103/PhysRevLett.112.210401} {\bibfield  {journal} {\bibinfo
  {journal} {Phys. Rev. Lett.}\ }\textbf {\bibinfo {volume} {112}},\ \bibinfo
  {pages} {210401}}\BibitemShut {NoStop}%
\bibitem [{\citenamefont {Girolami}\ \emph {et~al.}(2013)\citenamefont
  {Girolami}, \citenamefont {Tufarelli},\ and\ \citenamefont
  {Adesso}}]{Girolami2013}%
  \BibitemOpen
  \bibfield  {author} {\bibinfo {author} {\bibnamefont {Girolami},
  \bibfnamefont {D.}}, \bibinfo {author} {\bibfnamefont {T.}~\bibnamefont
  {Tufarelli}}, \ and\ \bibinfo {author} {\bibfnamefont {G.}~\bibnamefont
  {Adesso}}} (\bibinfo {year} {2013}),\ \href {\doibase
  10.1103/PhysRevLett.110.240402} {\bibfield  {journal} {\bibinfo  {journal}
  {Phys. Rev. Lett.}\ }\textbf {\bibinfo {volume} {110}},\ \bibinfo {pages}
  {240402}}\BibitemShut {NoStop}%
\bibitem [{\citenamefont {Goble}\ and\ \citenamefont
  {Trainor}(1965)}]{Goble1965}%
  \BibitemOpen
  \bibfield  {author} {\bibinfo {author} {\bibnamefont {Goble}, \bibfnamefont
  {D.~F.}}, \ and\ \bibinfo {author} {\bibfnamefont {L.~E.~H.}\ \bibnamefont
  {Trainor}}} (\bibinfo {year} {1965}),\ \href@noop {} {\bibfield  {journal}
  {\bibinfo  {journal} {Phys. Lett.}\ }\textbf {\bibinfo {volume} {18}},\
  \bibinfo {pages} {122}}\BibitemShut {NoStop}%
\bibitem [{\citenamefont {Goble}\ and\ \citenamefont
  {Trainor}(1966)}]{Goble1966}%
  \BibitemOpen
  \bibfield  {author} {\bibinfo {author} {\bibnamefont {Goble}, \bibfnamefont
  {D.~F.}}, \ and\ \bibinfo {author} {\bibfnamefont {L.~E.~H.}\ \bibnamefont
  {Trainor}}} (\bibinfo {year} {1966}),\ \href@noop {} {\bibfield  {journal}
  {\bibinfo  {journal} {Can. J. Phys.}\ }\textbf {\bibinfo {volume} {44}},\
  \bibinfo {pages} {27}}\BibitemShut {NoStop}%
\bibitem [{\citenamefont {Goble}\ and\ \citenamefont
  {Trainor}(1967)}]{Goble67}%
  \BibitemOpen
  \bibfield  {author} {\bibinfo {author} {\bibnamefont {Goble}, \bibfnamefont
  {D.~F.}}, \ and\ \bibinfo {author} {\bibfnamefont {L.~E.~H.}\ \bibnamefont
  {Trainor}}} (\bibinfo {year} {1967}),\ \href {\doibase
  10.1103/PhysRev.157.167} {\bibfield  {journal} {\bibinfo  {journal} {Phys.
  Rev.}\ }\textbf {\bibinfo {volume} {157}},\ \bibinfo {pages}
  {167}}\BibitemShut {NoStop}%
\bibitem [{\citenamefont {Gong}\ and\ \citenamefont {Tong}(2008)}]{Gong2008}%
  \BibitemOpen
  \bibfield  {author} {\bibinfo {author} {\bibnamefont {Gong}, \bibfnamefont
  {L.}}, \ and\ \bibinfo {author} {\bibfnamefont {P.}~\bibnamefont {Tong}}}
  (\bibinfo {year} {2008}),\ \href {\doibase 10.1103/PhysRevB.78.115114}
  {\bibfield  {journal} {\bibinfo  {journal} {Phys. Rev. B}\ }\textbf {\bibinfo
  {volume} {78}},\ \bibinfo {pages} {115114}}\BibitemShut {NoStop}%
\bibitem [{\citenamefont {G\"orlitz}\ \emph {et~al.}(2001)\citenamefont
  {G\"orlitz}, \citenamefont {Vogels}, \citenamefont {Leanhardt}, \citenamefont
  {Raman}, \citenamefont {Gustavson}, \citenamefont {Abo-Shaeer}, \citenamefont
  {Chikkatur}, \citenamefont {Gupta}, \citenamefont {Inouye}, \citenamefont
  {Rosenband},\ and\ \citenamefont {Ketterle}}]{Gorlitz2001}%
  \BibitemOpen
  \bibfield  {author} {\bibinfo {author} {\bibnamefont {G\"orlitz},
  \bibfnamefont {A.}}, \bibinfo {author} {\bibfnamefont {J.}~\bibnamefont
  {Vogels}}, \bibinfo {author} {\bibfnamefont {A.}~\bibnamefont {Leanhardt}},
  \bibinfo {author} {\bibfnamefont {C.}~\bibnamefont {Raman}}, \bibinfo
  {author} {\bibfnamefont {T.}~\bibnamefont {Gustavson}}, \bibinfo {author}
  {\bibfnamefont {J.}~\bibnamefont {Abo-Shaeer}}, \bibinfo {author}
  {\bibfnamefont {A.}~\bibnamefont {Chikkatur}}, \bibinfo {author}
  {\bibfnamefont {S.}~\bibnamefont {Gupta}}, \bibinfo {author} {\bibfnamefont
  {S.}~\bibnamefont {Inouye}}, \bibinfo {author} {\bibfnamefont
  {T.}~\bibnamefont {Rosenband}}, \ and\ \bibinfo {author} {\bibfnamefont
  {W.}~\bibnamefont {Ketterle}}} (\bibinfo {year} {2001}),\ \href@noop {}
  {\bibfield  {journal} {\bibinfo  {journal} {Phys. Rev. Lett.}\ }\textbf
  {\bibinfo {volume} {87}},\ \bibinfo {pages} {130402}}\BibitemShut {NoStop}%
\bibitem [{\citenamefont {Gottesman}(1996)}]{Gottesman96}%
  \BibitemOpen
  \bibfield  {author} {\bibinfo {author} {\bibnamefont {Gottesman},
  \bibfnamefont {D.}}} (\bibinfo {year} {1996}),\ \href@noop {} {\bibfield
  {journal} {\bibinfo  {journal} {Phys. Rev. A}\ }\textbf {\bibinfo {volume}
  {54}},\ \bibinfo {pages} {1862}}\BibitemShut {NoStop}%
\bibitem [{\citenamefont {Gottesman}\ \emph {et~al.}(2001)\citenamefont
  {Gottesman}, \citenamefont {Kitaev},\ and\ \citenamefont
  {Preskill}}]{gottesman_encoding_2001}%
  \BibitemOpen
  \bibfield  {author} {\bibinfo {author} {\bibnamefont {Gottesman},
  \bibfnamefont {D.}}, \bibinfo {author} {\bibfnamefont {A.}~\bibnamefont
  {Kitaev}}, \ and\ \bibinfo {author} {\bibfnamefont {J.}~\bibnamefont
  {Preskill}}} (\bibinfo {year} {2001}),\ \href {\doibase
  10.1103/PhysRevA.64.012310} {\bibfield  {journal} {\bibinfo  {journal} {Phys.
  Rev. A}\ }\textbf {\bibinfo {volume} {64}}~(\bibinfo {number} {1}),\ \bibinfo
  {pages} {012310}}\BibitemShut {NoStop}%
\bibitem [{\citenamefont {Grabowski}\ \emph {et~al.}(2011)\citenamefont
  {Grabowski}, \citenamefont {Ku\`s},\ and\ \citenamefont
  {Marmo}}]{Grabowski0}%
  \BibitemOpen
  \bibfield  {author} {\bibinfo {author} {\bibnamefont {Grabowski},
  \bibfnamefont {J.}}, \bibinfo {author} {\bibfnamefont {M.}~\bibnamefont
  {Ku\`s}}, \ and\ \bibinfo {author} {\bibfnamefont {G.}~\bibnamefont {Marmo}}}
  (\bibinfo {year} {2011}),\ \href@noop {} {\bibfield  {journal} {\bibinfo
  {journal} {J. Phys. A}\ }\textbf {\bibinfo {volume} {44}},\ \bibinfo {pages}
  {175302}}\BibitemShut {NoStop}%
\bibitem [{\citenamefont {Grabowski}\ \emph {et~al.}(2012)\citenamefont
  {Grabowski}, \citenamefont {Ku\`s},\ and\ \citenamefont
  {Marmo}}]{Grabowski1}%
  \BibitemOpen
  \bibfield  {author} {\bibinfo {author} {\bibnamefont {Grabowski},
  \bibfnamefont {J.}}, \bibinfo {author} {\bibfnamefont {M.}~\bibnamefont
  {Ku\`s}}, \ and\ \bibinfo {author} {\bibfnamefont {G.}~\bibnamefont {Marmo}}}
  (\bibinfo {year} {2012}),\ \href@noop {} {\bibfield  {journal} {\bibinfo
  {journal} {J. Phys. A}\ }\textbf {\bibinfo {volume} {45}},\ \bibinfo {pages}
  {105301}}\BibitemShut {NoStop}%
\bibitem [{\citenamefont {Greiner}\ \emph {et~al.}(2001)\citenamefont
  {Greiner}, \citenamefont {Bloch}, \citenamefont {Mandel}, \citenamefont
  {H\"ansch},\ and\ \citenamefont {Esslinger}}]{Greiner2001}%
  \BibitemOpen
  \bibfield  {author} {\bibinfo {author} {\bibnamefont {Greiner}, \bibfnamefont
  {M.}}, \bibinfo {author} {\bibfnamefont {I.}~\bibnamefont {Bloch}}, \bibinfo
  {author} {\bibfnamefont {O.}~\bibnamefont {Mandel}}, \bibinfo {author}
  {\bibfnamefont {T.}~\bibnamefont {H\"ansch}}, \ and\ \bibinfo {author}
  {\bibfnamefont {T.}~\bibnamefont {Esslinger}}} (\bibinfo {year} {2001}),\
  \href@noop {} {\bibfield  {journal} {\bibinfo  {journal} {Phys. Rev. Lett.}\
  }\textbf {\bibinfo {volume} {87}},\ \bibinfo {pages} {160405}}\BibitemShut
  {NoStop}%
\bibitem [{\citenamefont {Greschner}\ \emph {et~al.}(2013)\citenamefont
  {Greschner}, \citenamefont {Kolezhuk},\ and\ \citenamefont
  {Vekua}}]{Greschner2013}%
  \BibitemOpen
  \bibfield  {author} {\bibinfo {author} {\bibnamefont {Greschner},
  \bibfnamefont {S.}}, \bibinfo {author} {\bibfnamefont {A.~K.}\ \bibnamefont
  {Kolezhuk}}, \ and\ \bibinfo {author} {\bibfnamefont {T.}~\bibnamefont
  {Vekua}}} (\bibinfo {year} {2013}),\ \href {\doibase
  10.1103/PhysRevB.88.195101} {\bibfield  {journal} {\bibinfo  {journal} {Phys.
  Rev. B}\ }\textbf {\bibinfo {volume} {88}},\ \bibinfo {pages}
  {195101}}\BibitemShut {NoStop}%
\bibitem [{\citenamefont {Griffiths}(1969)}]{Griffiths1969}%
  \BibitemOpen
  \bibfield  {author} {\bibinfo {author} {\bibnamefont {Griffiths},
  \bibfnamefont {R.~B.}}} (\bibinfo {year} {1969}),\ \href {\doibase
  10.1103/PhysRevLett.23.17} {\bibfield  {journal} {\bibinfo  {journal} {Phys.
  Rev. Lett.}\ }\textbf {\bibinfo {volume} {23}},\ \bibinfo {pages}
  {17}}\BibitemShut {NoStop}%
\bibitem [{\citenamefont {Gross}\ \emph
  {et~al.}(2010{\natexlab{a}})\citenamefont {Gross}, \citenamefont {Zibold},
  \citenamefont {Nicklas}, \citenamefont {Est{\`e}ve},\ and\ \citenamefont
  {Oberthaler}}]{GrossN2010}%
  \BibitemOpen
  \bibfield  {author} {\bibinfo {author} {\bibnamefont {Gross}, \bibfnamefont
  {C.}}, \bibinfo {author} {\bibfnamefont {T.}~\bibnamefont {Zibold}}, \bibinfo
  {author} {\bibfnamefont {E.}~\bibnamefont {Nicklas}}, \bibinfo {author}
  {\bibfnamefont {J.}~\bibnamefont {Est{\`e}ve}}, \ and\ \bibinfo {author}
  {\bibfnamefont {M.~K.}\ \bibnamefont {Oberthaler}}} (\bibinfo {year}
  {2010}{\natexlab{a}}),\ \href {http://dx.doi.org/10.1038/nature08919}
  {\bibfield  {journal} {\bibinfo  {journal} {Nature}\ }\textbf {\bibinfo
  {volume} {464}}~(\bibinfo {number} {7292}),\ \bibinfo {pages}
  {1165}}\BibitemShut {NoStop}%
\bibitem [{\citenamefont {Gross}\ \emph
  {et~al.}(2010{\natexlab{b}})\citenamefont {Gross}, \citenamefont {Zibold},
  \citenamefont {Nicklas}, \citenamefont {amd M.~K.~Oberthaler},\ and\
  \citenamefont {Gross}}]{Oberthaler1}%
  \BibitemOpen
  \bibfield  {author} {\bibinfo {author} {\bibnamefont {Gross}, \bibfnamefont
  {C.}}, \bibinfo {author} {\bibfnamefont {T.}~\bibnamefont {Zibold}}, \bibinfo
  {author} {\bibfnamefont {E.}~\bibnamefont {Nicklas}}, \bibinfo {author}
  {\bibfnamefont {J.~E.}\ \bibnamefont {amd M.~K.~Oberthaler}}, \ and\ \bibinfo
  {author} {\bibfnamefont {C.}~\bibnamefont {Gross}}} (\bibinfo {year}
  {2010}{\natexlab{b}}),\ \href@noop {} {\bibfield  {journal} {\bibinfo
  {journal} {Nature}\ }\textbf {\bibinfo {volume} {464}},\ \bibinfo {pages}
  {1165}}\BibitemShut {NoStop}%
\bibitem [{\citenamefont {Gu}(2010)}]{Gu2010}%
  \BibitemOpen
  \bibfield  {author} {\bibinfo {author} {\bibnamefont {Gu}, \bibfnamefont
  {S.-J.}}} (\bibinfo {year} {2010}),\ \href@noop {} {\bibfield  {journal}
  {\bibinfo  {journal} {Int. J. Mod. Phys. B}\ }\textbf {\bibinfo {volume}
  {24}},\ \bibinfo {pages} {4371}}\BibitemShut {NoStop}%
\bibitem [{\citenamefont {Gu}\ \emph {et~al.}(2008)\citenamefont {Gu},
  \citenamefont {Kwok}, \citenamefont {Ning},\ and\ \citenamefont
  {Lin}}]{Gu2008}%
  \BibitemOpen
  \bibfield  {author} {\bibinfo {author} {\bibnamefont {Gu}, \bibfnamefont
  {S.-J.}}, \bibinfo {author} {\bibfnamefont {H.-M.}\ \bibnamefont {Kwok}},
  \bibinfo {author} {\bibfnamefont {W.-Q.}\ \bibnamefont {Ning}}, \ and\
  \bibinfo {author} {\bibfnamefont {H.-Q.}\ \bibnamefont {Lin}}} (\bibinfo
  {year} {2008}),\ \href {\doibase 10.1103/PhysRevB.77.245109} {\bibfield
  {journal} {\bibinfo  {journal} {Phys. Rev. B}\ }\textbf {\bibinfo {volume}
  {77}},\ \bibinfo {pages} {245109}}\BibitemShut {NoStop}%
\bibitem [{\citenamefont {Gu}\ and\ \citenamefont {Lin}(2009)}]{Gu2009}%
  \BibitemOpen
  \bibfield  {author} {\bibinfo {author} {\bibnamefont {Gu}, \bibfnamefont
  {S.-J.}}, \ and\ \bibinfo {author} {\bibfnamefont {H.-Q.}\ \bibnamefont
  {Lin}}} (\bibinfo {year} {2009}),\ \href@noop {} {\bibfield  {journal}
  {\bibinfo  {journal} {Europhys. Lett.}\ }\textbf {\bibinfo {volume} {87}},\
  \bibinfo {pages} {10003}}\BibitemShut {NoStop}%
\bibitem [{\citenamefont {Haag}(1992)}]{Haag0}%
  \BibitemOpen
  \bibfield  {author} {\bibinfo {author} {\bibnamefont {Haag}, \bibfnamefont
  {R.}}} (\bibinfo {year} {1992}),\ \href@noop {} {\emph {\bibinfo {title}
  {Local Quantum Physics}}}\ (\bibinfo  {publisher} {Springer},\ \bibinfo
  {address} {Heidelberg})\BibitemShut {NoStop}%
\bibitem [{\citenamefont {Hall}\ \emph {et~al.}(2012)\citenamefont {Hall},
  \citenamefont {Berry}, \citenamefont {Zwierz},\ and\ \citenamefont
  {Wiseman}}]{hall_universality_2012}%
  \BibitemOpen
  \bibfield  {author} {\bibinfo {author} {\bibnamefont {Hall}, \bibfnamefont
  {M.~J.~W.}}, \bibinfo {author} {\bibfnamefont {D.~W.}\ \bibnamefont {Berry}},
  \bibinfo {author} {\bibfnamefont {M.}~\bibnamefont {Zwierz}}, \ and\ \bibinfo
  {author} {\bibfnamefont {H.~M.}\ \bibnamefont {Wiseman}}} (\bibinfo {year}
  {2012}),\ \href@noop {} {\bibfield  {journal} {\bibinfo  {journal} {Physical
  Review A}\ }\textbf {\bibinfo {volume} {85}}~(\bibinfo {number}
  {4})}\BibitemShut {NoStop}%
\bibitem [{\citenamefont {Hall}\ and\ \citenamefont
  {Wiseman}(2012)}]{HallPRX2012}%
  \BibitemOpen
  \bibfield  {author} {\bibinfo {author} {\bibnamefont {Hall}, \bibfnamefont
  {M.~J.~W.}}, \ and\ \bibinfo {author} {\bibfnamefont {H.~M.}\ \bibnamefont
  {Wiseman}}} (\bibinfo {year} {2012}),\ \href {\doibase
  10.1103/PhysRevX.2.041006} {\bibfield  {journal} {\bibinfo  {journal} {Phys.
  Rev. X}\ }\textbf {\bibinfo {volume} {2}},\ \bibinfo {pages}
  {041006}}\BibitemShut {NoStop}%
\bibitem [{\citenamefont {Halvorson}\ and\ \citenamefont
  {Clifton}(2000)}]{Halvorson0}%
  \BibitemOpen
  \bibfield  {author} {\bibinfo {author} {\bibnamefont {Halvorson},
  \bibfnamefont {H.}}, \ and\ \bibinfo {author} {\bibfnamefont
  {R.}~\bibnamefont {Clifton}}} (\bibinfo {year} {2000}),\ \href@noop {}
  {\bibfield  {journal} {\bibinfo  {journal} {J. Math. Phys.}\ }\textbf
  {\bibinfo {volume} {41}},\ \bibinfo {pages} {1711}}\BibitemShut {NoStop}%
\bibitem [{\citenamefont {Haroche}(2013)}]{haroche_nobel_2013}%
  \BibitemOpen
  \bibfield  {author} {\bibinfo {author} {\bibnamefont {Haroche}, \bibfnamefont
  {S.}}} (\bibinfo {year} {2013}),\ \href {\doibase 10.1103/RevModPhys.85.1083}
  {\bibfield  {journal} {\bibinfo  {journal} {Rev. Mod. Phys.}\ }\textbf
  {\bibinfo {volume} {85}}~(\bibinfo {number} {3}),\ \bibinfo {pages}
  {1083}}\BibitemShut {NoStop}%
\bibitem [{\citenamefont {Haroche}\ and\ \citenamefont
  {Raimond}(2006)}]{Haroche0}%
  \BibitemOpen
  \bibfield  {author} {\bibinfo {author} {\bibnamefont {Haroche}, \bibfnamefont
  {S.}}, \ and\ \bibinfo {author} {\bibfnamefont {J.-M.}\ \bibnamefont
  {Raimond}}} (\bibinfo {year} {2006}),\ \href@noop {} {\emph {\bibinfo {title}
  {Exploring the Quantum: Atoms, Cavities and Photons}}}\ (\bibinfo
  {publisher} {Oxford University Press},\ \bibinfo {address}
  {Oxford})\BibitemShut {NoStop}%
\bibitem [{\citenamefont {Helstrom}(1969)}]{helstrom_quantum_1969}%
  \BibitemOpen
  \bibfield  {author} {\bibinfo {author} {\bibnamefont {Helstrom},
  \bibfnamefont {C.~W.}}} (\bibinfo {year} {1969}),\ \href@noop {} {\bibfield
  {journal} {\bibinfo  {journal} {J. Stat. Phys.}\ }\textbf {\bibinfo {volume}
  {1}}~(\bibinfo {number} {2}),\ \bibinfo {pages} {231}}\BibitemShut {NoStop}%
\bibitem [{\citenamefont {Helstrom}(1976)}]{Helstrom1976}%
  \BibitemOpen
  \bibfield  {author} {\bibinfo {author} {\bibnamefont {Helstrom},
  \bibfnamefont {C.~W.}}} (\bibinfo {year} {1976}),\ \href@noop {} {\emph
  {\bibinfo {title} {Quantum Detection and Estimation Theory}}}\ (\bibinfo
  {publisher} {Academic Press, New York})\BibitemShut {NoStop}%
\bibitem [{\citenamefont {Henderson}\ and\ \citenamefont
  {Vedral}(2001)}]{Henderson2001}%
  \BibitemOpen
  \bibfield  {author} {\bibinfo {author} {\bibnamefont {Henderson},
  \bibfnamefont {L.}}, \ and\ \bibinfo {author} {\bibfnamefont
  {V.}~\bibnamefont {Vedral}}} (\bibinfo {year} {2001}),\ \href
  {http://stacks.iop.org/0305-4470/34/i=35/a=315} {\bibfield  {journal}
  {\bibinfo  {journal} {J. Phys. A: Math. Gen.}\ }\textbf {\bibinfo {volume}
  {34}}~(\bibinfo {number} {35}),\ \bibinfo {pages} {6899}}\BibitemShut
  {NoStop}%
\bibitem [{\citenamefont {Higgins}\ \emph {et~al.}(2009)\citenamefont
  {Higgins}, \citenamefont {Berry}, \citenamefont {Bartlett}, \citenamefont
  {Mitchell}, \citenamefont {Wiseman},\ and\ \citenamefont
  {Pryde}}]{HigginsNJP2009}%
  \BibitemOpen
  \bibfield  {author} {\bibinfo {author} {\bibnamefont {Higgins}, \bibfnamefont
  {B.~L.}}, \bibinfo {author} {\bibfnamefont {D.~W.}\ \bibnamefont {Berry}},
  \bibinfo {author} {\bibfnamefont {S.~D.}\ \bibnamefont {Bartlett}}, \bibinfo
  {author} {\bibfnamefont {M.~W.}\ \bibnamefont {Mitchell}}, \bibinfo {author}
  {\bibfnamefont {H.~M.}\ \bibnamefont {Wiseman}}, \ and\ \bibinfo {author}
  {\bibfnamefont {G.~J.}\ \bibnamefont {Pryde}}} (\bibinfo {year} {2009}),\
  \href {http://stacks.iop.org/1367-2630/11/i=7/a=073023} {\bibfield  {journal}
  {\bibinfo  {journal} {New Journal of Physics}\ }\textbf {\bibinfo {volume}
  {11}}~(\bibinfo {number} {7}),\ \bibinfo {pages} {073023}}\BibitemShut
  {NoStop}%
\bibitem [{\citenamefont {Higgins}\ \emph {et~al.}(2007)\citenamefont
  {Higgins}, \citenamefont {Berry}, \citenamefont {Bartlett}, \citenamefont
  {Wiseman},\ and\ \citenamefont {Pryde}}]{Higgins07}%
  \BibitemOpen
  \bibfield  {author} {\bibinfo {author} {\bibnamefont {Higgins}, \bibfnamefont
  {B.~L.}}, \bibinfo {author} {\bibfnamefont {D.~W.}\ \bibnamefont {Berry}},
  \bibinfo {author} {\bibfnamefont {S.~D.}\ \bibnamefont {Bartlett}}, \bibinfo
  {author} {\bibfnamefont {H.~M.}\ \bibnamefont {Wiseman}}, \ and\ \bibinfo
  {author} {\bibfnamefont {G.~J.}\ \bibnamefont {Pryde}}} (\bibinfo {year}
  {2007}),\ \href@noop {} {\bibfield  {journal} {\bibinfo  {journal} {Nature}\
  }\textbf {\bibinfo {volume} {450}},\ \bibinfo {pages} {393}}\BibitemShut
  {NoStop}%
\bibitem [{\citenamefont {Hines}\ \emph {et~al.}(2003)\citenamefont {Hines},
  \citenamefont {McKenzie},\ and\ \citenamefont {Milburn}}]{Milburn0}%
  \BibitemOpen
  \bibfield  {author} {\bibinfo {author} {\bibnamefont {Hines}, \bibfnamefont
  {A.}}, \bibinfo {author} {\bibfnamefont {R.}~\bibnamefont {McKenzie}}, \ and\
  \bibinfo {author} {\bibfnamefont {G.}~\bibnamefont {Milburn}}} (\bibinfo
  {year} {2003}),\ \href@noop {} {\bibfield  {journal} {\bibinfo  {journal}
  {Phys. Rev. A}\ }\textbf {\bibinfo {volume} {67}},\ \bibinfo {pages}
  {013609}}\BibitemShut {NoStop}%
\bibitem [{\citenamefont {Hofheinz}\ \emph {et~al.}(2009)\citenamefont
  {Hofheinz}, \citenamefont {Wang}, \citenamefont {Ansmann}, \citenamefont
  {Bialczak}, \citenamefont {Lucero}, \citenamefont {Neeley}, \citenamefont
  {O'Connell}, \citenamefont {Sank}, \citenamefont {Wenner}, \citenamefont
  {Martinis},\ and\ \citenamefont {Cleland}}]{hofheinz_synthesizing_2009}%
  \BibitemOpen
  \bibfield  {author} {\bibinfo {author} {\bibnamefont {Hofheinz},
  \bibfnamefont {M.}}, \bibinfo {author} {\bibfnamefont {H.}~\bibnamefont
  {Wang}}, \bibinfo {author} {\bibfnamefont {M.}~\bibnamefont {Ansmann}},
  \bibinfo {author} {\bibfnamefont {R.~C.}\ \bibnamefont {Bialczak}}, \bibinfo
  {author} {\bibfnamefont {E.}~\bibnamefont {Lucero}}, \bibinfo {author}
  {\bibfnamefont {M.}~\bibnamefont {Neeley}}, \bibinfo {author} {\bibfnamefont
  {A.~D.}\ \bibnamefont {O'Connell}}, \bibinfo {author} {\bibfnamefont
  {D.}~\bibnamefont {Sank}}, \bibinfo {author} {\bibfnamefont {J.}~\bibnamefont
  {Wenner}}, \bibinfo {author} {\bibfnamefont {J.~M.}\ \bibnamefont
  {Martinis}}, \ and\ \bibinfo {author} {\bibfnamefont {A.~N.}\ \bibnamefont
  {Cleland}}} (\bibinfo {year} {2009}),\ \href {\doibase 10.1038/nature08005}
  {\bibfield  {journal} {\bibinfo  {journal} {Nature}\ }\textbf {\bibinfo
  {volume} {459}}~(\bibinfo {number} {7246}),\ \bibinfo {pages}
  {546}}\BibitemShut {NoStop}%
\bibitem [{\citenamefont {Holevo}(1982)}]{Holevo1982}%
  \BibitemOpen
  \bibfield  {author} {\bibinfo {author} {\bibnamefont {Holevo}, \bibfnamefont
  {A.~S.}}} (\bibinfo {year} {1982}),\ \href@noop {} {\emph {\bibinfo {title}
  {Probabilistic and Statistical Aspect of Quantum Theory}}}\ (\bibinfo
  {publisher} {North-Holland, Amsterdam})\BibitemShut {NoStop}%
\bibitem [{\citenamefont {Holland}\ \emph {et~al.}(2002)\citenamefont
  {Holland}, \citenamefont {Kok},\ and\ \citenamefont {Dowling}}]{Dowling2}%
  \BibitemOpen
  \bibfield  {author} {\bibinfo {author} {\bibnamefont {Holland}, \bibfnamefont
  {H.}}, \bibinfo {author} {\bibfnamefont {P.}~\bibnamefont {Kok}}, \ and\
  \bibinfo {author} {\bibfnamefont {J.}~\bibnamefont {Dowling}}} (\bibinfo
  {year} {2002}),\ \href@noop {} {\bibfield  {journal} {\bibinfo  {journal} {J.
  Mod. Opt.}\ }\textbf {\bibinfo {volume} {49}},\ \bibinfo {pages}
  {2325}}\BibitemShut {NoStop}%
\bibitem [{\citenamefont {Holland}\ and\ \citenamefont
  {Burnett}(1993)}]{HollandPRL1993}%
  \BibitemOpen
  \bibfield  {author} {\bibinfo {author} {\bibnamefont {Holland}, \bibfnamefont
  {M.~J.}}, \ and\ \bibinfo {author} {\bibfnamefont {K.}~\bibnamefont
  {Burnett}}} (\bibinfo {year} {1993}),\ \href {\doibase
  10.1103/PhysRevLett.71.1355} {\bibfield  {journal} {\bibinfo  {journal}
  {Phys. Rev. Lett.}\ }\textbf {\bibinfo {volume} {71}},\ \bibinfo {pages}
  {1355}}\BibitemShut {NoStop}%
\bibitem [{\citenamefont {Horodecki}\ \emph {et~al.}(2009)\citenamefont
  {Horodecki}, \citenamefont {Horodecki}, \citenamefont {Horodecki},\ and\
  \citenamefont {Horodecki}}]{Horodecki09}%
  \BibitemOpen
  \bibfield  {author} {\bibinfo {author} {\bibnamefont {Horodecki},
  \bibfnamefont {R.}}, \bibinfo {author} {\bibfnamefont {P.}~\bibnamefont
  {Horodecki}}, \bibinfo {author} {\bibfnamefont {M.}~\bibnamefont
  {Horodecki}}, \ and\ \bibinfo {author} {\bibfnamefont {K.}~\bibnamefont
  {Horodecki}}} (\bibinfo {year} {2009}),\ \href {\doibase
  10.1103/RevModPhys.81.865} {\bibfield  {journal} {\bibinfo  {journal} {Rev.
  Mod. Phys.}\ }\textbf {\bibinfo {volume} {81}}~(\bibinfo {number} {2}),\
  \bibinfo {pages} {865}}\BibitemShut {NoStop}%
\bibitem [{\citenamefont {Hosten}\ \emph {et~al.}(2016)\citenamefont {Hosten},
  \citenamefont {Krishnakumar}, \citenamefont {Engelsen},\ and\ \citenamefont
  {Kasevich}}]{HostenS2016}%
  \BibitemOpen
  \bibfield  {author} {\bibinfo {author} {\bibnamefont {Hosten}, \bibfnamefont
  {O.}}, \bibinfo {author} {\bibfnamefont {R.}~\bibnamefont {Krishnakumar}},
  \bibinfo {author} {\bibfnamefont {N.~J.}\ \bibnamefont {Engelsen}}, \ and\
  \bibinfo {author} {\bibfnamefont {M.~A.}\ \bibnamefont {Kasevich}}} (\bibinfo
  {year} {2016}),\ \href
  {http://science.sciencemag.org/content/352/6293/1552.abstract} {\bibfield
  {journal} {\bibinfo  {journal} {Science}\ }\textbf {\bibinfo {volume}
  {352}}~(\bibinfo {number} {6293}),\ \bibinfo {pages} {1552}}\BibitemShut
  {NoStop}%
\bibitem [{\citenamefont {Hu}\ \emph {et~al.}(2014)\citenamefont {Hu},
  \citenamefont {Xianlong},\ and\ \citenamefont {Liu}}]{Hu2014}%
  \BibitemOpen
  \bibfield  {author} {\bibinfo {author} {\bibnamefont {Hu}, \bibfnamefont
  {H.}}, \bibinfo {author} {\bibfnamefont {G.}~\bibnamefont {Xianlong}}, \ and\
  \bibinfo {author} {\bibfnamefont {X.-J.}\ \bibnamefont {Liu}}} (\bibinfo
  {year} {2014}),\ \href {\doibase 10.1103/PhysRevA.90.013622} {\bibfield
  {journal} {\bibinfo  {journal} {Phys. Rev. A}\ }\textbf {\bibinfo {volume}
  {90}},\ \bibinfo {pages} {013622}}\BibitemShut {NoStop}%
\bibitem [{\citenamefont {Huang}\ \emph {et~al.}(2015)\citenamefont {Huang},
  \citenamefont {Le~Jeannic}, \citenamefont {Ruaudel}, \citenamefont {Verma},
  \citenamefont {Shaw}, \citenamefont {Marsili}, \citenamefont {Nam},
  \citenamefont {Wu}, \citenamefont {Zeng}, \citenamefont {Jeong},
  \citenamefont {Filip}, \citenamefont {Morin},\ and\ \citenamefont
  {Laurat}}]{huang_optical_2015}%
  \BibitemOpen
  \bibfield  {author} {\bibinfo {author} {\bibnamefont {Huang}, \bibfnamefont
  {K.}}, \bibinfo {author} {\bibfnamefont {H.}~\bibnamefont {Le~Jeannic}},
  \bibinfo {author} {\bibfnamefont {J.}~\bibnamefont {Ruaudel}}, \bibinfo
  {author} {\bibfnamefont {V.}~\bibnamefont {Verma}}, \bibinfo {author}
  {\bibfnamefont {M.}~\bibnamefont {Shaw}}, \bibinfo {author} {\bibfnamefont
  {F.}~\bibnamefont {Marsili}}, \bibinfo {author} {\bibfnamefont
  {S.}~\bibnamefont {Nam}}, \bibinfo {author} {\bibfnamefont {E.}~\bibnamefont
  {Wu}}, \bibinfo {author} {\bibfnamefont {H.}~\bibnamefont {Zeng}}, \bibinfo
  {author} {\bibfnamefont {Y.-C.}\ \bibnamefont {Jeong}}, \bibinfo {author}
  {\bibfnamefont {R.}~\bibnamefont {Filip}}, \bibinfo {author} {\bibfnamefont
  {O.}~\bibnamefont {Morin}}, \ and\ \bibinfo {author} {\bibfnamefont
  {J.}~\bibnamefont {Laurat}}} (\bibinfo {year} {2015}),\ \href {\doibase
  10.1103/PhysRevLett.115.023602} {\bibfield  {journal} {\bibinfo  {journal}
  {Phys. Rev. Lett.}\ }\textbf {\bibinfo {volume} {115}}~(\bibinfo {number}
  {2}),\ \bibinfo {pages} {023602}}\BibitemShut {NoStop}%
\bibitem [{\citenamefont {Huang}\ \emph {et~al.}(2016)\citenamefont {Huang},
  \citenamefont {Macchiavello},\ and\ \citenamefont {Maccone}}]{Maccone2016}%
  \BibitemOpen
  \bibfield  {author} {\bibinfo {author} {\bibnamefont {Huang}, \bibfnamefont
  {Z.}}, \bibinfo {author} {\bibfnamefont {C.}~\bibnamefont {Macchiavello}}, \
  and\ \bibinfo {author} {\bibfnamefont {L.}~\bibnamefont {Maccone}}} (\bibinfo
  {year} {2016}),\ \href {\doibase 10.1103/PhysRevA.94.012101} {\bibfield
  {journal} {\bibinfo  {journal} {Phys. Rev. A}\ }\textbf {\bibinfo {volume}
  {94}},\ \bibinfo {pages} {012101}}\BibitemShut {NoStop}%
\bibitem [{\citenamefont {Huelga}\ \emph {et~al.}(1997)\citenamefont {Huelga},
  \citenamefont {Macchiavello}, \citenamefont {Pellizzari}, \citenamefont
  {Ekert}, \citenamefont {Plenio},\ and\ \citenamefont
  {Cirac}}]{huelga_improvement_1997}%
  \BibitemOpen
  \bibfield  {author} {\bibinfo {author} {\bibnamefont {Huelga}, \bibfnamefont
  {S.~F.}}, \bibinfo {author} {\bibfnamefont {C.}~\bibnamefont {Macchiavello}},
  \bibinfo {author} {\bibfnamefont {T.}~\bibnamefont {Pellizzari}}, \bibinfo
  {author} {\bibfnamefont {A.~K.}\ \bibnamefont {Ekert}}, \bibinfo {author}
  {\bibfnamefont {M.~B.}\ \bibnamefont {Plenio}}, \ and\ \bibinfo {author}
  {\bibfnamefont {J.~I.}\ \bibnamefont {Cirac}}} (\bibinfo {year} {1997}),\
  \href@noop {} {\bibfield  {journal} {\bibinfo  {journal} {Phys. Rev. Lett.}\
  }\textbf {\bibinfo {volume} {79}}~(\bibinfo {number} {20}),\ \bibinfo {pages}
  {3865{\textendash}3868}}\BibitemShut {NoStop}%
\bibitem [{\citenamefont {Hyllus}\ \emph {et~al.}(2012)\citenamefont {Hyllus},
  \citenamefont {Laskowski}, \citenamefont {Krischek}, \citenamefont
  {Schwemmer}, \citenamefont {Wieczorek}, \citenamefont {Weinfurter},
  \citenamefont {Pezz\`e},\ and\ \citenamefont {Smerzi}}]{Hyllus12}%
  \BibitemOpen
  \bibfield  {author} {\bibinfo {author} {\bibnamefont {Hyllus}, \bibfnamefont
  {P.}}, \bibinfo {author} {\bibfnamefont {W.}~\bibnamefont {Laskowski}},
  \bibinfo {author} {\bibfnamefont {R.}~\bibnamefont {Krischek}}, \bibinfo
  {author} {\bibfnamefont {C.}~\bibnamefont {Schwemmer}}, \bibinfo {author}
  {\bibfnamefont {W.}~\bibnamefont {Wieczorek}}, \bibinfo {author}
  {\bibfnamefont {H.}~\bibnamefont {Weinfurter}}, \bibinfo {author}
  {\bibfnamefont {L.}~\bibnamefont {Pezz\`e}}, \ and\ \bibinfo {author}
  {\bibfnamefont {A.}~\bibnamefont {Smerzi}}} (\bibinfo {year} {2012}),\ \href
  {\doibase 10.1103/PhysRevA.85.022321} {\bibfield  {journal} {\bibinfo
  {journal} {Phys. Rev. A}\ }\textbf {\bibinfo {volume} {85}},\ \bibinfo
  {pages} {022321}}\BibitemShut {NoStop}%
\bibitem [{\citenamefont {Inguscio}\ and\ \citenamefont
  {Fallani}(2013)}]{InguscioFallani}%
  \BibitemOpen
  \bibfield  {author} {\bibinfo {author} {\bibnamefont {Inguscio},
  \bibfnamefont {M.}}, \ and\ \bibinfo {author} {\bibfnamefont
  {L.}~\bibnamefont {Fallani}}} (\bibinfo {year} {2013}),\ \href@noop {} {\emph
  {\bibinfo {title} {{Atomic Physics: Precision Measurements and Ultracold
  Matter}}}}\ (\bibinfo  {publisher} {Oxford University Press})\BibitemShut
  {NoStop}%
\bibitem [{\citenamefont {Inguscio}\ \emph {et~al.}(2006)\citenamefont
  {Inguscio}, \citenamefont {Ketterle},\ and\ \citenamefont
  {Salomon}}]{Inguscio0}%
  \BibitemOpen
  \bibinfo {editor} {\bibnamefont {Inguscio}, \bibfnamefont {M.}}, \bibinfo
  {editor} {\bibfnamefont {W.}~\bibnamefont {Ketterle}}, \ and\ \bibinfo
  {editor} {\bibfnamefont {C.}~\bibnamefont {Salomon}},\ Eds. (\bibinfo {year}
  {2006}),\ \href@noop {} {\emph {\bibinfo {title} {Ultra-cold Fermi Gases}}}\
  (\bibinfo  {publisher} {IOS Press},\ \bibinfo {address}
  {Amsterdam})\BibitemShut {NoStop}%
\bibitem [{\citenamefont {Ino}\ and\ \citenamefont {Kohmoto}(2006)}]{Ino2006}%
  \BibitemOpen
  \bibfield  {author} {\bibinfo {author} {\bibnamefont {Ino}, \bibfnamefont
  {K.}}, \ and\ \bibinfo {author} {\bibfnamefont {M.}~\bibnamefont {Kohmoto}}}
  (\bibinfo {year} {2006}),\ \href {\doibase 10.1103/PhysRevB.73.205111}
  {\bibfield  {journal} {\bibinfo  {journal} {Phys. Rev. B}\ }\textbf {\bibinfo
  {volume} {73}},\ \bibinfo {pages} {205111}}\BibitemShut {NoStop}%
\bibitem [{\citenamefont {Invernizzi}\ \emph {et~al.}(2008)\citenamefont
  {Invernizzi}, \citenamefont {Korbman}, \citenamefont {Campos~Venuti},\ and\
  \citenamefont {Paris}}]{Invernizzi2008}%
  \BibitemOpen
  \bibfield  {author} {\bibinfo {author} {\bibnamefont {Invernizzi},
  \bibfnamefont {C.}}, \bibinfo {author} {\bibfnamefont {M.}~\bibnamefont
  {Korbman}}, \bibinfo {author} {\bibfnamefont {L.}~\bibnamefont
  {Campos~Venuti}}, \ and\ \bibinfo {author} {\bibfnamefont {M.~G.~A.}\
  \bibnamefont {Paris}}} (\bibinfo {year} {2008}),\ \href {\doibase
  10.1103/PhysRevA.78.042106} {\bibfield  {journal} {\bibinfo  {journal} {Phys.
  Rev. A}\ }\textbf {\bibinfo {volume} {78}},\ \bibinfo {pages}
  {042106}}\BibitemShut {NoStop}%
\bibitem [{\citenamefont {Invernizzi}\ \emph {et~al.}(2011)\citenamefont
  {Invernizzi}, \citenamefont {Paris},\ and\ \citenamefont
  {Pirandola}}]{Invernizzi11}%
  \BibitemOpen
  \bibfield  {author} {\bibinfo {author} {\bibnamefont {Invernizzi},
  \bibfnamefont {C.}}, \bibinfo {author} {\bibfnamefont {M.~G.~A.}\
  \bibnamefont {Paris}}, \ and\ \bibinfo {author} {\bibfnamefont
  {S.}~\bibnamefont {Pirandola}}} (\bibinfo {year} {2011}),\ \href@noop {}
  {\bibfield  {journal} {\bibinfo  {journal} {Phys. Rev. A}\ }\textbf {\bibinfo
  {volume} {84}},\ \bibinfo {pages} {022334}}\BibitemShut {NoStop}%
\bibitem [{\citenamefont {Jahnke}\ \emph {et~al.}(2011)\citenamefont {Jahnke},
  \citenamefont {Lan\'ery},\ and\ \citenamefont {Mahler}}]{PhysRevE.83.011109}%
  \BibitemOpen
  \bibfield  {author} {\bibinfo {author} {\bibnamefont {Jahnke}, \bibfnamefont
  {T.}}, \bibinfo {author} {\bibfnamefont {S.}~\bibnamefont {Lan\'ery}}, \ and\
  \bibinfo {author} {\bibfnamefont {G.}~\bibnamefont {Mahler}}} (\bibinfo
  {year} {2011}),\ \href {\doibase 10.1103/PhysRevE.83.011109} {\bibfield
  {journal} {\bibinfo  {journal} {Phys. Rev. E}\ }\textbf {\bibinfo {volume}
  {83}},\ \bibinfo {pages} {011109}}\BibitemShut {NoStop}%
\bibitem [{\citenamefont {Janke}\ \emph {et~al.}(2003)\citenamefont {Janke},
  \citenamefont {Johnston},\ and\ \citenamefont {Kenna}}]{Janke2003}%
  \BibitemOpen
  \bibfield  {author} {\bibinfo {author} {\bibnamefont {Janke}, \bibfnamefont
  {W.}}, \bibinfo {author} {\bibfnamefont {D.~A.}\ \bibnamefont {Johnston}}, \
  and\ \bibinfo {author} {\bibfnamefont {R.}~\bibnamefont {Kenna}}} (\bibinfo
  {year} {2003}),\ \href {\doibase 10.1103/PhysRevE.67.046106} {\bibfield
  {journal} {\bibinfo  {journal} {Phys. Rev. E}\ }\textbf {\bibinfo {volume}
  {67}},\ \bibinfo {pages} {046106}}\BibitemShut {NoStop}%
\bibitem [{\citenamefont {Janke}\ \emph {et~al.}(2002)\citenamefont {Janke},
  \citenamefont {Johnston},\ and\ \citenamefont {Malmini}}]{Janke2002}%
  \BibitemOpen
  \bibfield  {author} {\bibinfo {author} {\bibnamefont {Janke}, \bibfnamefont
  {W.}}, \bibinfo {author} {\bibfnamefont {D.~A.}\ \bibnamefont {Johnston}}, \
  and\ \bibinfo {author} {\bibfnamefont {R.~P. K.~C.}\ \bibnamefont {Malmini}}}
  (\bibinfo {year} {2002}),\ \href {\doibase 10.1103/PhysRevE.66.056119}
  {\bibfield  {journal} {\bibinfo  {journal} {Phys. Rev. E}\ }\textbf {\bibinfo
  {volume} {66}},\ \bibinfo {pages} {056119}}\BibitemShut {NoStop}%
\bibitem [{\citenamefont {Janyszek}(1986{\natexlab{a}})}]{Janyszek1986-2}%
  \BibitemOpen
  \bibfield  {author} {\bibinfo {author} {\bibnamefont {Janyszek},
  \bibfnamefont {H.}}} (\bibinfo {year} {1986}{\natexlab{a}}),\ \href@noop {}
  {\bibfield  {journal} {\bibinfo  {journal} {Rep. Math. Phys.}\ }\textbf
  {\bibinfo {volume} {24}},\ \bibinfo {pages} {11}}\BibitemShut {NoStop}%
\bibitem [{\citenamefont {Janyszek}(1986{\natexlab{b}})}]{Janyszek1986}%
  \BibitemOpen
  \bibfield  {author} {\bibinfo {author} {\bibnamefont {Janyszek},
  \bibfnamefont {H.}}} (\bibinfo {year} {1986}{\natexlab{b}}),\ \href@noop {}
  {\bibfield  {journal} {\bibinfo  {journal} {Rep. Math. Phys.}\ }\textbf
  {\bibinfo {volume} {24}},\ \bibinfo {pages} {1}}\BibitemShut {NoStop}%
\bibitem [{\citenamefont {Janyszek}(1990)}]{Janyszek1990}%
  \BibitemOpen
  \bibfield  {author} {\bibinfo {author} {\bibnamefont {Janyszek},
  \bibfnamefont {H.}}} (\bibinfo {year} {1990}),\ \href@noop {} {\bibfield
  {journal} {\bibinfo  {journal} {J. Phys. A}\ }\textbf {\bibinfo {volume}
  {23}},\ \bibinfo {pages} {477}}\BibitemShut {NoStop}%
\bibitem [{\citenamefont {Janyszek}\ and\ \citenamefont
  {Mruga\l{}a}(1989)}]{Janyszek1989}%
  \BibitemOpen
  \bibfield  {author} {\bibinfo {author} {\bibnamefont {Janyszek},
  \bibfnamefont {H.}}, \ and\ \bibinfo {author} {\bibfnamefont
  {R.}~\bibnamefont {Mruga\l{}a}}} (\bibinfo {year} {1989}),\ \href {\doibase
  10.1103/PhysRevA.39.6515} {\bibfield  {journal} {\bibinfo  {journal} {Phys.
  Rev. A}\ }\textbf {\bibinfo {volume} {39}},\ \bibinfo {pages}
  {6515}}\BibitemShut {NoStop}%
\bibitem [{\citenamefont {Janyszek}\ and\ \citenamefont
  {Mruga\l{}a}(1990)}]{Janyszek1990-2}%
  \BibitemOpen
  \bibfield  {author} {\bibinfo {author} {\bibnamefont {Janyszek},
  \bibfnamefont {H.}}, \ and\ \bibinfo {author} {\bibfnamefont
  {R.}~\bibnamefont {Mruga\l{}a}}} (\bibinfo {year} {1990}),\ \href@noop {}
  {\bibfield  {journal} {\bibinfo  {journal} {J. Phys. A}\ }\textbf {\bibinfo
  {volume} {23}},\ \bibinfo {pages} {467}}\BibitemShut {NoStop}%
\bibitem [{\citenamefont {Jaynes}(1957{\natexlab{a}})}]{Jaynes1957-1}%
  \BibitemOpen
  \bibfield  {author} {\bibinfo {author} {\bibnamefont {Jaynes}, \bibfnamefont
  {E.~T.}}} (\bibinfo {year} {1957}{\natexlab{a}}),\ \href {\doibase
  10.1103/PhysRev.106.620} {\bibfield  {journal} {\bibinfo  {journal} {Phys.
  Rev.}\ }\textbf {\bibinfo {volume} {106}},\ \bibinfo {pages}
  {620}}\BibitemShut {NoStop}%
\bibitem [{\citenamefont {Jaynes}(1957{\natexlab{b}})}]{Jaynes1957-2}%
  \BibitemOpen
  \bibfield  {author} {\bibinfo {author} {\bibnamefont {Jaynes}, \bibfnamefont
  {E.~T.}}} (\bibinfo {year} {1957}{\natexlab{b}}),\ \href {\doibase
  10.1103/PhysRev.108.171} {\bibfield  {journal} {\bibinfo  {journal} {Phys.
  Rev.}\ }\textbf {\bibinfo {volume} {108}},\ \bibinfo {pages}
  {171}}\BibitemShut {NoStop}%
\bibitem [{\citenamefont {Jevtic}\ \emph {et~al.}(2015)\citenamefont {Jevtic},
  \citenamefont {Newman}, \citenamefont {Rudolph},\ and\ \citenamefont
  {Stace}}]{PhysRevA.91.012331}%
  \BibitemOpen
  \bibfield  {author} {\bibinfo {author} {\bibnamefont {Jevtic}, \bibfnamefont
  {S.}}, \bibinfo {author} {\bibfnamefont {D.}~\bibnamefont {Newman}}, \bibinfo
  {author} {\bibfnamefont {T.}~\bibnamefont {Rudolph}}, \ and\ \bibinfo
  {author} {\bibfnamefont {T.~M.}\ \bibnamefont {Stace}}} (\bibinfo {year}
  {2015}),\ \href {\doibase 10.1103/PhysRevA.91.012331} {\bibfield  {journal}
  {\bibinfo  {journal} {Phys. Rev. A}\ }\textbf {\bibinfo {volume} {91}},\
  \bibinfo {pages} {012331}}\BibitemShut {NoStop}%
\bibitem [{\citenamefont {Jiang}(2014)}]{Jiang2014}%
  \BibitemOpen
  \bibfield  {author} {\bibinfo {author} {\bibnamefont {Jiang}, \bibfnamefont
  {Z.}}} (\bibinfo {year} {2014}),\ \href {\doibase 10.1103/PhysRevA.89.032128}
  {\bibfield  {journal} {\bibinfo  {journal} {Phys. Rev. A}\ }\textbf {\bibinfo
  {volume} {89}},\ \bibinfo {pages} {032128}}\BibitemShut {NoStop}%
\bibitem [{\citenamefont {Jones}\ \emph {et~al.}(2009)\citenamefont {Jones},
  \citenamefont {Karlen}, \citenamefont {Fitzsimons}, \citenamefont {Ardavan},
  \citenamefont {Benjamin}, \citenamefont {Briggs},\ and\ \citenamefont
  {Morton}}]{Jones2009}%
  \BibitemOpen
  \bibfield  {author} {\bibinfo {author} {\bibnamefont {Jones}, \bibfnamefont
  {J.~A.}}, \bibinfo {author} {\bibfnamefont {S.~D.}\ \bibnamefont {Karlen}},
  \bibinfo {author} {\bibfnamefont {J.}~\bibnamefont {Fitzsimons}}, \bibinfo
  {author} {\bibfnamefont {A.}~\bibnamefont {Ardavan}}, \bibinfo {author}
  {\bibfnamefont {S.~C.}\ \bibnamefont {Benjamin}}, \bibinfo {author}
  {\bibfnamefont {G.~A.~D.}\ \bibnamefont {Briggs}}, \ and\ \bibinfo {author}
  {\bibfnamefont {J.~J.~L.}\ \bibnamefont {Morton}}} (\bibinfo {year} {2009}),\
  \href {\doibase 10.1126/science.1170730} {\bibfield  {journal} {\bibinfo
  {journal} {Science}\ }\textbf {\bibinfo {volume} {324}}~(\bibinfo {number}
  {5931}),\ \bibinfo {pages} {1166}},\ \Eprint
  {http://arxiv.org/abs/http://science.sciencemag.org/content/324/5931/1166.full.pdf}
  {http://science.sciencemag.org/content/324/5931/1166.full.pdf} \BibitemShut
  {NoStop}%
\bibitem [{\citenamefont {Juffmann}\ \emph
  {et~al.}(2016{\natexlab{a}})\citenamefont {Juffmann}, \citenamefont
  {Klopfer}, \citenamefont {Frankort}, \citenamefont {Haslinger},\ and\
  \citenamefont {Kasevich}}]{juffmann_multi-pass_2016}%
  \BibitemOpen
  \bibfield  {author} {\bibinfo {author} {\bibnamefont {Juffmann},
  \bibfnamefont {T.}}, \bibinfo {author} {\bibfnamefont {B.~B.}\ \bibnamefont
  {Klopfer}}, \bibinfo {author} {\bibfnamefont {T.~L.~I.}\ \bibnamefont
  {Frankort}}, \bibinfo {author} {\bibfnamefont {P.}~\bibnamefont {Haslinger}},
  \ and\ \bibinfo {author} {\bibfnamefont {M.~A.}\ \bibnamefont {Kasevich}}}
  (\bibinfo {year} {2016}{\natexlab{a}}),\ \href {\doibase 10.1038/ncomms12858}
  {\bibfield  {journal} {\bibinfo  {journal} {Nature Communications}\ }\textbf
  {\bibinfo {volume} {7}},\ \bibinfo {pages} {12858}}\BibitemShut {NoStop}%
\bibitem [{\citenamefont {Juffmann}\ \emph
  {et~al.}(2016{\natexlab{b}})\citenamefont {Juffmann}, \citenamefont
  {Koppell}, \citenamefont {Klopfer}, \citenamefont {Ophus}, \citenamefont
  {Glaeser},\ and\ \citenamefont {Kasevich}}]{juffmann_multi-pass_2016e}%
  \BibitemOpen
  \bibfield  {author} {\bibinfo {author} {\bibnamefont {Juffmann},
  \bibfnamefont {T.}}, \bibinfo {author} {\bibfnamefont {S.~A.}\ \bibnamefont
  {Koppell}}, \bibinfo {author} {\bibfnamefont {B.~B.}\ \bibnamefont
  {Klopfer}}, \bibinfo {author} {\bibfnamefont {C.}~\bibnamefont {Ophus}},
  \bibinfo {author} {\bibfnamefont {R.}~\bibnamefont {Glaeser}}, \ and\
  \bibinfo {author} {\bibfnamefont {M.~A.}\ \bibnamefont {Kasevich}}} (\bibinfo
  {year} {2016}{\natexlab{b}}),\ \href {http://arxiv.org/abs/1612.04931}
  {\enquote {\bibinfo {title} {Multi-pass transmission electron microscopy},}\
  }\bibinfo {note} {ArXiv:1612.04931}\BibitemShut {NoStop}%
\bibitem [{\citenamefont {Kacprowicz}\ \emph {et~al.}(2010)\citenamefont
  {Kacprowicz}, \citenamefont {Demkowicz-Dobrza\'nski}, \citenamefont
  {Wasilewski}, \citenamefont {Banaszek},\ and\ \citenamefont
  {Walmsley}}]{Kacprowicz0}%
  \BibitemOpen
  \bibfield  {author} {\bibinfo {author} {\bibnamefont {Kacprowicz},
  \bibfnamefont {M.}}, \bibinfo {author} {\bibfnamefont {R.}~\bibnamefont
  {Demkowicz-Dobrza\'nski}}, \bibinfo {author} {\bibfnamefont {W.}~\bibnamefont
  {Wasilewski}}, \bibinfo {author} {\bibfnamefont {K.}~\bibnamefont
  {Banaszek}}, \ and\ \bibinfo {author} {\bibfnamefont {I.~A.}\ \bibnamefont
  {Walmsley}}} (\bibinfo {year} {2010}),\ \href@noop {} {\bibfield  {journal}
  {\bibinfo  {journal} {Nature Photonics}\ }\textbf {\bibinfo {volume} {4}},\
  \bibinfo {pages} {357}}\BibitemShut {NoStop}%
\bibitem [{\citenamefont {Kessler}\ \emph {et~al.}(2014)\citenamefont
  {Kessler}, \citenamefont {Lovchinsky}, \citenamefont {Sushkov},\ and\
  \citenamefont {Lukin}}]{kessler_quantum_2014}%
  \BibitemOpen
  \bibfield  {author} {\bibinfo {author} {\bibnamefont {Kessler}, \bibfnamefont
  {E.}}, \bibinfo {author} {\bibfnamefont {I.}~\bibnamefont {Lovchinsky}},
  \bibinfo {author} {\bibfnamefont {A.}~\bibnamefont {Sushkov}}, \ and\
  \bibinfo {author} {\bibfnamefont {M.}~\bibnamefont {Lukin}}} (\bibinfo {year}
  {2014}),\ \href {\doibase 10.1103/PhysRevLett.112.150802} {\bibfield
  {journal} {\bibinfo  {journal} {Phys. Rev. Lett.}\ }\textbf {\bibinfo
  {volume} {112}}~(\bibinfo {number} {15}),\ \bibinfo {pages}
  {150802}}\BibitemShut {NoStop}%
\bibitem [{\citenamefont {Ketterle}\ and\ \citenamefont {van
  Druten}(1982)}]{Ketterle1996}%
  \BibitemOpen
  \bibfield  {author} {\bibinfo {author} {\bibnamefont {Ketterle},
  \bibfnamefont {W.}}, \ and\ \bibinfo {author} {\bibfnamefont {N.~J.}\
  \bibnamefont {van Druten}}} (\bibinfo {year} {1982}),\ \href@noop {}
  {\bibfield  {journal} {\bibinfo  {journal} {Phys. Rev. A}\ }\textbf {\bibinfo
  {volume} {54}},\ \bibinfo {pages} {656}}\BibitemShut {NoStop}%
\bibitem [{\citenamefont {Keyl}\ \emph {et~al.}(2006)\citenamefont {Keyl},
  \citenamefont {Matsui}, \citenamefont {Schlingemann},\ and\ \citenamefont
  {Werner}}]{Werner6}%
  \BibitemOpen
  \bibfield  {author} {\bibinfo {author} {\bibnamefont {Keyl}, \bibfnamefont
  {M.}}, \bibinfo {author} {\bibfnamefont {T.}~\bibnamefont {Matsui}}, \bibinfo
  {author} {\bibfnamefont {D.}~\bibnamefont {Schlingemann}}, \ and\ \bibinfo
  {author} {\bibfnamefont {R.}~\bibnamefont {Werner}}} (\bibinfo {year}
  {2006}),\ \href@noop {} {\bibfield  {journal} {\bibinfo  {journal} {Rev.
  Math. Phys.}\ }\textbf {\bibinfo {volume} {18}},\ \bibinfo {pages}
  {935}}\BibitemShut {NoStop}%
\bibitem [{\citenamefont {Keyl}\ \emph {et~al.}(2003)\citenamefont {Keyl},
  \citenamefont {Schlingemann},\ and\ \citenamefont {Werner}}]{Werner4}%
  \BibitemOpen
  \bibfield  {author} {\bibinfo {author} {\bibnamefont {Keyl}, \bibfnamefont
  {M.}}, \bibinfo {author} {\bibfnamefont {D.}~\bibnamefont {Schlingemann}}, \
  and\ \bibinfo {author} {\bibfnamefont {R.}~\bibnamefont {Werner}}} (\bibinfo
  {year} {2003}),\ \href@noop {} {\bibfield  {journal} {\bibinfo  {journal}
  {Quant. Inf. Comput.}\ }\textbf {\bibinfo {volume} {3}},\ \bibinfo {pages}
  {281}}\BibitemShut {NoStop}%
\bibitem [{\citenamefont {Khorana}\ and\ \citenamefont
  {Douglass}(1965)}]{Khorana1965}%
  \BibitemOpen
  \bibfield  {author} {\bibinfo {author} {\bibnamefont {Khorana}, \bibfnamefont
  {B.~M.}}, \ and\ \bibinfo {author} {\bibfnamefont {D.~H.}\ \bibnamefont
  {Douglass}}} (\bibinfo {year} {1965}),\ \href {\doibase
  10.1103/PhysRev.138.A35} {\bibfield  {journal} {\bibinfo  {journal} {Phys.
  Rev.}\ }\textbf {\bibinfo {volume} {138}},\ \bibinfo {pages}
  {A35}}\BibitemShut {NoStop}%
\bibitem [{\citenamefont {Killoran}\ \emph {et~al.}(2014)\citenamefont
  {Killoran}, \citenamefont {Cramer},\ and\ \citenamefont {Plenio}}]{Plenio0}%
  \BibitemOpen
  \bibfield  {author} {\bibinfo {author} {\bibnamefont {Killoran},
  \bibfnamefont {N.}}, \bibinfo {author} {\bibfnamefont {M.}~\bibnamefont
  {Cramer}}, \ and\ \bibinfo {author} {\bibfnamefont {M.}~\bibnamefont
  {Plenio}}} (\bibinfo {year} {2014}),\ \href@noop {} {\bibfield  {journal}
  {\bibinfo  {journal} {Phys. Rev. Lett.}\ }\textbf {\bibinfo {volume} {112}},\
  \bibinfo {pages} {150501}}\BibitemShut {NoStop}%
\bibitem [{\citenamefont {Killoran}\ \emph {et~al.}(2016)\citenamefont
  {Killoran}, \citenamefont {Steinhoff},\ and\ \citenamefont
  {Plenio}}]{killoran2016converting}%
  \BibitemOpen
  \bibfield  {author} {\bibinfo {author} {\bibnamefont {Killoran},
  \bibfnamefont {N.}}, \bibinfo {author} {\bibfnamefont {F.~E.~S.}\
  \bibnamefont {Steinhoff}}, \ and\ \bibinfo {author} {\bibfnamefont {M.~B.}\
  \bibnamefont {Plenio}}} (\bibinfo {year} {2016}),\ \href {\doibase
  10.1103/PhysRevLett.116.080402} {\bibfield  {journal} {\bibinfo  {journal}
  {Phys. Rev. Lett.}\ }\textbf {\bibinfo {volume} {116}},\ \bibinfo {pages}
  {080402}}\BibitemShut {NoStop}%
\bibitem [{\citenamefont {Kitagawa}\ and\ \citenamefont
  {Ueda}(1993)}]{Kitagawa0}%
  \BibitemOpen
  \bibfield  {author} {\bibinfo {author} {\bibnamefont {Kitagawa},
  \bibfnamefont {M.}}, \ and\ \bibinfo {author} {\bibfnamefont
  {M.}~\bibnamefont {Ueda}}} (\bibinfo {year} {1993}),\ \href@noop {}
  {\bibfield  {journal} {\bibinfo  {journal} {Phys. Rev. A}\ }\textbf {\bibinfo
  {volume} {47}},\ \bibinfo {pages} {5138}}\BibitemShut {NoStop}%
\bibitem [{\citenamefont {Klaers}(2014)}]{Klaers2014}%
  \BibitemOpen
  \bibfield  {author} {\bibinfo {author} {\bibnamefont {Klaers}, \bibfnamefont
  {J.}}} (\bibinfo {year} {2014}),\ \href@noop {} {\bibfield  {journal}
  {\bibinfo  {journal} {J. Phys. B}\ }\textbf {\bibinfo {volume} {47}},\
  \bibinfo {pages} {243001}}\BibitemShut {NoStop}%
\bibitem [{\citenamefont {Knill}\ and\ \citenamefont
  {Laflamme}(1998)}]{knill_power_1998}%
  \BibitemOpen
  \bibfield  {author} {\bibinfo {author} {\bibnamefont {Knill}, \bibfnamefont
  {E.}}, \ and\ \bibinfo {author} {\bibfnamefont {R.}~\bibnamefont {Laflamme}}}
  (\bibinfo {year} {1998}),\ \href {\doibase 10.1103/PhysRevLett.81.5672}
  {\bibfield  {journal} {\bibinfo  {journal} {Phys. Rev. Lett.}\ }\textbf
  {\bibinfo {volume} {81}}~(\bibinfo {number} {25}),\ \bibinfo {pages}
  {5672}}\BibitemShut {NoStop}%
\bibitem [{\citenamefont {Knott}\ \emph {et~al.}(2016)\citenamefont {Knott},
  \citenamefont {Proctor}, \citenamefont {Hayes}, \citenamefont {Ralph},
  \citenamefont {Kok},\ and\ \citenamefont {Dunningham}}]{knott2016}%
  \BibitemOpen
  \bibfield  {author} {\bibinfo {author} {\bibnamefont {Knott}, \bibfnamefont
  {P.~A.}}, \bibinfo {author} {\bibfnamefont {T.~J.}\ \bibnamefont {Proctor}},
  \bibinfo {author} {\bibfnamefont {A.~J.}\ \bibnamefont {Hayes}}, \bibinfo
  {author} {\bibfnamefont {J.~F.}\ \bibnamefont {Ralph}}, \bibinfo {author}
  {\bibfnamefont {P.}~\bibnamefont {Kok}}, \ and\ \bibinfo {author}
  {\bibfnamefont {J.~A.}\ \bibnamefont {Dunningham}}} (\bibinfo {year}
  {2016}),\ \href@noop {} {\bibfield  {journal} {\bibinfo  {journal} {Phys.
  Rev. A}\ }\textbf {\bibinfo {volume} {94}},\ \bibinfo {pages}
  {062312}}\BibitemShut {NoStop}%
\bibitem [{\citenamefont {K\"ohl}\ and\ \citenamefont
  {Esslinger}(2006)}]{Kohl0}%
  \BibitemOpen
  \bibfield  {author} {\bibinfo {author} {\bibnamefont {K\"ohl}, \bibfnamefont
  {M.}}, \ and\ \bibinfo {author} {\bibfnamefont {T.}~\bibnamefont
  {Esslinger}}} (\bibinfo {year} {2006}),\ \href@noop {} {\bibfield  {journal}
  {\bibinfo  {journal} {Europhys. News}\ }\textbf {\bibinfo {volume} {37}},\
  \bibinfo {pages} {18}}\BibitemShut {NoStop}%
\bibitem [{\citenamefont {Ko\l{}ody\'{n}ski}\ and\ \citenamefont
  {Demkowicz-Dobrza\'{n}ski}(2010)}]{Kolodynski10}%
  \BibitemOpen
  \bibfield  {author} {\bibinfo {author} {\bibnamefont {Ko\l{}ody\'{n}ski},
  \bibfnamefont {J.}}, \ and\ \bibinfo {author} {\bibfnamefont
  {R.}~\bibnamefont {Demkowicz-Dobrza\'{n}ski}}} (\bibinfo {year} {2010}),\
  \href {\doibase 10.1103/PhysRevA.82.053804} {\bibfield  {journal} {\bibinfo
  {journal} {Phys. Rev. A}\ }\textbf {\bibinfo {volume} {82}}~(\bibinfo
  {number} {5}),\ \bibinfo {pages} {053804}}\BibitemShut {NoStop}%
\bibitem [{\citenamefont {Kominis}\ \emph {et~al.}(2003)\citenamefont
  {Kominis}, \citenamefont {Kornack}, \citenamefont {Allred},\ and\
  \citenamefont {Romalis}}]{KominisN2003}%
  \BibitemOpen
  \bibfield  {author} {\bibinfo {author} {\bibnamefont {Kominis}, \bibfnamefont
  {I.}}, \bibinfo {author} {\bibfnamefont {T.}~\bibnamefont {Kornack}},
  \bibinfo {author} {\bibfnamefont {J.}~\bibnamefont {Allred}}, \ and\ \bibinfo
  {author} {\bibfnamefont {M.}~\bibnamefont {Romalis}}} (\bibinfo {year}
  {2003}),\ \href@noop {} {\bibfield  {journal} {\bibinfo  {journal} {Nature}\
  }\textbf {\bibinfo {volume} {422}}~(\bibinfo {number} {6932}),\ \bibinfo
  {pages} {596}}\BibitemShut {NoStop}%
\bibitem [{\citenamefont {Korbicz}\ \emph {et~al.}(2005)\citenamefont
  {Korbicz}, \citenamefont {Cirac},\ and\ \citenamefont
  {Lewenstein}}]{Korbicz0}%
  \BibitemOpen
  \bibfield  {author} {\bibinfo {author} {\bibnamefont {Korbicz}, \bibfnamefont
  {J.}}, \bibinfo {author} {\bibfnamefont {J.}~\bibnamefont {Cirac}}, \ and\
  \bibinfo {author} {\bibfnamefont {M.}~\bibnamefont {Lewenstein}}} (\bibinfo
  {year} {2005}),\ \href@noop {} {\bibfield  {journal} {\bibinfo  {journal}
  {Phys. Rev. Lett.}\ }\textbf {\bibinfo {volume} {95}},\ \bibinfo {pages}
  {120502}}\BibitemShut {NoStop}%
\bibitem [{\citenamefont {Kraus}\ \emph {et~al.}(2009)\citenamefont {Kraus},
  \citenamefont {Wolf}, \citenamefont {Cirac},\ and\ \citenamefont
  {Giedke}}]{Kraus0}%
  \BibitemOpen
  \bibfield  {author} {\bibinfo {author} {\bibnamefont {Kraus}, \bibfnamefont
  {C.}}, \bibinfo {author} {\bibfnamefont {M.}~\bibnamefont {Wolf}}, \bibinfo
  {author} {\bibfnamefont {J.}~\bibnamefont {Cirac}}, \ and\ \bibinfo {author}
  {\bibfnamefont {G.}~\bibnamefont {Giedke}}} (\bibinfo {year} {2009}),\
  \href@noop {} {\bibfield  {journal} {\bibinfo  {journal} {Phys. Rev. A}\
  }\textbf {\bibinfo {volume} {79}},\ \bibinfo {pages} {012306}}\BibitemShut
  {NoStop}%
\bibitem [{\citenamefont {Krueger}(1968)}]{Krueger1968}%
  \BibitemOpen
  \bibfield  {author} {\bibinfo {author} {\bibnamefont {Krueger}, \bibfnamefont
  {D.~A.}}} (\bibinfo {year} {1968}),\ \href@noop {} {\bibfield  {journal}
  {\bibinfo  {journal} {Phys. Rev.}\ }\textbf {\bibinfo {volume} {172}},\
  \bibinfo {pages} {211}}\BibitemShut {NoStop}%
\bibitem [{\citenamefont {Laghaout}\ \emph {et~al.}(2013)\citenamefont
  {Laghaout}, \citenamefont {Neergaard-Nielsen}, \citenamefont {Rigas},
  \citenamefont {Kragh}, \citenamefont {Tipsmark},\ and\ \citenamefont
  {Andersen}}]{laghaout_amplification_2013}%
  \BibitemOpen
  \bibfield  {author} {\bibinfo {author} {\bibnamefont {Laghaout},
  \bibfnamefont {A.}}, \bibinfo {author} {\bibfnamefont {J.~S.}\ \bibnamefont
  {Neergaard-Nielsen}}, \bibinfo {author} {\bibfnamefont {I.}~\bibnamefont
  {Rigas}}, \bibinfo {author} {\bibfnamefont {C.}~\bibnamefont {Kragh}},
  \bibinfo {author} {\bibfnamefont {A.}~\bibnamefont {Tipsmark}}, \ and\
  \bibinfo {author} {\bibfnamefont {U.~L.}\ \bibnamefont {Andersen}}} (\bibinfo
  {year} {2013}),\ \href {\doibase 10.1103/PhysRevA.87.043826} {\bibfield
  {journal} {\bibinfo  {journal} {Phys. Rev. A}\ }\textbf {\bibinfo {volume}
  {87}}~(\bibinfo {number} {4}),\ \bibinfo {pages} {043826}}\BibitemShut
  {NoStop}%
\bibitem [{\citenamefont {Langen}\ \emph {et~al.}(2015)\citenamefont {Langen},
  \citenamefont {Erne}, \citenamefont {Geiger}, \citenamefont {Rauer},
  \citenamefont {Schweigier}, \citenamefont {Kuhnert}, \citenamefont
  {Rohringer}, \citenamefont {Mazets}, \citenamefont {Gasenzer},\ and\
  \citenamefont {Schmiedmayer}}]{Langen2015}%
  \BibitemOpen
  \bibfield  {author} {\bibinfo {author} {\bibnamefont {Langen}, \bibfnamefont
  {T.}}, \bibinfo {author} {\bibfnamefont {S.}~\bibnamefont {Erne}}, \bibinfo
  {author} {\bibfnamefont {R.}~\bibnamefont {Geiger}}, \bibinfo {author}
  {\bibfnamefont {B.}~\bibnamefont {Rauer}}, \bibinfo {author} {\bibfnamefont
  {T.}~\bibnamefont {Schweigier}}, \bibinfo {author} {\bibfnamefont
  {W.}~\bibnamefont {Kuhnert}}, \bibinfo {author} {\bibfnamefont
  {W.}~\bibnamefont {Rohringer}}, \bibinfo {author} {\bibfnamefont {I.~E.}\
  \bibnamefont {Mazets}}, \bibinfo {author} {\bibfnamefont {T.}~\bibnamefont
  {Gasenzer}}, \ and\ \bibinfo {author} {\bibfnamefont {J.}~\bibnamefont
  {Schmiedmayer}}} (\bibinfo {year} {2015}),\ \href@noop {} {\bibfield
  {journal} {\bibinfo  {journal} {Science}\ }\textbf {\bibinfo {volume}
  {348}},\ \bibinfo {pages} {207}}\BibitemShut {NoStop}%
\bibitem [{\citenamefont {Lanyon}\ \emph {et~al.}(2008)\citenamefont {Lanyon},
  \citenamefont {Barbieri}, \citenamefont {Almeida},\ and\ \citenamefont
  {White}}]{lanyon_experimental_2008}%
  \BibitemOpen
  \bibfield  {author} {\bibinfo {author} {\bibnamefont {Lanyon}, \bibfnamefont
  {B.~P.}}, \bibinfo {author} {\bibfnamefont {M.}~\bibnamefont {Barbieri}},
  \bibinfo {author} {\bibfnamefont {M.~P.}\ \bibnamefont {Almeida}}, \ and\
  \bibinfo {author} {\bibfnamefont {A.~G.}\ \bibnamefont {White}}} (\bibinfo
  {year} {2008}),\ \href {\doibase 10.1103/PhysRevLett.101.200501} {\bibfield
  {journal} {\bibinfo  {journal} {Phys. Rev. Lett.}\ }\textbf {\bibinfo
  {volume} {101}}~(\bibinfo {number} {20}),\
  10.1103/PhysRevLett.101.200501}\BibitemShut {NoStop}%
\bibitem [{\citenamefont {Laurenza}\ \emph {et~al.}(2017)\citenamefont
  {Laurenza}, \citenamefont {Lupo}, \citenamefont {Spedalieri}, \citenamefont
  {Braunstein},\ and\ \citenamefont {Pirandola}}]{Laurenza17}%
  \BibitemOpen
  \bibfield  {author} {\bibinfo {author} {\bibnamefont {Laurenza},
  \bibfnamefont {R.}}, \bibinfo {author} {\bibfnamefont {C.}~\bibnamefont
  {Lupo}}, \bibinfo {author} {\bibfnamefont {G.}~\bibnamefont {Spedalieri}},
  \bibinfo {author} {\bibfnamefont {S.~L.}\ \bibnamefont {Braunstein}}, \ and\
  \bibinfo {author} {\bibfnamefont {S.}~\bibnamefont {Pirandola}}} (\bibinfo
  {year} {2017}),\ \href@noop {} {\bibinfo  {journal} {arXiv preprint
  arXiv:1712.06603}\ }\BibitemShut {NoStop}%
\bibitem [{\citenamefont {Leanhardt}\ \emph {et~al.}(2003)\citenamefont
  {Leanhardt}, \citenamefont {Pasquini}, \citenamefont {Saba}, \citenamefont
  {Schirotzek}, \citenamefont {Shin}, \citenamefont {Kielpinski}, \citenamefont
  {Pritchard},\ and\ \citenamefont {Ketterle}}]{Leanhardt2003}%
  \BibitemOpen
\bibfield  {journal} {  }\bibfield  {author} {\bibinfo {author} {\bibnamefont
  {Leanhardt}, \bibfnamefont {A.~E.}}, \bibinfo {author} {\bibfnamefont
  {T.}~\bibnamefont {Pasquini}}, \bibinfo {author} {\bibfnamefont
  {M.}~\bibnamefont {Saba}}, \bibinfo {author} {\bibfnamefont {A.}~\bibnamefont
  {Schirotzek}}, \bibinfo {author} {\bibfnamefont {Y.}~\bibnamefont {Shin}},
  \bibinfo {author} {\bibfnamefont {D.}~\bibnamefont {Kielpinski}}, \bibinfo
  {author} {\bibfnamefont {D.~E.}\ \bibnamefont {Pritchard}}, \ and\ \bibinfo
  {author} {\bibfnamefont {W.}~\bibnamefont {Ketterle}}} (\bibinfo {year}
  {2003}),\ \href@noop {} {\bibfield  {journal} {\bibinfo  {journal} {Science}\
  }\textbf {\bibinfo {volume} {301}},\ \bibinfo {pages} {1513}}\BibitemShut
  {NoStop}%
\bibitem [{\citenamefont {Lee}\ \emph {et~al.}(2015)\citenamefont {Lee},
  \citenamefont {Cho},\ and\ \citenamefont {Choi}}]{lee_emergence_2015}%
  \BibitemOpen
  \bibfield  {author} {\bibinfo {author} {\bibnamefont {Lee}, \bibfnamefont
  {S.~K.}}, \bibinfo {author} {\bibfnamefont {J.}~\bibnamefont {Cho}}, \ and\
  \bibinfo {author} {\bibfnamefont {K.~S.}\ \bibnamefont {Choi}}} (\bibinfo
  {year} {2015}),\ \href@noop {} {\bibfield  {journal} {\bibinfo  {journal}
  {New J. Phys.}\ }\textbf {\bibinfo {volume} {17}},\ \bibinfo {pages}
  {113053}}\BibitemShut {NoStop}%
\bibitem [{\citenamefont {Leggett}(2001)}]{Leggett1}%
  \BibitemOpen
  \bibfield  {author} {\bibinfo {author} {\bibnamefont {Leggett}, \bibfnamefont
  {A.}}} (\bibinfo {year} {2001}),\ \href@noop {} {\bibfield  {journal}
  {\bibinfo  {journal} {Rev. Mod. Phys.}\ }\textbf {\bibinfo {volume} {73}},\
  \bibinfo {pages} {307}}\BibitemShut {NoStop}%
\bibitem [{\citenamefont {Leggett}(2006)}]{Leggett2}%
  \BibitemOpen
  \bibfield  {author} {\bibinfo {author} {\bibnamefont {Leggett}, \bibfnamefont
  {A.}}} (\bibinfo {year} {2006}),\ \href@noop {} {\emph {\bibinfo {title}
  {Quantum Liquids}}}\ (\bibinfo  {publisher} {Oxford University Press},\
  \bibinfo {address} {Oxford})\BibitemShut {NoStop}%
\bibitem [{\citenamefont {Lewenstein}\ \emph {et~al.}(2007)\citenamefont
  {Lewenstein}, \citenamefont {Sanpera}, \citenamefont {Ahufinger},
  \citenamefont {Damski}, \citenamefont {Sen},\ and\ \citenamefont
  {Sen}}]{Lewenstein0}%
  \BibitemOpen
  \bibfield  {author} {\bibinfo {author} {\bibnamefont {Lewenstein},
  \bibfnamefont {M.}}, \bibinfo {author} {\bibfnamefont {A.}~\bibnamefont
  {Sanpera}}, \bibinfo {author} {\bibfnamefont {V.}~\bibnamefont {Ahufinger}},
  \bibinfo {author} {\bibfnamefont {B.}~\bibnamefont {Damski}}, \bibinfo
  {author} {\bibfnamefont {A.}~\bibnamefont {Sen}}, \ and\ \bibinfo {author}
  {\bibfnamefont {U.}~\bibnamefont {Sen}}} (\bibinfo {year} {2007}),\
  \href@noop {} {\bibfield  {journal} {\bibinfo  {journal} {Adv. in Phys.}\
  }\textbf {\bibinfo {volume} {56}},\ \bibinfo {pages} {243}}\BibitemShut
  {NoStop}%
\bibitem [{\citenamefont {Li}\ \emph {et~al.}(2001)\citenamefont {Li},
  \citenamefont {Zeng}, \citenamefont {Liu},\ and\ \citenamefont {Long}}]{Li0}%
  \BibitemOpen
  \bibfield  {author} {\bibinfo {author} {\bibnamefont {Li}, \bibfnamefont
  {Y.}}, \bibinfo {author} {\bibfnamefont {B.}~\bibnamefont {Zeng}}, \bibinfo
  {author} {\bibfnamefont {X.}~\bibnamefont {Liu}}, \ and\ \bibinfo {author}
  {\bibfnamefont {G.}~\bibnamefont {Long}}} (\bibinfo {year} {2001}),\
  \href@noop {} {\bibfield  {journal} {\bibinfo  {journal} {Phys. Rev. A}\
  }\textbf {\bibinfo {volume} {64}},\ \bibinfo {pages} {054302}}\BibitemShut
  {NoStop}%
\bibitem [{\citenamefont {Linnemann}\ \emph {et~al.}(2016)\citenamefont
  {Linnemann}, \citenamefont {Strobel}, \citenamefont {Muessel}, \citenamefont
  {Schulz}, \citenamefont {Lewis-Swan}, \citenamefont {Kheruntsyan},\ and\
  \citenamefont {Oberthaler}}]{LinnemannPRL2016}%
  \BibitemOpen
  \bibfield  {author} {\bibinfo {author} {\bibnamefont {Linnemann},
  \bibfnamefont {D.}}, \bibinfo {author} {\bibfnamefont {H.}~\bibnamefont
  {Strobel}}, \bibinfo {author} {\bibfnamefont {W.}~\bibnamefont {Muessel}},
  \bibinfo {author} {\bibfnamefont {J.}~\bibnamefont {Schulz}}, \bibinfo
  {author} {\bibfnamefont {R.~J.}\ \bibnamefont {Lewis-Swan}}, \bibinfo
  {author} {\bibfnamefont {K.~V.}\ \bibnamefont {Kheruntsyan}}, \ and\ \bibinfo
  {author} {\bibfnamefont {M.~K.}\ \bibnamefont {Oberthaler}}} (\bibinfo {year}
  {2016}),\ \href {\doibase 10.1103/PhysRevLett.117.013001} {\bibfield
  {journal} {\bibinfo  {journal} {Phys. Rev. Lett.}\ }\textbf {\bibinfo
  {volume} {117}},\ \bibinfo {pages} {013001}}\BibitemShut {NoStop}%
\bibitem [{\citenamefont {Liu}\ and\ \citenamefont
  {Yuan}(2016)}]{1367-2630-18-9-093009}%
  \BibitemOpen
  \bibfield  {author} {\bibinfo {author} {\bibnamefont {Liu}, \bibfnamefont
  {J.}}, \ and\ \bibinfo {author} {\bibfnamefont {H.}~\bibnamefont {Yuan}}}
  (\bibinfo {year} {2016}),\ \href
  {http://stacks.iop.org/1367-2630/18/i=9/a=093009} {\bibfield  {journal}
  {\bibinfo  {journal} {New Journal of Physics}\ }\textbf {\bibinfo {volume}
  {18}}~(\bibinfo {number} {9}),\ \bibinfo {pages} {093009}}\BibitemShut
  {NoStop}%
\bibitem [{\citenamefont {Lloyd}(2008)}]{Lloyd08}%
  \BibitemOpen
  \bibfield  {author} {\bibinfo {author} {\bibnamefont {Lloyd}, \bibfnamefont
  {S.}}} (\bibinfo {year} {2008}),\ \href@noop {} {\bibfield  {journal}
  {\bibinfo  {journal} {Science}\ }\textbf {\bibinfo {volume} {321}},\ \bibinfo
  {pages} {1463}}\BibitemShut {NoStop}%
\bibitem [{\citenamefont {{Lo{\hskip 5pt}Franco}}\ and\ \citenamefont
  {Compagno}(2016)}]{Compagno}%
  \BibitemOpen
  \bibfield  {author} {\bibinfo {author} {\bibnamefont {{Lo{\hskip
  5pt}Franco}}, \bibfnamefont {R.}}, \ and\ \bibinfo {author} {\bibfnamefont
  {G.}~\bibnamefont {Compagno}}} (\bibinfo {year} {2016}),\ \href@noop {}
  {\bibfield  {journal} {\bibinfo  {journal} {Sci Rep}\ }\textbf {\bibinfo
  {volume} {20603}},\ \bibinfo {pages} {6}}\BibitemShut {NoStop}%
\bibitem [{\citenamefont {Lopaeva}\ \emph {et~al.}(2013)\citenamefont
  {Lopaeva}, \citenamefont {Berchera}, \citenamefont {Degiovanni},
  \citenamefont {Olivares}, \citenamefont {Brida},\ and\ \citenamefont
  {Genovese}}]{Lopaeva13}%
  \BibitemOpen
  \bibfield  {author} {\bibinfo {author} {\bibnamefont {Lopaeva}, \bibfnamefont
  {E.~D.}}, \bibinfo {author} {\bibfnamefont {I.~R.}\ \bibnamefont {Berchera}},
  \bibinfo {author} {\bibfnamefont {I.~P.}\ \bibnamefont {Degiovanni}},
  \bibinfo {author} {\bibfnamefont {S.}~\bibnamefont {Olivares}}, \bibinfo
  {author} {\bibfnamefont {G.}~\bibnamefont {Brida}}, \ and\ \bibinfo {author}
  {\bibfnamefont {M.}~\bibnamefont {Genovese}}} (\bibinfo {year} {2013}),\
  \href {\doibase 10.1103/PhysRevLett.110.153603} {\bibfield  {journal}
  {\bibinfo  {journal} {Phys. Rev. Lett.}\ }\textbf {\bibinfo {volume} {110}},\
  \bibinfo {pages} {153603}}\BibitemShut {NoStop}%
\bibitem [{\citenamefont {Lorenzo}\ \emph {et~al.}(2017)\citenamefont
  {Lorenzo}, \citenamefont {Marino}, \citenamefont {Plastina}, \citenamefont
  {Palma},\ and\ \citenamefont {Apollaro}}]{Lorenzo2017}%
  \BibitemOpen
  \bibfield  {author} {\bibinfo {author} {\bibnamefont {Lorenzo}, \bibfnamefont
  {S.}}, \bibinfo {author} {\bibfnamefont {J.}~\bibnamefont {Marino}}, \bibinfo
  {author} {\bibfnamefont {F.}~\bibnamefont {Plastina}}, \bibinfo {author}
  {\bibfnamefont {G.~M.}\ \bibnamefont {Palma}}, \ and\ \bibinfo {author}
  {\bibfnamefont {T.~J.~G.}\ \bibnamefont {Apollaro}}} (\bibinfo {year}
  {2017}),\ \href@noop {} {\bibfield  {journal} {\bibinfo  {journal} {Sci.
  Rep.}\ }\textbf {\bibinfo {volume} {7}},\ \bibinfo {pages}
  {5672}}\BibitemShut {NoStop}%
\bibitem [{\citenamefont {Luis}(2004)}]{LuisPL2004}%
  \BibitemOpen
  \bibfield  {author} {\bibinfo {author} {\bibnamefont {Luis}, \bibfnamefont
  {A.}}} (\bibinfo {year} {2004}),\ \href@noop {} {\bibfield  {journal}
  {\bibinfo  {journal} {Phys. Lett. A}\ }\textbf {\bibinfo {volume}
  {329}}~(\bibinfo {number} {1-2}),\ \bibinfo {pages} {8 }}\BibitemShut
  {NoStop}%
\bibitem [{\citenamefont {Luis}(2007)}]{LuisPR2007}%
  \BibitemOpen
  \bibfield  {author} {\bibinfo {author} {\bibnamefont {Luis}, \bibfnamefont
  {A.}}} (\bibinfo {year} {2007}),\ \href {\doibase 10.1103/PhysRevA.76.035801}
  {\bibfield  {journal} {\bibinfo  {journal} {Phys. Rev. A}\ }\textbf {\bibinfo
  {volume} {76}}~(\bibinfo {number} {3}),\ \bibinfo {eid} {035801}}\BibitemShut
  {NoStop}%
\bibitem [{\citenamefont {Lund}\ \emph {et~al.}(2004)\citenamefont {Lund},
  \citenamefont {Jeong}, \citenamefont {Ralph},\ and\ \citenamefont
  {Kim}}]{lund_conditional_2004}%
  \BibitemOpen
  \bibfield  {author} {\bibinfo {author} {\bibnamefont {Lund}, \bibfnamefont
  {A.~P.}}, \bibinfo {author} {\bibfnamefont {H.}~\bibnamefont {Jeong}},
  \bibinfo {author} {\bibfnamefont {T.~C.}\ \bibnamefont {Ralph}}, \ and\
  \bibinfo {author} {\bibfnamefont {M.~S.}\ \bibnamefont {Kim}}} (\bibinfo
  {year} {2004}),\ \href {\doibase 10.1103/PhysRevA.70.020101} {\bibfield
  {journal} {\bibinfo  {journal} {Phys. Rev. A}\ }\textbf {\bibinfo {volume}
  {70}}~(\bibinfo {number} {2}),\ \bibinfo {pages} {020101}}\BibitemShut
  {NoStop}%
\bibitem [{\citenamefont {Lupo}\ and\ \citenamefont
  {Pirandola}(2016)}]{Cosmo2016}%
  \BibitemOpen
  \bibfield  {author} {\bibinfo {author} {\bibnamefont {Lupo}, \bibfnamefont
  {C.}}, \ and\ \bibinfo {author} {\bibfnamefont {S.}~\bibnamefont
  {Pirandola}}} (\bibinfo {year} {2016}),\ \href {\doibase
  10.1103/PhysRevLett.117.190802} {\bibfield  {journal} {\bibinfo  {journal}
  {Phys. Rev. Lett.}\ }\textbf {\bibinfo {volume} {117}},\ \bibinfo {pages}
  {190802}}\BibitemShut {NoStop}%
\bibitem [{\citenamefont {Lupo}\ \emph {et~al.}(2013)\citenamefont {Lupo},
  \citenamefont {Pirandola}, \citenamefont {Giovannetti},\ and\ \citenamefont
  {Mancini}}]{Lupo13}%
  \BibitemOpen
  \bibfield  {author} {\bibinfo {author} {\bibnamefont {Lupo}, \bibfnamefont
  {C.}}, \bibinfo {author} {\bibfnamefont {S.}~\bibnamefont {Pirandola}},
  \bibinfo {author} {\bibfnamefont {V.}~\bibnamefont {Giovannetti}}, \ and\
  \bibinfo {author} {\bibfnamefont {S.}~\bibnamefont {Mancini}}} (\bibinfo
  {year} {2013}),\ \href@noop {} {\bibfield  {journal} {\bibinfo  {journal}
  {Phys. Rev. A}\ }\textbf {\bibinfo {volume} {87}},\ \bibinfo {pages}
  {062310}}\BibitemShut {NoStop}%
\bibitem [{\citenamefont {MacCormick}\ \emph {et~al.}(2016)\citenamefont
  {MacCormick}, \citenamefont {Bergamini}, \citenamefont {Mansell},
  \citenamefont {Cable},\ and\ \citenamefont {Modi}}]{PhysRevA.93.023805}%
  \BibitemOpen
  \bibfield  {author} {\bibinfo {author} {\bibnamefont {MacCormick},
  \bibfnamefont {C.}}, \bibinfo {author} {\bibfnamefont {S.}~\bibnamefont
  {Bergamini}}, \bibinfo {author} {\bibfnamefont {C.}~\bibnamefont {Mansell}},
  \bibinfo {author} {\bibfnamefont {H.}~\bibnamefont {Cable}}, \ and\ \bibinfo
  {author} {\bibfnamefont {K.}~\bibnamefont {Modi}}} (\bibinfo {year} {2016}),\
  \href {\doibase 10.1103/PhysRevA.93.023805} {\bibfield  {journal} {\bibinfo
  {journal} {Phys. Rev. A}\ }\textbf {\bibinfo {volume} {93}},\ \bibinfo
  {pages} {023805}}\BibitemShut {NoStop}%
\bibitem [{\citenamefont {Macieszczak}\ \emph {et~al.}(2014)\citenamefont
  {Macieszczak}, \citenamefont {Demkowicz-Dobrza{\'n}ski},\ and\ \citenamefont
  {Fraas}}]{macieszczak_optimal_2014}%
  \BibitemOpen
  \bibfield  {author} {\bibinfo {author} {\bibnamefont {Macieszczak},
  \bibfnamefont {K.}}, \bibinfo {author} {\bibfnamefont {R.}~\bibnamefont
  {Demkowicz-Dobrza{\'n}ski}}, \ and\ \bibinfo {author} {\bibfnamefont
  {M.}~\bibnamefont {Fraas}}} (\bibinfo {year} {2014}),\ \href
  {http://arxiv.org/abs/1311.5576} {\bibfield  {journal} {\bibinfo  {journal}
  {New J. Phys.}\ }\textbf {\bibinfo {volume} {16}},\ \bibinfo {pages}
  {113002}}\BibitemShut {NoStop}%
\bibitem [{\citenamefont {Mahmud}\ \emph {et~al.}(2014)\citenamefont {Mahmud},
  \citenamefont {Tiesinga},\ and\ \citenamefont {Johnson}}]{MahmudPRA2014}%
  \BibitemOpen
  \bibfield  {author} {\bibinfo {author} {\bibnamefont {Mahmud}, \bibfnamefont
  {K.~W.}}, \bibinfo {author} {\bibfnamefont {E.}~\bibnamefont {Tiesinga}}, \
  and\ \bibinfo {author} {\bibfnamefont {P.~R.}\ \bibnamefont {Johnson}}}
  (\bibinfo {year} {2014}),\ \href {\doibase 10.1103/PhysRevA.90.041602}
  {\bibfield  {journal} {\bibinfo  {journal} {Phys. Rev. A}\ }\textbf {\bibinfo
  {volume} {90}},\ \bibinfo {pages} {041602}}\BibitemShut {NoStop}%
\bibitem [{\citenamefont {Maldonado-Mundo}\ and\ \citenamefont
  {Luis}(2009)}]{Maldonado-MundoPRA2009}%
  \BibitemOpen
  \bibfield  {author} {\bibinfo {author} {\bibnamefont {Maldonado-Mundo},
  \bibfnamefont {D.}}, \ and\ \bibinfo {author} {\bibfnamefont
  {A.}~\bibnamefont {Luis}}} (\bibinfo {year} {2009}),\ \href {\doibase
  10.1103/PhysRevA.80.063811} {\bibfield  {journal} {\bibinfo  {journal} {Phys.
  Rev. A}\ }\textbf {\bibinfo {volume} {80}},\ \bibinfo {pages}
  {063811}}\BibitemShut {NoStop}%
\bibitem [{\citenamefont {Mancino}\ \emph {et~al.}(2016)\citenamefont
  {Mancino}, \citenamefont {Sbroscia}, \citenamefont {Gianani}, \citenamefont
  {Roccia},\ and\ \citenamefont {Barbieri}}]{mancino_quantum_2016}%
  \BibitemOpen
  \bibfield  {author} {\bibinfo {author} {\bibnamefont {Mancino}, \bibfnamefont
  {L.}}, \bibinfo {author} {\bibfnamefont {M.}~\bibnamefont {Sbroscia}},
  \bibinfo {author} {\bibfnamefont {I.}~\bibnamefont {Gianani}}, \bibinfo
  {author} {\bibfnamefont {E.}~\bibnamefont {Roccia}}, \ and\ \bibinfo {author}
  {\bibfnamefont {M.}~\bibnamefont {Barbieri}}} (\bibinfo {year} {2016}),\
  \href@noop {} {\enquote {\bibinfo {title} {Quantum {Simulation} of
  single-qubit thermometry using linear optics},}\ }\bibinfo {note}
  {ArXiv:1609.01590}\BibitemShut {NoStop}%
\bibitem [{\citenamefont {Mandel}\ and\ \citenamefont
  {Wolf}(1965)}]{MandelRMP}%
  \BibitemOpen
  \bibfield  {author} {\bibinfo {author} {\bibnamefont {Mandel}, \bibfnamefont
  {L.}}, \ and\ \bibinfo {author} {\bibfnamefont {E.}~\bibnamefont {Wolf}}}
  (\bibinfo {year} {1965}),\ \href {\doibase 10.1103/RevModPhys.37.231}
  {\bibfield  {journal} {\bibinfo  {journal} {Rev. Mod. Phys.}\ }\textbf
  {\bibinfo {volume} {37}},\ \bibinfo {pages} {231}}\BibitemShut {NoStop}%
\bibitem [{\citenamefont {Marian}\ and\ \citenamefont
  {Marian}(2012)}]{Marians12}%
  \BibitemOpen
  \bibfield  {author} {\bibinfo {author} {\bibnamefont {Marian}, \bibfnamefont
  {P.}}, \ and\ \bibinfo {author} {\bibfnamefont {T.~A.}\ \bibnamefont
  {Marian}}} (\bibinfo {year} {2012}),\ \href@noop {} {\bibfield  {journal}
  {\bibinfo  {journal} {Phys. Rev. A}\ }\textbf {\bibinfo {volume} {86}},\
  \bibinfo {pages} {022340}}\BibitemShut {NoStop}%
\bibitem [{\citenamefont {Marvian}\ and\ \citenamefont
  {Spekkens}(2016)}]{Marvian2016}%
  \BibitemOpen
  \bibfield  {author} {\bibinfo {author} {\bibnamefont {Marvian}, \bibfnamefont
  {I.}}, \ and\ \bibinfo {author} {\bibfnamefont {R.~W.}\ \bibnamefont
  {Spekkens}}} (\bibinfo {year} {2016}),\ \href {\doibase
  10.1103/PhysRevA.94.052324} {\bibfield  {journal} {\bibinfo  {journal} {Phys.
  Rev. A}\ }\textbf {\bibinfo {volume} {94}},\ \bibinfo {pages}
  {052324}}\BibitemShut {NoStop}%
\bibitem [{\citenamefont {Marzolino}(2013)}]{Marzolino0}%
  \BibitemOpen
  \bibfield  {author} {\bibinfo {author} {\bibnamefont {Marzolino},
  \bibfnamefont {U.}}} (\bibinfo {year} {2013}),\ \href@noop {} {\bibfield
  {journal} {\bibinfo  {journal} {Europhys. Lett.}\ }\textbf {\bibinfo {volume}
  {104}},\ \bibinfo {pages} {40004}}\BibitemShut {NoStop}%
\bibitem [{\citenamefont {Marzolino}\ and\ \citenamefont
  {Braun}(2013)}]{marzolino_precision_2013}%
  \BibitemOpen
  \bibfield  {author} {\bibinfo {author} {\bibnamefont {Marzolino},
  \bibfnamefont {U.}}, \ and\ \bibinfo {author} {\bibfnamefont
  {D.}~\bibnamefont {Braun}}} (\bibinfo {year} {2013}),\ \href {\doibase
  10.1103/PhysRevA.88.063609} {\bibfield  {journal} {\bibinfo  {journal} {Phys.
  Rev. A}\ }\textbf {\bibinfo {volume} {88}}~(\bibinfo {number} {6}),\ \bibinfo
  {pages} {063609}}\BibitemShut {NoStop}%
\bibitem [{\citenamefont {Marzolino}\ and\ \citenamefont
  {Braun}(2015)}]{Marzolino2015}%
  \BibitemOpen
  \bibfield  {author} {\bibinfo {author} {\bibnamefont {Marzolino},
  \bibfnamefont {U.}}, \ and\ \bibinfo {author} {\bibfnamefont
  {D.}~\bibnamefont {Braun}}} (\bibinfo {year} {2015}),\ \href {\doibase
  10.1103/PhysRevA.91.039902} {\bibfield  {journal} {\bibinfo  {journal} {Phys.
  Rev. A}\ }\textbf {\bibinfo {volume} {91}},\ \bibinfo {pages}
  {039902}}\BibitemShut {NoStop}%
\bibitem [{\citenamefont {Marzolino}\ and\ \citenamefont
  {Prosen}(2014)}]{Marzolino2014}%
  \BibitemOpen
  \bibfield  {author} {\bibinfo {author} {\bibnamefont {Marzolino},
  \bibfnamefont {U.}}, \ and\ \bibinfo {author} {\bibfnamefont
  {T.}~\bibnamefont {Prosen}}} (\bibinfo {year} {2014}),\ \href {\doibase
  10.1103/PhysRevA.90.062130} {\bibfield  {journal} {\bibinfo  {journal} {Phys.
  Rev. A}\ }\textbf {\bibinfo {volume} {90}},\ \bibinfo {pages}
  {062130}}\BibitemShut {NoStop}%
\bibitem [{\citenamefont {Marzolino}\ and\ \citenamefont
  {Prosen}(2016{\natexlab{a}})}]{Marzolino2016}%
  \BibitemOpen
  \bibfield  {author} {\bibinfo {author} {\bibnamefont {Marzolino},
  \bibfnamefont {U.}}, \ and\ \bibinfo {author} {\bibfnamefont
  {T.}~\bibnamefont {Prosen}}} (\bibinfo {year} {2016}{\natexlab{a}}),\ \href
  {\doibase 10.1103/PhysRevA.93.032306} {\bibfield  {journal} {\bibinfo
  {journal} {Phys. Rev. A}\ }\textbf {\bibinfo {volume} {93}},\ \bibinfo
  {pages} {032306}}\BibitemShut {NoStop}%
\bibitem [{\citenamefont {Marzolino}\ and\ \citenamefont
  {Prosen}(2016{\natexlab{b}})}]{Marzolino2016-2}%
  \BibitemOpen
  \bibfield  {author} {\bibinfo {author} {\bibnamefont {Marzolino},
  \bibfnamefont {U.}}, \ and\ \bibinfo {author} {\bibfnamefont
  {T.}~\bibnamefont {Prosen}}} (\bibinfo {year} {2016}{\natexlab{b}}),\
  \href@noop {} {\bibfield  {journal} {\bibinfo  {journal} {Phys. Rev. A}\
  }\textbf {\bibinfo {volume} {94}},\ \bibinfo {pages} {039903}}\BibitemShut
  {NoStop}%
\bibitem [{\citenamefont {Marzolino}\ and\ \citenamefont
  {Prosen}(2017)}]{Marzolino2017}%
  \BibitemOpen
  \bibfield  {author} {\bibinfo {author} {\bibnamefont {Marzolino},
  \bibfnamefont {U.}}, \ and\ \bibinfo {author} {\bibfnamefont {T.~c.~v.}\
  \bibnamefont {Prosen}}} (\bibinfo {year} {2017}),\ \href {\doibase
  10.1103/PhysRevB.96.104402} {\bibfield  {journal} {\bibinfo  {journal} {Phys.
  Rev. B}\ }\textbf {\bibinfo {volume} {96}},\ \bibinfo {pages}
  {104402}}\BibitemShut {NoStop}%
\bibitem [{\citenamefont {Matsuzaki}\ \emph {et~al.}(2011)\citenamefont
  {Matsuzaki}, \citenamefont {Benjamin},\ and\ \citenamefont
  {Fitzsimons}}]{PhysRevA.84.012103}%
  \BibitemOpen
  \bibfield  {author} {\bibinfo {author} {\bibnamefont {Matsuzaki},
  \bibfnamefont {Y.}}, \bibinfo {author} {\bibfnamefont {S.~C.}\ \bibnamefont
  {Benjamin}}, \ and\ \bibinfo {author} {\bibfnamefont {J.}~\bibnamefont
  {Fitzsimons}}} (\bibinfo {year} {2011}),\ \href {\doibase
  10.1103/PhysRevA.84.012103} {\bibfield  {journal} {\bibinfo  {journal} {Phys.
  Rev. A}\ }\textbf {\bibinfo {volume} {84}},\ \bibinfo {pages}
  {012103}}\BibitemShut {NoStop}%
\bibitem [{\citenamefont {Mehboudi}\ \emph {et~al.}(2016)\citenamefont
  {Mehboudi}, \citenamefont {Correa},\ and\ \citenamefont
  {Sanpera}}]{Mehboudi2016}%
  \BibitemOpen
  \bibfield  {author} {\bibinfo {author} {\bibnamefont {Mehboudi},
  \bibfnamefont {M.}}, \bibinfo {author} {\bibfnamefont {L.~A.}\ \bibnamefont
  {Correa}}, \ and\ \bibinfo {author} {\bibfnamefont {A.}~\bibnamefont
  {Sanpera}}} (\bibinfo {year} {2016}),\ \href {\doibase
  10.1103/PhysRevA.94.042121} {\bibfield  {journal} {\bibinfo  {journal} {Phys.
  Rev. A}\ }\textbf {\bibinfo {volume} {94}},\ \bibinfo {pages}
  {042121}}\BibitemShut {NoStop}%
\bibitem [{\citenamefont {Merkel}\ \emph {et~al.}(2013)\citenamefont {Merkel},
  \citenamefont {Gambetta}, \citenamefont {Smolin}, \citenamefont {Poletto},
  \citenamefont {C\'orcoles}, \citenamefont {Johnson}, \citenamefont {Ryan},\
  and\ \citenamefont {Steffen}}]{Merkel2013}%
  \BibitemOpen
  \bibfield  {author} {\bibinfo {author} {\bibnamefont {Merkel}, \bibfnamefont
  {S.~T.}}, \bibinfo {author} {\bibfnamefont {J.~M.}\ \bibnamefont {Gambetta}},
  \bibinfo {author} {\bibfnamefont {J.~A.}\ \bibnamefont {Smolin}}, \bibinfo
  {author} {\bibfnamefont {S.}~\bibnamefont {Poletto}}, \bibinfo {author}
  {\bibfnamefont {A.~D.}\ \bibnamefont {C\'orcoles}}, \bibinfo {author}
  {\bibfnamefont {B.~R.}\ \bibnamefont {Johnson}}, \bibinfo {author}
  {\bibfnamefont {C.~A.}\ \bibnamefont {Ryan}}, \ and\ \bibinfo {author}
  {\bibfnamefont {M.}~\bibnamefont {Steffen}}} (\bibinfo {year} {2013}),\ \href
  {\doibase 10.1103/PhysRevA.87.062119} {\bibfield  {journal} {\bibinfo
  {journal} {Phys. Rev. A}\ }\textbf {\bibinfo {volume} {87}},\ \bibinfo
  {pages} {062119}}\BibitemShut {NoStop}%
\bibitem [{\citenamefont {Micheli}\ \emph {et~al.}(2003)\citenamefont
  {Micheli}, \citenamefont {Jaksch}, \citenamefont {Cirac},\ and\ \citenamefont
  {Zoller}}]{Micheli0}%
  \BibitemOpen
  \bibfield  {author} {\bibinfo {author} {\bibnamefont {Micheli}, \bibfnamefont
  {A.}}, \bibinfo {author} {\bibfnamefont {D.}~\bibnamefont {Jaksch}}, \bibinfo
  {author} {\bibfnamefont {J.}~\bibnamefont {Cirac}}, \ and\ \bibinfo {author}
  {\bibfnamefont {P.}~\bibnamefont {Zoller}}} (\bibinfo {year} {2003}),\
  \href@noop {} {\bibfield  {journal} {\bibinfo  {journal} {Phys. Rev. A}\
  }\textbf {\bibinfo {volume} {67}},\ \bibinfo {pages} {013607}}\BibitemShut
  {NoStop}%
\bibitem [{\citenamefont {Mills}(1964)}]{Mills1964}%
  \BibitemOpen
  \bibfield  {author} {\bibinfo {author} {\bibnamefont {Mills}, \bibfnamefont
  {D.~L.}}} (\bibinfo {year} {1964}),\ \href@noop {} {\bibfield  {journal}
  {\bibinfo  {journal} {Phys. Rev.}\ }\textbf {\bibinfo {volume} {134}},\
  \bibinfo {pages} {A306}}\BibitemShut {NoStop}%
\bibitem [{\citenamefont {Miszczak}\ \emph {et~al.}(2009)\citenamefont
  {Miszczak}, \citenamefont {Pucha\l{}a}, \citenamefont {Horodecki},
  \citenamefont {Uhlmann},\ and\ \citenamefont
  {K.\.{Z}yczkowski}}]{Miszczak09}%
  \BibitemOpen
  \bibfield  {author} {\bibinfo {author} {\bibnamefont {Miszczak},
  \bibfnamefont {J.~A.}}, \bibinfo {author} {\bibfnamefont {Z.}~\bibnamefont
  {Pucha\l{}a}}, \bibinfo {author} {\bibfnamefont {P.}~\bibnamefont
  {Horodecki}}, \bibinfo {author} {\bibfnamefont {A.}~\bibnamefont {Uhlmann}},
  \ and\ \bibinfo {author} {\bibnamefont {K.\.{Z}yczkowski}}} (\bibinfo {year}
  {2009}),\ \href@noop {} {\bibfield  {journal} {\bibinfo  {journal} {Quantum
  Information and Computation}\ }\textbf {\bibinfo {volume} {9}},\ \bibinfo
  {pages} {0103}}\BibitemShut {NoStop}%
\bibitem [{\citenamefont {Mitchell}(2017)}]{MitchellQST2017}%
  \BibitemOpen
  \bibfield  {author} {\bibinfo {author} {\bibnamefont {Mitchell},
  \bibfnamefont {M.~W.}}} (\bibinfo {year} {2017}),\ \href {\doibase
  10.1088/2058-9565/aa80c0} {\bibfield  {journal} {\bibinfo  {journal} {Quantum
  Science and Technology}\ }\textbf {\bibinfo {volume} {2}}~(\bibinfo {number}
  {4}),\ \bibinfo {pages} {044005}}\BibitemShut {NoStop}%
\bibitem [{\citenamefont {Modi}\ \emph {et~al.}(2012)\citenamefont {Modi},
  \citenamefont {Brodutch}, \citenamefont {Cable}, \citenamefont {Paterek},\
  and\ \citenamefont {Vedral}}]{Modi2012}%
  \BibitemOpen
  \bibfield  {author} {\bibinfo {author} {\bibnamefont {Modi}, \bibfnamefont
  {K.}}, \bibinfo {author} {\bibfnamefont {A.}~\bibnamefont {Brodutch}},
  \bibinfo {author} {\bibfnamefont {H.}~\bibnamefont {Cable}}, \bibinfo
  {author} {\bibfnamefont {T.}~\bibnamefont {Paterek}}, \ and\ \bibinfo
  {author} {\bibfnamefont {V.}~\bibnamefont {Vedral}}} (\bibinfo {year}
  {2012}),\ \href {\doibase 10.1103/RevModPhys.84.1655} {\bibfield  {journal}
  {\bibinfo  {journal} {Rev. Mod. Phys.}\ }\textbf {\bibinfo {volume} {84}},\
  \bibinfo {pages} {1655}}\BibitemShut {NoStop}%
\bibitem [{\citenamefont {Modi}\ \emph {et~al.}(2011)\citenamefont {Modi},
  \citenamefont {Cable}, \citenamefont {Williamson},\ and\ \citenamefont
  {Vedral}}]{Modi2011}%
  \BibitemOpen
  \bibfield  {author} {\bibinfo {author} {\bibnamefont {Modi}, \bibfnamefont
  {K.}}, \bibinfo {author} {\bibfnamefont {H.}~\bibnamefont {Cable}}, \bibinfo
  {author} {\bibfnamefont {M.}~\bibnamefont {Williamson}}, \ and\ \bibinfo
  {author} {\bibfnamefont {V.}~\bibnamefont {Vedral}}} (\bibinfo {year}
  {2011}),\ \href {\doibase 10.1103/PhysRevX.1.021022} {\bibfield  {journal}
  {\bibinfo  {journal} {Phys. Rev. X}\ }\textbf {\bibinfo {volume} {1}},\
  \bibinfo {pages} {021022}}\BibitemShut {NoStop}%
\bibitem [{\citenamefont {Modi}\ \emph {et~al.}(2010)\citenamefont {Modi},
  \citenamefont {Paterek}, \citenamefont {Son}, \citenamefont {Vedral},\ and\
  \citenamefont {Williamson}}]{Modi10}%
  \BibitemOpen
  \bibfield  {author} {\bibinfo {author} {\bibnamefont {Modi}, \bibfnamefont
  {K.}}, \bibinfo {author} {\bibfnamefont {T.}~\bibnamefont {Paterek}},
  \bibinfo {author} {\bibfnamefont {W.}~\bibnamefont {Son}}, \bibinfo {author}
  {\bibfnamefont {V.}~\bibnamefont {Vedral}}, \ and\ \bibinfo {author}
  {\bibfnamefont {M.}~\bibnamefont {Williamson}}} (\bibinfo {year} {2010}),\
  \href {\doibase 10.1103/PhysRevLett.104.080501} {\bibfield  {journal}
  {\bibinfo  {journal} {Phys. Rev. Lett.}\ }\textbf {\bibinfo {volume} {104}},\
  \bibinfo {pages} {080501}}\BibitemShut {NoStop}%
\bibitem [{\citenamefont {Mohseni}\ \emph {et~al.}(2008)\citenamefont
  {Mohseni}, \citenamefont {Rezakhani},\ and\ \citenamefont
  {Lidar}}]{Mohseni2008}%
  \BibitemOpen
  \bibfield  {author} {\bibinfo {author} {\bibnamefont {Mohseni}, \bibfnamefont
  {M.}}, \bibinfo {author} {\bibfnamefont {A.~T.}\ \bibnamefont {Rezakhani}}, \
  and\ \bibinfo {author} {\bibfnamefont {D.~A.}\ \bibnamefont {Lidar}}}
  (\bibinfo {year} {2008}),\ \href {\doibase 10.1103/PhysRevA.77.032322}
  {\bibfield  {journal} {\bibinfo  {journal} {Phys. Rev. A}\ }\textbf {\bibinfo
  {volume} {77}},\ \bibinfo {pages} {032322}}\BibitemShut {NoStop}%
\bibitem [{\citenamefont {Monras}(2013)}]{Monras13}%
  \BibitemOpen
  \bibfield  {author} {\bibinfo {author} {\bibnamefont {Monras}, \bibfnamefont
  {A.}}} (\bibinfo {year} {2013}),\ \href@noop {} {\enquote {\bibinfo {title}
  {Phase space formalism for quantum estimation of gaussian states},}\ }\Eprint
  {http://arxiv.org/abs/ArXiv:1303.3682} {ArXiv:1303.3682} \BibitemShut
  {NoStop}%
\bibitem [{\citenamefont {Monras}\ and\ \citenamefont
  {Illuminati}(2011)}]{Monras11}%
  \BibitemOpen
  \bibfield  {author} {\bibinfo {author} {\bibnamefont {Monras}, \bibfnamefont
  {A.}}, \ and\ \bibinfo {author} {\bibfnamefont {F.}~\bibnamefont
  {Illuminati}}} (\bibinfo {year} {2011}),\ \href@noop {} {\bibfield  {journal}
  {\bibinfo  {journal} {Phys. Rev. A}\ }\textbf {\bibinfo {volume} {83}},\
  \bibinfo {pages} {012315}}\BibitemShut {NoStop}%
\bibitem [{\citenamefont {Monras}\ and\ \citenamefont
  {Paris}(2007)}]{Monras07}%
  \BibitemOpen
  \bibfield  {author} {\bibinfo {author} {\bibnamefont {Monras}, \bibfnamefont
  {A.}}, \ and\ \bibinfo {author} {\bibfnamefont {M.~G.~A.}\ \bibnamefont
  {Paris}}} (\bibinfo {year} {2007}),\ \href@noop {} {\bibfield  {journal}
  {\bibinfo  {journal} {Phys. Rev. Lett.}\ }\textbf {\bibinfo {volume} {98}},\
  \bibinfo {pages} {160401}}\BibitemShut {NoStop}%
\bibitem [{\citenamefont {Monroe}\ \emph {et~al.}(1996)\citenamefont {Monroe},
  \citenamefont {Meekho}, \citenamefont {King},\ and\ \citenamefont
  {Wineland}}]{Monroe96}%
  \BibitemOpen
  \bibfield  {author} {\bibinfo {author} {\bibnamefont {Monroe}, \bibfnamefont
  {C.}}, \bibinfo {author} {\bibfnamefont {D.~M.}\ \bibnamefont {Meekho}},
  \bibinfo {author} {\bibfnamefont {B.~E.}\ \bibnamefont {King}}, \ and\
  \bibinfo {author} {\bibfnamefont {D.~J.}\ \bibnamefont {Wineland}}} (\bibinfo
  {year} {1996}),\ \href@noop {} {\bibfield  {journal} {\bibinfo  {journal}
  {Science}\ }\textbf {\bibinfo {volume} {272}},\ \bibinfo {pages}
  {1131}}\BibitemShut {NoStop}%
\bibitem [{\citenamefont {Montina}\ and\ \citenamefont
  {Arecchi}(1998)}]{PhysRevA.58.3472}%
  \BibitemOpen
  \bibfield  {author} {\bibinfo {author} {\bibnamefont {Montina}, \bibfnamefont
  {A.}}, \ and\ \bibinfo {author} {\bibfnamefont {F.~T.}\ \bibnamefont
  {Arecchi}}} (\bibinfo {year} {1998}),\ \href {\doibase
  10.1103/PhysRevA.58.3472} {\bibfield  {journal} {\bibinfo  {journal} {Phys.
  Rev. A}\ }\textbf {\bibinfo {volume} {58}},\ \bibinfo {pages}
  {3472}}\BibitemShut {NoStop}%
\bibitem [{\citenamefont {Moritz}\ \emph {et~al.}(2003)\citenamefont {Moritz},
  \citenamefont {St\"oferle}, \citenamefont {K\"ohl},\ and\ \citenamefont
  {Esslinger}}]{Moritz2003}%
  \BibitemOpen
  \bibfield  {author} {\bibinfo {author} {\bibnamefont {Moritz}, \bibfnamefont
  {H.}}, \bibinfo {author} {\bibfnamefont {T.}~\bibnamefont {St\"oferle}},
  \bibinfo {author} {\bibfnamefont {M.}~\bibnamefont {K\"ohl}}, \ and\ \bibinfo
  {author} {\bibfnamefont {T.}~\bibnamefont {Esslinger}}} (\bibinfo {year}
  {2003}),\ \href {\doibase 10.1103/PhysRevLett.91.250402} {\bibfield
  {journal} {\bibinfo  {journal} {Phys. Rev. Lett.}\ }\textbf {\bibinfo
  {volume} {91}},\ \bibinfo {pages} {250402}}\BibitemShut {NoStop}%
\bibitem [{\citenamefont {Moriya}(2006)}]{Moriya1}%
  \BibitemOpen
  \bibfield  {author} {\bibinfo {author} {\bibnamefont {Moriya}, \bibfnamefont
  {H.}}} (\bibinfo {year} {2006}),\ \href@noop {} {\bibfield  {journal}
  {\bibinfo  {journal} {J. Phys. A}\ }\textbf {\bibinfo {volume} {39}},\
  \bibinfo {pages} {3753}}\BibitemShut {NoStop}%
\bibitem [{\citenamefont {Motamedifar}\ \emph {et~al.}(2013)\citenamefont
  {Motamedifar}, \citenamefont {Mahdavifar}, \citenamefont {Shayesteh},\ and\
  \citenamefont {Nemati}}]{Motamedifar2013}%
  \BibitemOpen
  \bibfield  {author} {\bibinfo {author} {\bibnamefont {Motamedifar},
  \bibfnamefont {M.}}, \bibinfo {author} {\bibfnamefont {S.}~\bibnamefont
  {Mahdavifar}}, \bibinfo {author} {\bibfnamefont {S.~F.}\ \bibnamefont
  {Shayesteh}}, \ and\ \bibinfo {author} {\bibfnamefont {S.}~\bibnamefont
  {Nemati}}} (\bibinfo {year} {2013}),\ \href@noop {} {\bibfield  {journal}
  {\bibinfo  {journal} {Phys. Scr.}\ }\textbf {\bibinfo {volume} {88}},\
  \bibinfo {pages} {015003}}\BibitemShut {NoStop}%
\bibitem [{\citenamefont {Mruga\l{}a}(1984)}]{Mrugala1984}%
  \BibitemOpen
  \bibfield  {author} {\bibinfo {author} {\bibnamefont {Mruga\l{}a},
  \bibfnamefont {R.}}} (\bibinfo {year} {1984}),\ \href@noop {} {\bibfield
  {journal} {\bibinfo  {journal} {Physica A}\ }\textbf {\bibinfo {volume}
  {125}},\ \bibinfo {pages} {631}}\BibitemShut {NoStop}%
\bibitem [{\citenamefont {Muessel}\ \emph {et~al.}(2014)\citenamefont
  {Muessel}, \citenamefont {Strobel}, \citenamefont {Linnemann}, \citenamefont
  {Hume},\ and\ \citenamefont {Oberthaler}}]{MuesselPRL2014}%
  \BibitemOpen
  \bibfield  {author} {\bibinfo {author} {\bibnamefont {Muessel}, \bibfnamefont
  {W.}}, \bibinfo {author} {\bibfnamefont {H.}~\bibnamefont {Strobel}},
  \bibinfo {author} {\bibfnamefont {D.}~\bibnamefont {Linnemann}}, \bibinfo
  {author} {\bibfnamefont {D.~B.}\ \bibnamefont {Hume}}, \ and\ \bibinfo
  {author} {\bibfnamefont {M.~K.}\ \bibnamefont {Oberthaler}}} (\bibinfo {year}
  {2014}),\ \href {\doibase 10.1103/PhysRevLett.113.103004} {\bibfield
  {journal} {\bibinfo  {journal} {Phys. Rev. Lett.}\ }\textbf {\bibinfo
  {volume} {113}},\ \bibinfo {pages} {103004}}\BibitemShut {NoStop}%
\bibitem [{\citenamefont {M\"uller}\ \emph {et~al.}(2010)\citenamefont
  {M\"uller}, \citenamefont {Zimmermann}, \citenamefont {Meineke},
  \citenamefont {Brantut}, \citenamefont {Esslinger},\ and\ \citenamefont
  {Moritz}}]{Muller2010}%
  \BibitemOpen
  \bibfield  {author} {\bibinfo {author} {\bibnamefont {M\"uller},
  \bibfnamefont {T.}}, \bibinfo {author} {\bibfnamefont {B.}~\bibnamefont
  {Zimmermann}}, \bibinfo {author} {\bibfnamefont {J.}~\bibnamefont {Meineke}},
  \bibinfo {author} {\bibfnamefont {J.-P.}\ \bibnamefont {Brantut}}, \bibinfo
  {author} {\bibfnamefont {T.}~\bibnamefont {Esslinger}}, \ and\ \bibinfo
  {author} {\bibfnamefont {H.}~\bibnamefont {Moritz}}} (\bibinfo {year}
  {2010}),\ \href {\doibase 10.1103/PhysRevLett.105.040401} {\bibfield
  {journal} {\bibinfo  {journal} {Phys. Rev. Lett.}\ }\textbf {\bibinfo
  {volume} {105}},\ \bibinfo {pages} {040401}}\BibitemShut {NoStop}%
\bibitem [{\citenamefont {Mullin}(1997)}]{Mullin1997}%
  \BibitemOpen
  \bibfield  {author} {\bibinfo {author} {\bibnamefont {Mullin}, \bibfnamefont
  {W.~J.}}} (\bibinfo {year} {1997}),\ \href@noop {} {\bibfield  {journal}
  {\bibinfo  {journal} {J. Low Temp. Phys.}\ }\textbf {\bibinfo {volume}
  {106}},\ \bibinfo {pages} {615}}\BibitemShut {NoStop}%
\bibitem [{\citenamefont {Mullin}\ and\ \citenamefont
  {Sakhel}(2012)}]{Mullin2012}%
  \BibitemOpen
  \bibfield  {author} {\bibinfo {author} {\bibnamefont {Mullin}, \bibfnamefont
  {W.~J.}}, \ and\ \bibinfo {author} {\bibfnamefont {A.~R.}\ \bibnamefont
  {Sakhel}}} (\bibinfo {year} {2012}),\ \href@noop {} {\bibfield  {journal}
  {\bibinfo  {journal} {J. Low Temp. Phys.}\ }\textbf {\bibinfo {volume}
  {166}},\ \bibinfo {pages} {125}}\BibitemShut {NoStop}%
\bibitem [{\citenamefont {Nagaoka}(1988)}]{Nagaoka88}%
  \BibitemOpen
  \bibfield  {author} {\bibinfo {author} {\bibnamefont {Nagaoka}, \bibfnamefont
  {H.}}} (\bibinfo {year} {1988}),\ in\ \href@noop {} {\emph {\bibinfo
  {booktitle} {Proc. Int. Symp. on Inform. Theory}}},\ Vol.~\bibinfo {volume}
  {98},\ p.\ \bibinfo {pages} {577}\BibitemShut {NoStop}%
\bibitem [{\citenamefont {Nair}(2011)}]{Nair11}%
  \BibitemOpen
  \bibfield  {author} {\bibinfo {author} {\bibnamefont {Nair}, \bibfnamefont
  {R.}}} (\bibinfo {year} {2011}),\ \href@noop {} {\bibfield  {journal}
  {\bibinfo  {journal} {Phys. Rev. A}\ }\textbf {\bibinfo {volume} {84}},\
  \bibinfo {pages} {032312}}\BibitemShut {NoStop}%
\bibitem [{\citenamefont {{Nair, R.}}\ and\ \citenamefont {{Tsang,
  T.}}(2016)}]{Nair2016}%
  \BibitemOpen
  \bibfield  {author} {\bibinfo {author} {\bibnamefont {{Nair, R.}},}, \ and\
  \bibinfo {author} {\bibnamefont {{Tsang, T.}}}} (\bibinfo {year} {2016}),\
  \href@noop {} {\bibfield  {journal} {\bibinfo  {journal} {Phys. Rev. Lett.}\
  }\textbf {\bibinfo {volume} {77}},\ \bibinfo {pages} {190801}}\BibitemShut
  {NoStop}%
\bibitem [{\citenamefont {Napolitano}\ \emph {et~al.}(2011)\citenamefont
  {Napolitano}, \citenamefont {Koschorreck}, \citenamefont {Dubost},
  \citenamefont {Behbood}, \citenamefont {Sewell},\ and\ \citenamefont
  {Mitchell}}]{NapolitanoN2011}%
  \BibitemOpen
  \bibfield  {author} {\bibinfo {author} {\bibnamefont {Napolitano},
  \bibfnamefont {M.}}, \bibinfo {author} {\bibfnamefont {M.}~\bibnamefont
  {Koschorreck}}, \bibinfo {author} {\bibfnamefont {B.}~\bibnamefont {Dubost}},
  \bibinfo {author} {\bibfnamefont {N.}~\bibnamefont {Behbood}}, \bibinfo
  {author} {\bibfnamefont {R.~J.}\ \bibnamefont {Sewell}}, \ and\ \bibinfo
  {author} {\bibfnamefont {M.~W.}\ \bibnamefont {Mitchell}}} (\bibinfo {year}
  {2011}),\ \href {\doibase 10.1038/nature09778} {\bibfield  {journal}
  {\bibinfo  {journal} {Nature}\ }\textbf {\bibinfo {volume} {471}}~(\bibinfo
  {number} {7339}),\ \bibinfo {pages} {486}}\BibitemShut {NoStop}%
\bibitem [{\citenamefont {Napolitano}\ and\ \citenamefont
  {Mitchell}(2010)}]{NapolitanoNJP2010}%
  \BibitemOpen
  \bibfield  {author} {\bibinfo {author} {\bibnamefont {Napolitano},
  \bibfnamefont {M.}}, \ and\ \bibinfo {author} {\bibfnamefont {M.~W.}\
  \bibnamefont {Mitchell}}} (\bibinfo {year} {2010}),\ \href {\doibase
  10.1088/1367-2630/12/9/093016} {\bibfield  {journal} {\bibinfo  {journal}
  {New J. Phys.}\ }\textbf {\bibinfo {volume} {12}}~(\bibinfo {number} {9}),\
  \bibinfo {pages} {093016}}\BibitemShut {NoStop}%
\bibitem [{\citenamefont {Narnhofer}(2004)}]{Narnhofer0}%
  \BibitemOpen
  \bibfield  {author} {\bibinfo {author} {\bibnamefont {Narnhofer},
  \bibfnamefont {H.}}} (\bibinfo {year} {2004}),\ \href@noop {} {\bibfield
  {journal} {\bibinfo  {journal} {Phys. Lett.}\ }\textbf {\bibinfo {volume}
  {A310}},\ \bibinfo {pages} {423}}\BibitemShut {NoStop}%
\bibitem [{\citenamefont {Neergaard-Nielsen}\ \emph {et~al.}(2006)\citenamefont
  {Neergaard-Nielsen}, \citenamefont {Nielsen}, \citenamefont {Hettich},
  \citenamefont {M{\o}lmer},\ and\ \citenamefont
  {Polzik}}]{neergaard-nielsen_generation_2006}%
  \BibitemOpen
  \bibfield  {author} {\bibinfo {author} {\bibnamefont {Neergaard-Nielsen},
  \bibfnamefont {J.~S.}}, \bibinfo {author} {\bibfnamefont {B.~M.}\
  \bibnamefont {Nielsen}}, \bibinfo {author} {\bibfnamefont {C.}~\bibnamefont
  {Hettich}}, \bibinfo {author} {\bibfnamefont {K.}~\bibnamefont {M{\o}lmer}},
  \ and\ \bibinfo {author} {\bibfnamefont {E.~S.}\ \bibnamefont {Polzik}}}
  (\bibinfo {year} {2006}),\ \href {\doibase 10.1103/PhysRevLett.97.083604}
  {\bibfield  {journal} {\bibinfo  {journal} {Phys. Rev. Lett.}\ }\textbf
  {\bibinfo {volume} {97}}~(\bibinfo {number} {8}),\ \bibinfo {pages}
  {083604}}\BibitemShut {NoStop}%
\bibitem [{\citenamefont {Ng}\ and\ \citenamefont {Dam}(2000)}]{Ng00}%
  \BibitemOpen
  \bibfield  {author} {\bibinfo {author} {\bibnamefont {Ng}, \bibfnamefont
  {Y.~J.}}, \ and\ \bibinfo {author} {\bibfnamefont {H.~v.}\ \bibnamefont
  {Dam}}} (\bibinfo {year} {2000}),\ \href {\doibase 10.1023/A:1003745212871}
  {\bibfield  {journal} {\bibinfo  {journal} {Foundations of Physics}\ }\textbf
  {\bibinfo {volume} {30}}~(\bibinfo {number} {5}),\ \bibinfo {pages}
  {795}}\BibitemShut {NoStop}%
\bibitem [{\citenamefont {Nichols}\ \emph {et~al.}(2016)\citenamefont
  {Nichols}, \citenamefont {Bromley}, \citenamefont {Correa},\ and\
  \citenamefont {Adesso}}]{Nichols2016}%
  \BibitemOpen
  \bibfield  {author} {\bibinfo {author} {\bibnamefont {Nichols}, \bibfnamefont
  {R.}}, \bibinfo {author} {\bibfnamefont {T.~R.}\ \bibnamefont {Bromley}},
  \bibinfo {author} {\bibfnamefont {L.~A.}\ \bibnamefont {Correa}}, \ and\
  \bibinfo {author} {\bibfnamefont {G.}~\bibnamefont {Adesso}}} (\bibinfo
  {year} {2016}),\ \href {\doibase 10.1103/PhysRevA.94.042101} {\bibfield
  {journal} {\bibinfo  {journal} {Phys. Rev. A}\ }\textbf {\bibinfo {volume}
  {94}},\ \bibinfo {pages} {042101}}\BibitemShut {NoStop}%
\bibitem [{\citenamefont {Nichols}\ \emph {et~al.}(2017)\citenamefont
  {Nichols}, \citenamefont {Liuzzo-Scorpo}, \citenamefont {Knott},\ and\
  \citenamefont {Adesso}}]{Liuzzo2017}%
  \BibitemOpen
  \bibfield  {author} {\bibinfo {author} {\bibnamefont {Nichols}, \bibfnamefont
  {R.}}, \bibinfo {author} {\bibfnamefont {P.}~\bibnamefont {Liuzzo-Scorpo}},
  \bibinfo {author} {\bibfnamefont {P.~A.}\ \bibnamefont {Knott}}, \ and\
  \bibinfo {author} {\bibfnamefont {G.}~\bibnamefont {Adesso}}} (\bibinfo
  {year} {2017}),\ \href@noop {} {\bibinfo  {journal} {Arxiv preprint
  arXiv:1711.09132}\ }\BibitemShut {NoStop}%
\bibitem [{\citenamefont {Nulton}\ and\ \citenamefont
  {Salamon}(1985)}]{Nulton1985}%
  \BibitemOpen
\bibfield  {journal} {  }\bibfield  {author} {\bibinfo {author} {\bibnamefont
  {Nulton}, \bibfnamefont {J.~D.}}, \ and\ \bibinfo {author} {\bibfnamefont
  {P.}~\bibnamefont {Salamon}}} (\bibinfo {year} {1985}),\ \href {\doibase
  10.1103/PhysRevA.31.2520} {\bibfield  {journal} {\bibinfo  {journal} {Phys.
  Rev. A}\ }\textbf {\bibinfo {volume} {31}},\ \bibinfo {pages}
  {2520}}\BibitemShut {NoStop}%
\bibitem [{\citenamefont {Okamoto}\ \emph {et~al.}(2012)\citenamefont
  {Okamoto}, \citenamefont {Iefuji}, \citenamefont {Oyama}, \citenamefont
  {Yamagata}, \citenamefont {Imai}, \citenamefont {Fujiwara},\ and\
  \citenamefont {Takeuchi}}]{okamoto_experimental_2012}%
  \BibitemOpen
  \bibfield  {author} {\bibinfo {author} {\bibnamefont {Okamoto}, \bibfnamefont
  {R.}}, \bibinfo {author} {\bibfnamefont {M.}~\bibnamefont {Iefuji}}, \bibinfo
  {author} {\bibfnamefont {S.}~\bibnamefont {Oyama}}, \bibinfo {author}
  {\bibfnamefont {K.}~\bibnamefont {Yamagata}}, \bibinfo {author}
  {\bibfnamefont {H.}~\bibnamefont {Imai}}, \bibinfo {author} {\bibfnamefont
  {A.}~\bibnamefont {Fujiwara}}, \ and\ \bibinfo {author} {\bibfnamefont
  {S.}~\bibnamefont {Takeuchi}}} (\bibinfo {year} {2012}),\ \href {\doibase
  10.1103/PhysRevLett.109.130404} {\bibfield  {journal} {\bibinfo  {journal}
  {Phys. Rev. Lett.}\ }\textbf {\bibinfo {volume} {109}}~(\bibinfo {number}
  {13}),\ \bibinfo {pages} {130404}}\BibitemShut {NoStop}%
\bibitem [{\citenamefont {Ollivier}\ and\ \citenamefont
  {Zurek}(2001)}]{Ollivier2001}%
  \BibitemOpen
  \bibfield  {author} {\bibinfo {author} {\bibnamefont {Ollivier},
  \bibfnamefont {H.}}, \ and\ \bibinfo {author} {\bibfnamefont {W.~H.}\
  \bibnamefont {Zurek}}} (\bibinfo {year} {2001}),\ \href {\doibase
  10.1103/PhysRevLett.88.017901} {\bibfield  {journal} {\bibinfo  {journal}
  {Phys. Rev. Lett.}\ }\textbf {\bibinfo {volume} {88}},\ \bibinfo {pages}
  {017901}}\BibitemShut {NoStop}%
\bibitem [{\citenamefont {Osborne}(1949)}]{Osborne1949}%
  \BibitemOpen
  \bibfield  {author} {\bibinfo {author} {\bibnamefont {Osborne}, \bibfnamefont
  {M.~F.~M.}}} (\bibinfo {year} {1949}),\ \href@noop {} {\bibfield  {journal}
  {\bibinfo  {journal} {Phys. Rev.}\ }\textbf {\bibinfo {volume} {76}},\
  \bibinfo {pages} {396}}\BibitemShut {NoStop}%
\bibitem [{\citenamefont {Oszmaniec}\ \emph {et~al.}(2016)\citenamefont
  {Oszmaniec}, \citenamefont {Augusiak}, \citenamefont {Gogolin}, \citenamefont
  {Ko\l{}ody\'{n}ski}, \citenamefont {Ac\'{\i}n},\ and\ \citenamefont
  {Lewenstein}}]{Oz16}%
  \BibitemOpen
  \bibfield  {author} {\bibinfo {author} {\bibnamefont {Oszmaniec},
  \bibfnamefont {M.}}, \bibinfo {author} {\bibfnamefont {R.}~\bibnamefont
  {Augusiak}}, \bibinfo {author} {\bibfnamefont {C.}~\bibnamefont {Gogolin}},
  \bibinfo {author} {\bibfnamefont {J.}~\bibnamefont {Ko\l{}ody\'{n}ski}},
  \bibinfo {author} {\bibfnamefont {A.}~\bibnamefont {Ac\'{\i}n}}, \ and\
  \bibinfo {author} {\bibfnamefont {M.}~\bibnamefont {Lewenstein}}} (\bibinfo
  {year} {2016}),\ \href {\doibase 10.1103/PhysRevX.6.041044} {\bibfield
  {journal} {\bibinfo  {journal} {Phys. Rev. X}\ }\textbf {\bibinfo {volume}
  {6}},\ \bibinfo {pages} {041044}}\BibitemShut {NoStop}%
\bibitem [{\citenamefont {Ourjoumtsev}\ \emph {et~al.}(2006)\citenamefont
  {Ourjoumtsev}, \citenamefont {Tualle-Brouri}, \citenamefont {Laurat},\ and\
  \citenamefont {Grangier}}]{ourjoumtsev_generating_2006}%
  \BibitemOpen
  \bibfield  {author} {\bibinfo {author} {\bibnamefont {Ourjoumtsev},
  \bibfnamefont {A.}}, \bibinfo {author} {\bibfnamefont {R.}~\bibnamefont
  {Tualle-Brouri}}, \bibinfo {author} {\bibfnamefont {J.}~\bibnamefont
  {Laurat}}, \ and\ \bibinfo {author} {\bibfnamefont {P.}~\bibnamefont
  {Grangier}}} (\bibinfo {year} {2006}),\ \href {\doibase
  10.1126/science.1122858} {\bibfield  {journal} {\bibinfo  {journal}
  {Science}\ }\textbf {\bibinfo {volume} {312}}~(\bibinfo {number} {5770}),\
  \bibinfo {pages} {83}}\BibitemShut {NoStop}%
\bibitem [{\citenamefont {Pang}\ and\ \citenamefont
  {Brun}(2014)}]{PhysRevA.90.022117}%
  \BibitemOpen
  \bibfield  {author} {\bibinfo {author} {\bibnamefont {Pang}, \bibfnamefont
  {S.}}, \ and\ \bibinfo {author} {\bibfnamefont {T.~A.}\ \bibnamefont {Brun}}}
  (\bibinfo {year} {2014}),\ \href {\doibase 10.1103/PhysRevA.90.022117}
  {\bibfield  {journal} {\bibinfo  {journal} {Phys. Rev. A}\ }\textbf {\bibinfo
  {volume} {90}},\ \bibinfo {pages} {022117}}\BibitemShut {NoStop}%
\bibitem [{\citenamefont {Pang}\ and\ \citenamefont
  {Brun}(2016)}]{PhysRevA.93.059901}%
  \BibitemOpen
  \bibfield  {author} {\bibinfo {author} {\bibnamefont {Pang}, \bibfnamefont
  {S.}}, \ and\ \bibinfo {author} {\bibfnamefont {T.~A.}\ \bibnamefont {Brun}}}
  (\bibinfo {year} {2016}),\ \href {\doibase 10.1103/PhysRevA.93.059901}
  {\bibfield  {journal} {\bibinfo  {journal} {Phys. Rev. A}\ }\textbf {\bibinfo
  {volume} {93}},\ \bibinfo {pages} {059901}}\BibitemShut {NoStop}%
\bibitem [{\citenamefont {Paraoanu}\ and\ \citenamefont
  {Scutaru}(2000)}]{Paraoanu2000}%
  \BibitemOpen
  \bibfield  {author} {\bibinfo {author} {\bibnamefont {Paraoanu},
  \bibfnamefont {G.}}, \ and\ \bibinfo {author} {\bibfnamefont
  {H.}~\bibnamefont {Scutaru}}} (\bibinfo {year} {2000}),\ \href@noop {}
  {\bibfield  {journal} {\bibinfo  {journal} {Physical Review A}\ }\textbf
  {\bibinfo {volume} {61}}~(\bibinfo {number} {2}),\ \bibinfo {pages}
  {022306}}\BibitemShut {NoStop}%
\bibitem [{\citenamefont {Paris}(2009)}]{Paris2009}%
  \BibitemOpen
  \bibfield  {author} {\bibinfo {author} {\bibnamefont {Paris}, \bibfnamefont
  {M.~G.~A.}}} (\bibinfo {year} {2009}),\ \href@noop {} {\bibfield  {journal}
  {\bibinfo  {journal} {Int. J. Quant. Inf.}\ }\textbf {\bibinfo {volume}
  {7}},\ \bibinfo {pages} {125}}\BibitemShut {NoStop}%
\bibitem [{\citenamefont {Paskauskas}\ and\ \citenamefont
  {You}(2001)}]{Paskauskas0}%
  \BibitemOpen
  \bibfield  {author} {\bibinfo {author} {\bibnamefont {Paskauskas},
  \bibfnamefont {R.}}, \ and\ \bibinfo {author} {\bibfnamefont
  {L.}~\bibnamefont {You}}} (\bibinfo {year} {2001}),\ \href@noop {} {\bibfield
   {journal} {\bibinfo  {journal} {Phys. Rev. A}\ }\textbf {\bibinfo {volume}
  {64}},\ \bibinfo {pages} {042310}}\BibitemShut {NoStop}%
\bibitem [{\citenamefont {Peres}(1993)}]{Peres93}%
  \BibitemOpen
  \bibfield  {author} {\bibinfo {author} {\bibnamefont {Peres}, \bibfnamefont
  {A.}}} (\bibinfo {year} {1993}),\ \href@noop {} {\emph {\bibinfo {title}
  {{Quantum Theory: Concepts and Methods}}}}\ (\bibinfo  {publisher} {Kluwer
  Academic Publishers},\ \bibinfo {address} {Dordrecht})\BibitemShut {NoStop}%
\bibitem [{\citenamefont {Pethick}\ and\ \citenamefont
  {Smith}(2004)}]{Pethick0}%
  \BibitemOpen
  \bibfield  {author} {\bibinfo {author} {\bibnamefont {Pethick}, \bibfnamefont
  {C.}}, \ and\ \bibinfo {author} {\bibfnamefont {H.}~\bibnamefont {Smith}}}
  (\bibinfo {year} {2004}),\ \href@noop {} {\emph {\bibinfo {title}
  {Bose-Einstein Condensation in Dilute Gases}}}\ (\bibinfo  {publisher}
  {Cambridge University Press},\ \bibinfo {address} {Cambridge})\BibitemShut
  {NoStop}%
\bibitem [{\citenamefont {Pezz\`e}\ and\ \citenamefont
  {Smerzi}(2009)}]{Smerzi0}%
  \BibitemOpen
  \bibfield  {author} {\bibinfo {author} {\bibnamefont {Pezz\`e}, \bibfnamefont
  {L.}}, \ and\ \bibinfo {author} {\bibfnamefont {A.}~\bibnamefont {Smerzi}}}
  (\bibinfo {year} {2009}),\ \href@noop {} {\bibfield  {journal} {\bibinfo
  {journal} {Phys. Rev. Lett.}\ }\textbf {\bibinfo {volume} {102}},\ \bibinfo
  {pages} {100401}}\BibitemShut {NoStop}%
\bibitem [{\citenamefont {Pezz\`e}\ and\ \citenamefont
  {Smerzi}(2014)}]{SmerziR}%
  \BibitemOpen
  \bibfield  {author} {\bibinfo {author} {\bibnamefont {Pezz\`e}, \bibfnamefont
  {L.}}, \ and\ \bibinfo {author} {\bibfnamefont {A.}~\bibnamefont {Smerzi}}}
  (\bibinfo {year} {2014}),\ \href@noop {} {\bibfield  {journal} {\bibinfo
  {journal} {Arxiv:1411.5164}\ }}\bibinfo {note} {Published in ''Atom
  Interferometry'', Proceedings of the International School of Physics ''Enrico
  Fermi'', Course 188, Varenna. Edited by G.M. Tino and M.A. Kasevich (IOS
  Press, Amsterdam, 2014). Page 691}\BibitemShut {NoStop}%
\bibitem [{\citenamefont {Pezz\`e}\ \emph {et~al.}(2016)\citenamefont
  {Pezz\`e}, \citenamefont {Smerzi}, \citenamefont {Oberthaler}, \citenamefont
  {Schmied},\ and\ \citenamefont {Treutlein}}]{pezze_non-classical_2016}%
  \BibitemOpen
  \bibfield  {author} {\bibinfo {author} {\bibnamefont {Pezz\`e}, \bibfnamefont
  {L.}}, \bibinfo {author} {\bibfnamefont {A.}~\bibnamefont {Smerzi}}, \bibinfo
  {author} {\bibfnamefont {M.~K.}\ \bibnamefont {Oberthaler}}, \bibinfo
  {author} {\bibfnamefont {R.}~\bibnamefont {Schmied}}, \ and\ \bibinfo
  {author} {\bibfnamefont {P.}~\bibnamefont {Treutlein}}} (\bibinfo {year}
  {2016}),\ \href {http://arxiv.org/abs/1609.01609} {\enquote {\bibinfo {title}
  {Non-classical states of atomic ensembles: fundamentals and applications in
  quantum metrology},}\ }\bibinfo {note} {ArXiv:1609.01609}\BibitemShut
  {NoStop}%
\bibitem [{\citenamefont {Pinel}\ \emph {et~al.}(2012)\citenamefont {Pinel},
  \citenamefont {Fade}, \citenamefont {Braun}, \citenamefont {Jian},
  \citenamefont {Treps},\ and\ \citenamefont {Fabre}}]{Pinel12}%
  \BibitemOpen
  \bibfield  {author} {\bibinfo {author} {\bibnamefont {Pinel}, \bibfnamefont
  {O.}}, \bibinfo {author} {\bibfnamefont {J.}~\bibnamefont {Fade}}, \bibinfo
  {author} {\bibfnamefont {D.}~\bibnamefont {Braun}}, \bibinfo {author}
  {\bibfnamefont {P.}~\bibnamefont {Jian}}, \bibinfo {author} {\bibfnamefont
  {N.}~\bibnamefont {Treps}}, \ and\ \bibinfo {author} {\bibfnamefont
  {C.}~\bibnamefont {Fabre}}} (\bibinfo {year} {2012}),\ \href@noop {}
  {\bibfield  {journal} {\bibinfo  {journal} {Phys. Rev. A}\ }\textbf {\bibinfo
  {volume} {85}},\ \bibinfo {pages} {010101}}\BibitemShut {NoStop}%
\bibitem [{\citenamefont {Pinel}\ \emph {et~al.}(2013)\citenamefont {Pinel},
  \citenamefont {Jian}, \citenamefont {Treps}, \citenamefont {Fabre},\ and\
  \citenamefont {Braun}}]{Pinel13}%
  \BibitemOpen
  \bibfield  {author} {\bibinfo {author} {\bibnamefont {Pinel}, \bibfnamefont
  {O.}}, \bibinfo {author} {\bibfnamefont {P.}~\bibnamefont {Jian}}, \bibinfo
  {author} {\bibfnamefont {N.}~\bibnamefont {Treps}}, \bibinfo {author}
  {\bibfnamefont {C.}~\bibnamefont {Fabre}}, \ and\ \bibinfo {author}
  {\bibfnamefont {D.}~\bibnamefont {Braun}}} (\bibinfo {year} {2013}),\ \href
  {\doibase 10.1103/PhysRevA.88.040102} {\bibfield  {journal} {\bibinfo
  {journal} {Phys. Rev. A}\ }\textbf {\bibinfo {volume} {88}},\ \bibinfo
  {pages} {040102}}\BibitemShut {NoStop}%
\bibitem [{\citenamefont {Pirandola}(2011)}]{Pirandola11}%
  \BibitemOpen
  \bibfield  {author} {\bibinfo {author} {\bibnamefont {Pirandola},
  \bibfnamefont {S.}}} (\bibinfo {year} {2011}),\ \href@noop {} {\bibfield
  {journal} {\bibinfo  {journal} {Phys. Rev. Lett.}\ }\textbf {\bibinfo
  {volume} {106}},\ \bibinfo {pages} {090504}}\BibitemShut {NoStop}%
\bibitem [{\citenamefont {Pirandola}(2016)}]{Pirandola16}%
  \BibitemOpen
  \bibfield  {author} {\bibinfo {author} {\bibnamefont {Pirandola},
  \bibfnamefont {S.}}} (\bibinfo {year} {2016}),\ \href@noop {} {\enquote
  {\bibinfo {title} {Capacities of repeater-assisted quantum communications},}\
  }\bibinfo {note} {Arxiv:1601.00966}\BibitemShut {NoStop}%
\bibitem [{\citenamefont {Pirandola}\ \emph {et~al.}(2017)\citenamefont
  {Pirandola}, \citenamefont {Laurenza}, \citenamefont {Ottaviani},\ and\
  \citenamefont {Banchi}}]{Pirandola15}%
  \BibitemOpen
  \bibfield  {author} {\bibinfo {author} {\bibnamefont {Pirandola},
  \bibfnamefont {S.}}, \bibinfo {author} {\bibfnamefont {R.}~\bibnamefont
  {Laurenza}}, \bibinfo {author} {\bibfnamefont {C.}~\bibnamefont {Ottaviani}},
  \ and\ \bibinfo {author} {\bibfnamefont {L.}~\bibnamefont {Banchi}}}
  (\bibinfo {year} {2017}),\ \href@noop {} {\bibfield  {journal} {\bibinfo
  {journal} {Nature Communications}\ }\textbf {\bibinfo {volume} {8}},\
  \bibinfo {pages} {15043}},\ \bibinfo {note} {arXiv:1510.08863
  (2015)}\BibitemShut {NoStop}%
\bibitem [{\citenamefont {Pirandola}\ and\ \citenamefont
  {Lloyd}(2008)}]{PirandolaLloyd08}%
  \BibitemOpen
  \bibfield  {author} {\bibinfo {author} {\bibnamefont {Pirandola},
  \bibfnamefont {S.}}, \ and\ \bibinfo {author} {\bibfnamefont
  {S.}~\bibnamefont {Lloyd}}} (\bibinfo {year} {2008}),\ \href@noop {}
  {\bibfield  {journal} {\bibinfo  {journal} {Phys. Rev. A}\ }\textbf {\bibinfo
  {volume} {78}},\ \bibinfo {pages} {012331}}\BibitemShut {NoStop}%
\bibitem [{\citenamefont {Pirandola}\ and\ \citenamefont
  {Lupo}(2017)}]{PirandolaLupo16}%
  \BibitemOpen
  \bibfield  {author} {\bibinfo {author} {\bibnamefont {Pirandola},
  \bibfnamefont {S.}}, \ and\ \bibinfo {author} {\bibfnamefont
  {C.}~\bibnamefont {Lupo}}} (\bibinfo {year} {2017}),\ \href@noop {}
  {\bibfield  {journal} {\bibinfo  {journal} {Phys. Rev. Lett.}\ }\textbf
  {\bibinfo {volume} {118}},\ \bibinfo {pages} {100502}}\BibitemShut {NoStop}%
\bibitem [{\citenamefont {Pirandola}\ \emph {et~al.}(2011)\citenamefont
  {Pirandola}, \citenamefont {Lupo}, \citenamefont {Giovannetti}, \citenamefont
  {Mancini},\ and\ \citenamefont {Braunstein}}]{Pirandola11b}%
  \BibitemOpen
  \bibfield  {author} {\bibinfo {author} {\bibnamefont {Pirandola},
  \bibfnamefont {S.}}, \bibinfo {author} {\bibfnamefont {C.}~\bibnamefont
  {Lupo}}, \bibinfo {author} {\bibfnamefont {V.}~\bibnamefont {Giovannetti}},
  \bibinfo {author} {\bibfnamefont {S.}~\bibnamefont {Mancini}}, \ and\
  \bibinfo {author} {\bibfnamefont {S.~L.}\ \bibnamefont {Braunstein}}}
  (\bibinfo {year} {2011}),\ \href@noop {} {\bibfield  {journal} {\bibinfo
  {journal} {New J. Phys.}\ }\textbf {\bibinfo {volume} {13}},\ \bibinfo
  {pages} {113012}}\BibitemShut {NoStop}%
\bibitem [{\citenamefont {Pitaevskii}\ and\ \citenamefont
  {Stringari}(2003)}]{Stringari0}%
  \BibitemOpen
  \bibfield  {author} {\bibinfo {author} {\bibnamefont {Pitaevskii},
  \bibfnamefont {L.}}, \ and\ \bibinfo {author} {\bibfnamefont
  {S.}~\bibnamefont {Stringari}}} (\bibinfo {year} {2003}),\ \href@noop {}
  {\emph {\bibinfo {title} {Bose-Einstein Condensation}}}\ (\bibinfo
  {publisher} {Oxford University Press})\BibitemShut {NoStop}%
\bibitem [{\citenamefont {Plastino}\ \emph {et~al.}(2009)\citenamefont
  {Plastino}, \citenamefont {Manzano},\ and\ \citenamefont
  {Dehesa}}]{Plastino0}%
  \BibitemOpen
  \bibfield  {author} {\bibinfo {author} {\bibnamefont {Plastino},
  \bibfnamefont {A.~R.}}, \bibinfo {author} {\bibfnamefont {D.}~\bibnamefont
  {Manzano}}, \ and\ \bibinfo {author} {\bibfnamefont {J.}~\bibnamefont
  {Dehesa}}} (\bibinfo {year} {2009}),\ \href@noop {} {\bibfield  {journal}
  {\bibinfo  {journal} {Europhys. Lett.}\ }\textbf {\bibinfo {volume} {86}},\
  \bibinfo {pages} {20005}}\BibitemShut {NoStop}%
\bibitem [{\citenamefont {Plenio}\ \emph {et~al.}(1999)\citenamefont {Plenio},
  \citenamefont {Huelga}, \citenamefont {Beige},\ and\ \citenamefont
  {Knight}}]{Plenio99}%
  \BibitemOpen
  \bibfield  {author} {\bibinfo {author} {\bibnamefont {Plenio}, \bibfnamefont
  {M.~B.}}, \bibinfo {author} {\bibfnamefont {S.~F.}\ \bibnamefont {Huelga}},
  \bibinfo {author} {\bibfnamefont {A.}~\bibnamefont {Beige}}, \ and\ \bibinfo
  {author} {\bibfnamefont {P.~L.}\ \bibnamefont {Knight}}} (\bibinfo {year}
  {1999}),\ \href@noop {} {\bibfield  {journal} {\bibinfo  {journal} {Phys.
  Rev. A}\ }\textbf {\bibinfo {volume} {59}}~(\bibinfo {number} {3}),\ \bibinfo
  {pages} {2468}}\BibitemShut {NoStop}%
\bibitem [{\citenamefont {Pope}\ \emph {et~al.}(2004)\citenamefont {Pope},
  \citenamefont {Wiseman},\ and\ \citenamefont
  {Langford}}]{PhysRevA.70.043812}%
  \BibitemOpen
  \bibfield  {author} {\bibinfo {author} {\bibnamefont {Pope}, \bibfnamefont
  {D.~T.}}, \bibinfo {author} {\bibfnamefont {H.~M.}\ \bibnamefont {Wiseman}},
  \ and\ \bibinfo {author} {\bibfnamefont {N.~K.}\ \bibnamefont {Langford}}}
  (\bibinfo {year} {2004}),\ \href {\doibase 10.1103/PhysRevA.70.043812}
  {\bibfield  {journal} {\bibinfo  {journal} {Phys. Rev. A}\ }\textbf {\bibinfo
  {volume} {70}},\ \bibinfo {pages} {043812}}\BibitemShut {NoStop}%
\bibitem [{\citenamefont {Popoviciu}(1935)}]{Popoviciu35}%
  \BibitemOpen
  \bibfield  {author} {\bibinfo {author} {\bibnamefont {Popoviciu},
  \bibfnamefont {T.}}} (\bibinfo {year} {1935}),\ \href@noop {} {\bibfield
  {journal} {\bibinfo  {journal} {Mathematica}\ }\textbf {\bibinfo {volume}
  {9}},\ \bibinfo {pages} {129}}\BibitemShut {NoStop}%
\bibitem [{\citenamefont {Proctor}\ \emph {et~al.}(2018)\citenamefont
  {Proctor}, \citenamefont {Knott},\ and\ \citenamefont
  {Dunningham}}]{proctor2018}%
  \BibitemOpen
  \bibfield  {author} {\bibinfo {author} {\bibnamefont {Proctor}, \bibfnamefont
  {T.~J.}}, \bibinfo {author} {\bibfnamefont {P.~A.}\ \bibnamefont {Knott}}, \
  and\ \bibinfo {author} {\bibfnamefont {J.~A.}\ \bibnamefont {Dunningham}}}
  (\bibinfo {year} {2018}),\ \href@noop {} {\bibfield  {journal} {\bibinfo
  {journal} {Phys. Rev. Lett.}\ }\textbf {\bibinfo {volume} {120}},\ \bibinfo
  {pages} {080501}}\BibitemShut {NoStop}%
\bibitem [{\citenamefont {Prokopenko}\ \emph {et~al.}(2011)\citenamefont
  {Prokopenko}, \citenamefont {Lizier}, \citenamefont {Obst},\ and\
  \citenamefont {Wang}}]{Prokopenko2011}%
  \BibitemOpen
  \bibfield  {author} {\bibinfo {author} {\bibnamefont {Prokopenko},
  \bibfnamefont {M.}}, \bibinfo {author} {\bibfnamefont {J.~T.}\ \bibnamefont
  {Lizier}}, \bibinfo {author} {\bibfnamefont {O.}~\bibnamefont {Obst}}, \ and\
  \bibinfo {author} {\bibfnamefont {X.~R.}\ \bibnamefont {Wang}}} (\bibinfo
  {year} {2011}),\ \href {\doibase 10.1103/PhysRevE.84.041116} {\bibfield
  {journal} {\bibinfo  {journal} {Phys. Rev. E}\ }\textbf {\bibinfo {volume}
  {84}},\ \bibinfo {pages} {041116}}\BibitemShut {NoStop}%
\bibitem [{\citenamefont {Prosen}(2015)}]{Prosen2015}%
  \BibitemOpen
  \bibfield  {author} {\bibinfo {author} {\bibnamefont {Prosen}, \bibfnamefont
  {T.}}} (\bibinfo {year} {2015}),\ \href@noop {} {\bibfield  {journal}
  {\bibinfo  {journal} {J. Phys. A}\ }\textbf {\bibinfo {volume} {48}},\
  \bibinfo {pages} {373001}}\BibitemShut {NoStop}%
\bibitem [{\citenamefont {Pustelny}\ \emph {et~al.}(2008)\citenamefont
  {Pustelny}, \citenamefont {Wojciechowski}, \citenamefont {Gring},
  \citenamefont {Kotyrba}, \citenamefont {Zachorowski},\ and\ \citenamefont
  {Gawlik}}]{PustelnyJAP2008}%
  \BibitemOpen
  \bibfield  {author} {\bibinfo {author} {\bibnamefont {Pustelny},
  \bibfnamefont {S.}}, \bibinfo {author} {\bibfnamefont {A.}~\bibnamefont
  {Wojciechowski}}, \bibinfo {author} {\bibfnamefont {M.}~\bibnamefont
  {Gring}}, \bibinfo {author} {\bibfnamefont {M.}~\bibnamefont {Kotyrba}},
  \bibinfo {author} {\bibfnamefont {J.}~\bibnamefont {Zachorowski}}, \ and\
  \bibinfo {author} {\bibfnamefont {W.}~\bibnamefont {Gawlik}}} (\bibinfo
  {year} {2008}),\ \href {\doibase 10.1063/1.2844494} {\bibfield  {journal}
  {\bibinfo  {journal} {Journal of Applied Physics}\ }\textbf {\bibinfo
  {volume} {103}}~(\bibinfo {number} {6}),\ \bibinfo {eid}
  {063108}}\BibitemShut {NoStop}%
\bibitem [{\citenamefont {Quan}\ and\ \citenamefont
  {Cucchietti}(2009)}]{Quan2009}%
  \BibitemOpen
  \bibfield  {author} {\bibinfo {author} {\bibnamefont {Quan}, \bibfnamefont
  {H.~T.}}, \ and\ \bibinfo {author} {\bibfnamefont {F.~M.}\ \bibnamefont
  {Cucchietti}}} (\bibinfo {year} {2009}),\ \href {\doibase
  10.1103/PhysRevE.79.031101} {\bibfield  {journal} {\bibinfo  {journal} {Phys.
  Rev. E}\ }\textbf {\bibinfo {volume} {79}},\ \bibinfo {pages}
  {031101}}\BibitemShut {NoStop}%
\bibitem [{\citenamefont {Ragy}\ and\ \citenamefont {Adesso}(2012)}]{Ragy2012}%
  \BibitemOpen
  \bibfield  {author} {\bibinfo {author} {\bibnamefont {Ragy}, \bibfnamefont
  {S.}}, \ and\ \bibinfo {author} {\bibfnamefont {G.}~\bibnamefont {Adesso}}}
  (\bibinfo {year} {2012}),\ \href {\doibase 10.1038/srep00651} {\bibfield
  {journal} {\bibinfo  {journal} {Scientific Reports}\ }\textbf {\bibinfo
  {volume} {2}},\ \bibinfo {pages} {651}}\BibitemShut {NoStop}%
\bibitem [{\citenamefont {Ragy}\ \emph {et~al.}(2016)\citenamefont {Ragy},
  \citenamefont {Jarzyna},\ and\ \citenamefont
  {Demkowicz-Dobrza\ifmmode~\acute{n}\else \'{n}\fi{}ski}}]{Ragy2016}%
  \BibitemOpen
  \bibfield  {author} {\bibinfo {author} {\bibnamefont {Ragy}, \bibfnamefont
  {S.}}, \bibinfo {author} {\bibfnamefont {M.}~\bibnamefont {Jarzyna}}, \ and\
  \bibinfo {author} {\bibfnamefont {R.}~\bibnamefont
  {Demkowicz-Dobrza\ifmmode~\acute{n}\else \'{n}\fi{}ski}}} (\bibinfo {year}
  {2016}),\ \href {\doibase 10.1103/PhysRevA.94.052108} {\bibfield  {journal}
  {\bibinfo  {journal} {Phys. Rev. A}\ }\textbf {\bibinfo {volume} {94}},\
  \bibinfo {pages} {052108}}\BibitemShut {NoStop}%
\bibitem [{\citenamefont {Raitz}\ \emph {et~al.}(2015)\citenamefont {Raitz},
  \citenamefont {Souza}, \citenamefont {Auccaise}, \citenamefont {Sarthour},\
  and\ \citenamefont {Oliveira}}]{Raitz2015}%
  \BibitemOpen
  \bibfield  {author} {\bibinfo {author} {\bibnamefont {Raitz}, \bibfnamefont
  {C.}}, \bibinfo {author} {\bibfnamefont {A.~M.}\ \bibnamefont {Souza}},
  \bibinfo {author} {\bibfnamefont {R.}~\bibnamefont {Auccaise}}, \bibinfo
  {author} {\bibfnamefont {R.~S.}\ \bibnamefont {Sarthour}}, \ and\ \bibinfo
  {author} {\bibfnamefont {I.~S.}\ \bibnamefont {Oliveira}}} (\bibinfo {year}
  {2015}),\ \href {\doibase 10.1007/s11128-014-0858-z} {\bibfield  {journal}
  {\bibinfo  {journal} {Quantum Information Processing}\ }\textbf {\bibinfo
  {volume} {14}}~(\bibinfo {number} {1}),\ \bibinfo {pages} {37}}\BibitemShut
  {NoStop}%
\bibitem [{\citenamefont {Ralph}(2002)}]{PhysRevA.65.042313}%
  \BibitemOpen
  \bibfield  {author} {\bibinfo {author} {\bibnamefont {Ralph}, \bibfnamefont
  {T.~C.}}} (\bibinfo {year} {2002}),\ \href {\doibase
  10.1103/PhysRevA.65.042313} {\bibfield  {journal} {\bibinfo  {journal} {Phys.
  Rev. A}\ }\textbf {\bibinfo {volume} {65}},\ \bibinfo {pages}
  {042313}}\BibitemShut {NoStop}%
\bibitem [{\citenamefont {Reeb}\ and\ \citenamefont
  {Wolf}(2015)}]{reeb_tight_2015}%
  \BibitemOpen
  \bibfield  {author} {\bibinfo {author} {\bibnamefont {Reeb}, \bibfnamefont
  {D.}}, \ and\ \bibinfo {author} {\bibfnamefont {M.~M.}\ \bibnamefont {Wolf}}}
  (\bibinfo {year} {2015}),\ \href {\doibase 10.1109/TIT.2014.2387822}
  {\bibfield  {journal} {\bibinfo  {journal} {IEEE Transactions on Information
  Theory}\ }\textbf {\bibinfo {volume} {61}}~(\bibinfo {number} {3}),\ \bibinfo
  {pages} {1458}}\BibitemShut {NoStop}%
\bibitem [{\citenamefont {Rehr}(1970)}]{Rehr1970}%
  \BibitemOpen
  \bibfield  {author} {\bibinfo {author} {\bibnamefont {Rehr}, \bibfnamefont
  {J.~J.}}} (\bibinfo {year} {1970}),\ \href@noop {} {\bibfield  {journal}
  {\bibinfo  {journal} {Phys. Rev. B}\ }\textbf {\bibinfo {volume} {1}},\
  \bibinfo {pages} {3160}}\BibitemShut {NoStop}%
\bibitem [{\citenamefont {Reichl}(1998)}]{Reichl1998}%
  \BibitemOpen
  \bibfield  {author} {\bibinfo {author} {\bibnamefont {Reichl}, \bibfnamefont
  {L.~E.}}} (\bibinfo {year} {1998}),\ \href@noop {} {\emph {\bibinfo {title}
  {A modern course in statistical physics}}}\ (\bibinfo  {publisher} {John
  Wiley \& Sons, Inc.})\BibitemShut {NoStop}%
\bibitem [{\citenamefont {Riedel}(2015)}]{PhysRevA.92.010101}%
  \BibitemOpen
  \bibfield  {author} {\bibinfo {author} {\bibnamefont {Riedel}, \bibfnamefont
  {C.~J.}}} (\bibinfo {year} {2015}),\ \href {\doibase
  10.1103/PhysRevA.92.010101} {\bibfield  {journal} {\bibinfo  {journal} {Phys.
  Rev. A}\ }\textbf {\bibinfo {volume} {92}},\ \bibinfo {pages}
  {010101}}\BibitemShut {NoStop}%
\bibitem [{\citenamefont {Riedel}\ \emph
  {et~al.}(2010{\natexlab{a}})\citenamefont {Riedel}, \citenamefont {B\"{o}hi},
  \citenamefont {Li}, \citenamefont {H\"{a}nsch}, \citenamefont {Sinatra},\
  and\ \citenamefont {Treutlein}}]{RiedelN2010}%
  \BibitemOpen
  \bibfield  {author} {\bibinfo {author} {\bibnamefont {Riedel}, \bibfnamefont
  {M.~F.}}, \bibinfo {author} {\bibfnamefont {P.}~\bibnamefont {B\"{o}hi}},
  \bibinfo {author} {\bibfnamefont {Y.}~\bibnamefont {Li}}, \bibinfo {author}
  {\bibfnamefont {T.~W.}\ \bibnamefont {H\"{a}nsch}}, \bibinfo {author}
  {\bibfnamefont {A.}~\bibnamefont {Sinatra}}, \ and\ \bibinfo {author}
  {\bibfnamefont {P.}~\bibnamefont {Treutlein}}} (\bibinfo {year}
  {2010}{\natexlab{a}}),\ \href {http://dx.doi.org/10.1038/nature08988}
  {\bibfield  {journal} {\bibinfo  {journal} {Nature}\ }\textbf {\bibinfo
  {volume} {464}}~(\bibinfo {number} {7292}),\ \bibinfo {pages}
  {1170}}\BibitemShut {NoStop}%
\bibitem [{\citenamefont {Riedel}\ \emph
  {et~al.}(2010{\natexlab{b}})\citenamefont {Riedel}, \citenamefont {B\"ohi},
  \citenamefont {Li}, \citenamefont {H\"ansch}, \citenamefont {Sinatra},
  \citenamefont {Treutlein},\ and\ \citenamefont {Riedel}}]{Treutlein10}%
  \BibitemOpen
  \bibfield  {author} {\bibinfo {author} {\bibnamefont {Riedel}, \bibfnamefont
  {M.~F.}}, \bibinfo {author} {\bibfnamefont {P.}~\bibnamefont {B\"ohi}},
  \bibinfo {author} {\bibfnamefont {Y.}~\bibnamefont {Li}}, \bibinfo {author}
  {\bibfnamefont {T.~W.}\ \bibnamefont {H\"ansch}}, \bibinfo {author}
  {\bibfnamefont {A.}~\bibnamefont {Sinatra}}, \bibinfo {author} {\bibfnamefont
  {P.}~\bibnamefont {Treutlein}}, \ and\ \bibinfo {author} {\bibfnamefont
  {M.}~\bibnamefont {Riedel}}} (\bibinfo {year} {2010}{\natexlab{b}}),\
  \href@noop {} {\bibfield  {journal} {\bibinfo  {journal} {Nature}\ }\textbf
  {\bibinfo {volume} {464}},\ \bibinfo {pages} {1170}}\BibitemShut {NoStop}%
\bibitem [{\citenamefont {Rigovacca}\ \emph {et~al.}(2015)\citenamefont
  {Rigovacca}, \citenamefont {Farace}, \citenamefont {De~Pasquale},\ and\
  \citenamefont {Giovannetti}}]{Rigovacca2015}%
  \BibitemOpen
  \bibfield  {author} {\bibinfo {author} {\bibnamefont {Rigovacca},
  \bibfnamefont {L.}}, \bibinfo {author} {\bibfnamefont {A.}~\bibnamefont
  {Farace}}, \bibinfo {author} {\bibfnamefont {A.}~\bibnamefont {De~Pasquale}},
  \ and\ \bibinfo {author} {\bibfnamefont {V.}~\bibnamefont {Giovannetti}}}
  (\bibinfo {year} {2015}),\ \href {\doibase 10.1103/PhysRevA.92.042331}
  {\bibfield  {journal} {\bibinfo  {journal} {Phys. Rev. A}\ }\textbf {\bibinfo
  {volume} {92}},\ \bibinfo {pages} {042331}}\BibitemShut {NoStop}%
\bibitem [{\citenamefont {Rigovacca}\ \emph {et~al.}(2017)\citenamefont
  {Rigovacca}, \citenamefont {Farace}, \citenamefont {Souza}, \citenamefont
  {De~Pasquale}, \citenamefont {Giovannetti},\ and\ \citenamefont
  {Adesso}}]{Rigovacca2017}%
  \BibitemOpen
  \bibfield  {author} {\bibinfo {author} {\bibnamefont {Rigovacca},
  \bibfnamefont {L.}}, \bibinfo {author} {\bibfnamefont {A.}~\bibnamefont
  {Farace}}, \bibinfo {author} {\bibfnamefont {L.~A.~M.}\ \bibnamefont
  {Souza}}, \bibinfo {author} {\bibfnamefont {A.}~\bibnamefont {De~Pasquale}},
  \bibinfo {author} {\bibfnamefont {V.}~\bibnamefont {Giovannetti}}, \ and\
  \bibinfo {author} {\bibfnamefont {G.}~\bibnamefont {Adesso}}} (\bibinfo
  {year} {2017}),\ \href {\doibase 10.1103/PhysRevA.95.052331} {\bibfield
  {journal} {\bibinfo  {journal} {Phys. Rev. A}\ }\textbf {\bibinfo {volume}
  {95}},\ \bibinfo {pages} {052331}}\BibitemShut {NoStop}%
\bibitem [{\citenamefont {Rivas}\ and\ \citenamefont
  {Luis}(2010)}]{RivasPRL2010}%
  \BibitemOpen
  \bibfield  {author} {\bibinfo {author} {\bibnamefont {Rivas}, \bibfnamefont
  {{\'A}.}}, \ and\ \bibinfo {author} {\bibfnamefont {A.}~\bibnamefont {Luis}}}
  (\bibinfo {year} {2010}),\ \href {\doibase 10.1103/PhysRevLett.105.010403}
  {\bibfield  {journal} {\bibinfo  {journal} {Phys. Rev. Lett.}\ }\textbf
  {\bibinfo {volume} {105}},\ \bibinfo {pages} {010403}}\BibitemShut {NoStop}%
\bibitem [{\citenamefont {Rivas}\ and\ \citenamefont
  {Luis}(2012)}]{RivasNJP2012}%
  \BibitemOpen
  \bibfield  {author} {\bibinfo {author} {\bibnamefont {Rivas}, \bibfnamefont
  {{\'A}.}}, \ and\ \bibinfo {author} {\bibfnamefont {A.}~\bibnamefont {Luis}}}
  (\bibinfo {year} {2012}),\ \href
  {http://stacks.iop.org/1367-2630/14/i=9/a=093052} {\bibfield  {journal}
  {\bibinfo  {journal} {New Journal of Physics}\ }\textbf {\bibinfo {volume}
  {14}}~(\bibinfo {number} {9}),\ \bibinfo {pages} {093052}}\BibitemShut
  {NoStop}%
\bibitem [{\citenamefont {Robertson}(1929)}]{RobertsonPR1929}%
  \BibitemOpen
  \bibfield  {author} {\bibinfo {author} {\bibnamefont {Robertson},
  \bibfnamefont {H.~P.}}} (\bibinfo {year} {1929}),\ \href {\doibase
  10.1103/PhysRev.34.163} {\bibfield  {journal} {\bibinfo  {journal} {Phys.
  Rev.}\ }\textbf {\bibinfo {volume} {34}},\ \bibinfo {pages}
  {163}}\BibitemShut {NoStop}%
\bibitem [{\citenamefont {Robillard}\ \emph {et~al.}(2008)\citenamefont
  {Robillard}, \citenamefont {Devos}, \citenamefont {Roch-Jeune},\ and\
  \citenamefont {Mante}}]{Robillard2008}%
  \BibitemOpen
  \bibfield  {author} {\bibinfo {author} {\bibnamefont {Robillard},
  \bibfnamefont {J.-F.}}, \bibinfo {author} {\bibfnamefont {A.}~\bibnamefont
  {Devos}}, \bibinfo {author} {\bibfnamefont {I.}~\bibnamefont {Roch-Jeune}}, \
  and\ \bibinfo {author} {\bibfnamefont {P.~A.}\ \bibnamefont {Mante}}}
  (\bibinfo {year} {2008}),\ \href {\doibase 10.1103/PhysRevB.78.064302}
  {\bibfield  {journal} {\bibinfo  {journal} {Phys. Rev. B}\ }\textbf {\bibinfo
  {volume} {78}},\ \bibinfo {pages} {064302}}\BibitemShut {NoStop}%
\bibitem [{\citenamefont {Roga}\ \emph {et~al.}(2015)\citenamefont {Roga},
  \citenamefont {Buono},\ and\ \citenamefont {Illuminati}}]{Roga2015}%
  \BibitemOpen
  \bibfield  {author} {\bibinfo {author} {\bibnamefont {Roga}, \bibfnamefont
  {W.}}, \bibinfo {author} {\bibfnamefont {D.}~\bibnamefont {Buono}}, \ and\
  \bibinfo {author} {\bibfnamefont {F.}~\bibnamefont {Illuminati}}} (\bibinfo
  {year} {2015}),\ \href {http://stacks.iop.org/1367-2630/17/i=1/a=013031}
  {\bibfield  {journal} {\bibinfo  {journal} {New Journal of Physics}\ }\textbf
  {\bibinfo {volume} {17}}~(\bibinfo {number} {1}),\ \bibinfo {pages}
  {013031}}\BibitemShut {NoStop}%
\bibitem [{\citenamefont {Roga}\ \emph {et~al.}(2014)\citenamefont {Roga},
  \citenamefont {Giampaolo},\ and\ \citenamefont {Illuminati}}]{Roga2014}%
  \BibitemOpen
  \bibfield  {author} {\bibinfo {author} {\bibnamefont {Roga}, \bibfnamefont
  {W.}}, \bibinfo {author} {\bibfnamefont {S.}~\bibnamefont {Giampaolo}}, \
  and\ \bibinfo {author} {\bibfnamefont {F.}~\bibnamefont {Illuminati}}}
  (\bibinfo {year} {2014}),\ \href@noop {} {\bibfield  {journal} {\bibinfo
  {journal} {J. Phys. A: Math. Theor.}\ }\textbf {\bibinfo {volume}
  {47}}~(\bibinfo {number} {36}),\ \bibinfo {pages} {365301}}\BibitemShut
  {NoStop}%
\bibitem [{\citenamefont {Roy}\ and\ \citenamefont
  {Braunstein}(2008)}]{RoyPRL2008}%
  \BibitemOpen
  \bibfield  {author} {\bibinfo {author} {\bibnamefont {Roy}, \bibfnamefont
  {S.~M.}}, \ and\ \bibinfo {author} {\bibfnamefont {S.~L.}\ \bibnamefont
  {Braunstein}}} (\bibinfo {year} {2008}),\ \href {\doibase
  10.1103/PhysRevLett.100.220501} {\bibfield  {journal} {\bibinfo  {journal}
  {Phys. Rev. Lett.}\ }\textbf {\bibinfo {volume} {100}},\ \bibinfo {pages}
  {220501}}\BibitemShut {NoStop}%
\bibitem [{\citenamefont {Ruppeiner}(1979)}]{Ruppeiner1979}%
  \BibitemOpen
  \bibfield  {author} {\bibinfo {author} {\bibnamefont {Ruppeiner},
  \bibfnamefont {G.}}} (\bibinfo {year} {1979}),\ \href {\doibase
  10.1103/PhysRevA.20.1608} {\bibfield  {journal} {\bibinfo  {journal} {Phys.
  Rev. A}\ }\textbf {\bibinfo {volume} {20}},\ \bibinfo {pages}
  {1608}}\BibitemShut {NoStop}%
\bibitem [{\citenamefont {Ruppeiner}(1981)}]{Ruppeiner1981}%
  \BibitemOpen
  \bibfield  {author} {\bibinfo {author} {\bibnamefont {Ruppeiner},
  \bibfnamefont {G.}}} (\bibinfo {year} {1981}),\ \href {\doibase
  10.1103/PhysRevA.24.488} {\bibfield  {journal} {\bibinfo  {journal} {Phys.
  Rev. A}\ }\textbf {\bibinfo {volume} {24}},\ \bibinfo {pages}
  {488}}\BibitemShut {NoStop}%
\bibitem [{\citenamefont {Ruppeiner}(1991)}]{Ruppeiner1991}%
  \BibitemOpen
  \bibfield  {author} {\bibinfo {author} {\bibnamefont {Ruppeiner},
  \bibfnamefont {G.}}} (\bibinfo {year} {1991}),\ \href {\doibase
  10.1103/PhysRevA.44.3583} {\bibfield  {journal} {\bibinfo  {journal} {Phys.
  Rev. A}\ }\textbf {\bibinfo {volume} {44}},\ \bibinfo {pages}
  {3583}}\BibitemShut {NoStop}%
\bibitem [{\citenamefont {Ruppeiner}(1995)}]{Ruppeiner1995}%
  \BibitemOpen
  \bibfield  {author} {\bibinfo {author} {\bibnamefont {Ruppeiner},
  \bibfnamefont {G.}}} (\bibinfo {year} {1995}),\ \href {\doibase
  10.1103/RevModPhys.67.605} {\bibfield  {journal} {\bibinfo  {journal} {Rev.
  Mod. Phys.}\ }\textbf {\bibinfo {volume} {67}},\ \bibinfo {pages}
  {605}}\BibitemShut {NoStop}%
\bibitem [{\citenamefont {Sacchi}(2005)}]{Sacchi05}%
  \BibitemOpen
  \bibfield  {author} {\bibinfo {author} {\bibnamefont {Sacchi}, \bibfnamefont
  {M.}}} (\bibinfo {year} {2005}),\ \href@noop {} {\bibfield  {journal}
  {\bibinfo  {journal} {Phys. Rev. A}\ }\textbf {\bibinfo {volume} {72}},\
  \bibinfo {pages} {014305}}\BibitemShut {NoStop}%
\bibitem [{\citenamefont {Sachdev}(1999)}]{Sachdev}%
  \BibitemOpen
  \bibfield  {author} {\bibinfo {author} {\bibnamefont {Sachdev}, \bibfnamefont
  {S.}}} (\bibinfo {year} {1999}),\ \href@noop {} {\emph {\bibinfo {title}
  {{Quantum Phase Transitions}}}}\ (\bibinfo  {publisher} {Cambridge University
  Press})\BibitemShut {NoStop}%
\bibitem [{\citenamefont {Sacramento}\ \emph {et~al.}(2011)\citenamefont
  {Sacramento}, \citenamefont {Paunkovi\ifmmode~\acute{c}\else \'{c}\fi{}},\
  and\ \citenamefont {Vieira}}]{Sacramento2011}%
  \BibitemOpen
  \bibfield  {author} {\bibinfo {author} {\bibnamefont {Sacramento},
  \bibfnamefont {P.~D.}}, \bibinfo {author} {\bibfnamefont {N.}~\bibnamefont
  {Paunkovi\ifmmode~\acute{c}\else \'{c}\fi{}}}, \ and\ \bibinfo {author}
  {\bibfnamefont {V.~R.}\ \bibnamefont {Vieira}}} (\bibinfo {year} {2011}),\
  \href {\doibase 10.1103/PhysRevA.84.062318} {\bibfield  {journal} {\bibinfo
  {journal} {Phys. Rev. A}\ }\textbf {\bibinfo {volume} {84}},\ \bibinfo
  {pages} {062318}}\BibitemShut {NoStop}%
\bibitem [{\citenamefont {Sakurai}(1994)}]{Sakurai0}%
  \BibitemOpen
  \bibfield  {author} {\bibinfo {author} {\bibnamefont {Sakurai}, \bibfnamefont
  {J.}}} (\bibinfo {year} {1994}),\ \href@noop {} {\emph {\bibinfo {title}
  {Modern Quantum Mechanics}}}\ (\bibinfo  {publisher} {Addison-Wesley},\
  \bibinfo {address} {Reading (MA)})\BibitemShut {NoStop}%
\bibitem [{\citenamefont {Salamon}\ \emph {et~al.}(1984)\citenamefont
  {Salamon}, \citenamefont {Nulton},\ and\ \citenamefont
  {Ihrig}}]{Salamon1984}%
  \BibitemOpen
  \bibfield  {author} {\bibinfo {author} {\bibnamefont {Salamon}, \bibfnamefont
  {P.}}, \bibinfo {author} {\bibfnamefont {J.}~\bibnamefont {Nulton}}, \ and\
  \bibinfo {author} {\bibfnamefont {E.}~\bibnamefont {Ihrig}}} (\bibinfo {year}
  {1984}),\ \href@noop {} {\bibfield  {journal} {\bibinfo  {journal} {J. Chem.
  Phys.}\ }\textbf {\bibinfo {volume} {80}},\ \bibinfo {pages}
  {436}}\BibitemShut {NoStop}%
\bibitem [{\citenamefont {Sanders}\ and\ \citenamefont
  {Milburn}(1995)}]{Sanders0}%
  \BibitemOpen
  \bibfield  {author} {\bibinfo {author} {\bibnamefont {Sanders}, \bibfnamefont
  {B.}}, \ and\ \bibinfo {author} {\bibfnamefont {G.}~\bibnamefont {Milburn}}}
  (\bibinfo {year} {1995}),\ \href@noop {} {\bibfield  {journal} {\bibinfo
  {journal} {Phys. Rev. Lett.}\ }\textbf {\bibinfo {volume} {75}},\ \bibinfo
  {pages} {2944}}\BibitemShut {NoStop}%
\bibitem [{\citenamefont {Sanner}\ \emph {et~al.}(2010)\citenamefont {Sanner},
  \citenamefont {Su}, \citenamefont {Keshet}, \citenamefont {Gommers},
  \citenamefont {Shin}, \citenamefont {Huang},\ and\ \citenamefont
  {Ketterle}}]{Sanner2010}%
  \BibitemOpen
  \bibfield  {author} {\bibinfo {author} {\bibnamefont {Sanner}, \bibfnamefont
  {C.}}, \bibinfo {author} {\bibfnamefont {E.~J.}\ \bibnamefont {Su}}, \bibinfo
  {author} {\bibfnamefont {A.}~\bibnamefont {Keshet}}, \bibinfo {author}
  {\bibfnamefont {R.}~\bibnamefont {Gommers}}, \bibinfo {author} {\bibfnamefont
  {Y.-i.}\ \bibnamefont {Shin}}, \bibinfo {author} {\bibfnamefont
  {W.}~\bibnamefont {Huang}}, \ and\ \bibinfo {author} {\bibfnamefont
  {W.}~\bibnamefont {Ketterle}}} (\bibinfo {year} {2010}),\ \href {\doibase
  10.1103/PhysRevLett.105.040402} {\bibfield  {journal} {\bibinfo  {journal}
  {Phys. Rev. Lett.}\ }\textbf {\bibinfo {volume} {105}},\ \bibinfo {pages}
  {040402}}\BibitemShut {NoStop}%
\bibitem [{\citenamefont {Sattler}(2011)}]{Sattler2011}%
  \BibitemOpen
  \bibfield  {author} {\bibinfo {author} {\bibnamefont {Sattler}, \bibfnamefont
  {K.~D.}}} (\bibinfo {year} {2011}),\ \href@noop {} {\emph {\bibinfo {title}
  {{Handbook of Nanophysics Principles and Methods}}}}\ (\bibinfo  {publisher}
  {CRC Press,Taylor \& Francis Group})\BibitemShut {NoStop}%
\bibitem [{\citenamefont {Schleich}(2001)}]{Schleich01}%
  \BibitemOpen
  \bibfield  {author} {\bibinfo {author} {\bibnamefont {Schleich},
  \bibfnamefont {W.}}} (\bibinfo {year} {2001}),\ \href@noop {} {\emph
  {\bibinfo {title} {Quantum Optics in Phase Space}}}\ (\bibinfo  {publisher}
  {Wiley-VCH Verlag Berlin GmbH},\ \bibinfo {address} {Berlin,
  Germany})\BibitemShut {NoStop}%
\bibitem [{\citenamefont {Schliemann}\ \emph {et~al.}(2001)\citenamefont
  {Schliemann}, \citenamefont {Cirac}, \citenamefont {Kus}, \citenamefont
  {Lewenstein},\ and\ \citenamefont {Loss}}]{Schliemann0}%
  \BibitemOpen
  \bibfield  {author} {\bibinfo {author} {\bibnamefont {Schliemann},
  \bibfnamefont {J.}}, \bibinfo {author} {\bibfnamefont {J.}~\bibnamefont
  {Cirac}}, \bibinfo {author} {\bibfnamefont {M.}~\bibnamefont {Kus}}, \bibinfo
  {author} {\bibfnamefont {M.}~\bibnamefont {Lewenstein}}, \ and\ \bibinfo
  {author} {\bibfnamefont {D.}~\bibnamefont {Loss}}} (\bibinfo {year} {2001}),\
  \href@noop {} {\bibfield  {journal} {\bibinfo  {journal} {Phys. Rev. A}\
  }\textbf {\bibinfo {volume} {64}},\ \bibinfo {pages} {022303}}\BibitemShut
  {NoStop}%
\bibitem [{\citenamefont {Schmitt}\ \emph {et~al.}(2014)\citenamefont
  {Schmitt}, \citenamefont {Damm}, \citenamefont {Dung}, \citenamefont
  {Vewinger}, \citenamefont {Klaers},\ and\ \citenamefont
  {Weitz}}]{Schmitt2014}%
  \BibitemOpen
  \bibfield  {author} {\bibinfo {author} {\bibnamefont {Schmitt}, \bibfnamefont
  {J.}}, \bibinfo {author} {\bibfnamefont {T.}~\bibnamefont {Damm}}, \bibinfo
  {author} {\bibfnamefont {D.}~\bibnamefont {Dung}}, \bibinfo {author}
  {\bibfnamefont {F.}~\bibnamefont {Vewinger}}, \bibinfo {author}
  {\bibfnamefont {J.}~\bibnamefont {Klaers}}, \ and\ \bibinfo {author}
  {\bibfnamefont {M.}~\bibnamefont {Weitz}}} (\bibinfo {year} {2014}),\ \href
  {\doibase 10.1103/PhysRevLett.112.030401} {\bibfield  {journal} {\bibinfo
  {journal} {Phys. Rev. Lett.}\ }\textbf {\bibinfo {volume} {112}},\ \bibinfo
  {pages} {030401}}\BibitemShut {NoStop}%
\bibitem [{\citenamefont {Schuch}\ \emph {et~al.}(2004)\citenamefont {Schuch},
  \citenamefont {Verstraete},\ and\ \citenamefont {Cirac}}]{Verstraete0}%
  \BibitemOpen
  \bibfield  {author} {\bibinfo {author} {\bibnamefont {Schuch}, \bibfnamefont
  {N.}}, \bibinfo {author} {\bibfnamefont {F.}~\bibnamefont {Verstraete}}, \
  and\ \bibinfo {author} {\bibfnamefont {J.}~\bibnamefont {Cirac}}} (\bibinfo
  {year} {2004}),\ \href@noop {} {\bibfield  {journal} {\bibinfo  {journal}
  {Phys. Rev. A}\ }\textbf {\bibinfo {volume} {70}},\ \bibinfo {pages}
  {042310}}\BibitemShut {NoStop}%
\bibitem [{\citenamefont {Scutaru}(1998)}]{Scutaru98}%
  \BibitemOpen
  \bibfield  {author} {\bibinfo {author} {\bibnamefont {Scutaru}, \bibfnamefont
  {H.}}} (\bibinfo {year} {1998}),\ \href@noop {} {\bibfield  {journal}
  {\bibinfo  {journal} {J. Phys. A}\ }\textbf {\bibinfo {volume} {31}},\
  \bibinfo {pages} {3659}}\BibitemShut {NoStop}%
\bibitem [{\citenamefont {Segovia}\ \emph {et~al.}(1999)\citenamefont
  {Segovia}, \citenamefont {Purdie}, \citenamefont {Hengsberger},\ and\
  \citenamefont {Baer}}]{Segovia1999}%
  \BibitemOpen
  \bibfield  {author} {\bibinfo {author} {\bibnamefont {Segovia}, \bibfnamefont
  {P.}}, \bibinfo {author} {\bibfnamefont {D.}~\bibnamefont {Purdie}}, \bibinfo
  {author} {\bibfnamefont {M.}~\bibnamefont {Hengsberger}}, \ and\ \bibinfo
  {author} {\bibfnamefont {Y.}~\bibnamefont {Baer}}} (\bibinfo {year} {1999}),\
  \href@noop {} {\bibfield  {journal} {\bibinfo  {journal} {Nature}\ }\textbf
  {\bibinfo {volume} {402}},\ \bibinfo {pages} {504}}\BibitemShut {NoStop}%
\bibitem [{\citenamefont {Sekatski}\ \emph {et~al.}(2017)\citenamefont
  {Sekatski}, \citenamefont {Skotiniotis}, \citenamefont
  {Ko{\l{}}ody{\'{n}}ski},\ and\ \citenamefont {D{\"{u}}r}}]{Sekatski2016}%
  \BibitemOpen
  \bibfield  {author} {\bibinfo {author} {\bibnamefont {Sekatski},
  \bibfnamefont {P.}}, \bibinfo {author} {\bibfnamefont {M.}~\bibnamefont
  {Skotiniotis}}, \bibinfo {author} {\bibfnamefont {J.}~\bibnamefont
  {Ko{\l{}}ody{\'{n}}ski}}, \ and\ \bibinfo {author} {\bibfnamefont
  {W.}~\bibnamefont {D{\"{u}}r}}} (\bibinfo {year} {2017}),\ \href {\doibase
  10.22331/q-2017-09-06-27} {\bibfield  {journal} {\bibinfo  {journal}
  {{Quantum}}\ }\textbf {\bibinfo {volume} {1}},\ \bibinfo {pages}
  {27}}\BibitemShut {NoStop}%
\bibitem [{\citenamefont {Serafini}(2012)}]{serafini_feedback_2012}%
  \BibitemOpen
  \bibfield  {author} {\bibinfo {author} {\bibnamefont {Serafini},
  \bibfnamefont {A.}}} (\bibinfo {year} {2012}),\ \href {\doibase
  10.5402/2012/275016} {\bibfield  {journal} {\bibinfo  {journal}
  {International Scholarly Research Notices}\ }\textbf {\bibinfo {volume}
  {2012}},\ \bibinfo {pages} {e275016}}\BibitemShut {NoStop}%
\bibitem [{\citenamefont {Sewell}\ \emph {et~al.}(2012)\citenamefont {Sewell},
  \citenamefont {Koschorreck}, \citenamefont {Napolitano}, \citenamefont
  {Dubost}, \citenamefont {Behbood},\ and\ \citenamefont
  {Mitchell}}]{SewellPRL2012}%
  \BibitemOpen
  \bibfield  {author} {\bibinfo {author} {\bibnamefont {Sewell}, \bibfnamefont
  {R.~J.}}, \bibinfo {author} {\bibfnamefont {M.}~\bibnamefont {Koschorreck}},
  \bibinfo {author} {\bibfnamefont {M.}~\bibnamefont {Napolitano}}, \bibinfo
  {author} {\bibfnamefont {B.}~\bibnamefont {Dubost}}, \bibinfo {author}
  {\bibfnamefont {N.}~\bibnamefont {Behbood}}, \ and\ \bibinfo {author}
  {\bibfnamefont {M.~W.}\ \bibnamefont {Mitchell}}} (\bibinfo {year} {2012}),\
  \href {\doibase 10.1103/PhysRevLett.109.253605} {\bibfield  {journal}
  {\bibinfo  {journal} {Phys. Rev. Lett.}\ }\textbf {\bibinfo {volume} {109}},\
  \bibinfo {pages} {253605}}\BibitemShut {NoStop}%
\bibitem [{\citenamefont {Sewell}\ \emph {et~al.}(2014)\citenamefont {Sewell},
  \citenamefont {Napolitano}, \citenamefont {Behbood}, \citenamefont
  {Colangelo}, \citenamefont {Martin~Ciurana},\ and\ \citenamefont
  {Mitchell}}]{SewellPRX2014}%
  \BibitemOpen
  \bibfield  {author} {\bibinfo {author} {\bibnamefont {Sewell}, \bibfnamefont
  {R.~J.}}, \bibinfo {author} {\bibfnamefont {M.}~\bibnamefont {Napolitano}},
  \bibinfo {author} {\bibfnamefont {N.}~\bibnamefont {Behbood}}, \bibinfo
  {author} {\bibfnamefont {G.}~\bibnamefont {Colangelo}}, \bibinfo {author}
  {\bibfnamefont {F.}~\bibnamefont {Martin~Ciurana}}, \ and\ \bibinfo {author}
  {\bibfnamefont {M.~W.}\ \bibnamefont {Mitchell}}} (\bibinfo {year} {2014}),\
  \href {\doibase 10.1103/PhysRevX.4.021045} {\bibfield  {journal} {\bibinfo
  {journal} {Phys. Rev. X}\ }\textbf {\bibinfo {volume} {4}},\ \bibinfo {pages}
  {021045}}\BibitemShut {NoStop}%
\bibitem [{\citenamefont {Shah}\ \emph {et~al.}(2010)\citenamefont {Shah},
  \citenamefont {Vasilakis},\ and\ \citenamefont {Romalis}}]{ShahPRL2010}%
  \BibitemOpen
  \bibfield  {author} {\bibinfo {author} {\bibnamefont {Shah}, \bibfnamefont
  {V.}}, \bibinfo {author} {\bibfnamefont {G.}~\bibnamefont {Vasilakis}}, \
  and\ \bibinfo {author} {\bibfnamefont {M.~V.}\ \bibnamefont {Romalis}}}
  (\bibinfo {year} {2010}),\ \href {\doibase 10.1103/PhysRevLett.104.013601}
  {\bibfield  {journal} {\bibinfo  {journal} {Phys. Rev. Lett.}\ }\textbf
  {\bibinfo {volume} {104}}~(\bibinfo {number} {1}),\ \bibinfo {pages}
  {013601}}\BibitemShut {NoStop}%
\bibitem [{\citenamefont {Shor}(1995)}]{Shor95}%
  \BibitemOpen
  \bibfield  {author} {\bibinfo {author} {\bibnamefont {Shor}, \bibfnamefont
  {P.~W.}}} (\bibinfo {year} {1995}),\ \href@noop {} {\bibfield  {journal}
  {\bibinfo  {journal} {Phys. Rev. A}\ }\textbf {\bibinfo {volume} {52}},\
  \bibinfo {pages} {R2493}}\BibitemShut {NoStop}%
\bibitem [{\citenamefont {S\"oderholm}\ \emph {et~al.}(2003)\citenamefont
  {S\"oderholm}, \citenamefont {Bj\"ork}, \citenamefont {Hessmo},\ and\
  \citenamefont {Inoue}}]{Soderholm0}%
  \BibitemOpen
  \bibfield  {author} {\bibinfo {author} {\bibnamefont {S\"oderholm},
  \bibfnamefont {J.}}, \bibinfo {author} {\bibfnamefont {G.}~\bibnamefont
  {Bj\"ork}}, \bibinfo {author} {\bibfnamefont {B.}~\bibnamefont {Hessmo}}, \
  and\ \bibinfo {author} {\bibfnamefont {S.}~\bibnamefont {Inoue}}} (\bibinfo
  {year} {2003}),\ \href {\doibase 10.1103/PhysRevA.67.053803} {\bibfield
  {journal} {\bibinfo  {journal} {Phys. Rev. A}\ }\textbf {\bibinfo {volume}
  {67}},\ \bibinfo {pages} {053803}}\BibitemShut {NoStop}%
\bibitem [{\citenamefont {Sommers}\ and\ \citenamefont
  {Zyczkowski}(2003)}]{Sommers2003}%
  \BibitemOpen
  \bibfield  {author} {\bibinfo {author} {\bibnamefont {Sommers}, \bibfnamefont
  {H.-J.}}, \ and\ \bibinfo {author} {\bibfnamefont {K.}~\bibnamefont
  {Zyczkowski}}} (\bibinfo {year} {2003}),\ \href
  {http://stacks.iop.org/0305-4470/36/i=39/a=308} {\bibfield  {journal}
  {\bibinfo  {journal} {J. Phys A: Math. Gen.}\ }\textbf {\bibinfo {volume}
  {36}}~(\bibinfo {number} {39}),\ \bibinfo {pages} {10083}}\BibitemShut
  {NoStop}%
\bibitem [{\citenamefont {Song}\ \emph {et~al.}(2012)\citenamefont {Song},
  \citenamefont {Rachel}, \citenamefont {Flindt}, \citenamefont {Klich},
  \citenamefont {Laflorencie},\ and\ \citenamefont {Le~Hur}}]{Song0}%
  \BibitemOpen
  \bibfield  {author} {\bibinfo {author} {\bibnamefont {Song}, \bibfnamefont
  {H.~F.}}, \bibinfo {author} {\bibfnamefont {S.}~\bibnamefont {Rachel}},
  \bibinfo {author} {\bibfnamefont {C.}~\bibnamefont {Flindt}}, \bibinfo
  {author} {\bibfnamefont {I.}~\bibnamefont {Klich}}, \bibinfo {author}
  {\bibfnamefont {N.}~\bibnamefont {Laflorencie}}, \ and\ \bibinfo {author}
  {\bibfnamefont {K.}~\bibnamefont {Le~Hur}}} (\bibinfo {year} {2012}),\ \href
  {\doibase 10.1103/PhysRevB.85.035409} {\bibfield  {journal} {\bibinfo
  {journal} {Phys. Rev. B}\ }\textbf {\bibinfo {volume} {85}},\ \bibinfo
  {pages} {035409}}\BibitemShut {NoStop}%
\bibitem [{\citenamefont {Sonin}(1969)}]{Sonin1969}%
  \BibitemOpen
  \bibfield  {author} {\bibinfo {author} {\bibnamefont {Sonin}, \bibfnamefont
  {E.~A.}}} (\bibinfo {year} {1969}),\ \href@noop {} {\bibfield  {journal}
  {\bibinfo  {journal} {Sov. Phys. JEPT}\ }\textbf {\bibinfo {volume} {29}},\
  \bibinfo {pages} {520}}\BibitemShut {NoStop}%
\bibitem [{\citenamefont {S\o{}rensen}\ \emph {et~al.}(2001)\citenamefont
  {S\o{}rensen}, \citenamefont {Duan}, \citenamefont {Cirac},\ and\
  \citenamefont {Zoller}}]{Sorensen0}%
  \BibitemOpen
  \bibfield  {author} {\bibinfo {author} {\bibnamefont {S\o{}rensen},
  \bibfnamefont {A.}}, \bibinfo {author} {\bibfnamefont {L.-M.}\ \bibnamefont
  {Duan}}, \bibinfo {author} {\bibfnamefont {J.}~\bibnamefont {Cirac}}, \ and\
  \bibinfo {author} {\bibfnamefont {P.}~\bibnamefont {Zoller}}} (\bibinfo
  {year} {2001}),\ \href@noop {} {\bibfield  {journal} {\bibinfo  {journal}
  {Nature}\ }\textbf {\bibinfo {volume} {409}},\ \bibinfo {pages}
  {63}}\BibitemShut {NoStop}%
\bibitem [{\citenamefont {Spedalieri}\ \emph {et~al.}(2016)\citenamefont
  {Spedalieri}, \citenamefont {Braunstein},\ and\ \citenamefont
  {Pirandola}}]{Spedalieri16}%
  \BibitemOpen
  \bibfield  {author} {\bibinfo {author} {\bibnamefont {Spedalieri},
  \bibfnamefont {G.}}, \bibinfo {author} {\bibfnamefont {S.~L.}\ \bibnamefont
  {Braunstein}}, \ and\ \bibinfo {author} {\bibfnamefont {S.}~\bibnamefont
  {Pirandola}}} (\bibinfo {year} {2016}),\ \href@noop {} {\enquote {\bibinfo
  {title} {Thermal quantum metrology},}\ }\bibinfo {note}
  {Arxiv:1602.05958}\BibitemShut {NoStop}%
\bibitem [{\citenamefont {Spedalieri}\ \emph {et~al.}(2013)\citenamefont
  {Spedalieri}, \citenamefont {Weedbrook},\ and\ \citenamefont
  {Pirandola}}]{Spedalieri13}%
  \BibitemOpen
  \bibfield  {author} {\bibinfo {author} {\bibnamefont {Spedalieri},
  \bibfnamefont {G.}}, \bibinfo {author} {\bibfnamefont {C.}~\bibnamefont
  {Weedbrook}}, \ and\ \bibinfo {author} {\bibfnamefont {S.}~\bibnamefont
  {Pirandola}}} (\bibinfo {year} {2013}),\ \href@noop {} {\bibfield  {journal}
  {\bibinfo  {journal} {J. Phys. A: Math. Theor.}\ }\textbf {\bibinfo {volume}
  {46}},\ \bibinfo {pages} {025304}}\BibitemShut {NoStop}%
\bibitem [{\citenamefont {Stace}(2010)}]{Stace2010}%
  \BibitemOpen
  \bibfield  {author} {\bibinfo {author} {\bibnamefont {Stace}, \bibfnamefont
  {T.~M.}}} (\bibinfo {year} {2010}),\ \href {\doibase
  10.1103/PhysRevA.82.011611} {\bibfield  {journal} {\bibinfo  {journal} {Phys.
  Rev. A}\ }\textbf {\bibinfo {volume} {82}},\ \bibinfo {pages} {011611}},\
  \bibinfo {note} {note that there are typos in equations (13) and (17) of
  \cite{Stace2010} which are corrected in equation \eqref{sensit.interf} of
  this review (privite communication with Thomas M. Stace).}\BibitemShut
  {Stop}%
\bibitem [{\citenamefont {Steane}(1996)}]{Steane96}%
  \BibitemOpen
  \bibfield  {author} {\bibinfo {author} {\bibnamefont {Steane}, \bibfnamefont
  {A.~M.}}} (\bibinfo {year} {1996}),\ \href@noop {} {\bibfield  {journal}
  {\bibinfo  {journal} {Phys. Rev. Lett.}\ }\textbf {\bibinfo {volume} {77}},\
  \bibinfo {pages} {793}}\BibitemShut {NoStop}%
\bibitem [{\citenamefont {Streltsov}\ \emph {et~al.}(2017)\citenamefont
  {Streltsov}, \citenamefont {Adesso},\ and\ \citenamefont
  {Plenio}}]{Streltsov2016C}%
  \BibitemOpen
  \bibfield  {author} {\bibinfo {author} {\bibnamefont {Streltsov},
  \bibfnamefont {A.}}, \bibinfo {author} {\bibfnamefont {G.}~\bibnamefont
  {Adesso}}, \ and\ \bibinfo {author} {\bibfnamefont {M.~B.}\ \bibnamefont
  {Plenio}}} (\bibinfo {year} {2017}),\ \href {\doibase
  10.1103/RevModPhys.89.041003} {\bibfield  {journal} {\bibinfo  {journal}
  {Rev. Mod. Phys.}\ }\textbf {\bibinfo {volume} {89}},\ \bibinfo {pages}
  {041003}}\BibitemShut {NoStop}%
\bibitem [{\citenamefont {Streltsov}\ \emph {et~al.}(2015)\citenamefont
  {Streltsov}, \citenamefont {Singh}, \citenamefont {Dhar}, \citenamefont
  {Bera},\ and\ \citenamefont {Adesso}}]{Singh2015}%
  \BibitemOpen
  \bibfield  {author} {\bibinfo {author} {\bibnamefont {Streltsov},
  \bibfnamefont {A.}}, \bibinfo {author} {\bibfnamefont {U.}~\bibnamefont
  {Singh}}, \bibinfo {author} {\bibfnamefont {H.~S.}\ \bibnamefont {Dhar}},
  \bibinfo {author} {\bibfnamefont {M.~N.}\ \bibnamefont {Bera}}, \ and\
  \bibinfo {author} {\bibfnamefont {G.}~\bibnamefont {Adesso}}} (\bibinfo
  {year} {2015}),\ \href {\doibase 10.1103/PhysRevLett.115.020403} {\bibfield
  {journal} {\bibinfo  {journal} {Phys. Rev. Lett.}\ }\textbf {\bibinfo
  {volume} {115}},\ \bibinfo {pages} {020403}}\BibitemShut {NoStop}%
\bibitem [{\citenamefont {Strocchi}(1985)}]{Strocchi0}%
  \BibitemOpen
  \bibfield  {author} {\bibinfo {author} {\bibnamefont {Strocchi},
  \bibfnamefont {F.}}} (\bibinfo {year} {1985}),\ \href@noop {} {\emph
  {\bibinfo {title} {Elements of Quantum mechanics of Infinite Systems}}}\
  (\bibinfo  {publisher} {World Scientific},\ \bibinfo {address}
  {Singapore})\BibitemShut {NoStop}%
\bibitem [{\citenamefont {Strocchi}(2008{\natexlab{a}})}]{Strocchi1}%
  \BibitemOpen
  \bibfield  {author} {\bibinfo {author} {\bibnamefont {Strocchi},
  \bibfnamefont {F.}}} (\bibinfo {year} {2008}{\natexlab{a}}),\ \href@noop {}
  {\emph {\bibinfo {title} {An Introduction to the Mathematical Structure of
  Quantum Mechanics}}},\ \bibinfo {edition} {2nd}\ ed.\ (\bibinfo  {publisher}
  {World Scientific},\ \bibinfo {address} {Singapore})\BibitemShut {NoStop}%
\bibitem [{\citenamefont {Strocchi}(2008{\natexlab{b}})}]{Strocchi2}%
  \BibitemOpen
  \bibfield  {author} {\bibinfo {author} {\bibnamefont {Strocchi},
  \bibfnamefont {F.}}} (\bibinfo {year} {2008}{\natexlab{b}}),\ \href@noop {}
  {\emph {\bibinfo {title} {Symmetry Breaking}}},\ \bibinfo {edition} {2nd}\
  ed.\ (\bibinfo  {publisher} {Springer},\ \bibinfo {address}
  {Heidelberg})\BibitemShut {NoStop}%
\bibitem [{\citenamefont {Strocchi}(2012)}]{Strocchi3}%
  \BibitemOpen
  \bibfield  {author} {\bibinfo {author} {\bibnamefont {Strocchi},
  \bibfnamefont {F.}}} (\bibinfo {year} {2012}),\ \href@noop {} {\bibfield
  {journal} {\bibinfo  {journal} {Eur. Phys. J. Plus}\ }\textbf {\bibinfo
  {volume} {127}},\ \bibinfo {pages} {12}}\BibitemShut {NoStop}%
\bibitem [{\citenamefont {Summers}\ and\ \citenamefont
  {Werner}(1985)}]{Werner1}%
  \BibitemOpen
  \bibfield  {author} {\bibinfo {author} {\bibnamefont {Summers}, \bibfnamefont
  {J.}}, \ and\ \bibinfo {author} {\bibfnamefont {R.}~\bibnamefont {Werner}}}
  (\bibinfo {year} {1985}),\ \href@noop {} {\bibfield  {journal} {\bibinfo
  {journal} {Phys. Lett.}\ }\textbf {\bibinfo {volume} {A110}},\ \bibinfo
  {pages} {257}}\BibitemShut {NoStop}%
\bibitem [{\citenamefont {Summers}\ and\ \citenamefont
  {Werner}(1987{\natexlab{a}})}]{Werner2}%
  \BibitemOpen
  \bibfield  {author} {\bibinfo {author} {\bibnamefont {Summers}, \bibfnamefont
  {J.}}, \ and\ \bibinfo {author} {\bibfnamefont {R.}~\bibnamefont {Werner}}}
  (\bibinfo {year} {1987}{\natexlab{a}}),\ \href@noop {} {\bibfield  {journal}
  {\bibinfo  {journal} {J. Math. Phys.}\ }\textbf {\bibinfo {volume} {28}},\
  \bibinfo {pages} {2440}}\BibitemShut {NoStop}%
\bibitem [{\citenamefont {Summers}\ and\ \citenamefont
  {Werner}(1987{\natexlab{b}})}]{Werner3}%
  \BibitemOpen
  \bibfield  {author} {\bibinfo {author} {\bibnamefont {Summers}, \bibfnamefont
  {J.}}, \ and\ \bibinfo {author} {\bibfnamefont {R.}~\bibnamefont {Werner}}}
  (\bibinfo {year} {1987}{\natexlab{b}}),\ \href@noop {} {\bibfield  {journal}
  {\bibinfo  {journal} {Comm. Math. Phys.}\ }\textbf {\bibinfo {volume}
  {110}},\ \bibinfo {pages} {247}}\BibitemShut {NoStop}%
\bibitem [{\citenamefont {Sun}\ \emph {et~al.}(2015)\citenamefont {Sun},
  \citenamefont {Kolezhuk},\ and\ \citenamefont {Vekua}}]{Sun2015}%
  \BibitemOpen
  \bibfield  {author} {\bibinfo {author} {\bibnamefont {Sun}, \bibfnamefont
  {G.}}, \bibinfo {author} {\bibfnamefont {A.~K.}\ \bibnamefont {Kolezhuk}}, \
  and\ \bibinfo {author} {\bibfnamefont {T.}~\bibnamefont {Vekua}}} (\bibinfo
  {year} {2015}),\ \href {\doibase 10.1103/PhysRevB.91.014418} {\bibfield
  {journal} {\bibinfo  {journal} {Phys. Rev. B}\ }\textbf {\bibinfo {volume}
  {91}},\ \bibinfo {pages} {014418}}\BibitemShut {NoStop}%
\bibitem [{\citenamefont {Suzuki}\ \emph {et~al.}(2006)\citenamefont {Suzuki},
  \citenamefont {Takeoka}, \citenamefont {Sasaki}, \citenamefont {Andersen},\
  and\ \citenamefont {Kannari}}]{suzuki_practical_2006}%
  \BibitemOpen
  \bibfield  {author} {\bibinfo {author} {\bibnamefont {Suzuki}, \bibfnamefont
  {S.}}, \bibinfo {author} {\bibfnamefont {M.}~\bibnamefont {Takeoka}},
  \bibinfo {author} {\bibfnamefont {M.}~\bibnamefont {Sasaki}}, \bibinfo
  {author} {\bibfnamefont {U.~L.}\ \bibnamefont {Andersen}}, \ and\ \bibinfo
  {author} {\bibfnamefont {F.}~\bibnamefont {Kannari}}} (\bibinfo {year}
  {2006}),\ \href {\doibase 10.1103/PhysRevA.73.042304} {\bibfield  {journal}
  {\bibinfo  {journal} {Phys. Rev. A}\ }\textbf {\bibinfo {volume}
  {73}}~(\bibinfo {number} {4}),\ \bibinfo {pages} {042304}}\BibitemShut
  {NoStop}%
\bibitem [{\citenamefont {Tacla}\ and\ \citenamefont
  {Caves}(2013)}]{TaclaNJP2013}%
  \BibitemOpen
  \bibfield  {author} {\bibinfo {author} {\bibnamefont {Tacla}, \bibfnamefont
  {A.~B.}}, \ and\ \bibinfo {author} {\bibfnamefont {C.~M.}\ \bibnamefont
  {Caves}}} (\bibinfo {year} {2013}),\ \href
  {http://stacks.iop.org/1367-2630/15/i=2/a=023008} {\bibfield  {journal}
  {\bibinfo  {journal} {New Journal of Physics}\ }\textbf {\bibinfo {volume}
  {15}}~(\bibinfo {number} {2}),\ \bibinfo {pages} {023008}}\BibitemShut
  {NoStop}%
\bibitem [{\citenamefont {Tan}\ \emph {et~al.}(2008)\citenamefont {Tan},
  \citenamefont {Erkmen}, \citenamefont {Giovannetti}, \citenamefont {Guha},
  \citenamefont {Lloyd}, \citenamefont {Maccone}, \citenamefont {Pirandola},\
  and\ \citenamefont {Shapiro}}]{Tan08}%
  \BibitemOpen
  \bibfield  {author} {\bibinfo {author} {\bibnamefont {Tan}, \bibfnamefont
  {S.-H.}}, \bibinfo {author} {\bibfnamefont {B.~I.}\ \bibnamefont {Erkmen}},
  \bibinfo {author} {\bibfnamefont {V.}~\bibnamefont {Giovannetti}}, \bibinfo
  {author} {\bibfnamefont {S.}~\bibnamefont {Guha}}, \bibinfo {author}
  {\bibfnamefont {S.}~\bibnamefont {Lloyd}}, \bibinfo {author} {\bibfnamefont
  {L.}~\bibnamefont {Maccone}}, \bibinfo {author} {\bibfnamefont
  {S.}~\bibnamefont {Pirandola}}, \ and\ \bibinfo {author} {\bibfnamefont
  {J.~H.}\ \bibnamefont {Shapiro}}} (\bibinfo {year} {2008}),\ \href@noop {}
  {\bibfield  {journal} {\bibinfo  {journal} {Phys. Rev. Lett.}\ }\textbf
  {\bibinfo {volume} {101}},\ \bibinfo {pages} {253601}}\BibitemShut {NoStop}%
\bibitem [{\citenamefont {Tham}\ \emph {et~al.}(2016)\citenamefont {Tham},
  \citenamefont {Ferretti}, \citenamefont {Sadashivan},\ and\ \citenamefont
  {Steinberg}}]{tham_simulating_2016}%
  \BibitemOpen
  \bibfield  {author} {\bibinfo {author} {\bibnamefont {Tham}, \bibfnamefont
  {W.~K.}}, \bibinfo {author} {\bibfnamefont {H.}~\bibnamefont {Ferretti}},
  \bibinfo {author} {\bibfnamefont {A.~V.}\ \bibnamefont {Sadashivan}}, \ and\
  \bibinfo {author} {\bibfnamefont {A.~M.}\ \bibnamefont {Steinberg}}}
  (\bibinfo {year} {2016}),\ \href
  {https://www.ncbi.nlm.nih.gov/pmc/articles/PMC5156908/} {\bibfield  {journal}
  {\bibinfo  {journal} {Scientific Reports}\ }\textbf {\bibinfo {volume} {6}},\
  \bibinfo {pages} {38822}}\BibitemShut {NoStop}%
\bibitem [{\citenamefont {{The {LIGO} Scientific
  Collaboration}}(2011)}]{LIGONP2011}%
  \BibitemOpen
  \bibfield  {author} {\bibinfo {author} {\bibnamefont {{The {LIGO} Scientific
  Collaboration}},}} (\bibinfo {year} {2011}),\ \href
  {http://dx.doi.org/10.1038/nphys2083} {\bibfield  {journal} {\bibinfo
  {journal} {Nat. Phys.}\ }\textbf {\bibinfo {volume} {7}}~(\bibinfo {number}
  {12}),\ \bibinfo {pages} {962}}\BibitemShut {NoStop}%
\bibitem [{\citenamefont {Thesberg}\ and\ \citenamefont
  {S\o{}rensen}(2011)}]{Thesberg2011}%
  \BibitemOpen
  \bibfield  {author} {\bibinfo {author} {\bibnamefont {Thesberg},
  \bibfnamefont {M.}}, \ and\ \bibinfo {author} {\bibfnamefont {E.~S.}\
  \bibnamefont {S\o{}rensen}}} (\bibinfo {year} {2011}),\ \href {\doibase
  10.1103/PhysRevB.84.224435} {\bibfield  {journal} {\bibinfo  {journal} {Phys.
  Rev. B}\ }\textbf {\bibinfo {volume} {84}},\ \bibinfo {pages}
  {224435}}\BibitemShut {NoStop}%
\bibitem [{\citenamefont {Thirring}(2002)}]{Thirring0}%
  \BibitemOpen
  \bibfield  {author} {\bibinfo {author} {\bibnamefont {Thirring},
  \bibfnamefont {W.}}} (\bibinfo {year} {2002}),\ \href@noop {} {\emph
  {\bibinfo {title} {Quantum Mathematical Physics: Atoms, Molecules and Large
  Systems}}}\ (\bibinfo  {publisher} {Springer},\ \bibinfo {address}
  {Berlin})\BibitemShut {NoStop}%
\bibitem [{\citenamefont {Tichy}\ \emph {et~al.}(2013)\citenamefont {Tichy},
  \citenamefont {de~Melo}, \citenamefont {Ku\'s}, \citenamefont {Mintert},\
  and\ \citenamefont {Buchleitner}}]{Buchleitner0}%
  \BibitemOpen
  \bibfield  {author} {\bibinfo {author} {\bibnamefont {Tichy}, \bibfnamefont
  {M.}}, \bibinfo {author} {\bibfnamefont {F.}~\bibnamefont {de~Melo}},
  \bibinfo {author} {\bibfnamefont {M.}~\bibnamefont {Ku\'s}}, \bibinfo
  {author} {\bibfnamefont {F.}~\bibnamefont {Mintert}}, \ and\ \bibinfo
  {author} {\bibfnamefont {A.}~\bibnamefont {Buchleitner}}} (\bibinfo {year}
  {2013}),\ \href@noop {} {\bibfield  {journal} {\bibinfo  {journal} {Fortschr.
  Phys.}\ }\textbf {\bibinfo {volume} {61}},\ \bibinfo {pages}
  {225}}\BibitemShut {NoStop}%
\bibitem [{\citenamefont {Tilma}\ \emph {et~al.}(2010)\citenamefont {Tilma},
  \citenamefont {Hamaji}, \citenamefont {Munro},\ and\ \citenamefont
  {Nemoto}}]{TilmaPRA2010}%
  \BibitemOpen
  \bibfield  {author} {\bibinfo {author} {\bibnamefont {Tilma}, \bibfnamefont
  {T.}}, \bibinfo {author} {\bibfnamefont {S.}~\bibnamefont {Hamaji}}, \bibinfo
  {author} {\bibfnamefont {W.~J.}\ \bibnamefont {Munro}}, \ and\ \bibinfo
  {author} {\bibfnamefont {K.}~\bibnamefont {Nemoto}}} (\bibinfo {year}
  {2010}),\ \href {\doibase 10.1103/PhysRevA.81.022108} {\bibfield  {journal}
  {\bibinfo  {journal} {Phys. Rev. A}\ }\textbf {\bibinfo {volume} {81}},\
  \bibinfo {pages} {022108}}\BibitemShut {NoStop}%
\bibitem [{\citenamefont {T\'oth}(2012)}]{Toth12}%
  \BibitemOpen
  \bibfield  {author} {\bibinfo {author} {\bibnamefont {T\'oth}, \bibfnamefont
  {G.}}} (\bibinfo {year} {2012}),\ \href {\doibase 10.1103/PhysRevA.85.022322}
  {\bibfield  {journal} {\bibinfo  {journal} {Phys. Rev. A}\ }\textbf {\bibinfo
  {volume} {85}},\ \bibinfo {pages} {022322}}\BibitemShut {NoStop}%
\bibitem [{\citenamefont {T\'oth}\ and\ \citenamefont
  {Apellaniz}(2014)}]{toth_quantum_2014}%
  \BibitemOpen
  \bibfield  {author} {\bibinfo {author} {\bibnamefont {T\'oth}, \bibfnamefont
  {G.}}, \ and\ \bibinfo {author} {\bibfnamefont {I.}~\bibnamefont
  {Apellaniz}}} (\bibinfo {year} {2014}),\ \href {\doibase
  10.1088/1751-8113/47/42/424006} {\bibfield  {journal} {\bibinfo  {journal}
  {J. Phys. A: Math. Theor.}\ }\textbf {\bibinfo {volume} {47}}~(\bibinfo
  {number} {42}),\ \bibinfo {pages} {424006}}\BibitemShut {NoStop}%
\bibitem [{\citenamefont {T\'oth}\ \emph {et~al.}(2009)\citenamefont {T\'oth},
  \citenamefont {Knapp}, \citenamefont {G\"uhne},\ and\ \citenamefont
  {Briegel}}]{Briegel0}%
  \BibitemOpen
  \bibfield  {author} {\bibinfo {author} {\bibnamefont {T\'oth}, \bibfnamefont
  {G.}}, \bibinfo {author} {\bibfnamefont {C.}~\bibnamefont {Knapp}}, \bibinfo
  {author} {\bibfnamefont {O.}~\bibnamefont {G\"uhne}}, \ and\ \bibinfo
  {author} {\bibfnamefont {H.}~\bibnamefont {Briegel}}} (\bibinfo {year}
  {2009}),\ \href@noop {} {\bibfield  {journal} {\bibinfo  {journal} {Phys.
  Rev. A}\ }\textbf {\bibinfo {volume} {79}},\ \bibinfo {pages}
  {042334}}\BibitemShut {NoStop}%
\bibitem [{\citenamefont {van Trees}(2001)}]{van_trees_detection_2001}%
  \BibitemOpen
  \bibfield  {author} {\bibinfo {author} {\bibnamefont {van Trees},
  \bibfnamefont {H.~L.}}} (\bibinfo {year} {2001}),\ \href@noop {} {\emph
  {\bibinfo {title} {Detection, {Estimation}, and {Modulation} {Theory}}}},\
  \bibinfo {edition} {part i}\ ed.\ (\bibinfo  {publisher}
  {Wiley-Interscience},\ \bibinfo {address} {New York})\BibitemShut {NoStop}%
\bibitem [{\citenamefont {Tsang}\ \emph {et~al.}(2016)\citenamefont {Tsang},
  \citenamefont {Nair},\ and\ \citenamefont {Lu}}]{Mankei2016}%
  \BibitemOpen
  \bibfield  {author} {\bibinfo {author} {\bibnamefont {Tsang}, \bibfnamefont
  {M.}}, \bibinfo {author} {\bibfnamefont {R.}~\bibnamefont {Nair}}, \ and\
  \bibinfo {author} {\bibfnamefont {X.-M.}\ \bibnamefont {Lu}}} (\bibinfo
  {year} {2016}),\ \href {\doibase 10.1103/PhysRevX.6.031033} {\bibfield
  {journal} {\bibinfo  {journal} {Phys. Rev. X}\ }\textbf {\bibinfo {volume}
  {6}},\ \bibinfo {pages} {031033}}\BibitemShut {NoStop}%
\bibitem [{\citenamefont {Tsuda}\ and\ \citenamefont
  {Matsumoto}(2005)}]{Tsuda2005}%
  \BibitemOpen
  \bibfield  {author} {\bibinfo {author} {\bibnamefont {Tsuda}, \bibfnamefont
  {Y.}}, \ and\ \bibinfo {author} {\bibfnamefont {K.}~\bibnamefont
  {Matsumoto}}} (\bibinfo {year} {2005}),\ \href@noop {} {\bibfield  {journal}
  {\bibinfo  {journal} {Journal of Physics A: Mathematical and General}\
  }\textbf {\bibinfo {volume} {38}},\ \bibinfo {pages} {1593}}\BibitemShut
  {NoStop}%
\bibitem [{\citenamefont {Tzeng}\ \emph {et~al.}(2008)\citenamefont {Tzeng},
  \citenamefont {Hung}, \citenamefont {Chen},\ and\ \citenamefont
  {Yang}}]{Tzeng2008-2}%
  \BibitemOpen
  \bibfield  {author} {\bibinfo {author} {\bibnamefont {Tzeng}, \bibfnamefont
  {Y.-C.}}, \bibinfo {author} {\bibfnamefont {H.-H.}\ \bibnamefont {Hung}},
  \bibinfo {author} {\bibfnamefont {Y.-C.}\ \bibnamefont {Chen}}, \ and\
  \bibinfo {author} {\bibfnamefont {M.-F.}\ \bibnamefont {Yang}}} (\bibinfo
  {year} {2008}),\ \href {\doibase 10.1103/PhysRevA.77.062321} {\bibfield
  {journal} {\bibinfo  {journal} {Phys. Rev. A}\ }\textbf {\bibinfo {volume}
  {77}},\ \bibinfo {pages} {062321}}\BibitemShut {NoStop}%
\bibitem [{\citenamefont {Uys}\ and\ \citenamefont {Meystre}(2007)}]{Meystre0}%
  \BibitemOpen
  \bibfield  {author} {\bibinfo {author} {\bibnamefont {Uys}, \bibfnamefont
  {H.}}, \ and\ \bibinfo {author} {\bibfnamefont {P.}~\bibnamefont {Meystre}}}
  (\bibinfo {year} {2007}),\ \href@noop {} {\bibfield  {journal} {\bibinfo
  {journal} {Phys. Rev. A}\ }\textbf {\bibinfo {volume} {76}},\ \bibinfo
  {pages} {013804}}\BibitemShut {NoStop}%
\bibitem [{\citenamefont {Vasilakis}\ \emph {et~al.}(2011)\citenamefont
  {Vasilakis}, \citenamefont {Shah},\ and\ \citenamefont
  {Romalis}}]{VasilakisPRL2011}%
  \BibitemOpen
  \bibfield  {author} {\bibinfo {author} {\bibnamefont {Vasilakis},
  \bibfnamefont {G.}}, \bibinfo {author} {\bibfnamefont {V.}~\bibnamefont
  {Shah}}, \ and\ \bibinfo {author} {\bibfnamefont {M.~V.}\ \bibnamefont
  {Romalis}}} (\bibinfo {year} {2011}),\ \href {\doibase
  10.1103/PhysRevLett.106.143601} {\bibfield  {journal} {\bibinfo  {journal}
  {Phys. Rev. Lett.}\ }\textbf {\bibinfo {volume} {106}},\ \bibinfo {pages}
  {143601}}\BibitemShut {NoStop}%
\bibitem [{\citenamefont {\ifmmode \check{Z}\else
  \v{Z}\fi{}nidari\ifmmode~\check{c}\else \v{c}\fi{}}(2012)}]{Znidaric2012}%
  \BibitemOpen
  \bibfield  {author} {\bibinfo {author} {\bibnamefont {\ifmmode \check{Z}\else
  \v{Z}\fi{}nidari\ifmmode~\check{c}\else \v{c}\fi{}}, \bibfnamefont {M.}}}
  (\bibinfo {year} {2012}),\ \href {\doibase 10.1103/PhysRevA.85.012324}
  {\bibfield  {journal} {\bibinfo  {journal} {Phys. Rev. A}\ }\textbf {\bibinfo
  {volume} {85}},\ \bibinfo {pages} {012324}}\BibitemShut {NoStop}%
\bibitem [{\citenamefont {Vedral}(2003)}]{Vedral0}%
  \BibitemOpen
  \bibfield  {author} {\bibinfo {author} {\bibnamefont {Vedral}, \bibfnamefont
  {V.}}} (\bibinfo {year} {2003}),\ \href@noop {} {\bibfield  {journal}
  {\bibinfo  {journal} {Central Europ. J. Phys.}\ }\textbf {\bibinfo {volume}
  {2}},\ \bibinfo {pages} {289}}\BibitemShut {NoStop}%
\bibitem [{\citenamefont {Verch}\ and\ \citenamefont {Werner}(2005)}]{Werner5}%
  \BibitemOpen
  \bibfield  {author} {\bibinfo {author} {\bibnamefont {Verch}, \bibfnamefont
  {R.}}, \ and\ \bibinfo {author} {\bibfnamefont {R.}~\bibnamefont {Werner}}}
  (\bibinfo {year} {2005}),\ \href@noop {} {\bibfield  {journal} {\bibinfo
  {journal} {Rev. Math. Phys.}\ }\textbf {\bibinfo {volume} {17}},\ \bibinfo
  {pages} {545}}\BibitemShut {NoStop}%
\bibitem [{\citenamefont {Vlastakis}\ \emph {et~al.}(2013)\citenamefont
  {Vlastakis}, \citenamefont {Kirchmair}, \citenamefont {Leghtas},
  \citenamefont {Nigg}, \citenamefont {Frunzio}, \citenamefont {Girvin},
  \citenamefont {Mirrahimi}, \citenamefont {Devoret},\ and\ \citenamefont
  {Schoelkopf}}]{vlastakis_deterministically_2013}%
  \BibitemOpen
  \bibfield  {author} {\bibinfo {author} {\bibnamefont {Vlastakis},
  \bibfnamefont {B.}}, \bibinfo {author} {\bibfnamefont {G.}~\bibnamefont
  {Kirchmair}}, \bibinfo {author} {\bibfnamefont {Z.}~\bibnamefont {Leghtas}},
  \bibinfo {author} {\bibfnamefont {S.~E.}\ \bibnamefont {Nigg}}, \bibinfo
  {author} {\bibfnamefont {L.}~\bibnamefont {Frunzio}}, \bibinfo {author}
  {\bibfnamefont {S.~M.}\ \bibnamefont {Girvin}}, \bibinfo {author}
  {\bibfnamefont {M.}~\bibnamefont {Mirrahimi}}, \bibinfo {author}
  {\bibfnamefont {M.~H.}\ \bibnamefont {Devoret}}, \ and\ \bibinfo {author}
  {\bibfnamefont {R.~J.}\ \bibnamefont {Schoelkopf}}} (\bibinfo {year}
  {2013}),\ \href {\doibase 10.1126/science.1243289} {\bibfield  {journal}
  {\bibinfo  {journal} {Science}\ }\textbf {\bibinfo {volume} {342}}~(\bibinfo
  {number} {6158}),\ \bibinfo {pages} {607}}\BibitemShut {NoStop}%
\bibitem [{\citenamefont {Vogel}\ and\ \citenamefont
  {Sperling}(2014)}]{Vogel2014}%
  \BibitemOpen
  \bibfield  {author} {\bibinfo {author} {\bibnamefont {Vogel}, \bibfnamefont
  {W.}}, \ and\ \bibinfo {author} {\bibfnamefont {J.}~\bibnamefont {Sperling}}}
  (\bibinfo {year} {2014}),\ \href {\doibase 10.1103/PhysRevA.89.052302}
  {\bibfield  {journal} {\bibinfo  {journal} {Phys. Rev. A}\ }\textbf {\bibinfo
  {volume} {89}},\ \bibinfo {pages} {052302}}\BibitemShut {NoStop}%
\bibitem [{\citenamefont {Vogels}\ \emph {et~al.}(2002)\citenamefont {Vogels},
  \citenamefont {Xu}, \citenamefont {Raman}, \citenamefont {Abo-Shaeer},\ and\
  \citenamefont {Ketterle}}]{Vogels2002}%
  \BibitemOpen
  \bibfield  {author} {\bibinfo {author} {\bibnamefont {Vogels}, \bibfnamefont
  {J.~M.}}, \bibinfo {author} {\bibfnamefont {K.}~\bibnamefont {Xu}}, \bibinfo
  {author} {\bibfnamefont {C.}~\bibnamefont {Raman}}, \bibinfo {author}
  {\bibfnamefont {J.~R.}\ \bibnamefont {Abo-Shaeer}}, \ and\ \bibinfo {author}
  {\bibfnamefont {W.}~\bibnamefont {Ketterle}}} (\bibinfo {year} {2002}),\
  \href {\doibase 10.1103/PhysRevLett.88.060402} {\bibfield  {journal}
  {\bibinfo  {journal} {Phys. Rev. Lett.}\ }\textbf {\bibinfo {volume} {88}},\
  \bibinfo {pages} {060402}}\BibitemShut {NoStop}%
\bibitem [{\citenamefont {Vourdas}\ and\ \citenamefont
  {Dunningham}(2005)}]{Vourdas0}%
  \BibitemOpen
  \bibfield  {author} {\bibinfo {author} {\bibnamefont {Vourdas}, \bibfnamefont
  {A.}}, \ and\ \bibinfo {author} {\bibfnamefont {J.~A.}\ \bibnamefont
  {Dunningham}}} (\bibinfo {year} {2005}),\ \href@noop {} {\bibfield  {journal}
  {\bibinfo  {journal} {Phys. Rev. A}\ }\textbf {\bibinfo {volume} {71}},\
  \bibinfo {pages} {013809}}\BibitemShut {NoStop}%
\bibitem [{\citenamefont {\v{S}afr\'anek}\ and\ \citenamefont
  {Fuentes}(2016)}]{Safranek2016}%
  \BibitemOpen
  \bibfield  {author} {\bibinfo {author} {\bibnamefont {\v{S}afr\'anek},
  \bibfnamefont {D.}}, \ and\ \bibinfo {author} {\bibfnamefont
  {I.}~\bibnamefont {Fuentes}}} (\bibinfo {year} {2016}),\ \href {\doibase
  10.1103/PhysRevA.94.062313} {\bibfield  {journal} {\bibinfo  {journal} {Phys.
  Rev. A}\ }\textbf {\bibinfo {volume} {94}},\ \bibinfo {pages}
  {062313}}\BibitemShut {NoStop}%
\bibitem [{\citenamefont {\v{S}afr\'anek}\ \emph {et~al.}(2015)\citenamefont
  {\v{S}afr\'anek}, \citenamefont {Lee},\ and\ \citenamefont
  {Fuentes}}]{Safranek2015}%
  \BibitemOpen
  \bibfield  {author} {\bibinfo {author} {\bibnamefont {\v{S}afr\'anek},
  \bibfnamefont {D.}}, \bibinfo {author} {\bibfnamefont {A.~R.}\ \bibnamefont
  {Lee}}, \ and\ \bibinfo {author} {\bibfnamefont {I.}~\bibnamefont {Fuentes}}}
  (\bibinfo {year} {2015}),\ \href
  {http://stacks.iop.org/1367-2630/17/i=7/a=073016} {\bibfield  {journal}
  {\bibinfo  {journal} {New Journal of Physics}\ }\textbf {\bibinfo {volume}
  {17}}~(\bibinfo {number} {7}),\ \bibinfo {pages} {073016}}\BibitemShut
  {NoStop}%
\bibitem [{\citenamefont {\ifmmode~\check{S}\else
  \v{S}\fi{}afr\'anek}(2017)}]{safranek_discontinuities_2016}%
  \BibitemOpen
  \bibfield  {author} {\bibinfo {author} {\bibnamefont {\ifmmode~\check{S}\else
  \v{S}\fi{}afr\'anek}, \bibfnamefont {D.}}} (\bibinfo {year} {2017}),\ \href
  {\doibase 10.1103/PhysRevA.95.052320} {\bibfield  {journal} {\bibinfo
  {journal} {Phys. Rev. A}\ }\textbf {\bibinfo {volume} {95}},\ \bibinfo
  {pages} {052320}}\BibitemShut {NoStop}%
\bibitem [{\citenamefont {Wakui}\ \emph {et~al.}(2007)\citenamefont {Wakui},
  \citenamefont {Takahashi}, \citenamefont {Furusawa},\ and\ \citenamefont
  {Sasaki}}]{wakui_photon_2007}%
  \BibitemOpen
  \bibfield  {author} {\bibinfo {author} {\bibnamefont {Wakui}, \bibfnamefont
  {K.}}, \bibinfo {author} {\bibfnamefont {H.}~\bibnamefont {Takahashi}},
  \bibinfo {author} {\bibfnamefont {A.}~\bibnamefont {Furusawa}}, \ and\
  \bibinfo {author} {\bibfnamefont {M.}~\bibnamefont {Sasaki}}} (\bibinfo
  {year} {2007}),\ \href {\doibase 10.1364/OE.15.003568} {\bibfield  {journal}
  {\bibinfo  {journal} {Optics Express}\ }\textbf {\bibinfo {volume}
  {15}}~(\bibinfo {number} {6}),\ \bibinfo {pages} {3568}}\BibitemShut
  {NoStop}%
\bibitem [{\citenamefont {Wallis}(1996)}]{Wallis1996}%
  \BibitemOpen
  \bibfield  {author} {\bibinfo {author} {\bibnamefont {Wallis}, \bibfnamefont
  {H.}}} (\bibinfo {year} {1996}),\ \href@noop {} {\bibfield  {journal}
  {\bibinfo  {journal} {Quantum Semiclass. Opt.}\ }\textbf {\bibinfo {volume}
  {8}},\ \bibinfo {pages} {727}}\BibitemShut {NoStop}%
\bibitem [{\citenamefont {Wang}\ and\ \citenamefont {Sanders}(2003)}]{Wang0}%
  \BibitemOpen
  \bibfield  {author} {\bibinfo {author} {\bibnamefont {Wang}, \bibfnamefont
  {X.}}, \ and\ \bibinfo {author} {\bibfnamefont {B.}~\bibnamefont {Sanders}}}
  (\bibinfo {year} {2003}),\ \href@noop {} {\bibfield  {journal} {\bibinfo
  {journal} {Phys. Rev. A}\ }\textbf {\bibinfo {volume} {68}},\ \bibinfo
  {pages} {012101}}\BibitemShut {NoStop}%
\bibitem [{\citenamefont {Weedbrook}\ \emph {et~al.}(2012)\citenamefont
  {Weedbrook}, \citenamefont {Pirandola}, \citenamefont {Garcia-Patron},
  \citenamefont {Cerf}, \citenamefont {Ralph}, \citenamefont {Shapiro},\ and\
  \citenamefont {Lloyd}}]{Weedbrook12}%
  \BibitemOpen
  \bibfield  {author} {\bibinfo {author} {\bibnamefont {Weedbrook},
  \bibfnamefont {C.}}, \bibinfo {author} {\bibfnamefont {S.}~\bibnamefont
  {Pirandola}}, \bibinfo {author} {\bibfnamefont {R.}~\bibnamefont
  {Garcia-Patron}}, \bibinfo {author} {\bibfnamefont {N.~J.}\ \bibnamefont
  {Cerf}}, \bibinfo {author} {\bibfnamefont {T.~C.}\ \bibnamefont {Ralph}},
  \bibinfo {author} {\bibfnamefont {J.~H.}\ \bibnamefont {Shapiro}}, \ and\
  \bibinfo {author} {\bibfnamefont {S.}~\bibnamefont {Lloyd}}} (\bibinfo {year}
  {2012}),\ \href@noop {} {\bibfield  {journal} {\bibinfo  {journal} {Rev. Mod.
  Phys.}\ }\textbf {\bibinfo {volume} {84}},\ \bibinfo {pages}
  {621}}\BibitemShut {NoStop}%
\bibitem [{\citenamefont {Weedbrook}\ \emph {et~al.}(2016)\citenamefont
  {Weedbrook}, \citenamefont {Pirandola}, \citenamefont {Thompson},
  \citenamefont {Vedral},\ and\ \citenamefont {Gu}}]{Weedbrook2016}%
  \BibitemOpen
  \bibfield  {author} {\bibinfo {author} {\bibnamefont {Weedbrook},
  \bibfnamefont {C.}}, \bibinfo {author} {\bibfnamefont {S.}~\bibnamefont
  {Pirandola}}, \bibinfo {author} {\bibfnamefont {J.}~\bibnamefont {Thompson}},
  \bibinfo {author} {\bibfnamefont {V.}~\bibnamefont {Vedral}}, \ and\ \bibinfo
  {author} {\bibfnamefont {M.}~\bibnamefont {Gu}}} (\bibinfo {year} {2016}),\
  \href@noop {} {\bibfield  {journal} {\bibinfo  {journal} {New J. Phys.}\
  }\textbf {\bibinfo {volume} {18}},\ \bibinfo {pages} {043027}}\BibitemShut
  {NoStop}%
\bibitem [{\citenamefont {Weinhold}(1974)}]{Weinhold1974}%
  \BibitemOpen
  \bibfield  {author} {\bibinfo {author} {\bibnamefont {Weinhold},
  \bibfnamefont {F.}}} (\bibinfo {year} {1974}),\ \href@noop {} {\bibfield
  {journal} {\bibinfo  {journal} {J. Chem. Phys.}\ }\textbf {\bibinfo {volume}
  {63}},\ \bibinfo {pages} {2488}}\BibitemShut {NoStop}%
\bibitem [{\citenamefont {Weiss}(1999)}]{Weiss99}%
  \BibitemOpen
  \bibfield  {author} {\bibinfo {author} {\bibnamefont {Weiss}, \bibfnamefont
  {U.}}} (\bibinfo {year} {1999}),\ \href@noop {} {\emph {\bibinfo {title}
  {{Quantum Dissipative Systems}}}},\ Vol.\ \bibinfo {volume} {2nd edition}\
  (\bibinfo  {publisher} {World Scientific},\ \bibinfo {address}
  {Singapore})\BibitemShut {NoStop}%
\bibitem [{\citenamefont {Wheatley}\ \emph {et~al.}(2010)\citenamefont
  {Wheatley}, \citenamefont {Berry}, \citenamefont {Yonezawa}, \citenamefont
  {Nakane}, \citenamefont {Arao}, \citenamefont {Pope}, \citenamefont {Ralph},
  \citenamefont {Wiseman}, \citenamefont {Furusawa},\ and\ \citenamefont
  {Huntington}}]{PhysRevLett.104.093601}%
  \BibitemOpen
  \bibfield  {author} {\bibinfo {author} {\bibnamefont {Wheatley},
  \bibfnamefont {T.~A.}}, \bibinfo {author} {\bibfnamefont {D.~W.}\
  \bibnamefont {Berry}}, \bibinfo {author} {\bibfnamefont {H.}~\bibnamefont
  {Yonezawa}}, \bibinfo {author} {\bibfnamefont {D.}~\bibnamefont {Nakane}},
  \bibinfo {author} {\bibfnamefont {H.}~\bibnamefont {Arao}}, \bibinfo {author}
  {\bibfnamefont {D.~T.}\ \bibnamefont {Pope}}, \bibinfo {author}
  {\bibfnamefont {T.~C.}\ \bibnamefont {Ralph}}, \bibinfo {author}
  {\bibfnamefont {H.~M.}\ \bibnamefont {Wiseman}}, \bibinfo {author}
  {\bibfnamefont {A.}~\bibnamefont {Furusawa}}, \ and\ \bibinfo {author}
  {\bibfnamefont {E.~H.}\ \bibnamefont {Huntington}}} (\bibinfo {year}
  {2010}),\ \href {\doibase 10.1103/PhysRevLett.104.093601} {\bibfield
  {journal} {\bibinfo  {journal} {Phys. Rev. Lett.}\ }\textbf {\bibinfo
  {volume} {104}},\ \bibinfo {pages} {093601}}\BibitemShut {NoStop}%
\bibitem [{\citenamefont {Wigner}(1957)}]{wigner_relativistic_1957}%
  \BibitemOpen
  \bibfield  {author} {\bibinfo {author} {\bibnamefont {Wigner}, \bibfnamefont
  {E.~P.}}} (\bibinfo {year} {1957}),\ \href {\doibase
  10.1103/RevModPhys.29.255} {\bibfield  {journal} {\bibinfo  {journal} {Rev.
  Mod. Phys.}\ }\textbf {\bibinfo {volume} {29}}~(\bibinfo {number} {3}),\
  \bibinfo {pages} {255}}\BibitemShut {NoStop}%
\bibitem [{\citenamefont {Wimberger}(2016)}]{Wimberger2016}%
  \BibitemOpen
  \bibfield  {author} {\bibinfo {author} {\bibnamefont {Wimberger},
  \bibfnamefont {S.}}} (\bibinfo {year} {2016}),\ \href@noop {} {\bibfield
  {journal} {\bibinfo  {journal} {Phil. Trans. R. Soc. A}\ }\textbf {\bibinfo
  {volume} {374}},\ \bibinfo {pages} {20150153}}\BibitemShut {NoStop}%
\bibitem [{\citenamefont {Wineland}(2013)}]{wineland_nobel_2013}%
  \BibitemOpen
  \bibfield  {author} {\bibinfo {author} {\bibnamefont {Wineland},
  \bibfnamefont {D.~J.}}} (\bibinfo {year} {2013}),\ \href {\doibase
  10.1103/RevModPhys.85.1103} {\bibfield  {journal} {\bibinfo  {journal} {Rev.
  Mod. Phys.}\ }\textbf {\bibinfo {volume} {85}}~(\bibinfo {number} {3}),\
  \bibinfo {pages} {1103}}\BibitemShut {NoStop}%
\bibitem [{\citenamefont {Wineland}\ \emph {et~al.}(1994)\citenamefont
  {Wineland}, \citenamefont {Bollinger}, \citenamefont {Itano},\ and\
  \citenamefont {Heinzen}}]{Wineland0}%
  \BibitemOpen
  \bibfield  {author} {\bibinfo {author} {\bibnamefont {Wineland},
  \bibfnamefont {D.~J.}}, \bibinfo {author} {\bibfnamefont {J.~J.}\
  \bibnamefont {Bollinger}}, \bibinfo {author} {\bibfnamefont {W.~M.}\
  \bibnamefont {Itano}}, \ and\ \bibinfo {author} {\bibfnamefont {D.~J.}\
  \bibnamefont {Heinzen}}} (\bibinfo {year} {1994}),\ \href {\doibase
  10.1103/PhysRevA.50.67} {\bibfield  {journal} {\bibinfo  {journal} {Phys.
  Rev. A}\ }\textbf {\bibinfo {volume} {50}},\ \bibinfo {pages}
  {67}}\BibitemShut {NoStop}%
\bibitem [{\citenamefont {Wiseman}\ and\ \citenamefont
  {Vaccaro}(2003)}]{Wiseman0}%
  \BibitemOpen
  \bibfield  {author} {\bibinfo {author} {\bibnamefont {Wiseman}, \bibfnamefont
  {H.}}, \ and\ \bibinfo {author} {\bibfnamefont {J.}~\bibnamefont {Vaccaro}}}
  (\bibinfo {year} {2003}),\ \href@noop {} {\bibfield  {journal} {\bibinfo
  {journal} {Phys. Rev. Lett.}\ }\textbf {\bibinfo {volume} {91}},\ \bibinfo
  {pages} {097902}}\BibitemShut {NoStop}%
\bibitem [{\citenamefont {Wiseman}(1995)}]{wiseman_adaptive_1995}%
  \BibitemOpen
  \bibfield  {author} {\bibinfo {author} {\bibnamefont {Wiseman}, \bibfnamefont
  {H.~M.}}} (\bibinfo {year} {1995}),\ \href {\doibase
  10.1103/PhysRevLett.75.4587} {\bibfield  {journal} {\bibinfo  {journal}
  {Phys. Rev. Lett.}\ }\textbf {\bibinfo {volume} {75}}~(\bibinfo {number}
  {25}),\ \bibinfo {pages} {4587}}\BibitemShut {NoStop}%
\bibitem [{\citenamefont {Wiseman}\ \emph {et~al.}(2003)\citenamefont
  {Wiseman}, \citenamefont {Bartlett},\ and\ \citenamefont
  {Vaccaro}}]{wiseman_ferreting_2003}%
  \BibitemOpen
  \bibfield  {author} {\bibinfo {author} {\bibnamefont {Wiseman}, \bibfnamefont
  {H.~M.}}, \bibinfo {author} {\bibfnamefont {S.~D.}\ \bibnamefont {Bartlett}},
  \ and\ \bibinfo {author} {\bibfnamefont {J.~A.}\ \bibnamefont {Vaccaro}}}
  (\bibinfo {year} {2003}),\ \href {http://arxiv.org/abs/quant-ph/0309046}
  {\bibinfo  {journal} {arXiv:quant-ph/0309046}\ }\BibitemShut {NoStop}%
\bibitem [{\citenamefont {Wiseman}\ and\ \citenamefont
  {Milburn}(2009)}]{wiseman_quantum_2009}%
  \BibitemOpen
\bibfield  {journal} {  }\bibfield  {author} {\bibinfo {author} {\bibnamefont
  {Wiseman}, \bibfnamefont {H.~M.}}, \ and\ \bibinfo {author} {\bibfnamefont
  {G.~J.}\ \bibnamefont {Milburn}}} (\bibinfo {year} {2009}),\ \href@noop {}
  {\emph {\bibinfo {title} {Quantum Measurement and Control}}},\ \bibinfo
  {edition} {1st}\ ed.\ (\bibinfo  {publisher} {Cambridge University
  Press})\BibitemShut {NoStop}%
\bibitem [{\citenamefont {Woolley}\ \emph {et~al.}(2008)\citenamefont
  {Woolley}, \citenamefont {Milburn},\ and\ \citenamefont
  {Caves}}]{WoolleyNJP2008}%
  \BibitemOpen
  \bibfield  {author} {\bibinfo {author} {\bibnamefont {Woolley}, \bibfnamefont
  {M.~J.}}, \bibinfo {author} {\bibfnamefont {G.~J.}\ \bibnamefont {Milburn}},
  \ and\ \bibinfo {author} {\bibfnamefont {C.~M.}\ \bibnamefont {Caves}}}
  (\bibinfo {year} {2008}),\ \href {\doibase 10.1088/1367-2630/10/12/125018}
  {\bibfield  {journal} {\bibinfo  {journal} {New Journal of Physics}\ }\textbf
  {\bibinfo {volume} {10}}~(\bibinfo {number} {12}),\ \bibinfo {pages}
  {125018}}\BibitemShut {NoStop}%
\bibitem [{\citenamefont {Yan}\ \emph {et~al.}(2016)\citenamefont {Yan},
  \citenamefont {Huang},\ and\ \citenamefont {Wang}}]{Yan2016}%
  \BibitemOpen
  \bibfield  {author} {\bibinfo {author} {\bibnamefont {Yan}, \bibfnamefont
  {Z.}}, \bibinfo {author} {\bibfnamefont {P.-W.}\ \bibnamefont {Huang}}, \
  and\ \bibinfo {author} {\bibfnamefont {Z.}~\bibnamefont {Wang}}} (\bibinfo
  {year} {2016}),\ \href {\doibase 10.1103/PhysRevB.93.085138} {\bibfield
  {journal} {\bibinfo  {journal} {Phys. Rev. B}\ }\textbf {\bibinfo {volume}
  {93}},\ \bibinfo {pages} {085138}}\BibitemShut {NoStop}%
\bibitem [{\citenamefont {Yang}\ \emph {et~al.}(2008)\citenamefont {Yang},
  \citenamefont {Gu}, \citenamefont {Sun},\ and\ \citenamefont
  {Lin}}]{Yang2008}%
  \BibitemOpen
  \bibfield  {author} {\bibinfo {author} {\bibnamefont {Yang}, \bibfnamefont
  {S.}}, \bibinfo {author} {\bibfnamefont {S.-J.}\ \bibnamefont {Gu}}, \bibinfo
  {author} {\bibfnamefont {C.-P.}\ \bibnamefont {Sun}}, \ and\ \bibinfo
  {author} {\bibfnamefont {H.-Q.}\ \bibnamefont {Lin}}} (\bibinfo {year}
  {2008}),\ \href {\doibase 10.1103/PhysRevA.78.012304} {\bibfield  {journal}
  {\bibinfo  {journal} {Phys. Rev. A}\ }\textbf {\bibinfo {volume} {78}},\
  \bibinfo {pages} {012304}}\BibitemShut {NoStop}%
\bibitem [{\citenamefont {Yao}\ and\ \citenamefont {Kivelson}(2007)}]{Yao2007}%
  \BibitemOpen
  \bibfield  {author} {\bibinfo {author} {\bibnamefont {Yao}, \bibfnamefont
  {H.}}, \ and\ \bibinfo {author} {\bibfnamefont {S.~A.}\ \bibnamefont
  {Kivelson}}} (\bibinfo {year} {2007}),\ \href {\doibase
  10.1103/PhysRevLett.99.247203} {\bibfield  {journal} {\bibinfo  {journal}
  {Phys. Rev. Lett.}\ }\textbf {\bibinfo {volume} {99}},\ \bibinfo {pages}
  {247203}}\BibitemShut {NoStop}%
\bibitem [{\citenamefont {Yonezawa}\ \emph {et~al.}(2012)\citenamefont
  {Yonezawa}, \citenamefont {Nakane}, \citenamefont {Wheatley}, \citenamefont
  {Iwasawa}, \citenamefont {Takeda}, \citenamefont {Arao}, \citenamefont
  {Ohki}, \citenamefont {Tsumura}, \citenamefont {Berry}, \citenamefont
  {Ralph}, \citenamefont {Wiseman}, \citenamefont {Huntington},\ and\
  \citenamefont {Furusawa}}]{yonezawa_quantum-enhanced_2012}%
  \BibitemOpen
  \bibfield  {author} {\bibinfo {author} {\bibnamefont {Yonezawa},
  \bibfnamefont {H.}}, \bibinfo {author} {\bibfnamefont {D.}~\bibnamefont
  {Nakane}}, \bibinfo {author} {\bibfnamefont {T.~A.}\ \bibnamefont
  {Wheatley}}, \bibinfo {author} {\bibfnamefont {K.}~\bibnamefont {Iwasawa}},
  \bibinfo {author} {\bibfnamefont {S.}~\bibnamefont {Takeda}}, \bibinfo
  {author} {\bibfnamefont {H.}~\bibnamefont {Arao}}, \bibinfo {author}
  {\bibfnamefont {K.}~\bibnamefont {Ohki}}, \bibinfo {author} {\bibfnamefont
  {K.}~\bibnamefont {Tsumura}}, \bibinfo {author} {\bibfnamefont {D.~W.}\
  \bibnamefont {Berry}}, \bibinfo {author} {\bibfnamefont {T.~C.}\ \bibnamefont
  {Ralph}}, \bibinfo {author} {\bibfnamefont {H.~M.}\ \bibnamefont {Wiseman}},
  \bibinfo {author} {\bibfnamefont {E.~H.}\ \bibnamefont {Huntington}}, \ and\
  \bibinfo {author} {\bibfnamefont {A.}~\bibnamefont {Furusawa}}} (\bibinfo
  {year} {2012}),\ \href {\doibase 10.1126/science.1225258} {\bibfield
  {journal} {\bibinfo  {journal} {Science}\ }\textbf {\bibinfo {volume}
  {337}}~(\bibinfo {number} {6101}),\ \bibinfo {pages} {1514}}\BibitemShut
  {NoStop}%
\bibitem [{\citenamefont {You}\ and\ \citenamefont {He}(2015)}]{You2015}%
  \BibitemOpen
  \bibfield  {author} {\bibinfo {author} {\bibnamefont {You}, \bibfnamefont
  {W.-L.}}, \ and\ \bibinfo {author} {\bibfnamefont {L.}~\bibnamefont {He}}}
  (\bibinfo {year} {2015}),\ \href@noop {} {\bibfield  {journal} {\bibinfo
  {journal} {J. Phys.: Condens. Matter}\ }\textbf {\bibinfo {volume} {27}},\
  \bibinfo {pages} {205601}}\BibitemShut {NoStop}%
\bibitem [{\citenamefont {You}\ \emph {et~al.}(2007)\citenamefont {You},
  \citenamefont {Li},\ and\ \citenamefont {Gu}}]{You2007}%
  \BibitemOpen
  \bibfield  {author} {\bibinfo {author} {\bibnamefont {You}, \bibfnamefont
  {W.-L.}}, \bibinfo {author} {\bibfnamefont {Y.-W.}\ \bibnamefont {Li}}, \
  and\ \bibinfo {author} {\bibfnamefont {S.-J.}\ \bibnamefont {Gu}}} (\bibinfo
  {year} {2007}),\ \href {\doibase 10.1103/PhysRevE.76.022101} {\bibfield
  {journal} {\bibinfo  {journal} {Phys. Rev. E}\ }\textbf {\bibinfo {volume}
  {76}},\ \bibinfo {pages} {022101}}\BibitemShut {NoStop}%
\bibitem [{\citenamefont {Yousefjani}\ \emph {et~al.}(2017)\citenamefont
  {Yousefjani}, \citenamefont {Nichols}, \citenamefont {Salimi},\ and\
  \citenamefont {Adesso}}]{Youse2017}%
  \BibitemOpen
  \bibfield  {author} {\bibinfo {author} {\bibnamefont {Yousefjani},
  \bibfnamefont {R.}}, \bibinfo {author} {\bibfnamefont {R.}~\bibnamefont
  {Nichols}}, \bibinfo {author} {\bibfnamefont {S.}~\bibnamefont {Salimi}}, \
  and\ \bibinfo {author} {\bibfnamefont {G.}~\bibnamefont {Adesso}}} (\bibinfo
  {year} {2017}),\ \href {\doibase 10.1103/PhysRevA.95.062307} {\bibfield
  {journal} {\bibinfo  {journal} {Phys. Rev. A}\ }\textbf {\bibinfo {volume}
  {95}},\ \bibinfo {pages} {062307}}\BibitemShut {NoStop}%
\bibitem [{\citenamefont {Y.Shi}(2004)}]{Shi0}%
  \BibitemOpen
  \bibfield  {author} {\bibinfo {author} {\bibnamefont {Y.Shi},}} (\bibinfo
  {year} {2004}),\ \href@noop {} {\bibfield  {journal} {\bibinfo  {journal} {J.
  Phys. A}\ }\textbf {\bibinfo {volume} {37}},\ \bibinfo {pages}
  {6807}}\BibitemShut {NoStop}%
\bibitem [{\citenamefont {Yuan}(2016)}]{Yuan2016}%
  \BibitemOpen
  \bibfield  {author} {\bibinfo {author} {\bibnamefont {Yuan}, \bibfnamefont
  {H.}}} (\bibinfo {year} {2016}),\ \href {\doibase
  10.1103/PhysRevLett.117.160801} {\bibfield  {journal} {\bibinfo  {journal}
  {Phys. Rev. Lett.}\ }\textbf {\bibinfo {volume} {117}},\ \bibinfo {pages}
  {160801}}\BibitemShut {NoStop}%
\bibitem [{\citenamefont {Yuan}\ and\ \citenamefont {Fung}(2015)}]{Yuan2015}%
  \BibitemOpen
  \bibfield  {author} {\bibinfo {author} {\bibnamefont {Yuan}, \bibfnamefont
  {H.}}, \ and\ \bibinfo {author} {\bibfnamefont {C.-H.~F.}\ \bibnamefont
  {Fung}}} (\bibinfo {year} {2015}),\ \href {\doibase
  10.1103/PhysRevLett.115.110401} {\bibfield  {journal} {\bibinfo  {journal}
  {Phys. Rev. Lett.}\ }\textbf {\bibinfo {volume} {115}},\ \bibinfo {pages}
  {110401}}\BibitemShut {NoStop}%
\bibitem [{\citenamefont {Yukalov}(2009)}]{Yukalov0}%
  \BibitemOpen
  \bibfield  {author} {\bibinfo {author} {\bibnamefont {Yukalov}, \bibfnamefont
  {V.}}} (\bibinfo {year} {2009}),\ \href@noop {} {\bibfield  {journal}
  {\bibinfo  {journal} {Laser Physics}\ }\textbf {\bibinfo {volume} {19}},\
  \bibinfo {pages} {1}}\BibitemShut {NoStop}%
\bibitem [{\citenamefont {Yukawa}\ \emph {et~al.}(2013)\citenamefont {Yukawa},
  \citenamefont {Miyata}, \citenamefont {Mizuta}, \citenamefont {Yonezawa},
  \citenamefont {Marek}, \citenamefont {Filip},\ and\ \citenamefont
  {Furusawa}}]{yukawa_generating_2013}%
  \BibitemOpen
  \bibfield  {author} {\bibinfo {author} {\bibnamefont {Yukawa}, \bibfnamefont
  {M.}}, \bibinfo {author} {\bibfnamefont {K.}~\bibnamefont {Miyata}}, \bibinfo
  {author} {\bibfnamefont {T.}~\bibnamefont {Mizuta}}, \bibinfo {author}
  {\bibfnamefont {H.}~\bibnamefont {Yonezawa}}, \bibinfo {author}
  {\bibfnamefont {P.}~\bibnamefont {Marek}}, \bibinfo {author} {\bibfnamefont
  {R.}~\bibnamefont {Filip}}, \ and\ \bibinfo {author} {\bibfnamefont
  {A.}~\bibnamefont {Furusawa}}} (\bibinfo {year} {2013}),\ \href {\doibase
  10.1364/OE.21.005529} {\bibfield  {journal} {\bibinfo  {journal} {Optics
  Express}\ }\textbf {\bibinfo {volume} {21}}~(\bibinfo {number} {5}),\
  \bibinfo {pages} {5529}}\BibitemShut {NoStop}%
\bibitem [{\citenamefont {Yurke}(1986)}]{Yurke0}%
  \BibitemOpen
  \bibfield  {author} {\bibinfo {author} {\bibnamefont {Yurke}, \bibfnamefont
  {B.}}} (\bibinfo {year} {1986}),\ \href@noop {} {\bibfield  {journal}
  {\bibinfo  {journal} {Phys. Rev. Lett.}\ }\textbf {\bibinfo {volume} {56}},\
  \bibinfo {pages} {1515}}\BibitemShut {NoStop}%
\bibitem [{\citenamefont {Yurke}\ \emph {et~al.}(1986)\citenamefont {Yurke},
  \citenamefont {McCall},\ and\ \citenamefont {Klauder}}]{Klauder0}%
  \BibitemOpen
  \bibfield  {author} {\bibinfo {author} {\bibnamefont {Yurke}, \bibfnamefont
  {B.}}, \bibinfo {author} {\bibfnamefont {S.}~\bibnamefont {McCall}}, \ and\
  \bibinfo {author} {\bibfnamefont {J.}~\bibnamefont {Klauder}}} (\bibinfo
  {year} {1986}),\ \href@noop {} {\bibfield  {journal} {\bibinfo  {journal}
  {Phys. Rev. A}\ }\textbf {\bibinfo {volume} {33}},\ \bibinfo {pages}
  {4033}}\BibitemShut {NoStop}%
\bibitem [{\citenamefont {Zanardi}\ \emph
  {et~al.}(2007{\natexlab{a}})\citenamefont {Zanardi}, \citenamefont
  {Campos~Venuti},\ and\ \citenamefont {Giorda}}]{Zanardi2007-3}%
  \BibitemOpen
  \bibfield  {author} {\bibinfo {author} {\bibnamefont {Zanardi}, \bibfnamefont
  {P.}}, \bibinfo {author} {\bibfnamefont {L.}~\bibnamefont {Campos~Venuti}}, \
  and\ \bibinfo {author} {\bibfnamefont {P.}~\bibnamefont {Giorda}}} (\bibinfo
  {year} {2007}{\natexlab{a}}),\ \href {\doibase 10.1103/PhysRevA.76.062318}
  {\bibfield  {journal} {\bibinfo  {journal} {Phys. Rev. A}\ }\textbf {\bibinfo
  {volume} {76}},\ \bibinfo {pages} {062318}}\BibitemShut {NoStop}%
\bibitem [{\citenamefont {Zanardi}\ and\ \citenamefont
  {Paunkovi\ifmmode~\acute{c}\else \'{c}\fi{}}(2006)}]{Zanardi2006}%
  \BibitemOpen
  \bibfield  {author} {\bibinfo {author} {\bibnamefont {Zanardi}, \bibfnamefont
  {P.}}, \ and\ \bibinfo {author} {\bibfnamefont {N.}~\bibnamefont
  {Paunkovi\ifmmode~\acute{c}\else \'{c}\fi{}}}} (\bibinfo {year} {2006}),\
  \href {\doibase 10.1103/PhysRevE.74.031123} {\bibfield  {journal} {\bibinfo
  {journal} {Phys. Rev. E}\ }\textbf {\bibinfo {volume} {74}},\ \bibinfo
  {pages} {031123}}\BibitemShut {NoStop}%
\bibitem [{\citenamefont {Zanardi}\ \emph
  {et~al.}(2007{\natexlab{b}})\citenamefont {Zanardi}, \citenamefont
  {Cozzini},\ and\ \citenamefont {Giorda}}]{Zanardi2007}%
  \BibitemOpen
  \bibfield  {author} {\bibinfo {author} {\bibnamefont {Zanardi}, \bibfnamefont
  {P.}}, \bibinfo {author} {\bibfnamefont {M.}~\bibnamefont {Cozzini}}, \ and\
  \bibinfo {author} {\bibfnamefont {P.}~\bibnamefont {Giorda}}} (\bibinfo
  {year} {2007}{\natexlab{b}}),\ \href@noop {} {\bibinfo  {journal} {J. Stat.
  Phys.}\ ,\ \bibinfo {pages} {L02002}}\BibitemShut {NoStop}%
\bibitem [{\citenamefont {Zanardi}\ \emph
  {et~al.}(2007{\natexlab{c}})\citenamefont {Zanardi}, \citenamefont {Giorda},\
  and\ \citenamefont {Cozzini}}]{Zanardi2007-2}%
  \BibitemOpen
\bibfield  {journal} {  }\bibfield  {author} {\bibinfo {author} {\bibnamefont
  {Zanardi}, \bibfnamefont {P.}}, \bibinfo {author} {\bibfnamefont
  {P.}~\bibnamefont {Giorda}}, \ and\ \bibinfo {author} {\bibfnamefont
  {M.}~\bibnamefont {Cozzini}}} (\bibinfo {year} {2007}{\natexlab{c}}),\ \href
  {\doibase 10.1103/PhysRevLett.99.100603} {\bibfield  {journal} {\bibinfo
  {journal} {Phys. Rev. Lett.}\ }\textbf {\bibinfo {volume} {99}},\ \bibinfo
  {pages} {100603}}\BibitemShut {NoStop}%
\bibitem [{\citenamefont {Zanardi}\ \emph {et~al.}(2004)\citenamefont
  {Zanardi}, \citenamefont {Lidar},\ and\ \citenamefont {Lloyd}}]{Zanardi0}%
  \BibitemOpen
  \bibfield  {author} {\bibinfo {author} {\bibnamefont {Zanardi}, \bibfnamefont
  {P.}}, \bibinfo {author} {\bibfnamefont {D.}~\bibnamefont {Lidar}}, \ and\
  \bibinfo {author} {\bibfnamefont {S.}~\bibnamefont {Lloyd}}} (\bibinfo {year}
  {2004}),\ \href@noop {} {\bibfield  {journal} {\bibinfo  {journal} {Phys.
  Rev. Lett.}\ }\textbf {\bibinfo {volume} {92}},\ \bibinfo {pages}
  {060402}}\BibitemShut {NoStop}%
\bibitem [{\citenamefont {Zanardi}\ \emph {et~al.}(2008)\citenamefont
  {Zanardi}, \citenamefont {Paris},\ and\ \citenamefont
  {Campos~Venuti}}]{Zanardi2008}%
  \BibitemOpen
  \bibfield  {author} {\bibinfo {author} {\bibnamefont {Zanardi}, \bibfnamefont
  {P.}}, \bibinfo {author} {\bibfnamefont {M.~G.~A.}\ \bibnamefont {Paris}}, \
  and\ \bibinfo {author} {\bibfnamefont {L.}~\bibnamefont {Campos~Venuti}}}
  (\bibinfo {year} {2008}),\ \href {\doibase 10.1103/PhysRevA.78.042105}
  {\bibfield  {journal} {\bibinfo  {journal} {Phys. Rev. A}\ }\textbf {\bibinfo
  {volume} {78}},\ \bibinfo {pages} {042105}}\BibitemShut {NoStop}%
\bibitem [{\citenamefont {Zeng}\ \emph {et~al.}(2015)\citenamefont {Zeng},
  \citenamefont {Chen}, \citenamefont {Zhou},\ and\ \citenamefont
  {Wen}}]{Zeng2015}%
  \BibitemOpen
  \bibfield  {author} {\bibinfo {author} {\bibnamefont {Zeng}, \bibfnamefont
  {B.}}, \bibinfo {author} {\bibfnamefont {X.}~\bibnamefont {Chen}}, \bibinfo
  {author} {\bibfnamefont {D.-L.}\ \bibnamefont {Zhou}}, \ and\ \bibinfo
  {author} {\bibfnamefont {X.-G.}\ \bibnamefont {Wen}}} (\bibinfo {year}
  {2015}),\ \href@noop {} {\enquote {\bibinfo {title} {{Quantum Information
  Meets Quantum Matter -- From Quantum Entanglement to Topological Phase in
  Many-Body Systems}},}\ }\bibinfo {note} {ArXiv:1508.02595}\BibitemShut
  {NoStop}%
\bibitem [{\citenamefont {Zhang}\ and\ \citenamefont
  {Sarovar}(2015)}]{Zhang2015}%
  \BibitemOpen
  \bibfield  {author} {\bibinfo {author} {\bibnamefont {Zhang}, \bibfnamefont
  {J.}}, \ and\ \bibinfo {author} {\bibfnamefont {M.}~\bibnamefont {Sarovar}}}
  (\bibinfo {year} {2015}),\ \href {\doibase 10.1103/PhysRevA.91.052121}
  {\bibfield  {journal} {\bibinfo  {journal} {Phys. Rev. A}\ }\textbf {\bibinfo
  {volume} {91}},\ \bibinfo {pages} {052121}}\BibitemShut {NoStop}%
\bibitem [{\citenamefont {Zhang}\ \emph {et~al.}(2015)\citenamefont {Zhang},
  \citenamefont {Mouradian}, \citenamefont {Wong},\ and\ \citenamefont
  {Shapiro}}]{Zhang15}%
  \BibitemOpen
  \bibfield  {author} {\bibinfo {author} {\bibnamefont {Zhang}, \bibfnamefont
  {Z.}}, \bibinfo {author} {\bibfnamefont {S.}~\bibnamefont {Mouradian}},
  \bibinfo {author} {\bibfnamefont {F.~N.~C.}\ \bibnamefont {Wong}}, \ and\
  \bibinfo {author} {\bibfnamefont {J.~H.}\ \bibnamefont {Shapiro}}} (\bibinfo
  {year} {2015}),\ \href@noop {} {\bibfield  {journal} {\bibinfo  {journal}
  {Phys. Rev. Lett.}\ }\textbf {\bibinfo {volume} {114}},\ \bibinfo {pages}
  {110506}}\BibitemShut {NoStop}%
\bibitem [{\citenamefont {Zhang}\ \emph {et~al.}(2013)\citenamefont {Zhang},
  \citenamefont {Tengner}, \citenamefont {Zhong}, \citenamefont {Wong},\ and\
  \citenamefont {Shapiro}}]{Zhang13}%
  \BibitemOpen
  \bibfield  {author} {\bibinfo {author} {\bibnamefont {Zhang}, \bibfnamefont
  {Z.}}, \bibinfo {author} {\bibfnamefont {M.}~\bibnamefont {Tengner}},
  \bibinfo {author} {\bibfnamefont {T.}~\bibnamefont {Zhong}}, \bibinfo
  {author} {\bibfnamefont {F.~N.~C.}\ \bibnamefont {Wong}}, \ and\ \bibinfo
  {author} {\bibfnamefont {J.~H.}\ \bibnamefont {Shapiro}}} (\bibinfo {year}
  {2013}),\ \href@noop {} {\bibfield  {journal} {\bibinfo  {journal} {Phys.
  Rev. Lett.}\ }\textbf {\bibinfo {volume} {111}},\ \bibinfo {pages}
  {010501}}\BibitemShut {NoStop}%
\bibitem [{\citenamefont {Zhao}\ and\ \citenamefont {Zhou}(2009)}]{Zhao2009}%
  \BibitemOpen
  \bibfield  {author} {\bibinfo {author} {\bibnamefont {Zhao}, \bibfnamefont
  {J.-H.}}, \ and\ \bibinfo {author} {\bibfnamefont {H.-Q.}\ \bibnamefont
  {Zhou}}} (\bibinfo {year} {2009}),\ \href {\doibase
  10.1103/PhysRevB.80.014403} {\bibfield  {journal} {\bibinfo  {journal} {Phys.
  Rev. B}\ }\textbf {\bibinfo {volume} {80}},\ \bibinfo {pages}
  {014403}}\BibitemShut {NoStop}%
\bibitem [{\citenamefont {Zhou}\ and\ \citenamefont
  {Barjaktarevi\v{c}}(2008)}]{Zhou2008}%
  \BibitemOpen
  \bibfield  {author} {\bibinfo {author} {\bibnamefont {Zhou}, \bibfnamefont
  {H.-Q.}}, \ and\ \bibinfo {author} {\bibfnamefont {J.~P.}\ \bibnamefont
  {Barjaktarevi\v{c}}}} (\bibinfo {year} {2008}),\ \href@noop {} {\bibfield
  {journal} {\bibinfo  {journal} {J. Phys. A}\ }\textbf {\bibinfo {volume}
  {41}},\ \bibinfo {pages} {412001}}\BibitemShut {NoStop}%
\bibitem [{\citenamefont {Zhou}\ \emph {et~al.}(2008)\citenamefont {Zhou},
  \citenamefont {Zhao},\ and\ \citenamefont {Li}}]{Zhou2008-2}%
  \BibitemOpen
  \bibfield  {author} {\bibinfo {author} {\bibnamefont {Zhou}, \bibfnamefont
  {H.-Q.}}, \bibinfo {author} {\bibfnamefont {J.}~\bibnamefont {Zhao}}, \ and\
  \bibinfo {author} {\bibfnamefont {B.}~\bibnamefont {Li}}} (\bibinfo {year}
  {2008}),\ \href@noop {} {\bibfield  {journal} {\bibinfo  {journal} {J. Phys.
  A}\ }\textbf {\bibinfo {volume} {41}},\ \bibinfo {pages}
  {492002}}\BibitemShut {NoStop}%
\bibitem [{\citenamefont {Zobay}\ and\ \citenamefont
  {Garraway}(2004)}]{Zobay2004}%
  \BibitemOpen
  \bibfield  {author} {\bibinfo {author} {\bibnamefont {Zobay}, \bibfnamefont
  {O.}}, \ and\ \bibinfo {author} {\bibfnamefont {B.~M.}\ \bibnamefont
  {Garraway}}} (\bibinfo {year} {2004}),\ \href@noop {} {\bibfield  {journal}
  {\bibinfo  {journal} {Phys. Rev. A}\ }\textbf {\bibinfo {volume} {69}},\
  \bibinfo {pages} {023605}}\BibitemShut {NoStop}%
\bibitem [{\citenamefont {Zunkovic}\ and\ \citenamefont
  {Prosen}(2010)}]{Zunkovic2010}%
  \BibitemOpen
  \bibfield  {author} {\bibinfo {author} {\bibnamefont {Zunkovic},
  \bibfnamefont {B.}}, \ and\ \bibinfo {author} {\bibfnamefont
  {T.}~\bibnamefont {Prosen}}} (\bibinfo {year} {2010}),\ \href@noop {}
  {\bibfield  {journal} {\bibinfo  {journal} {J. Stat. Mech.}\ }\textbf
  {\bibinfo {volume} {1008}},\ \bibinfo {pages} {P08016}}\BibitemShut {NoStop}%
\bibitem [{\citenamefont {Zurek}(2001)}]{zurek_sub-planck_2001}%
  \BibitemOpen
  \bibfield  {author} {\bibinfo {author} {\bibnamefont {Zurek}, \bibfnamefont
  {W.~H.}}} (\bibinfo {year} {2001}),\ \href {\doibase 10.1038/35089017}
  {\bibfield  {journal} {\bibinfo  {journal} {Nature}\ }\textbf {\bibinfo
  {volume} {412}}~(\bibinfo {number} {6848}),\ \bibinfo {pages}
  {712}}\BibitemShut {NoStop}%
\bibitem [{\citenamefont {Zwierz}\ \emph {et~al.}(2010)\citenamefont {Zwierz},
  \citenamefont {P\'{e}rez-Delgado},\ and\ \citenamefont
  {Kok}}]{ZwierzPRL2010}%
  \BibitemOpen
  \bibfield  {author} {\bibinfo {author} {\bibnamefont {Zwierz}, \bibfnamefont
  {M.}}, \bibinfo {author} {\bibfnamefont {C.}~\bibnamefont
  {P\'{e}rez-Delgado}}, \ and\ \bibinfo {author} {\bibfnamefont
  {P.}~\bibnamefont {Kok}}} (\bibinfo {year} {2010}),\ \href
  {http://link.aps.org/doi/10.1103/PhysRevLett.105.180402} {\bibfield
  {journal} {\bibinfo  {journal} {Phys Rev Lett}\ }\textbf {\bibinfo {volume}
  {105}}~(\bibinfo {number} {18})}\BibitemShut {NoStop}%
\bibitem [{\citenamefont {Zwierz}\ \emph {et~al.}(2011)\citenamefont {Zwierz},
  \citenamefont {P\'erez-Delgado},\ and\ \citenamefont
  {Kok}}]{ZweirzPRL2010erratum}%
  \BibitemOpen
  \bibfield  {author} {\bibinfo {author} {\bibnamefont {Zwierz}, \bibfnamefont
  {M.}}, \bibinfo {author} {\bibfnamefont {C.~A.}\ \bibnamefont
  {P\'erez-Delgado}}, \ and\ \bibinfo {author} {\bibfnamefont {P.}~\bibnamefont
  {Kok}}} (\bibinfo {year} {2011}),\ \href {\doibase
  10.1103/PhysRevLett.107.059904} {\bibfield  {journal} {\bibinfo  {journal}
  {Phys. Rev. Lett.}\ }\textbf {\bibinfo {volume} {107}},\ \bibinfo {pages}
  {059904}}\BibitemShut {NoStop}%
\end{thebibliography}%

\end{document}